\documentclass{book}\pdfoutput=1%
\usepackage{amssymb}
\usepackage{amsfonts}
\usepackage{amsmath}
\usepackage{eurosym}
\usepackage{fancyhdr}
\usepackage[pdftex]{graphicx}%
\providecommand{\U}[1]{\protect\rule{.1in}{.1in}}

\begin{document}

\frontmatter
\title{LECTURES ON PROBABILITY, ENTROPY, AND STATISTICAL PHYSICS }
\author{Ariel Caticha\\Department of Physics, University at Albany -- SUNY}
\date{}
\maketitle
\tableofcontents

\bigskip\newpage

\thispagestyle{empty}

\chapter*{Preface\addcontentsline{toc}{chapter}{Preface}}

\markboth{PREFACE}{PREFACE}Science consists in using information about the
world for the purpose of predicting, explaining, understanding, and/or
controlling phenomena of interest. The basic difficulty is that the available
information is usually insufficient to attain any of those goals with certainty.

In these lectures we will be concerned with the problem of inductive
inference, that is, the problem of reasoning under conditions of incomplete
information. Is there a general\ method for handling uncertainty? Or, at
least, are there rules that could in principle be followed by an ideally
rational agent when discussing scientific matters? What makes one statement
more plausible than another? How much more plausible? And then, when new
information is acquired how does it change its mind? Or, to put it
differently, are there rules for learning? Are there rules for processing
information that are objective and consistent? Are they unique? And, come to
think of it, what, after all, is information? It is clear that data
\textquotedblleft contains\textquotedblright\ or \textquotedblleft
conveys\textquotedblright\ information, but what does this precisely mean? Can
information be conveyed in other ways? Is information some sort of physical
fluid that can be contained or transported? Is information \emph{physical}?
Can we measure amounts of information? Do we need to?

Our goal is to develop the main tools for inductive inference -- probability
and entropy -- and to illustrate their use in physics. To be specific we will
concentrate on examples borrowed from the foundations of classical statistical
physics, but this is not meant to reflect a limitation of these inductive
methods, which, as far as we can tell at present are of universal
applicability. It is just that statistical mechanics is rather special in that
it provides us with the first examples of fundamental laws of physics that can
be derived as examples of inductive inference. Perhaps all laws of physics can
be derived in this way.

The level of these lectures is somewhat uneven. Some topics are fairly
advanced -- the subject of recent research -- while some other topics are very
elementary. I can give two related reasons for including the latter. First,
the standard education of physicists includes a very limited study of
probability and even of entropy -- maybe just a little about errors in a
laboratory course, or maybe a couple of lectures as a brief mathematical
prelude to statistical mechanics. The result is a widespread misconception
that these subjects are trivial and unproblematic -- that the real problems of
theoretical physics lie elsewhere, and that if your experimental data require
analysis, then you have done the wrong experiment. Which brings me to the
second reason. It would be very surprising to find that the interpretations of
probability and of entropy turned out to bear no relation to our understanding
of statistical mechanics and quantum mechanics. Indeed, if the only notion of
probability at your disposal is that of a frequency in a large number of
trials you might be led to think that the ensembles of statistical mechanics
must be real, and to regard their absence as an urgent problem demanding an
immediate solution -- perhaps an ergodic solution. You might also be led to
think that similar ensembles are needed in quantum theory and therefore that
quantum theory requires the existence of an ensemble of parallel universes.
Similarly, if the only notion of entropy available to you is derived from
thermodynamics, you might end up thinking that entropy is a physical quantity
related to heat and disorder, that it can be measured in the lab, and that
therefore has little or no relevance beyond statistical mechanics.

It is very worthwhile to revisit the elementary basics not because they are
easy -- they are not -- but because they are fundamental.

Many are the subjects that I have left out but wish I had included in these
lectures. Some relate to inference proper -- the assignment of priors,
information geometry, model selection, and the theory of questions or
inductive inquiry -- while others deal with applications to the foundations of
both classical and quantum physics. As a provisional remedy at the very end I
provide a short and very biased list of suggestions for further reading.

\noindent\textbf{Acknowledgements:} The points of view expressed here reflect
much that I have learned from discussions with many colleagues and friends: C.
Cafaro, N. Caticha, V. Dose, R. Fischer, A. Garrett, A. Giffin, M. Grendar, K.
Knuth, R. Preuss, C. Rodr\'{\i}guez, J. Skilling, and C.-Y. Tseng. I hope they
will not judge these lectures by those few instances where we have not yet
managed to reach agreement. I would also like to express my special thanks to
Julio Stern and to the organizers of MaxEnt 2008 for their encouragement to
pull these notes together into some sort of printable form.

\bigskip

\medskip Albany, May 2008.

\mainmatter

\chapter{Inductive Inference}

The process of drawing conclusions from available information is called
inference. When the available information is sufficient to make unequivocal,
unique assessments of truth we speak of making deductions: on the basis of a
certain piece of information we deduce that a certain proposition is true. The
method of reasoning leading to deductive inferences is called logic.
Situations where the available information is insufficient to reach such
certainty lie outside the realm of logic. In these cases we speak of making a
probable inference, and the method of reasoning is probability theory.
Alternative names are `inductive inference' and `inductive logic'. The word
`induction' refers to the process of using limited information about a few
special cases to draw conclusions about more general situations.

\section{Probability}

The question of the meaning and interpretation of the concept of probability
has long been controversial. Needless to say the interpretations offered by
various schools are at least partially successful or else they would already
have been discarded. But the different interpretations are not equivalent.
They lead people to ask different questions and to pursue their research in
different directions. Some questions may become essential and urgent under one
interpretation while totally irrelevant under another. And perhaps even more
important: under different interpretations equations can be used differently
and this can lead to different predictions.

Historically the \emph{frequentist} interpretation has been the most popular:
the probability of a random event is given by the relative number of
occurrences of the event in a sufficiently large number of identical and
independent trials. The appeal of this interpretation is that it seems to
provide an empirical method to estimate probabilities by counting over the set
of trials -- an ensemble. The magnitude of a probability is obtained solely
from the observation of many repeated trials and does not depend on any
feature or characteristic of the observers. Probabilities interpreted in this
way have been called \emph{objective}. This view dominated the fields of
statistics and physics for most of the 19th and 20th centuries (see,
\emph{e.g.}, [von Mises 57]).

One disadvantage of the frequentist approach has to do with matters of rigor:
what precisely does one mean by `random'? If the trials are sufficiently
identical, shouldn't one always obtain the same outcome? Also, if the
interpretation is to be validated on the basis of its operational, empirical
value, how large should the number of trials be? Unfortunately, the answers to
these questions are neither easy nor free from controversy. By the time the
tentative answers have reached a moderately acceptable level of sophistication
the intuitive appeal of this interpretation has long been lost. In the end, it
seems the frequentist interpretation is most useful when left a bit vague.

A more serious objection is the following. In the frequentist approach the
notion of an ensemble of trials is central. In cases where there is a natural
ensemble (tossing a coin, or a die, spins in a lattice, etc.) the frequency
interpretation seems natural enough. But for many other problems the
construction of an ensemble is at best highly artificial. For example,
consider the probability of there being life in Mars. Are we to imagine an
ensemble of Mars planets and solar systems? In these cases the ensemble would
be purely hypothetical. It offers no possibility of an empirical determination
of a relative frequency and this defeats the original goal of providing an
objective operational interpretation of probabilities as frequencies. In yet
other problems there is no ensemble at all: consider the probability that the
$n$th digit of the number $\pi$ be 7. Are we to imagine alternative universes
with different values for the number $\pi$? It is clear that there a number of
interesting problems where one suspects the notion of probability could be
quite useful but which nevertheless lie outside the domain of the frequentist approach.

According to the \emph{Bayesian }interpretations, which can be traced back to
Bernoulli and Laplace, but have only achieved popularity in the last few
decades, a probability reflects the confidence, the degree of belief of an
individual in the truth of a proposition. These probabilities are said to be
\emph{Bayesian} because of the central role played by Bayes' theorem -- a
theorem which is actually due to Laplace. This approach enjoys several
advantages. One is that the difficulties associated with attempting to
pinpoint the precise meaning of the word `random' can be avoided. Bayesian
probabilities are not restricted to repeatable events; they allow us to reason
in a consistent and rational manner about unique, singular events. Thus, in
going from the frequentist to the Bayesian interpretations the domain of
applicability and therefore the usefulness of the concept of probability is
considerably enlarged.

The crucial aspect of Bayesian probabilities is that different individuals may
have different degrees of belief in the truth of the very same proposition, a
fact that is described by referring to Bayesian probabilities as being
\emph{subjective}. This term is somewhat misleading because there are (at
least) two views on this matter, one is the so-called subjective Bayesian or
\emph{personalistic} view (see, \emph{e.g.}, [Savage 72, Howson Urbach 93,
Jeffrey 04]), and the other is the \emph{objective} Bayesian view (see
\emph{e.g.} [Jeffreys 39, Jaynes 85, 03, Lucas 70]). For an excellent
introduction with a philosophical perspective see [Hacking 01]. According to
the subjective view, two reasonable individuals faced with the same evidence,
the same information, can legitimately differ in their confidence in the truth
of a proposition and may therefore assign different probabilities. Subjective
Bayesians accept that an individual can change his or her beliefs, merely on
the basis of introspection, reasoning, or even revelation.

At the other end of the Bayesian spectrum, the objective Bayesian view
considers the theory of probability as an extension of logic. It is said then
that a probability measures a degree of \emph{rational} belief. It is assumed
that the objective Bayesian has thought so long and hard about how
probabilities are assigned that no further reasoning will induce a revision of
beliefs except when confronted with new information. In an ideal situation two
different individuals will, on the basis of the same information, assign the
same probabilities.

Whether Bayesian probabilities are subjective or objective is still a matter
of dispute. Our position is that they lie somewhere in between. Probabilities
will always retain a \textquotedblleft subjective\textquotedblright\ element
because translating information into probabilities involves judgments and
different people will inevitably judge differently. On the other hand, not all
probability assignments are equally useful and it is plausible that what makes
some assignments better than others is that they represent or reflect some
objective feature of the world. One might even say that what makes them better
is that they provide a better guide to the \textquotedblleft
truth\textquotedblright. Thus, probabilities can be characterized by both
subjective and objective elements and, ultimately, it is their objectivity
that makes probabilities useful.

In fact we shall see that while the subjective element in probabilities can
never be completely eliminated, the rules for processing information, that is,
the rules for updating probabilities, are themselves quite objective. This
means that the new information can be objectively processed and incorporated
into our posterior probabilities. Thus, it is quite possible to continuously
suppress the subjective elements while enhancing the objective elements as we
process more and more information.

\section{Inductive reasoning}

We discussed how the study of macroscopic systems requires a general theory to
allow us to carry out inferences on the basis of incomplete information and
our first step should be to inquire what this theory or language for inference
should be. The principle of reasoning that we will follow is simple,
compelling, and quite common in science [Skilling 89]:\noindent\

\begin{description}
\item \qquad If a general theory exists, it must apply to special cases.

\item \qquad If a certain special case happens to be known then this knowledge
can be used to constrain the general theory: all candidate theories that fail
to reproduce the known example are discarded.

\item \qquad If a sufficient number of special cases is known then the general
theory might be completely determined.
\end{description}

\noindent The method allows us to extrapolate from a few special cases where
we know what to expect, to more general cases where we did not. This is a
method for induction, for generalization. Of course, it may happen that there
are too many constraints, in which case there is no general theory that
reproduces them all.

Philosophers have a name for such a method: they call it \emph{eliminative
induction} [Earman 92]. On the negative side, the \textbf{Principle of
Eliminative Induction} (PEI), like any other form of induction, is not
guaranteed to work. On the positive side, the PEI adds an interesting twist to
Popper's scientific methodology. According to Popper scientific theories can
never be proved right, they can only be proved false; a theory is corroborated
only to the extent that all attempts at falsifying it have failed. Eliminative
induction is fully compatible with Popper's notions but the point of view is
just the opposite. Instead of focusing on \emph{failure} to falsify one
focuses on \emph{success}: it is the successful falsification of all rival
theories that corroborates the surviving one. The advantage is that one
acquires a more explicit understanding of why competing theories are eliminated.

This inductive method will be used several times. First in chapter 2 to show
that if a general theory of inference exists, then it must coincide with the
usual theory of probability. \noindent In other words, we will show that
degrees of belief, those measures of plausibility that we require to do
inference, should be manipulated and calculated using the ordinary rules of
the calculus of probabilities and \emph{therefore} that probabilities
\emph{can} be interpreted as degrees of belief [Cox 46, Jaynes 57a, 03].

But with this achievement, enormous as it is, we do not yet reach our final
goal. The problem is that what the rules of probability theory will allow us
to do is to assign probabilities to some \textquotedblleft
complex\textquotedblright\ propositions on the basis of the probabilities that
have been previously assigned to other, perhaps more \textquotedblleft
elementary\textquotedblright\ propositions. The issue of how to assign
probabilities to the elementary propositions is not addressed.

Historically the first partial solution to this problem was suggested by James
Bernoulli (1713). The idea is simple: in those situations where there are
several alternatives that can be enumerated and counted, and where one has no
reason to favor one over another, the alternatives should be deemed equally
probable. The equality of the degrees of belief reflects the symmetry of one's
state of knowledge or, rather, of ignorance. This mode of reasoning has been
called the `\emph{Principle of Insufficient Reason}' and is usually associated
with the name of Laplace (1812).

The principle has been particularly successful in dealing with situations
where there is some positive, \emph{sufficient} reason\emph{\ }to suspect that
the various alternatives should be considered equally likely. For example, in
certain games of chance the symmetry among possible outcomes is attained\ on
purpose, by construction. These games are special because they are
deliberately designed so that information about previous outcomes is
irrelevant to the prediction of future outcomes and the symmetry of our state
of ignorance about the future is very robust.

The range of applications of Laplace's principle is, however, limited. There
are situations where it is not clear what `equally likely' means. For example,
it might not be possible to count the alternatives or maybe the possible
outcomes are distributed over continuous ranges. Also, there are situations
where there is information leading one to prefer some alternatives over
others; how can such information be incorporated in a systematic way? One
needs a method that generalizes Laplace's principle.

Progress toward this goal came from an unexpected direction. While
investigating the capacity of communication channels to transmit information
Shannon came to appreciate the need for a quantitative measure of the notion
of \textquotedblleft amount of missing information\textquotedblright\ or the
\textquotedblleft amount of uncertainty\textquotedblright\ in a probability
distribution. In 1948 he succeeded in finding such a measure and thereby
initiated the field of information theory [Shannon 48].

As we will see in chapter 4 Shannon's argument is a second application of the
induction principle above: A general theory, if it exists at all, must apply
to special cases. He argued that in order to qualify as a measure of ignorance
or of missing information a quantity $S$ would have to satisfy some reasonable
conditions -- the Shannon axioms -- and these conditions were sufficiently
constraining to determine the quantity $S$ uniquely: There is only one way to
measure the amount of uncertainty in a probability distribution. \noindent It
was rather surprising that the expression that Shannon obtained for $S$ in
communication theory coincided with expressions that had previously been used
by Boltzmann and by Gibbs to represent entropy in the very different context
of statistical mechanics and thermodynamics. This coincidence led Shannon to
choose the name `entropy' for his quantity $S $. Somewhat later, however,
Brillouin and Jaynes realized that the similarity of Shannon's entropy with
Gibbs' entropy could not be a mere coincidence and thus began a process that
would radically alter our understanding of the thermodynamical entropy of
Clausius. [Brillouin 52, Jaynes 57b]

The crucial contribution of Jaynes was the insight that the Shannon derivation
was not limited to information in communication channels, but that the same
mathematics can be applied to information in general. It establishes a basis
for a general method of inference that includes Laplace's principle of
insufficient reason as a special case. In fact, it became clear that on a
purely intuitive basis Boltzmann and Gibbs had already found and had made
extensive use of this method in statistical mechanics.

With the Boltzmann-Gibbs-Jaynes method we can revisit the question of how to
assign those probabilities that will be used as the starting point for the
calculation of all others. The answer is simple: among all possible
probability distributions that satisfy the constraints implied by the limited
available information we select that particular distribution that reflects
maximum ignorance about those aspects of the problem about which nothing is
known. What else could we do? It seems this is the only intellectually honest
way to proceed. And the procedure is mathematically clear: since ignorance is
measured by entropy the desired probability distribution is obtained by
maximizing the entropy subject to whatever conditions are known to constrain
the system. This is called the \emph{Method of Maximum Entropy} and it is
usually abbreviated as MaxEnt.

But the procedure is not without its problems. These may, to some, seem
relatively minor, but one may reasonably argue that any problem of principle
is necessarily a major problem. For example, the Shannon axioms refer to
discrete probability distributions rather than continuous ones, and
generalizing his measure of uncertainty is not altogether straightforward.
Another, perhaps more serious problem, is that the axioms themselves may be
self-evident to some but not to others: do the Shannon axioms really codify
what we mean by uncertainty? Are there other measures of uncertainty? Indeed,
others have been proposed. Thus, despite its obvious success, in the eyes of
many, the MaxEnt method remains controversial and several variations on its
justification have been proposed.

In chapter 6 we present an extension of the method of maximum entropy (which
we will abbreviate ME to distinguish it from the older MaxEnt) which derives
from the work of Shore and Johnson. They point out what is perhaps the main
drawback of the Shannon-Jaynes approach: it is indirect. First one finds how
to measure amount of uncertainty and then one argues that the only unbiased
way to incorporate information into a probability distribution is to maximize
this measure subject to constraints. The procedure can be challenged by
arguing that, even granted that entropy measures something, how sure can we be
this something is uncertainty, ignorance? Shore and Johnson argue that what
one really wants is a consistent method to process information directly,
without detours that invoke questionable measures of uncertainty.

A third application of the general inductive method -- a general theory, if it
exists at all, must apply to special cases [Skilling 88] -- yields the desired
procedure: There is a unique method to update from an old set of beliefs
codified in a prior probability distribution into a new set of beliefs
described by a new, posterior distribution when the information available is
in the form of a constraint on the family of acceptable posteriors. The
updated posterior distribution is that of maximum \textquotedblleft
relative\textquotedblright\ entropy. The axioms of the ME method are,
hopefully, more self-evident: They reflect the conviction that what was
learned in the past is important and should not be frivolously ignored. The
chosen posterior distribution should coincide with the prior as closely as
possible and one should only update those aspects of one's beliefs for which
corrective new evidence has been supplied. Furthermore, since the new axioms
do not tell us what and how to update, they merely tell us what not to update,
they have the added bonus of maximizing objectivity -- there are many ways to
change something but only one way to keep it the same. [Caticha 03,Caticha
Giffin 06, Caticha 07]

This alternative justification for the method of maximum entropy turns out to
be directly applicable to continuous distributions, and it establishes the
value of the concept of entropy irrespective of its interpretation in terms of
heat, or disorder, or uncertainty. In this approach \emph{entropy is purely a
tool for consistent reasoning; strictly, it needs no interpretation}. Perhaps
this is the reason why the meaning of entropy has turned out to be such an
elusive concept.

\chapter{Probability}

Our goal is to establish the theory of probability as the general theory for
reasoning on the basis of incomplete information. This requires us to tackle
two different problems. The first problem is to figure out how to achieve a
quantitative description of a state of knowledge. Once this is settled we
address the second problem of how to update from one state of knowledge to
another when new information becomes available.

Throughout we will assume that the subject matter -- the set of statements the
truth of which we want to assess -- has been clearly specified. This question
of what it that we are actually talking about is much less trivial than it
might appear at first sight.\footnote{Consider the example of quantum
mechanics: Are we talking about particles, or about experimental setups, or
both? Are we talking about position variables, or about momenta, or both? Or
neither? Is it the position of the particles or the position of the
detectors?} Nevertheless, it will not be discussed further.

The first problem, that of describing or characterizing a state of knowledge,
requires that we quantify the degree to which we believe each proposition in
the set is true. The most basic feature of these beliefs is that they form an
interconnected web that must be internally consistent. The idea is that in
general the strengths of one's beliefs in some propositions are constrained by
one's beliefs in other propositions; beliefs are not independent of each
other. For example, the belief in the truth of a certain statement $a$ is
strongly constrained by the belief in the truth of its negation, not-$a$: the
more I believe in one, the less I believe in the other. As we will see below,
the basic desiderata for such a scheme, which are expressed in the Cox axioms,
[Cox 46] lead to a unique formalism in which degrees of belief are related to
each other using the standard rules of probability theory. Then we explore
some of the consequences. For experiments that can be repeated indefinitely
one recovers standard results, such as the law of large numbers, and the
connection between probability and frequency.

The second problem, that of updating from one consistent web of beliefs to
another when new information becomes available, will be addressed for the
special case that the information is in the form of data. The basic updating
strategy reflects the conviction that what we learned in the past is valuable,
that the web of beliefs should only be revised to the extent required by the
data. We will see that this principle of \emph{minimal updating} leads to the
uniquely natural rule that is widely known as Bayes' theorem. (More general
kinds of information can also be processed using the minimal updating
principle but they require a more sophisticated tool, namely relative entropy.
This topic will be extensively explored later.) As an illustration of the
enormous power of Bayes' rule we will briefly explore its application to data analysis.

\section{Consistent reasoning: degrees of belief}

We discussed how the study of physical systems in general requires a theory of
inference on the basis of incomplete information. Here we will show that
\emph{a general theory of inference, if it exists at all, coincides with the
usual theory of probability}. \noindent We will show that the quantitative
measures of plausibility or \emph{degrees} \emph{of belief} that we introduce
as tools for reasoning should be manipulated and calculated using the ordinary
rules of the calculus of probabilities. \emph{Therefore} probabilities
\emph{can }be interpreted as degrees of belief.

The procedure we follow differs in one remarkable way from the traditional way
of setting up physical theories. Normally one starts with the mathematical
formalism, and then one proceeds to try to figure out what the formalism could
possibly mean, one tries to append an interpretation to it. This is a very
difficult problem; historically it has affected not only statistical physics
-- what is the meaning of probabilities and of entropy -- but also quantum
theory -- what is the meaning of wave functions and amplitudes. Here we
proceed in the opposite order, we first decide what we are talking about,
degrees of belief or plausibility (we use the two expressions interchangeably)
and then we \emph{design} rules to manipulate them; we design the formalism,
we construct it to suit our purposes. The advantage of this approach is that
the issue of meaning, of interpretation, is settled from the start.

Before we proceed further it may be important to emphasize that the degrees of
belief discussed here are those held by an idealized rational agent that would
not be subject to the practical limitations under which we humans operate. We
discuss degrees of \emph{rational} belief and not the irrational and
inconsistent beliefs that real humans seem to hold. We are concerned with the
ideal optimal standard of rationality that we humans ought to attain at least
when discussing scientific\ matters.

Any suitable measure of belief must allow us to represent the fact that given
any two statements $a$ and $b$ one must be able to describe the fact that
either $a$ is more plausible than $b$, or $a$ is less plausible than $b$, or
else $a$ and $b$ are equally plausible. That this is possible is implicit in
what we mean by `plausibility'. Thus we can order assertions according to
increasing plausibility: if a statement $a$ is more plausible than $b$, and
$b$ is itself more plausible than another statement $c$, then $a$ is more
plausible than $c$. Since any transitive ordering, such as the one just
described, can be represented with real numbers, we are led to the following requirement:

\begin{description}
\item \qquad\emph{Degrees of \ rational belief (or, as we shall later call
them, probabilities) are represented by real numbers.}
\end{description}

\noindent The next and most crucial requirement is that whenever a degree of
belief can be computed in two different ways the two results must agree.

\begin{description}
\item \qquad\emph{The assignment of degrees of rational belief must be
consistent.}
\end{description}

\noindent Otherwise we could get entangled in confusing paradoxes: by
following one computational path we could decide that a statement $a$ is more
plausible than a statement $b$, but if we were to follow a different path we
could conclude the opposite. Consistency is the crucial requirement that
eliminates vagueness and transforms our general qualitative statements into
precise quantitative ones.

Our general theory of inference is constructed using the inductive method
described in the previous chapter: If a general theory exists, then it must
reproduce the right answers in those special cases where the answers happen to
be known; these special cases constrain the general theory; given enough such
constraints, the general theory is fully determined\emph{.}

Before we write down the special cases that will play the role of the axioms
of probability theory we should introduce a convenient notation. A degree of
plausibility is a real number that we will assign to a statement $a$ on the
basis of some information that we have and will obviously depend on what that
information actually is. A common kind of information takes the form of
another statement $b$ which is asserted to be true. Therefore, a degree of
plausibility is a real number assigned to two statements $a$ and $b$, rather
than just one. Our notation should reflect this. Let $P(a|b)$ denote the
plausibility that statement $a$ is true provided we know $b$ to be true.
$P(a|b)$ is read `the degree of plausibility (or, later, the probability) of
$a$ given $b$'. $P(a|b)$ is commonly called a conditional probability (the
probability of $a$ given that condition $b$ holds). When $b$ turns out to be
false, we shall regard $P(a|b)$ as undefined. Although the notation $P(a|b)$
is quite convenient we will not always use it; we will often just write
$P\left(  a\right)  $ omitting the statement $b$, or we might even just write
$P $. It is, however, important to realize that degrees of belief and
probabilities are always conditional on something even if that something is
not explicitly stated.

More notation: For every statement $a$ there exists its negation not-$a$,
which will be denoted with a prime, $a^{\prime}{}$. If $a$ is true, then
$a^{\prime}$ is false and vice versa. Given two statements $a_{1}$ and $a_{2}
$ we can form their \emph{conjunction} `$a_{1}$ and $a_{2}$' which we will
denote it as $a_{1}a_{2}$. The conjunction is true if and only if both $a_{1}
$ and $a_{2}$ are true. Given $a_{1}$ and $a_{2}$, we can also form their
\emph{disjunction} `$a_{1}$ or $a_{2}$'. The disjunction will be denoted by
$a_{1}+a_{2}$ and it is true when either $a_{1}$ or $a_{2}$ or both are true;
it is false when both $a_{1}$ and $a_{2}$ are false.

Now we proceed to state the axioms [Cox 46, Jaynes 03].

\section{The Cox Axioms}

The degrees of belief or plausibility we assign to a statement $a$ and to its
negation $a^{\prime}$ are not independent of each other. The more plausible
one is, the less plausible the other becomes; if one increases we expect the
other to decrease and vice-versa. This is expressed by our first axiom.

\begin{description}
\item[Axiom 1.] The plausibility of not-$a$ is a monotonic function of the
plausibility of $a$,
\end{description}%

\begin{equation}
P(a^{\prime}|b)=f\left(  P(a|b)\right)  .\label{Cox axiom 1}%
\end{equation}
\noindent At this point we do not know the precise relation between $P(a|b)$
and $P(a^{\prime}|b)$, we only know that some such function $f$ must exist.

The second axiom expresses the fact that a measure of plausibility for a
complex statement such as the conjunction \textquotedblleft$a_{1}$ and $a_{2}%
$\textquotedblright, must somehow depend on the separate plausibilities of
$a_{1}$ and of $a_{2}$. We consider it \textquotedblleft
self-evident\textquotedblright\ that the plausibility that both $a_{1}$ and
$a_{2}$\ are simultaneously true, $P(a_{1}a_{2}|b)$, can be analyzed in
stages: In order for $a_{1}a_{2}$ to be true it must first be the case that
$a_{1}$ is itself true. Thus, $P(a_{1}a_{2}|b)$ must depend on $P(a_{1}|b)$.
Furthermore, once we have established that $a_{1}$ is in fact true, in order
for $a_{1}a_{2}$ to be true, it must be the case that $a_{2}$ is also true.
Thus, $P(a_{1}a_{2}|b)$ must depend on $P(a_{2}|a_{1}b)$ as well. This
argument is carried out in more detail in [Tribus 69]. Therefore, our second
axiom is

\begin{description}
\item[\textbf{Axiom 2.}] The plausibility $P(a_{1}a_{2}|b)$ of a conjunction
$a_{1}a_{2}$, is determined once we specify the plausibility $P(a_{1}|b)$ of
$a_{1}$ and the plausibility $P(a_{2}|a_{1}b)$ of $a_{2}$ given $a_{1}$.
\end{description}

\noindent What this means is that $P(a_{1}a_{2}|b)$ must be calculable in
terms of $P(a_{1}|b)$ and $P(a_{2}|a_{1}b)$: the second axiom asserts that
there exists a function $g$ such that
\begin{equation}
P(a_{1}a_{2}|b)=g\left(  P(a_{1}|b),P(a_{2}|a_{1}b)\right)
.\label{Cox axiom 2}%
\end{equation}
Remarkably this is all we need! Note the \emph{qualitative} nature of these
axioms: what is being asserted is the existence of some unspecified functions
$f$ and $g$ and not their specific quantitative mathematical forms.
Furthermore, note that the same $f$ and $g$ apply to any and all propositions.
This reflects our desire to construct a single theory of universal
applicability. It also means that the axioms represent a huge number of known
special cases.

At this point the functions $f$ and $g$ are unknown, but they are not
arbitrary. In fact, as we shall see below, the requirement of consistency is
very constraining. For example, notice that since $a_{1}a_{2}=a_{2}a_{1}$, in
\ref{Cox axiom 2} the roles of $a_{1}$ and $a_{2}$ may be interchanged,
\begin{equation}
P(a_{1}a_{2}|b)=g\left(  P(a_{2}|b),P(a_{1}|a_{2}b)\right)  .
\end{equation}
Consistency requires that
\begin{equation}
g\left(  P(a_{1}|b),P(a_{2}|a_{1}b)\right)  =g\left(  P(a_{2}|b),P(a_{1}%
|a_{2}b)\right)  .\label{constr g1}%
\end{equation}
We will have to check that this is indeed the case. As a second example, since
$a^{\prime\prime}=a$, it must be the case that
\begin{equation}
P(a|b)=P(a^{\prime\prime}|b)=f\left(  P(a^{\prime}|b)\right)  =f\left[
f\left(  P(a|b)\right)  \right]  .
\end{equation}
The plausibility $P(a|b)$ is just a number, call it $u$, this can be written as%

\begin{equation}
f\left(  f\left(  u\right)  \right)  =u\text{ .}\label{constr f1}%
\end{equation}
These two constraints are not at this point helpful in fixing the functions
$f$ and $g$. But the following one is. \quad

\section{Regraduation: the Product Rule}

\subsection{Cox's first theorem}

A consistency constraint that follows from the associativity property of the
conjunction goes a long way toward fixing the acceptable forms of the function
$g$. The constraint is obtained by noting that since $\left(  ab\right)
c=a\left(  bc\right)  $, we have two ways to compute $P\left(  abc|d\right)
$. Starting from \
\begin{equation}
P\left[  \left(  ab\right)  c|d\right]  =P\left[  a\left(  bc\right)
|d\right]  \text{,}%
\end{equation}
we get\newline%
\begin{equation}
g\left[  P\left(  ab|d\right)  ,P\left(  c|abd\right)  \right]  =g\left[
P\left(  a|d\right)  ,P\left(  bc|ad\right)  \right]
\end{equation}
and\newline%
\begin{equation}
g\left[  g\left(  P\left(  a|d\right)  ,P\left(  b|ad\right)  \right)
,P\left(  c|abd\right)  \right]  =g\left[  P\left(  a|d\right)  ,g\left(
P\left(  b|ad\right)  ,P\left(  c|bad\right)  \right)  \right]  .
\end{equation}
\newline Writing $P\left(  a|d\right)  =u$, $P\left(  b|ad\right)  =v$, and
$P\left(  c|abd\right)  =w$, the \textquotedblleft
associativity\textquotedblright\ constraint is
\begin{equation}
g\left(  g(u,v),w\right)  =g\left(  u,g(v,w)\right)  .\label{constr g2}%
\end{equation}

It is quite obvious that the functional equation eq.(\ref{constr g2}) has an
infinity of solutions. Indeed, by direct substitution one can easily check
that functions of the form
\begin{equation}
g(u,v))=G^{-1}\left[  G(u)G(v)\right]
\end{equation}
are solutions for any invertible (and therefore monotonic) function $G(u)$.
What is not so easy to prove is that this is the general solution.

\noindent\textbf{Associativity Theorem:} Given any function $g(u,v)$ that
satisfies the associativity constraint, eq.(\ref{constr g2}), one can
construct another monotonic function $G(u)$ such that
\begin{equation}
G(g(u,v))=G(u)G(v).\label{regraduation1}%
\end{equation}
Cox's proof of this theorem is somewhat lengthy and is relegated to the next subsection.

The significance of this result becomes apparent when one rewrites it as
\begin{equation}
G\left[  P(ab|c)\right]  =G\left[  P(a|c)\right]  G\left[  P(b|ac)\right]
\label{regraduation2}%
\end{equation}
and realizes that there was nothing particularly special about the original
assignment of real numbers $P(a|c)$, $P(b|ac)$, and so on. Their only purpose
was to provide us with a ranking, an ordering of propositions according to how
plausible they are. Since the function $G(u)$ is monotonic, the same ordering
can be achieved using a new set positive numbers
\begin{equation}
p(a|c)\overset{\operatorname*{def}}{=}G\left[  P(a|c)\right]  ,\quad
p(b|ac)\overset{\operatorname*{def}}{=}G\left[  P(b|ac)\right]
,...\label{regraduation3}%
\end{equation}
instead of the old. The advantage of using these `regraduated' plausibilities
is that the plausibility of $ab$ can be calculated in terms of the
plausibilities of $a$ and of $b$ given $a$ in a particularly simple way: it is
just their product. Thus, while the new numbers are neither more nor less
correct than the old, they are just considerably more convenient. The theorem
can be rephrased as follows.

\noindent\textbf{Cox's First Regraduation Theorem:} Once a consistent
representation of the ordering of propositions according to their degree of
plausibility has been set up by assigning a real number $P(a|b)$ to each pair
of propositions $a$ and $b$ one can always find another equivalent
representation by assigning positive numbers $p(a|c)$ that satisfy the product
rule
\begin{equation}
p(ab|c)=p(a|c)p(b|ac).\label{product rule}%
\end{equation}

Perhaps one can make the logic behind this regraduation a little bit clearer
by considering the somewhat analogous situation of introducing the quantity
temperature as a measure of degree of \textquotedblleft
hotness\textquotedblright. Clearly any acceptable measure of \textquotedblleft
hotness\textquotedblright\ must reflect its transitivity -- if $a$ is hotter
than $b$ and $b$ is hotter than $c$ then $a$ is hotter than $c$; thus,
temperatures are represented by real numbers. But the temperature scales are
so far arbitrary. While many temperature scales may serve equally well the
purpose of ordering systems according to their hotness, there is one choice --
the absolute or Kelvin scale -- that turns out to be considerably more
convenient because it simplifies the mathematical formalism. Switching from an
arbitrary temperature scale to the Kelvin scale is one instance of a
convenient regraduation. (The details of temperature regraduation are given in
chapter 3.)

On the basis of plain common sense one would have expected $g(u,v)$ to be
monotonic in both its arguments. Consider a change in the first argument
$P(a_{1}|b)$ while holding the second $P(a_{2}|a_{1}b)$ fixed. Since a
strengthening the belief in $a_{1}$ can only strengthen the belief in
$a_{1}a_{2}$ we require that a change in $P(a_{1}|b)$ should yield a change in
$P(a_{1}a_{2}|b)$ of the \emph{same} sign. It is therefore a reassuring check
that the product rule eq.(\ref{product rule}) behaves as expected.

\subsection{Proof of the Associativity Theorem}

Understanding the proof that eq.(\ref{regraduation1}) is the general solution
of the associativity constraint, eq.(\ref{constr g2}), is not necessary for
understanding other topics in this book. This section may be skipped on a
first reading. The proof given below, due to Cox, takes advantage of the fact
that our interest is not just to find the most general solution but rather
that we want the most general solution under the restricted circumstance that
the function $g$ is to be used for the purpose of inference. This allows us to
impose additional constraints on $g$.

We will assume that the functions $g$ are continuous and twice differentiable.
Indeed inference is quantified common sense and if the function $g$ had turned
out to be non-differentiable serious doubts would be cast on the legitimacy of
the whole scheme. Furthermore, common sense also requires that $g(u,v)$ be
monotonic increasing in both its arguments. Consider a change in the first
argument $P(a_{1}|b)$ while holding the second $P(a_{2}|a_{1}b)$ fixed. Since
a strengthening of one's belief in $a_{1}$ must be reflected in a
corresponding strengthening in ones's belief in $a_{1}a_{2}$ we require that a
change in $P(a_{1}|b)$ should yield a change in $P(a_{1}a_{2}|b)$ of the
\emph{same} sign. An analogous line of reasoning leads one to impose that
$g(u,v)$ must be monotonic increasing in the second argument as well,
\begin{equation}
\frac{\partial g\left(  u,v\right)  }{\partial u}\geq0\quad\text{and}%
\quad\frac{\partial g\left(  u,v\right)  }{\partial v}\geq0.
\end{equation}

Let
\begin{equation}
r\overset{\operatorname*{def}}{=}g\left(  u,v\right)  \quad\text{and}\quad
s\overset{\operatorname*{def}}{=}g\left(  v,w\right)  ,\label{a11}%
\end{equation}
and
\begin{equation}
g_{1}(u,v)\overset{\operatorname*{def}}{=}\frac{\partial g\left(  u,v\right)
}{\partial u}\geq0\quad\text{and\quad}g_{2}(u,v)\overset{\operatorname*{def}%
}{=}\frac{\partial g\left(  u,v\right)  }{\partial v}\geq0.
\end{equation}
Then eq.(\ref{constr g2}) and its derivatives with respect to $u$ and $v$ are
\begin{equation}
g\left(  r,w\right)  =g\left(  u,s\right)  ,\label{a1}%
\end{equation}%
\begin{equation}
g_{1}(r,w)g_{1}(u,v)=g_{1}(u,s),
\end{equation}
and
\begin{equation}
g_{1}(r,w)g_{2}(u,v)=g_{2}(u,s)g_{1}(v,w).
\end{equation}
\newline Eliminating $g_{1}(r,w)$ from these last two equations we get
\begin{equation}
K(u,v)=K(u,s)g_{1}(v,w).\label{g2}%
\end{equation}
\newline where
\begin{equation}
K(u,v)=\frac{g_{2}(u,v)}{g_{1}(u,v)}.\label{g3}%
\end{equation}
\newline Multiplying eq.(\ref{g2}) by $K(v,w)$ we get
\begin{equation}
K(u,v)K(v,w)=K(u,s)g_{2}(v,w)\text{ }\label{g4}%
\end{equation}
\newline Differentiating the right hand side of eq.(\ref{g4}) with respect to
$v$ and comparing with the derivative of eq.(\ref{g2}) with respect to $w$, we
have
\begin{equation}
\frac{\partial}{\partial v}\left(  K\left(  u,s\right)  g_{2}\left(
v,w\right)  \right)  =\frac{\partial}{\partial w}\left(  K\left(  u,s\right)
g_{1}\left(  v,w\right)  \right)  =\frac{\partial}{\partial w}\left(  K\left(
u,v\right)  \right)  =0.
\end{equation}
\newline Therefore
\begin{equation}
\frac{\partial}{\partial v}\left(  K\left(  u,v\right)  K\left(  v,w\right)
\right)  =0,
\end{equation}
or,
\begin{equation}
\frac{1}{K\left(  u,v\right)  }\frac{\partial K\left(  u,v\right)  }{\partial
v}=-\frac{1}{K\left(  v,w\right)  }\frac{\partial K\left(  v,w\right)
}{\partial v}\overset{\operatorname*{def}}{=}h\left(  v\right)  .\label{a5}%
\end{equation}
\newline Integrate using the fact that $K\geq0$ because both $g_{1}$ and
$g_{2}$ are positive, we get
\begin{equation}
K(u,v)=K(u,0)\,\exp\int_{0}^{v}h(v^{\prime})dv^{\prime},
\end{equation}
\newline and also
\begin{equation}
K\left(  v,w\right)  =K\left(  0,w\right)  \,\,\exp-\int_{0}^{v}h(v^{\prime
})dv^{\prime},
\end{equation}
\newline so that
\begin{equation}
K\left(  u,v\right)  =\alpha\,\frac{H\left(  u\right)  }{H\left(  v\right)  },
\end{equation}
where $\alpha=K(0,0)$ is a constant and $H(u)$ is the \emph{positive} function
\newline%
\begin{equation}
H(u)\overset{\operatorname*{def}}{=}\exp\,\left[  -\int_{0}^{u}h(u^{\prime
})du^{\prime}\right]  \geq0.\label{a6}%
\end{equation}
\newline On substituting back into eqs.(\ref{g2}) and (\ref{g4}) we get
\begin{equation}
g_{1}(v,w)=\,\frac{H(s)}{H(v)}\quad\quad\text{and}\quad\quad g_{2}%
(v,w)=\alpha\,\frac{H(s)}{H(w)}.\label{a7}%
\end{equation}
\newline

Next, use $s=g(v,w)$, so that
\begin{equation}
ds=g_{1}(v,w)dv+g_{2}(v,w)dw.
\end{equation}
Substituting (\ref{a7}) we get \ \
\begin{equation}
\frac{ds}{H(s)}=\frac{dv}{H(v)}+\alpha\frac{dw}{H(w)}.
\end{equation}
This is easily integrated. Let
\begin{equation}
G\left(  u\right)  =G\left(  0\right)  \,\exp\left(  \int_{0}^{u}%
\frac{du^{\prime}}{H(u^{\prime})}\right)  ,\label{a8}%
\end{equation}
so that $du/H(u)=dG(u)/G(u)$. Then
\begin{equation}
G\left(  g\left(  v,w\right)  \right)  =G\left(  v\right)  G^{{}\alpha}\left(
w\right)  ,
\end{equation}
where a multiplicative constant of integration has been absorbed into the
constant $G\left(  0\right)  $. Applying this function $G$ twice in
eq.(\ref{constr g2}) we obtain
\begin{equation}
G(u)G^{{}\alpha}(v)G^{{}\alpha}(w)=G(u)G^{{}\alpha}(v)G^{{}\alpha^{2}}(w),
\end{equation}
so that $\alpha=1$,
\begin{equation}
G\left(  g\left(  v,w\right)  \right)  =G\left(  v\right)  G\left(  w\right)
,\label{a9}%
\end{equation}
(The second possibility $\alpha=0$ is discarded because it leads to $g(u,v)=u$
which is not useful for inference.) This completes our proof
eq.(\ref{regraduation1}) is the general solution of eq.(\ref{constr g2}):
Given any $g(u,v)$ that satisfies eq.(\ref{constr g2}) one can construct the
corresponding $G(u)$ using eqs.(\ref{g3}), (\ref{a5}), (\ref{a6}), and
(\ref{a8}). Furthermore, since $G(u)$ is an exponential its sign is dictated
by the constant $G\left(  0\right)  $ which is positive because the right hand
side of eq.(\ref{a9}) is positive. Finally, since $H(u)\geq0$, eq. (\ref{a6}),
the regraduating function $G(u)$ is a monotonic function of its variable $u$.

\subsection{Setting the range of degrees of belief}

Degrees of belief range from the extreme of total certainty that an assertion
is true to the opposite extreme of total certainty that it is false. What
numerical values should we assign to these extremes?

Let $p_{T}$ and $p_{F}$ be the numerical values assigned to the (regraduated)
plausibilities of propositions which are known to be true and false
respectively. Notice that the extremes should be unique. There is a single
$p_{T}$ and a single $p_{F}$. The possibility of assigning two different
numerical values, for example $p_{T1}$ and $p_{T2}$, to propositions known to
be true is ruled out by our desire that degrees of plausibility be ordered.

The philosophy behind regraduation is to seek the most convenient
representation of degrees of belief in terms of real numbers. In particular,
we would like our regraduated plausibilities to reflect the fact that if $b$
is known to be true then we believe in $ab$ to precisely the same extent as we
believe in $a$, no more and no less. This is expressed by%

\begin{equation}
p(ab|b)=p(a|b)\,.\label{pt1}%
\end{equation}
On the other hand, using the product rule eq.(\ref{product rule}) we get
\begin{equation}
p(ab|b)=p(b|b)p(a|bb)=p_{T}p(a|b)\,.\label{pt2}%
\end{equation}
Comparing eqs.(\ref{pt1}) and (\ref{pt2}) we get
\begin{equation}
p_{T}=1
\end{equation}
Thus, the value of $p_{T}$ is assigned so that eq.(\ref{pt1}) holds:

\begin{description}
\item \qquad\emph{Belief that }$a$ \emph{is true is represented by }%
$p(a)=1$\emph{.}
\end{description}

For the other extreme value, $p_{F}$, which represents impossibility, consider
the plausibility of $ab^{\prime}$ given $b$. Using the product rule we have
\begin{equation}
p(ab^{\prime}|b)=p(a|b)p(b^{\prime}|ab)\,.
\end{equation}
But $p(ab^{\prime}|b)=p_{F}$ and $p(b^{\prime}|ab)=p_{F}$. Therefore \
\begin{equation}
p_{F}=p(a|b)\,p_{F}.
\end{equation}
\ Again, this should hold for arbitrary $a$. Therefore either $p_{F}=0$ or
$\infty$, either value is fine. (The value $-\infty$ is not allowed; negative
values of $p(a|b)$ would lead to an inconsistency.) We can either choose
plausibilities in the range $[0,1]$ so that a higher $p$ reflects a higher
degree of belief or, alternatively, we can choose `implausibilities' in the
range $[1,\infty)$ so that a higher $p$ reflects a lower degree of belief.
Both alternatives are equally consistent and correct. The usual convention is
to choose the former.

\begin{description}
\item \qquad\emph{Belief that }$a$ \emph{is false is represented by }$p(a)=0
$\emph{.}
\end{description}

The numerical values assigned to $p_{T}$ and $p_{F}$ follow from a
particularly convenient regraduation that led to the product rule. Other
possibilities are, of course, legitimate. Instead of eq.(\ref{regraduation3})
we could for example have regraduated plausibilities according to
$p(a|c)\overset{\operatorname*{def}}{=}CG\left[  P(a|c)\right]  $ where $C$ is
some constant. Then the product rule would read $Cp(ab|c)=p(a|c)p(b|ac)$ and
the analysis of the previous paragraphs would have led us to $p_{T}=C$ and
$p_{F}=0$ or $\infty$. The choice $C=100$ is quite common; it is implicit in
many colloquial uses of the notion of probability, as for example, when one
says `I am 100\% sure that...'. Notice, incidentally, that within a
frequentist interpretation most such statements would be meaningless.

\section{Further regraduation: the Sum Rule}

\subsection{Cox's second theorem}

Having restricted the form of $g$ considerably we next study the function $f$
by requiring its compatibility with $g$. It is here that we make use of the
constraints (\ref{constr g1}) and (\ref{constr f1}) that we had found earlier.

Consider plausibilities $P$ that have gone through a first process of
regraduation so that the product rule holds,
\begin{equation}
P(ab|c)=P(a|c)P(b|ac)=P(a|c)f\left(  P(b^{\prime}|ac)\right) \label{f1}%
\end{equation}
but $P(ab^{\prime}|c)=P(a|c)P(b^{\prime}|ac)$, then
\begin{equation}
P(ab|c)=P(a|c)f\left(  \frac{P(ab^{\prime}|c)}{P(a|c)}\right)  .
\end{equation}
But $P(ab|c)$ is symmetric in $ab=ba$. Therefore
\begin{equation}
P(a|c)f\left(  \frac{P(ab^{\prime}|c)}{P(a|c)}\right)  =P(b|c)f\left(
\frac{P(a^{\prime}b|c)}{P(b|c)}\right)  .
\end{equation}
This must hold irrespective of the choice of $a$, $b$, and $c$. In particular
suppose that $b^{\prime}=ad$. On the left hand side $P(ab^{\prime
}|c)=P(b^{\prime}|c)$ because $aa=a$. On the right hand side, to simplify
$P(a^{\prime}b|c)$ we note that $a^{\prime}b^{\prime}=a^{\prime}ad$ is false
and that $a^{\prime}b^{\prime}=(a+b)^{\prime}$. (In order for $a+b$ to be
false it must be the case that both $a$ is false and $b$ is false.) Therefore
$a+b$ is true: either $a$ is true or $b$ is true. If $b$ is true then
$a^{\prime}b=a^{\prime}$. If $a$ is true both $a^{\prime}$ and $a^{\prime}b$
are false which means that we also get $a^{\prime}b=a^{\prime}$. Therefore on
the right hand side $P(a^{\prime}b|c)=P(a^{\prime}|c)$ and we get
\begin{equation}
P(a|c)f\left(  \frac{f\left(  P(b|c)\right)  }{P(a|c)}\right)  =P(b|c)f\left(
\frac{f\left(  P(a|c)\right)  }{P(b|c)}\right)  .
\end{equation}
Writing $P\left(  a|c\right)  =u$, and $P\left(  b|c\right)  =v$, and
$P\left(  c|abd\right)  =w$, the \textquotedblleft
compatibility\textquotedblright\ constraint is
\begin{equation}
uf\left(  \frac{f\left(  v\right)  }{u}\right)  =vf\left(  \frac{f\left(
u\right)  }{v}\right)  .\label{ffuv}%
\end{equation}
We had earlier seen that certainty is represented by $1$ and impossibility by
$0$. Note that when $u=1$, using $f(1)=0$ and $f(0)=1$, we obtain $f[f(v)]=v$.
Thus, eq.(\ref{constr f1}) is a special case of (\ref{ffuv}).

\noindent\textbf{Compatibility Theorem: }The function $f(u)$ that satisfies
the compatibility constraint eq.(\ref{ffuv}) is
\begin{equation}
f(u)=\left(  1-u^{{}\alpha}\right)  ^{{}1/\alpha}\quad\text{or}\quad
u^{{}\alpha}+f^{{}\alpha}(u)=1.\label{fu}%
\end{equation}
where $\alpha$ is a constant.

\noindent It is easy to show that eq.(\ref{fu}) is a solution -- just
substitute. What is considerably more difficult is to show that it is the
general solution. The proof is given in the next subsection.

As a result of the first theorem we can consider both $u$ and $f(u)$ positive.
Therefore, for $\alpha>0$ impossibility must be represented by $0$, while for
$\alpha<0$ impossibility should be represented by $\infty$.

The significance of the solution for $f$ becomes clear when eq.(\ref{fu}) is
rewritten as
\begin{equation}
\left[  P(a|b)\right]  ^{{}\alpha}+\left[  P(a^{\prime}|b)\right]  ^{{}\alpha
}=1,\label{falpha}%
\end{equation}
and the product rule eq.(\ref{f1}) is raised to the same power $\alpha$,
\begin{equation}
\left[  P(ab|c)\right]  ^{{}\alpha}=\left[  P(a|c)\right]  ^{{}\alpha}\left[
P(b|ac)\right]  ^{{}\alpha}.
\end{equation}
This shows that, having regraduated plausibilities once, we can simplify the
solution (\ref{falpha}) considerably by regraduating a second time, while
still preserving the product rule. This second regraduation is
\begin{equation}
p(a|b)\overset{\operatorname*{def}}{=}\left[  P(a|b)\right]  ^{{}\alpha}.
\end{equation}

\noindent\textbf{Cox's Second Regraduation Theorem:} Once a consistent
representation of the ordering of propositions according to their degree of
plausibility has been set up in such a way that the product rule holds, one
can regraduate further and find an equivalent and more convenient
representation that assigns plausibilities $p(a|b)$ satisfying both the sum
rule,
\begin{equation}
p(a|b)+p(a^{\prime}|b)=1,\label{sr1}%
\end{equation}
and the product rule,
\begin{equation}
p(ab|c)=p(a|c)p(b|ac).\label{pr1}%
\end{equation}

These new, conveniently regraduated degrees of plausibility will be called
\emph{probabilities}, positive numbers in the interval $[0,1]$ with certainty
represented by $1$ and impossibility by $0$. From now on there is no need to
refer to plausibilities again; both notations, lower case $p$ as well as upper
case $P$ will be used to refer to the regraduated probabilities.

\subsection{Proof of the Compatibility Theorem}

The contents of this section is not essential to understanding other topics in
this book. It may be skipped on a first reading.

Just as in our previous consideration of the constraint imposed by
associativity on the function $g$, since the function $f$ is to be used for
the purpose of inference we can assume that it is continuous and twice
differentiable. Furthermore, once we have gone through the first stage of
regraduation, and plausibilities satisfy the product rule
eq.(\ref{product rule}), common sense also requires that the function $f(u)$
be monotonic decreasing,
\[
\frac{df(u)}{du}\leq0\quad\text{for}\quad0\leq u\leq1\,,
\]
with extreme values such that $f(0)=1$ and $f(1)=0$.

The first step is to transform the functional equation (\ref{ffuv}) into an
ordinary differential equation. Let
\begin{equation}
r\overset{\operatorname*{def}}{=}\frac{f\left(  v\right)  }{u}\quad
\text{and}\quad s\overset{\operatorname*{def}}{=}\frac{f\left(  u\right)  }%
{v}.
\end{equation}
and substitute into eq.(\ref{ffuv}),
\begin{equation}
uf\left(  r\right)  =vf\left(  s\right)  .\tag{2.45}%
\end{equation}
Next differentiate eq.(\ref{ffuv}) with respect to $u$, to $v$, and to $u$ and
$v$, to get (here primes denote derivatives)%
\begin{equation}
f(r)-rf^{\prime}(r)=f^{\prime}(s)f^{\prime}(u),\label{f2}%
\end{equation}%
\begin{equation}
f(s)-sf^{\prime}(s)=f^{\prime}(r)f^{\prime}(v),\label{f3}%
\end{equation}
and
\begin{equation}
\frac{s}{v}f^{\prime\prime}(s)f^{\prime}(u)=\frac{r}{u}f^{\prime\prime
}(r)f^{\prime}(v).\label{f4}%
\end{equation}
Multiply eq.(\ref{ffuv}) by eq.(\ref{f4}),
\begin{equation}
sf^{\prime\prime}(s)f^{\prime}(u)f(s)=rf^{\prime\prime}(r)f^{\prime}(v)f(r),
\end{equation}
and use eqs.(\ref{f2}) and (\ref{f3}) to eliminate $f^{\prime}(u)$ and
$f^{\prime}(v)$. After rearranging one gets,
\begin{equation}
\frac{sf^{\prime\prime}(s)f(s)}{f^{\prime}(s)\left[  f(s)-sf^{\prime
}(s)\right]  }=\frac{rf^{\prime\prime}(r)f(r)}{f^{\prime}(r)\left[
f(r)-rf^{\prime}(r)\right]  }.
\end{equation}
Since the left side does not depend on $r$, neither must the right side; both
sides must actually be constant. Call this constant $k$. Thus, the problem is
reduced to a differential equation,
\begin{equation}
rf^{\prime\prime}(r)f(r)=kf^{\prime}(r)\left[  f(r)-rf^{\prime}(r)\right]  .
\end{equation}
Multiplying by $dr/rff^{\prime}$ gives
\begin{equation}
\frac{df^{\prime}}{f^{\prime}}=k\left(  \frac{dr}{r}-\frac{df}{f}\right)  .
\end{equation}
Integrating twice gives
\begin{equation}
f(r)=\left(  Ar^{{}\alpha}+B\right)  ^{1/\alpha},
\end{equation}
where $A$ and $B$ are integration constants and $\alpha=1+k$. Substituting
back into eq.(\ref{ffuv}) allows us, after some simple algebra to determine
one of the integration constants, $B=A^{2}$, while substituting into
eq.(\ref{constr f1}) yields the other, $A=-1$. This concludes the proof.

\section{Some remarks on the sum and product rules}

\subsection{On meaning, ignorance and randomness}

The product and sum rules can be used as the starting point for a theory of
probability: Quite independently of what probabilities could possibly mean, we
can develop a formalism of real numbers (measures) that are manipulated
according to eqs.(\ref{sr1}) and (\ref{pr1}). This is the approach taken by
Kolmogorov. The advantage is mathematical clarity and rigor. The disadvantage,
of course, is that in actual applications the issue of meaning, of
interpretation, turns out to be important because it affects how and why
probabilities are used.

The advantage of the approach due to Cox is that the issue of meaning is
clarified from the start: the theory was designed to apply to degrees of
belief. Consistency requires that these numbers be manipulated according to
the rules of probability theory. This is all we need. There is no reference to
measures of sets or large ensembles of trials or even to random variables.
This is remarkable: it means that we can apply the powerful methods of
probability theory to thinking and reasoning about problems where nothing
random is going on, and to single events for which the notion of an ensemble
is either absurd or at best highly contrived and artificial. Thus, probability
theory is \emph{the} method for consistent reasoning in situations where the
information available might be insufficient to reach certainty: probability is
\emph{the} tool for dealing with uncertainty and ignorance.

This interpretation is not in conflict with the common view that probabilities
are associated with randomness. It may, of course, happen that there is an
unknown influence that affects the system in unpredictable ways and that there
is a good reason why this influence remains unknown, namely, it is so
complicated that the information necessary to characterize it cannot be
supplied. Such an influence we call `random'. Thus, being random is just one
among many possible reasons why a quantity might be uncertain or unknown.

\subsection{The general sum rule}

From the sum and product rules, eqs.(\ref{sr1}) and (\ref{pr1}) we can easily
deduce a third one:

\noindent\textbf{Theorem: }The probability of a disjunction (or) is given by
the sum rule%

\begin{equation}
p(a+b|c)=p(a|c)+p(b|c)-p(ab|c).\label{sr2}%
\end{equation}

\noindent The proof is straightforward. Use $(a+b)^{\prime}=a^{\prime
}b^{\prime}$, (for $a+b$ to be false both $a$ and $b$ must be false) then
\[
p\left(  a+b|c\right)  =1-p\left(  a^{\prime}b^{\prime}|c\right)  =1-p\left(
a^{\prime}|c\right)  p\left(  b^{\prime}|a^{\prime}c\right)  =
\]%
\[
1-p\left(  a^{\prime}|c\right)  \left(  1-p\left(  b|a^{\prime}c\right)
\right)  =p\left(  a|c\right)  +p\left(  a^{\prime}b|c\right)  =p\left(
a|c\right)  +p\left(  b|c\right)  p\left(  a^{\prime}|bc\right)  =
\]%
\[
p\left(  a|c\right)  +p\left(  b|c\right)  \left(  1-p\left(  a|bc\right)
\right)  =p(a|c)+p(b|c)-p(ab|c).
\]

These theorems are rather obvious on the basis of the interpretation of a
probability as a frequency or as the measure of a set. This is conveyed
graphically in a very clear way by Venn diagrams (see fig.2.1).%

\begin{figure}
[t]
\begin{center}
\includegraphics[
trim=1.671967in 1.672411in 1.672966in 1.673161in,
natheight=7.499600in,
natwidth=9.999800in,
height=2.3125in,
width=3.6876in
]%
{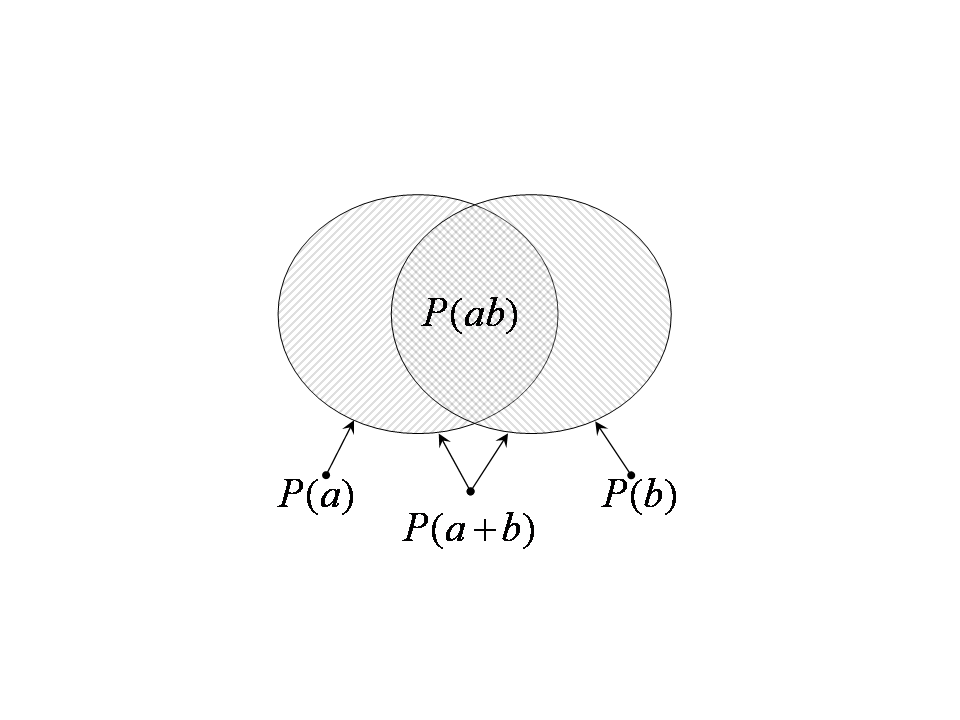}%
\caption{Venn diagram showing $P(a)$, $P(b) $, $P(ab)$ and $P(a+b)$.}%
\end{center}
\end{figure}

\subsection{Independent and mutually exclusive events}

In special cases the sum and product rules can be rewritten in various useful
ways. Two statements or events $a$ and $b$ are said to be \emph{independent}
if the probability of one is not altered by information about the truth of the
other. More specifically, event $a$ is independent of $b$ (given $c$) if
\begin{equation}
p\left(  a|bc\right)  =p\left(  a|c\right)  ~.
\end{equation}
For independent events the product rule simplifies to
\begin{equation}
p(ab|c)=p(a|c)p(b|c)\quad\text{or}\quad p(ab)=p(a)p(b)~.
\end{equation}
The symmetry of these expressions implies that $p\left(  b|ac\right)
=p\left(  b|c\right)  $ as well: if $a$ is independent of $b$, then $b$ is
independent of $a$.

Two statements or events $a_{1}$ and $a_{2}$ are \emph{mutually exclusive}
given $b$ if they cannot be true simultaneously, i.e., $p(a_{1}a_{2}|b)=0$.
Notice that neither $p(a_{1}|b)$ nor $p(a_{2}|b)$ need vanish. For mutually
exclusive events the sum rule simplifies to
\begin{equation}
p(a_{1}+a_{2}|b)=p(a_{1}|b)+p(a_{2}|b).
\end{equation}
The generalization to many mutually exclusive statements $a_{1},a_{2}%
,\ldots,a_{n}$ (mutually exclusive given $b$) is immediate,
\begin{equation}
p(a_{1}+a_{2}+\cdots+a_{n}|b)=%
{\textstyle\sum\limits_{i=1}^{n}}
\,p(a_{i}|b)~.
\end{equation}

If one of the statements $a_{1},a_{2},\ldots,a_{n}$ is necessarily true, i.e.,
they cover all possibilities, they are said to be \emph{exhaustive}. Then
their conjunction is necessarily true, $a_{1}+a_{2}+\cdots+a_{n}=\top$, so
that
\begin{equation}
p(a_{1}+a_{2}+\cdots+a_{n}|b)=1.
\end{equation}
If, in addition to being exhaustive, the statements $a_{1},a_{2},\ldots,a_{n}$
are also mutually exclusive then
\begin{equation}%
{\textstyle\sum\limits_{i=1}^{n}}
\,p(a_{i})=1~.\label{eme a}%
\end{equation}
A useful generalization involving the probabilities $\,p(a_{i}|b)$ conditional
on any arbitrary proposition $b$ is
\begin{equation}%
{\textstyle\sum\limits_{i=1}^{n}}
\,p(a_{i}|b)=1~.\label{eme b}%
\end{equation}
The proof is straightforward:
\begin{equation}
p(b)=p(b\top)=%
{\textstyle\sum\limits_{i=1}^{n}}
\,p(ba_{i})=p(b)%
{\textstyle\sum\limits_{i=1}^{n}}
p(a_{i}|b)~.
\end{equation}

\subsection{Marginalization}

Once we decide that it is legitimate to quantify degrees of belief by real
numbers $p$ the problem becomes how do we assign these numbers. The sum and
product rules show how we should assign probabilities to some statements once
probabilities have been assigned to others. Here is an important example of
how this works.

We want to assign a probability to a particular statement $b$. Let
$a_{1},a_{2},\ldots,a_{n}$ be mutually exclusive and exhaustive statements and
suppose that the probabilities of the conjunctions $ba_{j}$ are known. We want
to calculate $p(b)$ given the joint probabilities $p(ba_{j})$. The solution is
straightforward: sum $p(ba_{j})$ over all $a_{j}$s, use the product rule, and
eq.(\ref{eme b}) to get
\begin{equation}%
{\textstyle\sum\limits_{j}}
p(ba_{j})=p(b)%
{\textstyle\sum\limits_{j}}
p(a_{j}|b)=p(b)~.
\end{equation}
This procedure, called marginalization, is quite useful when we want to
eliminate uninteresting variables $a$ so we can concentrate on those variables
$b$ that really matter to us. The distribution $p(b)$ is referred to as the
marginal of the joint distribution $p(ab)$.

For a second use of formulas such as these suppose that we happen to know the
conditional probabilities $p(b|a)$. When $a$ is known we can make good
inferences about $b$, but what can we tell about $b$ when we are uncertain
about the actual value of $a$? Then we proceed as follows. Use of the sum and
product rules gives
\begin{equation}
p(b)=%
{\textstyle\sum\limits_{j}}
p(ba_{j})=%
{\textstyle\sum\limits_{j}}
p(b|a_{j})p(a_{j})~.
\end{equation}
This is quite reasonable: the probability of $b$ is the probability we would
assign if the value of $a$ were precisely known, averaged over all $a$s. The
assignment $p(b)$ clearly depends on how uncertain we are about the value of
$a$. In the extreme case when we are totally certain that $a$ takes the
particular value $a_{k}$ we have $p(a_{j})=\delta_{jk}$ and we recover
$p(b)=p(b|a_{k})$ as expected.

\section{The expected value}

Suppose we know that a quantity $x$ can take values $x_{i}$ with probabilities
$p_{i}$. Sometimes we need an estimate for the quantity $x$. What should we
choose? It seems reasonable that those values $x_{i}$ that have larger $p_{i}$
should have a dominant contribution to $x$. We therefore make the following
reasonable choice: The expected value of the quantity $x$ is$\,$denoted by
$\langle x\rangle$ and is given by
\begin{equation}
\langle x\rangle\overset{\operatorname*{def}}{=}%
{\textstyle\sum\limits_{i}}
p_{i}\,x_{i}\text{ .}\label{expected x}%
\end{equation}
\newline The term `expected' value is not always an appropriate one because
$\langle x\rangle$ may not be one of the actually allowed values $x_{i}$ and,
therefore, it is not a value we would expect. The expected value of a die toss
is $(1+\cdots+6)/6=3.5$ which is not an allowed result.

Using the average $\langle x\rangle$ as an estimate for the expected value of
$x$ is reasonable, but it is also somewhat arbitrary. Alternative estimates
are possible; for example, one could have chosen the value for which the
probability is maximum -- this is called the `mode'. This raises two questions.

The first question is whether $\langle x\rangle$ is a good estimate. If the
probability distribution is sharply peaked all the values of $x$ that have
appreciable probabilities are close to each other and to $\langle x\rangle$.
Then $\langle x\rangle$ is a good estimate. \ But if the distribution is broad
the actual value of $x$ may deviate from $\langle x\rangle$ considerably. To
describe quantitatively how large this deviation might be we need to describe
how broad the probability distribution is.

A convenient measure of the width of the distribution is the root mean square
(\emph{rms}) deviation defined by
\begin{equation}
\Delta x\overset{\operatorname*{def}}{=}\left\langle \left(  x-\langle
x\rangle\right)  ^{2}\right\rangle ^{1/2}.
\end{equation}
\newline The quantity $\Delta x$ is also called the standard deviation, its
square $(\Delta x)^{2}$ is called the variance. For historical reasons it is
common to refer to the `variance of $x$' but this is misleading because it
suggests that $x$ itself could vary; $\Delta x$ refers to our knowledge about
$x$.

If $\Delta x\ll\langle x\rangle$ then $x$ will not deviate much from $\langle
x\rangle$ and we expect $\langle x\rangle$ to be a good estimate.

The definition of $\Delta x$ is somewhat arbitrary. It is dictated both by
common sense and by convenience. Alternatively we could have chosen to define
the width of the distribution as $\left\langle \left\vert x-\langle
x\rangle\right\vert \right\rangle $ or $\langle\left(  x-\langle
x\rangle\right)  ^{4}\rangle^{1/4}$ but these definitions are less convenient
for calculations.

Now that we have a way of deciding whether $\langle x\rangle$ is a good
estimate for $x$ we may raise a second question: Is there such a thing as the
\textquotedblleft best\textquotedblright\ estimate for $x$? Consider another
estimate $x^{\prime}$. We expect $x^{\prime}$ to be accurate provided the
deviations from it are small, i.e., $\langle\left(  x-x^{\prime}\right)
^{2}\rangle$ is small. The best $x^{\prime}$ is that for which its variance is
a minimum
\begin{equation}
\left.  \frac{d}{dx^{\prime}}\langle\left(  x-x^{\prime}\right)  ^{2}%
\rangle\right\vert _{x^{\prime}{}_{\text{best}}}=0\text{,}%
\end{equation}
\newline which implies $x^{\prime}{}_{\text{best}}=\langle x\rangle$.
Conclusion: $\langle x\rangle$ is the best estimate for $x$ when by
\textquotedblleft best\textquotedblright\ we mean the one with the smallest
variance. But other choices are possible, for example, had we actually decided
to minimize the width $\left\langle \left\vert x-x^{\prime}\right\vert
\right\rangle $ the best estimate would have been the median, $x^{\prime}%
{}_{\text{best}}=x_{m}$, a value such that $\operatorname{Prob}(x<x_{m}%
)=\operatorname{Prob}(x>x_{m})=1/2$.

We conclude this section by mentioning two important identities that will be
repeatedly used in what follows. The first is that the average deviation from
the mean vanishes,%

\begin{equation}
\left\langle x-\langle x\rangle\right\rangle =0,
\end{equation}
because deviations from the mean are just as likely to be positive and
negative. The second useful identity is
\begin{equation}
\left\langle \left(  x-\langle x\rangle\right)  ^{2}\right\rangle =\langle
x^{2}\rangle-\langle x\rangle^{2}.
\end{equation}
The proofs are trivial -- just use the definition (\ref{expected x}).

\section{The binomial distribution}

Suppose the probability of a certain event $\alpha$ is $p$. The probability of
$\alpha$ not happening is $1-p$. Using the theorems discussed earlier we can
obtain the probability that $\alpha$ happens $m$ times in $N$ independent
trials. The probability that $\alpha$ happens in the first $m$ trials and
not-$\alpha$ or $\alpha^{\prime}$ happens in the subsequent $N-m$ trials is,
using the product rule for independent events, $p^{m}(1-p)^{N-m}$. But this is
only one particular ordering of the $m$ \ $\alpha$s and the $N-m$
\ $\alpha^{\prime}$s. There are
\begin{equation}
\frac{N!}{m!(N-m)!}=\binom{N}{m}%
\end{equation}
such orderings. Therefore, using the sum rule for mutually exclusive events,
the probability of $m$ $\alpha$s in $N$ independent trials irrespective of the
particular order of $\alpha$s and $\tilde{\alpha}$s is%

\begin{equation}
P(m|N,p)=\binom{N}{m}p^{m}(1-p)^{N-m}.\label{binomial}%
\end{equation}
\newline This is called the binomial distribution.

Using the binomial theorem (hence the name of the distribution) one can show
these probabilities are correctly normalized:
\begin{equation}
\sum_{m=0}^{N}P(m|N,p)=\sum_{m=0}^{N}\binom{N}{m}p^{m}(1-p)^{N-m}=\left(
p+(1-p)\right)  ^{N}=1.
\end{equation}
\newline The range of applicability of this distribution is enormous. Whenever
trials are independent of each other (i.e., the outcome of one trial has no
influence on the outcome of another, or alternatively, knowing the outcome of
one trial provides us with no information about the possible outcomes of
another) the distribution is binomial. Independence is the crucial feature.

The expected number of $\alpha$s is
\[
\langle m\rangle=\sum_{m=0}^{N}m\,P(m|N,p)=\sum_{m=0}^{N}m\text{ }\binom{N}%
{m}p^{m}(1-p)^{N-m}.
\]
\newline This sum over $m$ is complicated. The following elegant trick is
useful. Consider the sum
\[
S(p,q)=\sum_{m=0}^{N}m\text{ }\binom{N}{m}p^{m}q^{N-m},
\]
\newline where $p$ and $q$ are independent variables. After we calculate $S$
we will replace $q$ by $1-p$ to obtain the desired result, $\langle
m\rangle=S(p,1-p)$. The calculation of $S$ is easy if we note that
$m\,p^{m}=p\,\frac{\partial}{\partial p}p^{m}$. Then, using the binomial
theorem
\[
S(p,q)=p\,\frac{\partial}{\partial p}\,\sum_{m=0}^{N}\binom{N}{m}p^{m}%
q^{N-m}=p\,\frac{\partial}{\partial p}\left(  p+q\right)  ^{N}=Np\left(
p+q\right)  ^{N-1}.
\]
\newline Replacing $q$ by $1-p$ we obtain our best estimate for the expected
number of $\alpha$s
\begin{equation}
\langle m\rangle=Np\,.\label{binomial expectation}%
\end{equation}
\newline This is the best estimate, but how good is it? To answer we need to
calculate $\Delta m$. The variance is
\[
\left(  \Delta m\right)  ^{2}=\left\langle \left(  m-\langle m\rangle\right)
^{2}\right\rangle =\langle m^{2}\rangle-\langle m\rangle^{2}\text{,}%
\]
\newline which requires we calculate $\langle m^{2}\rangle$,
\[
\langle m^{2}\rangle=\sum_{m=0}^{N}m^{2}P(m|N,p)=\sum_{m=0}^{N}m^{2}\binom
{N}{m}p^{m}(1-p)^{N-m}.
\]
\newline We can use the same trick we used before to get $\langle m\rangle$:
\[
S^{\prime}(p,q)=\sum_{m=0}^{N}m^{2}\binom{N}{m}p^{m}q^{N-m}=p\,\frac{\partial
}{\partial p}\,\left(  p\,\frac{\partial}{\partial p}\,\left(  p+q\right)
^{N}\right)  .
\]
\newline Therefore,
\begin{equation}
\langle m^{2}\rangle=(Np)^{2}+Np(1-p)\text{,}%
\end{equation}
\newline and the final result for the \emph{rms} deviation $\Delta m$ is
\begin{equation}
\Delta m=\sqrt{Np\left(  1-p\right)  }.\label{binomial variance}%
\end{equation}
\newline Now we can address the question of how good an estimate $\langle
m\rangle$ is. Notice that $\Delta m$ grows with $N$. This might seem to
suggest that our estimate of $m$ gets worse for large $N$ but this is not
quite true because $\langle m\rangle$ also grows with $N$. The ratio
\begin{equation}
\frac{\Delta m}{\langle m\rangle}=\sqrt{\frac{\left(  1-p\right)  }{Np}%
}\propto\frac{1}{N^{1/2}}\text{,}\label{binomial relat uncert}%
\end{equation}
\newline shows that while both the estimate $\langle m\rangle$ and its
uncertainty $\Delta m$ grow with $N$, the relative uncertainty decreases.

\section[The law of large numbers]{Probability vs. frequency: the law of large
numbers}

Notice that the \textquotedblleft frequency\textquotedblright\ $f=m/N$ of
$\alpha$s obtained in one $N$-trial sequence is not equal to $p$. For one
given fixed value of $p$, the frequency $f$ can take any one of the values
$0/N,1/N,2/N,\ldots N/N$. What is equal to $p$ is not the frequency itself but
its expected value. Using eq.(\ref{binomial expectation}),
\begin{equation}
\langle f\rangle=\langle\frac{m}{N}\rangle=p~.
\end{equation}
$\,$

For large $N$ the distribution is quite narrow and the probability that the
observed frequency of $\alpha$s differs from $p$ tends to zero as
$N\rightarrow\infty$. Using eq.(\ref{binomial variance}),
\begin{equation}
\Delta f=\Delta\left(  \frac{m}{N}\right)  =\frac{\Delta m}{N}=\sqrt
{\frac{p\left(  1-p\right)  }{N}}\propto\frac{1}{N^{1/2}}.
\end{equation}

The same ideas are more precisely conveyed by a theorem due to Bernoulli known
as the `weak law of large numbers'. A simple proof of the theorem involves an
inequality due to Tchebyshev. Let $\rho\left(  x\right)  dx$ be the
probability that a variable $X$ lies in the range between $x$ and $x+dx$,
\[
P\left(  x<X<x+dx\right)  =\rho\left(  x\right)  dx.
\]
\ \ The variance of $X$ satisfies
\[
\left(  \Delta x\right)  ^{2}=\int\left(  x-\langle x\rangle\right)  ^{2}%
\rho\left(  x\right)  \,dx\geq\int_{\left\vert x-\langle x\rangle\right\vert
\geq\varepsilon}\left(  x-\langle x\rangle\right)  ^{2}\rho\left(  x\right)
\,dx\,,
\]
where $\varepsilon$ is an arbitrary constant. Replacing $\left(  x-\langle
x\rangle\right)  ^{2}$ by its least value $\varepsilon^{2}$ gives \newline%
\[
\left(  \Delta x\right)  ^{2}\geq\varepsilon^{2}\,\int_{\left\vert x-\langle
x\rangle\right\vert \geq\varepsilon}\,\rho\left(  x\right)  \,dx=\varepsilon
^{2}\text{ }P\left(  \left\vert x-\langle x\rangle\right\vert \geq
\varepsilon\right)  ,
\]
which is Tchebyshev's inequality, \newline%
\begin{equation}
P\left(  \left\vert x-\langle x\rangle\right\vert \geq\varepsilon\right)
\leq\left(  \frac{\Delta x}{\varepsilon}\right)  ^{2}\text{ }%
.\label{Tchebyshev}%
\end{equation}

Next we prove Bernoulli's theorem, the weak law of large numbers. First a
special case. Let $p$ be the probability of outcome $\alpha$ in an experiment
$E$, $P\left(  \alpha|E\right)  =p$. In a sequence of $N$ independent
repetitions of $E$ the probability of $m$ outcomes $\alpha$ is binomial.
Substituting \
\[
\langle f\rangle=p\,\,\,\,\,\,\,\,\text{and}\,\,\,\,\,\,\,\left(  \Delta
f\right)  ^{2}=\frac{p\left(  1-p\right)  }{N}\text{ }%
\]
\newline into Tchebyshev's inequality we get Bernoulli's theorem,
\begin{equation}
P\left(  \left\vert f-p\right\vert \geq\varepsilon\,|E^{N}\right)  \leq
\frac{p\left(  1-p\right)  }{N\varepsilon^{2}}\text{ .}\label{LLN a}%
\end{equation}
\newline Therefore, the probability that the observed frequency $f$ is
appreciably different from $p$ tends to zero as $N\rightarrow\infty$. Or
equivalently: for any small $\varepsilon$, the probability that the observed
frequency $f=m/N$ lies in the interval between $p-\varepsilon/2$ and
$p+\varepsilon/2$ tends to unity as $N\rightarrow\infty$.

In the mathematical/statistical literature this result is commonly stated in
the form
\begin{equation}
f\longrightarrow p\quad\text{\emph{in probability}.}\label{LLN b}%
\end{equation}
The qualifying words `in probability' are crucial: we are not saying that the
observed $f$ tends to $p$ for large $N$. What vanishes for large $N$ is not
the difference $f-p$ itself, but rather the \emph{probability} that
$\left\vert f-p\right\vert $ is larger than a certain (small) amount.

Thus, probabilities and frequencies are not the same thing but they are
related to each other. Since $\left\langle f\right\rangle =p$, one might
perhaps be tempted to define the probability $p$ in terms of the expected
frequency $\left\langle f\right\rangle $, but this does not work either. The
problem is that the notion of expected value already presupposes that the
concept of probability has been defined previously. The definition of a
probability in terms of expected values is unsatisfactory because it is circular.

The law of large numbers is easily generalized beyond the binomial
distribution. Consider the average
\begin{equation}
x=\frac{1}{N}%
{\textstyle\sum\limits_{r=1}^{N}}
x_{r}~,
\end{equation}
where $x_{1},\ldots,x_{N}$ are $N$ independent variables with the same mean
$\left\langle x_{r}\right\rangle =\mu$ and variance $\operatorname*{var}%
(x_{r})=\left(  \Delta x_{r}\right)  ^{2}=\sigma^{2}$. (In the previous
discussion leading to eq.(\ref{LLN a}) each variable $x_{r}$ is either $1$ or
$0$ according to whether outcome $\alpha$ happens or not in the $r$th
repetition of experiment $E$.)

To apply Tchebyshev's inequality, eq.(\ref{Tchebyshev}), we need the mean and
the variance of $x$. Clearly,
\begin{equation}
\left\langle x\right\rangle =\frac{1}{N}%
{\textstyle\sum\limits_{r=1}^{N}}
\left\langle x_{r}\right\rangle =\frac{1}{N}N\mu=\mu~.
\end{equation}
Furthermore, since the $x_{r}$ are independent, their variances are additive.
For example,
\begin{equation}
\operatorname*{var}(x_{1}+x_{2})=\operatorname*{var}(x_{1}%
)+\operatorname*{var}(x_{2})~.
\end{equation}
(Prove it.) Therefore,
\begin{equation}
\operatorname*{var}(x)=%
{\textstyle\sum\limits_{r=1}^{N}}
\operatorname*{var}(\frac{x_{r}}{N})=N\left(  \frac{\sigma}{N}\right)
^{2}=\frac{\sigma^{2}}{N}~.
\end{equation}
Tchebyshev's inequality now gives,
\begin{equation}
P\left(  \left\vert x-\mu\right\vert \geq\varepsilon|E^{N}\right)  \leq
\frac{\sigma^{2}}{N\varepsilon^{2}}\label{LLN c}%
\end{equation}
so that for any $\varepsilon>0$
\begin{equation}
\lim_{N\rightarrow\infty}\,P\left(  \left\vert x-\mu\right\vert \geq
\varepsilon|E^{N}\right)  =0\quad\text{or}\quad\lim_{N\rightarrow\infty
}\,P\left(  \left\vert x-\mu\right\vert \leq\varepsilon|E^{N}\right)
=1\,,\label{LLN d}%
\end{equation}
or
\begin{equation}
x\longrightarrow\mu\quad\text{\emph{in probability}.}\label{LLN e}%
\end{equation}
Again, what vanishes for large $N$ is not the difference $x-\mu$ itself, but
rather the \emph{probability} that $\left\vert x-\mu\right\vert $ is larger
than any given small amount.

\section{The Gaussian distribution}

The Gaussian distribution is quite remarkable, it applies to a wide variety of
problems such as the distribution of errors affecting experimental data, the
distribution of velocities of molecules in gases and liquids, the distribution
of fluctuations of thermodynamical quantities, and so on and on. One suspects
that a deeply fundamental reason must exist for its wide applicability.
Somehow the Gaussian distribution manages to codify the information that
happens to be relevant for prediction in a wide variety of problems. The
Central Limit Theorem discussed below provides an explanation.

\subsection{The de Moivre-Laplace theorem}

The Gaussian distribution turns out to be a special case of the binomial
distribution. It applies to situations when the number $N$ of trials and the
expected number of $\alpha$s, $\langle m\rangle=Np$, are both very large
(i.e., $N$ large, $p$ not too small).

To find an analytical expression for the Gaussian distribution we note that
when $N$ is large the binomial distribution, \ %

\[
P(m|N,p)=\frac{N!}{m!(N-m)!}\,p^{m}(1-p)^{N-m}\text{,}%
\]
\newline is very sharply peaked: $P(m|N,p)$ is essentially zero unless $m$ is
very close to $\langle m\rangle=Np$. This suggests that to find a good
approximation for $P$ we need to pay special attention to a very small range
of $m$ and this can be done following the usual approach of a Taylor
expansion. A problem is immediately apparent: if a small change in $m$
produces a small change in $P$ then we only need to keep the first few terms,
but in our case $P$ is a very sharp function. To reproduce this kind of
behavior we need a huge number of terms in the series expansion which is
impractical. Having diagnosed the problem one can easily find a cure: instead
of finding a Taylor expansion for the rapidly varying $P$, one finds an
expansion for $\log\,P$ which varies much more smoothly.

Let us therefore expand $\log\,P$ about its maximum at $m_{0}$, the location
of which is at this point still unknown. The first few terms are
\[
\log P=\left.  \log P\right\vert _{m_{0}}+\left.  \frac{d\log P}%
{dm}\right\vert _{m_{0}}\left(  m-m_{0}\right)  +\frac{1}{2}\left.
\frac{d^{2}\log P}{dm^{2}}\right\vert _{m_{0}}\left(  m-m_{0}\right)
^{2}+\ldots,
\]
where
\[
\log P=\log N!-\log m!-\log\left(  N-m\right)  !+m\,\log p+\left(  N-m\right)
\,\log\left(  1-p\right)  .\,
\]
What is a derivative with respect to an integer? For large $m$ the function
$\log m!$ varies so slowly (relative to the huge value of $\log m!$ itself)
that we may consider $m$ to be a continuous variable. Then
\begin{equation}
\frac{d\log m!}{dm}\approx\frac{\log m!-\log\left(  m-1\right)  !}{1}%
=\log\frac{m!}{\left(  m-1\right)  !}=\log m~.
\end{equation}
\newline Integrating one obtains a very useful approximation -- called the
Stirling approximation -- for the logarithm of a large factorial
\[
\log m!\approx\int_{0}^{m}\log x\,dx=\left.  \left(  x\log\,x-x\right)
\right\vert _{0}^{m}=m\log\,m-m.
\]
A somewhat better expression which includes the next term in the Stirling
expansion is \newline%
\begin{equation}
\log m!\approx m\log m-m+\frac{1}{2}\log2\pi m+\ldots\text{ }%
\end{equation}
\ \ Notice that the third term is much smaller than the first two; the first
two terms are of order $m$ while the last is of order $\log m$. For
$m=10^{23}$, $\log m$ is only $55.3$. \

The derivatives in the Taylor expansion are
\[
\frac{d\log P}{dm}=-\log m+\log\left(  n-m\right)  +\log p-\log\left(
1-p\right)  =\log\frac{p(N-m)}{m(1-p)},
\]
and
\[
\frac{d^{2}\log P}{dm^{2}}=-\frac{1}{m}-\frac{1}{N-m}=\frac{-N}{m(N-m)}.
\]
\newline To find the value $m_{0}$ where $P$ is maximum set $d\log P/dm=0$.
This gives $m_{0}=Np=\langle m\rangle$, and substituting into the second
derivative of $\log P$ we get
\[
\left.  \frac{d^{2}\log P}{dm^{2}}\right\vert _{\left\langle m\right\rangle
}=-\frac{1}{Np\left(  1-p\right)  }=-\frac{1}{\left(  \Delta m\right)  ^{2}}.
\]
Therefore\newline%
\[
\log P=\log P\left(  \langle m\rangle\right)  -\frac{\left(  m-\langle
m\rangle\right)  ^{2}}{2\,\left(  \Delta m\right)  ^{2}}+\ldots
\]
or
\[
P(m)=P\left(  \langle m\rangle\right)  \,\exp\left[  -\frac{\left(  m-\langle
m\rangle\right)  ^{2}}{2\,\left(  \Delta m\right)  ^{2}}\right]  .
\]
The remaining unknown constant $P\left(  \langle m\rangle\right)  $ can be
evaluated by requiring that the distribution $P(m)$ be properly normalized,
that is
\[
1=\sum_{m=0}^{N}P(m)\approx\int_{0}^{N}P(x)\,dx\approx\int_{-\infty}^{\infty
}P(x)\,dx.
\]
\ Using
\[
\int_{-\infty}^{\infty}e^{-\alpha x^{2}}dx=\sqrt{\frac{\pi}{\alpha}},
\]
we get\newline%
\[
P\left(  \langle m\rangle\right)  =\frac{1}{\sqrt{2\pi\left(  \Delta m\right)
^{2}}}\,.
\]
\newline Thus, the expression for the Gaussian distribution with mean $\langle
m\rangle$ and \emph{rms} deviation $\Delta m$ is \
\begin{equation}
P(m)=\frac{1}{\sqrt{2\pi\left(  \Delta m\right)  ^{2}}}\,\exp\left[
-\frac{\left(  m-\langle m\rangle\right)  ^{2}}{2\,\left(  \Delta m\right)
^{2}}\right]  .\label{Gaussian a}%
\end{equation}
It can be rewritten as a probability for the frequency $f=m/N$ using
$\left\langle m\right\rangle =Np$ and $\left(  \Delta m\right)  ^{2}=Np\left(
1-p\right)  $. The probability that $f$ lies in the small range $df=1/N$ is
\begin{equation}
p(f)df=\frac{1}{\sqrt{2\pi\sigma_{N}^{2}}}\,\exp\left[  -\frac{\left(
f-p\right)  ^{2}}{2\,\sigma_{N}^{2}}\right]  df~,\label{Gaussian b}%
\end{equation}
where $\sigma_{N}^{2}=p(1-p)/N$.

To appreciate the significance of the theorem consider a macroscopic variable
$x$ built up by adding a large number of small contributions, $x=\sum
_{n=1}^{N}\xi_{n}$, where the $\xi_{n}$ are statistically independent. We
assume that each $\xi_{n}$ takes the value $\varepsilon$ with probability $p$,
and the value $0$ with probability $1-p$. Then the probability that $x$ takes
the value $m\varepsilon$ is given by the binomial distribution $P(m|N,p)$. For
large $N$ the probability that $x$ lies in the small range $m\varepsilon\pm
dx/2$ where $dx=\varepsilon$ is
\begin{equation}
p(x)dx=\frac{1}{\sqrt{2\pi\left(  \Delta x\right)  ^{2}}}\,\exp\left[
-\frac{\left(  x-\left\langle x\right\rangle \right)  ^{2}}{2\,\left(  \Delta
x\right)  ^{2}}\right]  dx~,\label{Gaussian c}%
\end{equation}
where $\left\langle x\right\rangle =Np\varepsilon$ and $\left(  \Delta
x\right)  ^{2}=Np(1-p)\varepsilon^{2}$. Thus, \emph{the Gaussian distribution
arises whenever we have a quantity that is the result of adding a large number
of small independent contributions}. The derivation above assumes that the
microscopic contributions are discrete (binomial, either $0$ or $\varepsilon
$), and identically distributed but, as shown in the next section, both of
these conditions can be relaxed.

\subsection{The Central Limit Theorem}

Consider the average
\begin{equation}
x=\frac{1}{N}%
{\textstyle\sum\limits_{r=1}^{N}}
x_{r}~,
\end{equation}
of $N$ independent variables $x_{1},\ldots,x_{N}$. Our goal is to calculate
the probability of $x$ in the limit of large $N$. Let $p_{r}(x_{r})$ be the
probability distribution for the $r$th variable with
\begin{equation}
\left\langle x_{r}\right\rangle =\mu_{r}\quad\text{and}\quad\left(  \Delta
x_{r}\right)  ^{2}=\sigma_{r}^{2}~.
\end{equation}
The probability density for $x$ is given by the integral
\begin{equation}
P\left(  x\right)  =%
{\textstyle\int}
dx_{1}\ldots dx_{N}~p_{1}(x_{1})\ldots p_{N}(x_{N})\,\delta\left(  x-\frac
{1}{N}%
{\textstyle\sum\limits_{r=1}^{N}}
x_{r}\right)  ~.\label{central lim a}%
\end{equation}
(This is just an exercise in the sum and product rules.) To calculate
$P\left(  x\right)  $ introduce the averages
\begin{equation}
\bar{\mu}\overset{\operatorname*{def}}{=}\frac{1}{N}%
{\textstyle\sum\limits_{r=1}^{N}}
\mu_{r}\quad\text{and}\quad\bar{\sigma}^{2}\overset{\operatorname*{def}}%
{=}\frac{1}{N}%
{\textstyle\sum\limits_{r=1}^{N}}
\sigma_{r}^{2}~,
\end{equation}
and consider the distribution for the variable $x-\bar{\mu}$ which is
$\Pr(x-\bar{\mu})=P(x)$. It's Fourier transform,
\begin{align*}
F(k)  & =\int dx\,\Pr(x-\bar{\mu})e^{ik(x-\bar{\mu})}=\int
dx\,P(x)e^{ik(x-\bar{\mu})}\\
& =\int dx_{1}\ldots dx_{N}~p_{1}(x_{1})\ldots p_{N}(x_{N})\,\exp\,\left[
\frac{ik}{N}%
{\textstyle\sum\limits_{r=1}^{N}}
\left(  x_{r}-\mu_{r}\right)  \right]  ,
\end{align*}
can be rearranged into a product%
\begin{equation}
F(k)=\left[  \int dx_{1}\,p_{1}(x_{1})e^{i\frac{k}{N}\left(  x_{1}-\mu
_{1}\right)  }\right]  \ldots\left[  \int dx_{N}\,p_{N}(x_{N})e^{i\frac{k}%
{N}\left(  x_{N}-\mu_{N}\right)  }\right]  ~.\label{central lim convolution}%
\end{equation}

The Fourier transform $f(k)$ of a distribution $p(\xi)$ has many interesting
and useful properties. For example,
\begin{equation}
f(k)=\int d\xi\,p(\xi)e^{ik\xi}=\left\langle e^{ik\xi}\right\rangle ~,
\end{equation}
and the series expansion of the exponential gives
\begin{equation}
f(k)=\left\langle
{\textstyle\sum\limits_{n=0}^{\infty}}
\,\frac{\left(  ik\xi\right)  ^{n}}{n!}~\right\rangle =%
{\textstyle\sum\limits_{n=0}^{\infty}}
\,\frac{\left(  ik\right)  ^{n}}{n!}\left\langle \xi^{n}\right\rangle ~.
\end{equation}
In words, the coefficients of the Taylor expansion of $f(k)$ give all the
moments of $p(\xi)$. The Fourier transform $f(k)$ is called the \emph{moment
generating function} and also the \emph{characteristic function} of the distribution.

Going back to our calculation of $P(x)$, eq.(\ref{central lim a}), its Fourier
transform, eq.(\ref{central lim convolution}) is,
\begin{equation}
F(k)=%
{\textstyle\prod\limits_{r=1}^{N}}
\,f_{r}(\frac{k}{N})~,
\end{equation}
where
\begin{align}
f_{r}(\frac{k}{N})  & =\int dx_{r}\,p_{r}(x_{r})e^{i\frac{k}{N}\left(
x_{r}-\mu_{r}\right)  }\nonumber\\
& =1+i\frac{k}{N}\left\langle x_{r}-\mu_{r}\right\rangle -\frac{k^{2}}{2N^{2}%
}\left\langle \left(  x_{r}-\mu_{r}\right)  ^{2}\right\rangle +\ldots
\nonumber\\
& =1-\frac{k^{2}\sigma_{r}^{2}}{2N^{2}}+O\left(  \frac{k^{3}}{N^{3}}\right)
~.\label{central limit 1}%
\end{align}
For a sufficiently large $N$ this can be written as
\begin{equation}
f_{r}(\frac{k}{N})\longrightarrow\exp\left(  -\frac{k^{2}\sigma_{r}^{2}%
}{2N^{2}}\right)  ~.
\end{equation}
so that
\begin{equation}
F(k)=\exp\left(  -\frac{k^{2}}{2N^{2}}%
{\textstyle\sum\limits_{r=1}^{N}}
\sigma_{r}^{2}\right)  =\exp\left(  -\frac{k^{2}\bar{\sigma}^{2}}{2N}\right)
~.\label{central limit 2}%
\end{equation}
Finally, taking the inverse Fourier transform, we obtain the desired result,
which is called the central limit theorem
\begin{equation}
\Pr(x-\bar{\mu})=P(x)=\frac{1}{\sqrt{2\pi\bar{\sigma}^{2}/N}}\exp\left(
-\frac{(x-\bar{\mu})^{2}}{2\bar{\sigma}^{2}/N}\right)  ~.
\end{equation}

To conclude we comment on its significance. We have shown that almost
independently of the form of the distributions $p_{r}\left(  x_{r}\right)  $
the distribution of the average $x$ is Gaussian centered at $\bar{\mu}$ with
standard deviation $\bar{\sigma}^{2}/N$. Not only the $p_{r}\left(
x_{r}\right)  $ need not be binomial, they do not even have to be equal to
each other. This helps to explain the widespread applicability of the Gaussian
distribution: it applies to almost any `macro-variable' (such as $x$) that
results from adding a large number of independent `micro-variables' (such as
$x_{r}/N$).

But there are restrictions; although very common, Gaussian distributions do
not obtain always. A careful look at the derivation above shows the crucial
step was taken in eqs.(\ref{central limit 1}) and (\ref{central limit 2})
where we neglected the contributions of the third and higher moments. Earlier
we mentioned that the success of Gaussian distributions is due to the fact
that they codify the information that happens to be relevant to the particular
phenomenon under consideration. Now we see what that relevant information
might be: it is contained in the first two moments, the mean and the variance
-- Gaussian distributions are successful when third and higher moments are
irrelevant. (This can be stated more precisely in terms as the so-called
Lyapunov condition.)

Later we shall approach this same problem from the point of view of the method
of maximum entropy and there we will show that, indeed, the Gaussian
distribution can be derived as the distribution that codifies information
about the mean and the variance while remaining maximally ignorant about
everything else.

\section{Updating probabilities: Bayes' rule}

Now that we have solved the problem of how to represent a state of knowledge
as a consistent web of interconnected beliefs we can address the problem of
updating from one consistent web of beliefs to another when new information
becomes available. We will only consider those special situations where the
information to be processed is in the form of data.

Specifically the problem is to update our beliefs about $\theta$ (either a
single parameter or many) on the basis of data $x$ (either a single number or
several) and of a known relation between $\theta$ and $x$. The updating
consists of replacing the \emph{prior} probability distribution $p(\theta)$
that represents our beliefs before the data is processed, by a
\emph{posterior} distribution $p_{\text{new}}(\theta)$ that applies after the
data has been processed.

\subsection{Formulating the problem}

We must first describe the state of our knowledge before the data has been
collected or, if the data has already been collected, before we have taken it
into account. At this stage of the game not only we do not know $\theta$, we
do not know $x$ either. As mentioned above, in order to infer $\theta$ from
$x$ we must also know how these two quantities are related to each other.
Without this information one cannot proceed further. Fortunately we usually
know enough about the physics of an experiment that if $\theta$ were known we
would have a fairly good idea of what values of $x$ to expect. For example,
given a value $\theta$ for the charge of the electron, we can calculate the
velocity $x$ of an oil drop in Millikan's experiment, add some uncertainty in
the form of Gaussian noise and we have a very reasonable estimate of the
conditional distribution $p(x|\theta)$. The distribution $p(x|\theta)$ is
called the \emph{sampling} distribution and also (less appropriately) the
\emph{likelihood}. We will assume it is known.

We should emphasize that the crucial information about how $x$ is related to
$\theta$ is contained in the functional form of the distribution $p(x|\theta)$
-- say, whether it is a Gaussian or a Cauchy distribution -- and not in the
actual values of the arguments $x$ and $\theta$ which are, at this point,
still unknown.

Thus, to describe the web of prior beliefs we must know the prior $p(\theta) $
and also the sampling distribution $p(x|\theta)$. This means that we must know
the full joint distribution,
\begin{equation}
p(\theta,x)=p(\theta)p(x|\theta)~.\label{web old}%
\end{equation}
This is very important: we must be clear about what we are talking about.\ The
relevant universe of discourse is neither the space $\Theta$ of possible
parameters $\theta$ nor is it the space $\mathcal{X}$ of possible data $x$. It
is rather the product space $\Theta\times\mathcal{X}$ and the probability
distributions that concern us are the joint distributions $p(\theta,x)$.

Next we collect data: the observed value turns out to be $X$. Our goal is to
use this information to update to a web of posterior beliefs represented by a
new joint distribution $p_{\text{new}}(\theta,x)$. How shall we choose
$p_{\text{new}}(\theta,x)$? The new data tells us that the value of $x$ is now
known to be $X$. Therefore, the new web of beliefs must be such that
\begin{equation}
p_{\text{new}}(x)=%
{\textstyle\int}
d\theta\,p_{\text{new}}(\theta,x)=\delta(x-X)~.\label{web new a}%
\end{equation}
(For simplicity we have here assumed that $x$ is a continuous variable; had
$x$ been discrete Dirac $\delta$s would be replaced by Kronecker $\delta$s.)
This is all we know but it is not sufficient to determine $p_{\text{new}%
}(\theta,x)$. Apart from the general requirement that the new web of beliefs
must be internally consistent there is nothing in any of our previous
considerations that induces us to prefer one consistent web over another. A
new principle is needed.

\subsection{Minimal updating: Bayes' rule}

The basic updating strategy that we adopt below reflects the conviction that
what we have learned in the past, the prior knowledge, is a valuable resource
that should not be squandered. Prior beliefs should be revised only when this
is demanded by the new information; the new web of beliefs should coincide
with the old one as much as possible. We propose to adopt the following

\begin{description}
\item \qquad\emph{Principle of Minimal Updating }(PMU):\ The web of beliefs
needs to be revised \emph{only} to the extent required by the new data.
\end{description}

\noindent This seems so reasonable and natural that an explicit statement may
seem superfluous. The important point, however, is that \emph{it is not
logically necessary}. We could update in many other ways that preserve both
internal consistency and consistency with the new information.

As we saw above the new data, eq.(\ref{web new a}), does not fully determine
the joint distribution
\begin{equation}
p_{\text{new}}(\theta,x)=p_{\text{new}}(x)p_{\text{new}}(\theta|x)=\delta
(x-X)p_{\text{new}}(\theta|x)\,.
\end{equation}
All distributions of the form
\begin{equation}
p_{\text{new}}(\theta,x)=\delta(x-X)p_{\text{new}}(\theta|X)~,~
\end{equation}
where $p_{\text{new}}(\theta|X)$ remains arbitrary is compatible with the
newly acquired data. We still need to assign $p_{\text{new}}(\theta|X)$. It is
at this point that we invoke the PMU. We stipulate that no further revision is
needed and set%
\begin{equation}
p_{\text{new}}(\theta|X)=p_{\text{old}}(\theta|X)=p(\theta|X)~.
\end{equation}
Therefore, the web of posterior beliefs is described by
\begin{equation}
p_{\text{new}}(\theta,x)=\delta(x-X)p(\theta|X)~.\label{web new b}%
\end{equation}
The posterior probability $p_{\text{new}}(\theta)$ is
\begin{equation}
p_{\text{new}}(\theta)=%
{\textstyle\int}
dx\,p_{\text{new}}(\theta,x)=%
{\textstyle\int}
dx\,\delta(x-X)p(\theta|X)~,
\end{equation}
or,
\begin{equation}
p_{\text{new}}(\theta)=p(\theta|X)~.\label{Bayes rule}%
\end{equation}
In words, the \emph{posterior probability equals the prior conditional
probability} of $\theta$ given $X$. This result, known as Bayes' rule, is
extremely reasonable: we \emph{maintain} those beliefs about $\theta$ that are
consistent with the data values $X$ that turned out to be true. Data values
that were not observed are discarded because they are now known to be false.
`Maintain' is the key word: it reflects the PMU in action.

Using the product rule
\begin{equation}
p(\theta,X)=p(\theta)p(X|\theta)=p(X)p(\theta|X)~,\label{Bayes th}%
\end{equation}
Bayes' rule can be written as
\begin{equation}
p_{\text{new}}(\theta)=p(\theta)\frac{p(X|\theta)}{p(X)}~.\label{Bayes rule b}%
\end{equation}

\noindent\textbf{Remark:} Bayes' rule is usually written in the form
\begin{equation}
p(\theta|X)=p(\theta)\frac{p(X|\theta)}{p(X)}~,\label{Bayes theorem}%
\end{equation}
and called Bayes' theorem. This formula is very simple; perhaps it is too
simple. It is just a restatement of the product rule, eq.(\ref{Bayes th}), and
therefore it is a simple consequence of the \emph{internal} consistency of the
\emph{prior} web of beliefs. The drawback of this formula is that the left
hand side is not the \emph{posterior} but rather the \emph{prior}
\emph{conditional} probability; it obscures the fact that an additional
principle -- the PMU -- was needed for updating.

The interpretation of Bayes' rule is straightforward: according to
eq.(\ref{Bayes rule b}) the posterior distribution $p_{\text{new}}(\theta)$
gives preference to those values of $\theta$ that were previously preferred as
described by the prior $p(\theta)$, but this is now modulated by the
likelihood factor $p(X|\theta)$ in such a way as to enhance our preference for
values of $\theta$ that make the data more likely, less surprising. The factor
in the denominator $p(X)$ which is the prior probability of the data is given
by%
\begin{equation}
p(X)=%
{\textstyle\int}
p(\theta)p(X|\theta)\,d\theta~,
\end{equation}
and plays the role of a normalization constant for the posterior distribution
$p_{\text{new}}(\theta)$. It does not help to discriminate one value of
$\theta$ from another because it affects all values of $\theta$ equally and is
therefore not important except, as we shall see later in this chapter, in
problems of model selection.

Neither the rule, eq.(\ref{Bayes rule}), nor the theorem,
eq.(\ref{Bayes theorem}), was ever actually written down by Bayes. The person
who first explicitly stated the theorem and, more importantly, who first
realized its deep significance was Laplace.

\subsubsection*{Example: is there life on Mars?}

Suppose we are interested in whether there is life on Mars or not. How is the
probability that there is life on Mars altered by new data indicating the
presence of water on Mars. Let $\theta=$`There is life on Mars'. The prior
information includes the fact $I=\,$`All known life forms require water'. The
new data is that $X=\,$`There is water on Mars'. Let us look at Bayes' rule.
We can't say much about $p\left(  X|I\right)  $ but whatever its value it is
definitely less than 1. On the other hand $p\left(  X|\theta I\right)
\approx1$. Therefore the factor multiplying the prior is larger than 1. Our
belief in the truth of $\theta$ is strengthened by the new data $X$. This is
just common sense, but notice that this kind of probabilistic reasoning cannot
be carried out if one adheres to a strictly frequentist interpretation --
there is no set of trials. The name `Bayesian probabilities' given to `degrees
of belief' originates in the fact that it is only under this interpretation
that the full power of Bayes' rule can be exploited.

\subsubsection*{Example: testing positive for a rare disease}

Suppose you are tested for a disease, say cancer, and the test turns out to be
positive. Suppose further that the test is said to be 99\% accurate. Should
you panic? It may be wise to proceed with caution.

One should start by explaining that `99\% accurate' means that when the test
is applied to people known to have cancer the result is positive 99\% of the
time, and when applied to people known to be healthy, the result is negative
99\% of the time. We express this accuracy as $p(y|c)=A=0.99$ and
$p(n|\tilde{c})=A=0.99$ ($y$ and $n$ stand for `positive' and `negative', $c$
and $\tilde{c}$ stand for `cancer' or `no cancer'). There is a 1\% probability
of false positives, $p(y|\tilde{c})=1-A$, and a 1\% probability of false
negatives, $p(n|c)=1-A$.

On the other hand, what we really want to know is $p_{\text{new}}(c)=p(c|y)$,
the probability of having cancer given that you tested positive. This is not
the same as the probability of testing positive given that you have cancer,
$p(y|c)$; the two probabilities are not the same thing! So there might be some
hope. The connection between what we want, $p(c|y)$, and what we know,
$p(y|c)$, is given by Bayes' theorem,
\[
p(c|y)=\frac{p(c)p(y|c)}{p(y)}~.
\]

An important virtue of Bayes' rule is that it doesn't just tell you how to
process information; it also tells you what information you should seek. In
this case one should find $p(c)$, the probability of having cancer
irrespective of being tested positive or negative. Suppose you inquire and
find that the incidence of cancer in the general population is 1\%; this means
that $p(c)=0.01$. Thus,
\[
p(c|y)=\frac{p(c)A}{p(y)}%
\]

One also needs to know $p(y)$, the probability of the test being positive
irrespective of whether the person has cancer or not. To obtain $p(y)$ use
\[
p(\tilde{c}|y)=\frac{p(\tilde{c})p(y|\tilde{c})}{p(y)}=\frac{\left(
1-p(c)\right)  \left(  1-A\right)  }{p(y)}~,
\]
and $p(c|y)+p(\tilde{c}|y)=1$ which leads to our final answer
\begin{equation}
p(c|y)=\frac{p(c)A}{p(c)A+\left(  1-p(c)\right)  \left(  1-A\right)  }~.
\end{equation}
For an accuracy $A=0.99$ and an incidence $p(c)=0.01$ we get $p(c|y)=50\%$
which is not nearly as bad as one might have originally feared. Should one
dismiss the information provided by the test as misleading? No. Note that the
probability of having cancer prior to the test was 1\% and on learning the
test result this was raised all the way up to 50\%. Note also that when the
disease is really rare, $p(c)\rightarrow0$, we still get $p(c|y)\rightarrow0$
even when the test is quite accurate. This means that for rare diseases most
positive tests turn out to be false positives.

We conclude that both the prior and the data contain important information;
neither should be neglected.

\noindent\textbf{Remark:} The previous discussion illustrates a mistake that
is common in verbal discussions: if $h$ denotes a hypothesis and $e$ is some
evidence, it is quite obvious that we should not confuse $p(e|h)$ with
$p(h|e)$. However, when expressed verbally the distinction is not nearly as
obvious. For example, in a criminal trial jurors might be told that if the
defendant were guilty (the hypothesis) the probability of some observed
evidence would be large, and the jurors might easily be misled into concluding
that given the evidence the probability is high that the defendant is guilty.
Lawyers call this the \emph{prosecutor's fallacy.}

\subsubsection*{Example: uncertain data}

As before we want to update from a prior joint distribution $p(\theta
,x)=p(x)p(\theta|x)$ to a posterior joint distribution $p_{\text{new}}%
(\theta,x)=p_{\text{new}}(x)p_{\text{new}}(\theta|x)$ when information becomes
available. When the information is data $X$ that precisely fixes the value of
$x$, we impose that $p_{\text{new}}(x)=\delta(x-X)$. The remaining unknown
$p_{\text{new}}(\theta|x)$ is determined by invoking the PMU: no further
updating is needed. This fixes $p_{\text{new}}(\theta|x)$ to be the old
$p(\theta|x)$ and yields Bayes' rule.

It may happen, however, that there is a measurement error. The data $X$ that
was actually observed does not constrain the value of $x$ completely. To be
explicit let us assume that the remaining uncertainty in $x$ is well
understood: the observation $X$ constrains our beliefs about $x$ to a
distribution $P_{X}(x)$ that happens to be known. $P_{X}(x)$ could, for
example, be a Gaussian distribution centered at $X$, with some known standard
deviation $\sigma$.

This information is incorporated into the posterior distribution,
$p_{\text{new}}(\theta,x)=p_{\text{new}}(x)p_{\text{new}}(\theta|x)$, by
imposing that $p_{\text{new}}(x)=P_{X}(x)$. The remaining conditional
distribution is, as before, determined by invoking the PMU,
\begin{equation}
p_{\text{new}}(\theta|x)=p_{\text{old}}(\theta|x)=p(\theta|x)~,
\end{equation}
and therefore, the joint posterior is
\begin{equation}
p_{\text{new}}(\theta,x)=P_{X}(x)p(\theta|x)~.
\end{equation}
Marginalizing over the uncertain $x$ yields the new posterior for $\theta$,
\begin{equation}
p_{\text{new}}(\theta)=%
{\textstyle\int}
dx\,P_{X}(x)p(\theta|x)~.
\end{equation}
This generalization of Bayes' rule is sometimes called Jeffrey's
conditionalization rule.

Incidentally, this is an example of updating that shows that \emph{it is not
always the case that information comes purely in the form of data }$X$\emph{.}
In the derivation above there clearly is some information in the observed
value $X$ and some information in the particular functional form of the
distribution $P_{X}(x)$, whether it is a Gaussian or some other distribution.

The common element in our previous derivation of Bayes' rule and in the
present derivation of Jeffrey's rule is that in both cases the information
being processed is a constraint on the allowed posterior marginal
distributions $p_{\text{new}}(x)$. Later we shall see (chapter 5) how the
updating rules can be generalized still further to apply to even more general constraints.

\subsection{Multiple experiments, sequential updating}

The problem here is to update our beliefs about $\theta$ on the basis of data
$x_{1},x_{2},\ldots,x_{n}$ obtained in a sequence of experiments. The
relations between $\theta$ and the variables $x_{i}$ are given through known
sampling distributions. We will assume that the experiments are independent
but they need not be identical. When the experiments are not independent it is
more appropriate to refer to them as being performed is a single more complex
experiment the outcome of which is a set of numbers $\{x_{1},\ldots,x_{n}\}$.

For simplicity we deal with just two identical experiments. The prior web of
beliefs is described by the joint distribution,
\begin{equation}
p(x_{1},x_{2},\theta)=p(\theta)\,p(x_{1}|\theta)p(x_{2}|\theta)=p(x_{1}%
)p(\theta|x_{1})p(x_{2}|\theta)~,\label{mult a}%
\end{equation}
where we have used independence, $p(x_{2}|\theta,x_{1})=p(x_{2}|\theta)$.

The first experiment yields the data $x_{1}=X_{1}$. Bayes' rule gives the
updated distribution for $\theta$ as
\begin{equation}
p_{1}(\theta)=p(\theta|X_{1})=p(\theta)\frac{p(X_{1}|\theta)}{p(X_{1}%
)}~.\label{mult b}%
\end{equation}
The second experiment yields the data $x_{2}=X_{2}$ and requires a second
application of Bayes' rule. The posterior $p_{1}(\theta)$ in eq.(\ref{mult b})
now plays the role of the prior and the new posterior distribution for
$\theta$ is
\begin{equation}
p_{12}(\theta)=p_{1}(\theta|X_{2})=p_{1}(\theta)\frac{p(X_{2}|\theta)}%
{p_{1}(X_{2})}~.\label{mult c}%
\end{equation}

We have explicitly followed the update from $p(\theta)$ to $p_{1}(\theta)$ to
$p_{12}(\theta)$. It is straightforward to show that the same result is
obtained if the data from both experiments were processed simultaneously,
\begin{equation}
p_{12}(\theta)=p(\theta|X_{1},X_{2})=p(\theta)\frac{p(X_{1},X_{2}|\theta
)}{p(X_{1},X_{2})}~.\label{mult d}%
\end{equation}
Indeed, using eq.(\ref{mult a}) and (\ref{mult b}), this last equation can be
rewritten as
\begin{equation}
p_{12}(\theta)=p(\theta)\frac{p(X_{1}|\theta)}{p(X_{1})}\frac{p(X_{2}|\theta
)}{p(X_{2}|X_{1})}=p_{1}(\theta)\frac{p(X_{2}|\theta)}{p(X_{2}|X_{1})}\,~,
\end{equation}
and it remains to show that $p(X_{2}|X_{1})=p_{1}(X_{2})$. This last step is
straightforward; use eq.(\ref{mult c}) and (\ref{mult b}):
\begin{align}
p_{1}(X_{2})  & =%
{\textstyle\int}
p_{1}(\theta)p(X_{2}|\theta)d\theta=%
{\textstyle\int}
p(\theta)\frac{p(X_{1}|\theta)}{p(X_{1})}p(X_{2}|\theta)d\theta\nonumber\\
& =%
{\textstyle\int}
\frac{p(X_{1},X_{2},\theta)}{p(X_{1})}d\theta=p(X_{2}|X_{1})~.
\end{align}

From the symmetry of eq.(\ref{mult d}) it is clear that the same posterior
$p_{12}(\theta)$ is obtained irrespective of the order that the data $X_{1}$
and $X_{2}$ are processed. The commutativity of Bayesian updating follows from
the special circumstance that the information conveyed by one experiment does
not revise or render obsolete the information conveyed by the other
experiment. As we generalize our methods of inference for processing other
kinds of information that do interfere with each other (and therefore one may
render the other obsolete) we should not expect, much less demand, that
commutativity will continue to hold.

\subsection{Remarks on priors}

Let us return to the question of the extent to which probabilities incorporate
subjective and objective elements. We have seen that Bayes' rule allows us to
update from prior to posterior distributions. The posterior distributions
incorporate the presumably objective information contained in the data plus
whatever earlier beliefs had been codified into the prior. To the extent that
the Bayes updating rule is itself unique one can claim that the posterior is
\textquotedblleft more objective\textquotedblright\ than the prior. As we
update more and more we should expect that our probabilities should reflect
more and more the input data and less and less the original subjective prior
distribution. In other words, some subjectivity is unavoidable at the
beginning of an inference chain, but it can be gradually suppressed as more
and more information is processed.

The problem of choosing the first prior in the inference chain is a difficult
one. We will tackle it in several different ways. Later in this chapter, as we
introduce some elementary notions of data analysis, we will address it in the
standard way: just make a \textquotedblleft reasonable\textquotedblright%
\ guess -- whatever that might mean. With experience and intuition this seems
to work well. But when addressing new problems we have neither experience nor
intuition and guessing is risky. We would like to develop more systematic ways
to proceed. Indeed it can be shown that certain types of prior information
(for example, symmetries and/or other constraints) can be objectively
translated into a prior once we have developed the appropriate tools --
entropy and geometry. (See e.g. [Caticha Preuss 04] and references therein.)

Our immediate goal here is, first, to remark on the dangerous consequences of
extreme degrees of belief, and then to prove our previous intuitive assertion
that the accumulation of data will swamp the original prior and render it irrelevant.

\subsubsection{Dangerous extremes: the prejudiced mind}

The consistency of Bayes' rule can be checked for the extreme cases of
certainty and impossibility: Let $B$ describe any background information. If
$p\left(  \theta|B\right)  =1$, then $\theta B=B$ and $p(X|\theta B)=p(X|B)$,
so that Bayes' rule gives
\begin{equation}
p_{\text{new}}(\theta|B)=p(\theta|B)\frac{p(X|\theta B)}{p(X|B)}=1~.
\end{equation}
A similar argument can be carried through in the case of impossibility: If
$p\left(  \theta|B\right)  =0$, then $p_{\text{new}}\left(  \theta|B\right)
=0$. Conclusion: if we are absolutely certain about the truth of $\theta$,
acquiring data $X$ will have absolutely no effect on our opinions; the new
data is worthless.

This should serve as a warning to the dangers of erroneously assigning a
probability of $1$ or of $0$: since no amount of data could sway us from our
prior beliefs we may decide we did not need to collect the data in the first
place. If you are absolutely sure that Jupiter has no moons, you may either
decide that it is not necessary to look through the telescope, or, if you do
look and you see some little bright spots, you will probably decide the spots
are mere optical illusions. Extreme degrees of belief are dangerous: a truly
prejudiced mind does not, and indeed, \emph{cannot} question its own beliefs.

\subsubsection{Lots of data overwhelms the prior}

As more and more data is accumulated according to the sequential updating
described earlier one would expect that the continuous inflow of information
will eventually render irrelevant whatever prior information we might have had
at the start. This is indeed the case: unless we have assigned a pathological
prior -- all we need is a prior that is smooth where the likelihood is large
-- after a large number of experiments the posterior becomes essentially
independent of the prior.

Consider $N$ independent repetitions of a certain experiment $E$ that yield
the data $X=\{X_{1}\ldots X_{N}\}$. The corresponding likelihood is
\begin{equation}
p(X|\theta)=%
{\textstyle\prod\limits_{r=1}^{N}}
p(X_{r}|\theta)~,
\end{equation}
and the posterior distribution $p_{\text{new}}(\theta)$ is
\begin{equation}
p(\theta|X)=\frac{p(\theta)}{p(X)}p(X|\theta)=\frac{p(\theta)}{p(X)}%
{\textstyle\prod\limits_{r=1}^{N}}
p(X_{r}|\theta)~.
\end{equation}

To investigate the extent to which the data $X$ supports the particular value
$\theta_{1}$ rather than any other value $\theta_{2}$ it is convenient to
study the ratio
\begin{equation}
\frac{p(\theta_{1}|X)}{p(\theta_{2}|X)}=\frac{p(\theta_{1})}{p(\theta_{2}%
)}R(X)~,
\end{equation}
where we introduce the likelihood ratios,
\begin{equation}
R(X)\overset{\operatorname*{def}}{=}%
{\textstyle\prod\limits_{r=1}^{N}}
R_{r}(X_{r})\quad\text{and}\quad R_{r}(X_{r})\overset{\operatorname*{def}}%
{=}\frac{p(X_{r}|\theta_{1})}{p(X_{r}|\theta_{2})}~.
\end{equation}
We want to prove the following theorem: Barring two trivial exceptions, for
any arbitrarily large positive $\Lambda$, we have
\begin{equation}
\lim_{N\rightarrow\infty}P\left(  R(X)>\Lambda|\theta_{1}\right)  =1
\end{equation}
or, in other words,
\begin{equation}
\text{given }\theta_{1}\text{,\quad}R(X)\longrightarrow\infty\quad
\text{\emph{in probability.}}%
\end{equation}

The significance of the theorem is that as data accumulates a rational person
becomes more and more convinced of the truth -- in this case the true value is
$\theta_{1}$ -- and this happens essentially irrespective of the prior
$p(\theta)$.

The theorem fails in two cases: first, when the prior $p(\theta_{1})$
vanishes, in which case probabilities conditional on $\theta_{1}$ are
meaningless, and second, when $p(X_{r}|\theta_{1})=p(X_{r}|\theta_{2})$ for
all $X_{r}$ which describes an experiment $E$ that is flawed because it cannot
distinguish between $\theta_{1}$ and $\theta_{2}$.

The proof of the theorem is an application of the weak law of large numbers.
Consider the quantity
\begin{equation}
\frac{1}{N}\log R(X)=\frac{1}{N}%
{\textstyle\sum\limits_{r=1}^{N}}
\log R_{r}(X_{r})
\end{equation}
Since the variables $\log R_{r}(X_{r})$ are independent, eq.(\ref{LLN d})
gives
\begin{equation}
\lim_{N\rightarrow\infty}\,P\left(  \left\vert \frac{1}{N}\log R(X)-K(\theta
_{1},\theta_{2})\right\vert \leq\varepsilon|\theta_{1}\right)  =1
\end{equation}
where $\varepsilon$ is any small positive number and
\begin{align}
K(\theta_{1},\theta_{2})  & =\left\langle \frac{1}{N}\log R(X)|\theta
_{1}\right\rangle \nonumber\\
& =%
{\textstyle\sum\limits_{X_{r}}}
p(X_{r}|\theta_{1})\log R_{r}(X_{r})~.
\end{align}
In other words,
\begin{equation}
\text{given }\theta_{1}\text{,\quad}e^{N(K-\varepsilon)}\leq R(X)\leq
e^{N(K+\varepsilon)}\quad\text{\emph{in probability.}}%
\end{equation}
In Chapter 4 we will prove that $K(\theta_{1},\theta_{2})\geq0$ with equality
if and only if the two distributions $p(X_{r}|\theta_{1})$ and $p(X_{r}%
|\theta_{2})$ are identical, which is precisely the second of the two trivial
exceptions we explicitly avoid. Thus $K(\theta_{1},\theta_{2})>0$, and this
concludes the proof.

We see here the first appearance of a quantity,
\begin{equation}
K(\theta_{1},\theta_{2})=+%
{\textstyle\sum\limits_{X_{r}}}
p(X_{r}|\theta_{1})\log\frac{p(X_{r}|\theta_{1})}{p(X_{r}|\theta_{2})}~,
\end{equation}
that will prove to be central in later discussions. When multiplied by $-1$,
the quantity $-K(\theta_{1},\theta_{2})$ is called the \emph{relative
entropy},\footnote{Other names include relative information, directed
divergence, and Kullback-Leibler distance.} that is the entropy of
$p(X_{r}|\theta_{1})$ \emph{relative} to $p(X_{r}|\theta_{2})$.\emph{\ }It can
be interpreted as a measure of the extent that the distribution $p(X_{r}%
|\theta_{1})$ can be distinguished from $p(X_{r}|\theta_{2})$.\emph{\ }

\section{Examples from data analysis}

To illustrate the use of Bayes' theorem as a tool to process information when
the information is in the form of data we consider some elementary examples
from the field of data analysis. (For detailed treatments that are friendly to
physicists see e.g. [Sivia Skilling 06, Gregory 05].)

\subsection{Parameter estimation}

Suppose the probability for the quantity $x$ depends on certain parameters
$\theta$, $p=p(x|\theta)$. Although most of the discussion here can be carried
out for an arbitrary function $p$ it is best to be specific and focus on the
important case of a Gaussian distribution,
\begin{equation}
p(x|\mu,\sigma)=\frac{1}{\sqrt{2\pi\sigma^{2}}}\exp\left(  -\frac{(x-\mu)^{2}%
}{2\sigma^{2}}\right)  ~.\label{Gaussian PE}%
\end{equation}
The objective is to estimate the parameters $\theta=(\mu,\sigma)$ on the basis
of a set of data $X=\left\{  X_{1},\ldots X_{N}\right\}  $. We assume the
measurements are statistically independent of each other and use Bayes'
theorem to get
\begin{equation}
p(\mu,\sigma|X)=\frac{p(\mu,\sigma)}{p\left(  X\right)  }%
{\textstyle\prod\limits_{i=1}^{N}}
p(X_{i}|\mu,\sigma)~.\label{Bayes GPE 1}%
\end{equation}
Independence is important in practice because it leads to considerable
practical simplifications but it is not essential: instead of $N$ independent
measurements each providing a single datum we would have a single complex
experiment that provides $N$ non-independent data.

Looking at eq.(\ref{Bayes GPE 1}) we see that a more precise formulation of
the same problem is the following. We want to estimate certain parameters
$\theta$, in our case $\mu$ and $\sigma$, from repeated measurements of the
quantity $x$ on the basis of \emph{several} pieces of information. The most
obvious is

\begin{enumerate}
\item The information contained in the actual values of the collected data $X
$.
\end{enumerate}

\noindent Almost equally obvious (at least to those who are comfortable with
the Bayesian interpretation of probabilities) is

\begin{enumerate}
\item[2.] The information about the parameters that is codified into the prior
distribution $p(\theta)$.
\end{enumerate}

\noindent Where and how this prior information was obtained is not relevant at
this point; it could have resulted from previous experiments, or from other
background knowledge about the problem. The only relevant part is whatever
ended up being distilled into $p(\theta)$.

The last piece of information is not always explicitly recognized; it is

\begin{enumerate}
\item[3.] The information that is codified into the functional form of the
`sampling' distribution $p(X|\theta)$.
\end{enumerate}

\noindent If we are to estimate parameters $\theta$ on the basis of
measurements of a quantity $x$ it is clear that we must know how $\theta$ and
$x$ are related to each other. Notice that item 3 refers to the
\emph{functional form} -- whether the distribution is Gaussian as opposed to
Poisson or binomial or something else -- and not to the actual values of the
data $X$ which is what is taken into account in item 1. The nature of the
relation in $p(X|\theta)$ is in general statistical but it could also be
completely deterministic. For example, when $X$ is a known function of
$\theta$, say $X=f(\theta)$, we have $p(X|\theta)=\delta\left[  X-f(\theta
)\right]  $. In this latter case there is no need for Bayes' rule.

Eq. (\ref{Bayes GPE 1}) is rewritten as
\begin{equation}
p(\mu,\sigma|X)=\frac{p(\mu,\sigma)}{p\left(  X\right)  }\frac{1}{\left(
2\pi\sigma^{2}\right)  ^{N/2}}\exp\left[  -%
{\textstyle\sum\limits_{i=1}^{N}}
\frac{\left(  X_{i}-\mu\right)  ^{2}}{2\sigma^{2}}\right] \label{Bayes GPE 2}%
\end{equation}
Introducing the sample average $\bar{X}$ and sample variance $s^{2}$,
\begin{equation}
\bar{X}=\frac{1}{N}%
{\textstyle\sum\limits_{i=1}^{N}}
X_{i}\quad\text{and}\quad s^{2}=\frac{1}{N}%
{\textstyle\sum\limits_{i=1}^{N}}
\left(  X_{i}-\bar{X}\right)  ^{2}~,\label{sample avgs}%
\end{equation}
eq.(\ref{Bayes GPE 2}) becomes
\begin{equation}
p(\mu,\sigma|X)=\frac{p(\mu,\sigma)}{p\left(  X\right)  }\frac{1}{\left(
2\pi\sigma^{2}\right)  ^{N/2}}\exp\left[  -\frac{\left(  \mu-\bar{X}\right)
^{2}+s^{2}}{2\sigma^{2}/N}\right]  ~.\label{Bayes GPE 3}%
\end{equation}
It is interesting that the data appears here only in the particular
combination in eq.(\ref{sample avgs}) -- different sets of data characterized
by the same $\bar{X}$ and $s^{2}$ lead to the same inference about $\mu$ and
$\sigma$. (As discussed earlier the factor $p\left(  X\right)  $ is not
relevant here since it can be absorbed into the normalization of the posterior
$p(\mu,\sigma|X)$.)

Eq. (\ref{Bayes GPE 3}) incorporates the information described in items 1 and
3 above. The prior distribution, item 2, remains to be specified. Let us start
by considering the simple case where the value of $\sigma$ is actually known.
Then $p(\mu,\sigma)=p(\mu)\delta(\sigma-\sigma_{0})$ and the goal is to
estimate $\mu$. Bayes' theorem is now written as
\begin{align}
p(\mu|X)  & =\frac{p(\mu)}{p\left(  X\right)  }\frac{1}{\left(  2\pi\sigma
_{0}^{2}\right)  ^{N/2}}\exp\left[  -%
{\textstyle\sum\limits_{i=1}^{N}}
\frac{\left(  X_{i}-\mu\right)  ^{2}}{2\sigma_{0}^{2}}\right]
\label{Bayes GPE 4}\\
& =~\frac{p(\mu)}{p\left(  X\right)  }\frac{1}{\left(  2\pi\sigma_{0}%
^{2}\right)  ^{N/2}}\exp\left[  -\frac{\left(  \mu-\bar{X}\right)  ^{2}+s^{2}%
}{2\sigma_{0}^{2}/N}\right] \nonumber\\
& \propto p(\mu)\exp\left[  -\frac{\left(  \mu-\bar{X}\right)  ^{2}}%
{2\sigma_{0}^{2}/N}\right]  .\label{Bayes GPE 4b}%
\end{align}
Suppose further that we know nothing about $\mu$; it could have any value.
This state of extreme ignorance is represented by a very broad distribution
that we take as essentially uniform within some large range; $\mu$ is just as
likely to have one value as another. For $p(\mu)\sim\operatorname*{const}$ the
posterior distribution is Gaussian, with mean given by the sample average
$\bar{x}$, and variance $\sigma_{0}^{2}/N.$ The best estimate for the value of
$\mu$ is the sample average and the uncertainty is the standard deviation.
This is usually expressed in the form
\begin{equation}
\mu=\bar{X}\pm\frac{\sigma_{0}}{\sqrt{N}}~.
\end{equation}
Note that the estimate of $\mu$ from $N$ measurements has a much smaller error
than the estimate from just one measurement; the individual measurements are
plagued with errors but they tend to cancel out in the sample average.

In the case of very little prior information -- the uniform prior -- we have
recovered the same results as in the standard non-Bayesian data analysis
approach. The real difference arises when prior information is available: the
non-Bayesian approach can't deal with it and can only proceed by ignoring it.
On the other hand, within the Bayesian approach prior information is easily
taken into account. For example, if we know on the basis of other physical
considerations that $\mu$ has to be positive we assign $p(\mu)=0$ for $\mu<0$
and we calculate the estimate of $\mu$ from the truncated Gaussian in
eq.(\ref{Bayes GPE 4b}).

A slightly more complicated case arises when the value of $\sigma$ is not
known. Let us assume again that our ignorance of both $\mu$ and $\sigma$ is
quite extreme and choose a uniform prior,
\begin{equation}
p(\mu,\sigma)\propto\left\{
\begin{array}
[c]{ccc}%
C & \text{for} & \sigma>0\\
0 & \  & \text{otherwise.}%
\end{array}
\right.
\end{equation}
Another popular choice is a prior that is uniform in $\mu$ and in $\log\sigma
$. When there is a considerable amount of data the two choices lead to
practically the same conclusions but we see that there is an important
question here: what do we mean by the word `uniform'? Uniform in terms of
which variable? $\sigma$, or $\sigma^{2}$, or $\log\sigma\,$? Later we shall
have much more to say about this misleadingly innocuous question.

To estimate $\mu$ we return to eq.(\ref{Bayes GPE 2}) or (\ref{Bayes GPE 3}).
For the purpose of estimating $\mu$ the variable $\sigma$ is an uninteresting
nuisance which, as discussed in section 2.5.4, is eliminated through
marginalization,
\begin{align}
p(\mu|X)  & =%
{\textstyle\int\limits_{0}^{\infty}}
d\sigma\,p(\mu,\sigma|X)\\
& \propto%
{\textstyle\int\limits_{0}^{\infty}}
d\sigma\,\frac{1}{\sigma^{N}}\exp\left[  -\frac{\left(  \mu-\bar{X}\right)
^{2}+s^{2}}{2\sigma^{2}/N}\right]  ~.
\end{align}
Change variables to $t=1/\sigma$, then
\begin{equation}
p(\mu|X)\propto%
{\textstyle\int\limits_{0}^{\infty}}
dt\,t^{N-2}\exp\left[  -\frac{t^{2}}{2}N\left(  \left(  \mu-\bar{X}\right)
^{2}+s^{2}\right)  \right]  ~.
\end{equation}
Repeated integrations by parts lead to
\begin{equation}
p(\mu|X)\propto\left[  N\left(  \left(  \mu-\bar{X}\right)  ^{2}+s^{2}\right)
\right]  ^{-\frac{N-1}{2}}~,\label{Student t}%
\end{equation}
which is called the \emph{Student-t} distribution. Since the distribution is
symmetric the estimate for $\mu$ is easy to get,
\begin{equation}
\left\langle \mu\right\rangle =\bar{X}~.
\end{equation}
The posterior $p(\mu|X)$ is a Lorentzian-like function raised to some power.
As the number of data grows, say $N\gtrsim10$, the tails of the distribution
are suppressed and $p(\mu|X)$ approaches a Gaussian. To obtain an error bar in
the estimate $\mu=\bar{X}$ we can estimate the variance of $\mu$ using the
following trick. Note that for the Gaussian in eq.(\ref{Gaussian PE}),
\begin{equation}
\left.  \frac{d^{2}}{dx^{2}}\log\,p(x|\mu,\sigma)\right\vert _{x_{\max}%
}=-\frac{1}{\sigma^{2}}~.
\end{equation}
Therefore, to the extent that eq.(\ref{Student t}) approximates a Gaussian, we
can write
\begin{equation}
\left(  \Delta\mu\right)  ^{2}\approx\left[  -\left.  \frac{d^{2}}{d\mu^{2}%
}\log\,p(\mu|X)\right\vert _{\mu_{\max}}\right]  ^{-1}=\frac{s^{2}}{N-1}~.
\end{equation}
(This explains the famous factor of $N-1$. As we can see it is not a
particularly fundamental result; it follows from approximations that are
meaningful only for large $N$.)

We can also estimate $\sigma$ directly from the data. This requires that we
marginalize over $\mu$,
\begin{align}
p(\sigma|X)  & =%
{\textstyle\int\limits_{-\infty}^{\infty}}
d\mu\,p(\mu,\sigma|X)\\
& \propto\frac{1}{\sigma^{N}}\exp\left[  -\frac{Ns^{2}}{2\sigma^{2}}\right]
{\textstyle\int\limits_{-\infty}^{\infty}}
d\mu\,\exp\left[  -\frac{\left(  \mu-\bar{X}\right)  ^{2}}{2\sigma^{2}%
/N}\right]  ~.
\end{align}
The Gaussian integral over $\mu$ is $\left(  2\pi\sigma^{2}/N\right)
^{1/2}\propto\sigma$ and therefore
\begin{equation}
p(\sigma|X)\propto\frac{1}{\sigma^{N-1}}\exp\left[  -\frac{Ns^{2}}{2\sigma
^{2}}\right]  ~.
\end{equation}
As an estimate for $\sigma$ we can use the value where the distribution is
maximized,
\begin{equation}
\sigma_{\max}=\sqrt{\frac{N}{N-1}s^{2}}~,
\end{equation}
which agrees with our previous estimate of $\left(  \Delta\mu\right)  ^{2}$,
\begin{equation}
\frac{\sigma_{\max}^{2}}{N}=\frac{s^{2}}{N-1}~.
\end{equation}
An error bar for $\sigma$ itself can be obtained using the previous trick
(provided $N$ is large enough) of taking a second derivative of $\log\,p.$ The
result is
\begin{equation}
\sigma=\sigma_{\max}\pm\frac{\sigma_{\max}}{\sqrt{2\left(  N-1\right)  }}~.
\end{equation}

\subsection{Curve fitting}

The problem of fitting a curve to a set of data points is a problem of
parameter estimation. There are no new issues of principle to be resolved. In
practice, however, it can be considerably more complicated than the simple
cases discussed in the previous paragraphs.

The problem is as follows. The observed data is in the form of pairs $\left(
X_{i},Y_{i}\right)  $ with $i=1,\ldots N$ and we believe that the true $y$s
are related to the $X$s through a function $y_{i}=f_{\theta}(x_{i})$ which
depends on several parameters $\theta$. The goal is to estimate the parameters
$\theta$ and the complication is that the measured values of $y$ are afflicted
by experimental errors,
\begin{equation}
Y_{i}=f_{\theta}(X_{i})+\varepsilon_{i}~.
\end{equation}
For simplicity we assume that the probability of the error $\varepsilon_{i}$
is Gaussian with mean $\left\langle \varepsilon_{i}\right\rangle =0$ and that
the variances $\left\langle \varepsilon_{i}^{2}\right\rangle =\sigma^{2}$ are
known and the same for all data pairs. We also assume that there are no errors
affecting the $X$s. A more realistic account might have to reconsider these assumptions.

The sampling distribution is
\begin{equation}
p(Y|\theta)=%
{\textstyle\prod\limits_{i=1}^{N}}
~p(Y_{i}|\theta)~,
\end{equation}
where
\begin{equation}
p(Y_{i}|\theta)=\frac{1}{\sqrt{2\pi\sigma^{2}}}\exp\left(  -\frac
{(Y_{i}-f_{\theta}(X_{i}))^{2}}{2\sigma^{2}}\right)  ~.
\end{equation}
Bayes' theorem gives,
\begin{equation}
p(\theta|Y)\propto p(\theta)\exp\left[  -%
{\textstyle\sum\limits_{i=1}^{N}}
\frac{(Y_{i}-f_{\theta}(X_{i}))^{2}}{2\sigma^{2}}\right]  ~.
\end{equation}

As an example, suppose that we are trying to fit a straight line through data
points
\begin{equation}
f(x)=a+bx~,
\end{equation}
and suppose further that being ignorant about the values of $\theta=(a,b)$ we
choose $p(\theta)=p(a,b)\sim\operatorname*{const}$, then
\begin{equation}
p(a,b|Y)\propto\exp\left[  -%
{\textstyle\sum\limits_{i=1}^{N}}
\frac{(Y_{i}-a-bX_{i})^{2}}{2\sigma^{2}}\right]  ~.
\end{equation}
A good estimate of $a$ and $b$ is the value that maximizes the posterior
distribution, which as we see, is equivalent to using the method of least
squares. But this Bayesian analysis, simple as it is, can already give us
more: from $p(a,b|Y)$ we can also estimate the uncertainties $\Delta a$ and
$\Delta b$ which lies beyond the scope of least squares.

\subsection{Model selection}

Suppose we are trying to fit a curve $y=f_{\theta}(x)$ through data points
$\left(  X_{i},Y_{i}\right)  $, $i=1,\ldots N$. How do we choose the function
$f_{\theta}$? To be specific let $f_{\theta}$ be a polynomial of order $n$,
\begin{equation}
f_{\theta}(x)=\theta_{0}+\theta_{1}x+\ldots+\,\theta_{n}x^{n}~,
\end{equation}
the techniques of the previous section allow us to estimate the parameters
$\theta_{0},\ldots,\,\theta_{n}$ but how do we decide the order $n$? Should we
fit a straight or a quadratic line? It is not obvious. Having more parameters
means that we will be able to achieve a closer fit to the data, which is good,
but we might also be fitting the noise, which is bad. The same problem arises
when the data shows peaks and we want to estimate their location, their width,
and \emph{their number}; could there be an additional peak hiding in the
noise? Are we just fitting noise, or does the data really support one
additional peak?

We say these are `problems of model selection'. To appreciate how important
they can be consider replacing the modestly unassuming word `model' by the
more impressive sounding word `theory'. Given two competing theories, which
one does the data support best? What is at stake is nothing less than the
foundation of experimental science.

On the basis of data $X$ we want to select one model among several competing
candidates labeled by $m=1,2,\ldots$ Suppose model $m$ is defined in terms of
some parameters $\theta_{m}=\left\{  \theta_{m1},\theta_{m2},\ldots\right\}  $
and their relation to the data $X$ is contained in the sampling distribution
$p(X|m,\theta_{m})$. The extent to which the data supports model $m$,
\emph{i.e.}, the probability of model $m$ given the data, is given by Bayes'
theorem,
\begin{equation}
p(m|X)=\frac{p(m)}{p(X)}p(X|m)~,
\end{equation}
where $p(m)$ is the prior for the model. The factor $p(X|m)$, which is the
prior probability for the data given the model, plays the role of a
likelihood. It is often called the `evidence'. This is not altogether
appropriate because the meaning of $p(X|m)$ is already given as
\textquotedblleft the prior probability of the data.\textquotedblright\ There
is nothing more to be said about it. Calling it the `evidence' can only
mislead us by suggesting interpretations and therefore uses that go beyond and
could conceivably be in conflict with its probability meaning.\footnote{A
similar problem occurs when $\left\langle x\right\rangle $ is called the
\textquotedblleft expected\textquotedblright\ value. It misleads us into
thinking that $\left\langle x\right\rangle $ is the value we should expect,
which is not necessarily true.} After this warning, we follow standard
practice. The \textquotedblleft evidence\textquotedblright\ is calculated
from
\begin{equation}
p(X|m)=\int d\theta_{m}~p(X,\theta_{m}|m)=\int d\theta_{m}~p(\theta
_{m}|m)~p(X|m,\theta_{m})~.
\end{equation}
Therefore
\begin{equation}
p(m|X)\propto p(m)\int d\theta_{m}~p(\theta_{m}|m)p(X|m,\theta_{m}%
)~.\label{model selection}%
\end{equation}
Thus, the problem is solved, at least in principle, once the priors $p(m)$ and
$p(\theta_{m}|m)$ are assigned. Of course, the practical problem of
calculating the multi-dimensional integrals can still be quite formidable.

No further progress is possible without making specific choices for the
various functions in eq.(\ref{model selection}) but we can offer some
qualitative comments. When comparing two models, $m_{1}$ and $m_{2}$, it is
fairly common to argue that a priori we have no reason to prefer one over the
other and therefore we assign the same prior probability $p(m_{1})=p(m_{2})$.
(Of course this is not always justified. Particularly in the case of theories
that claim to be fundamental people usually have very strong prior prejudices
favoring one theory against the other. Be that as it may, let us proceed.)

Suppose the prior $p(\theta_{m}|m)$ represents a uniform distribution over the
parameter space. Since
\begin{equation}
\int d\theta_{m}~p(\theta_{m}|m)=1\quad\text{then}\quad p(\theta_{m}%
|m)\approx\frac{1}{V_{m}}~,
\end{equation}
where $V_{m}$ is the `volume' of the parameter space. Suppose further that
$p(X|m,\theta_{m})$ has a single peak of height $L_{\text{max}}$ spread out
over a region of `volume' $\delta\theta_{m}$. The value $\theta_{m}$ where
$p(X|m,\theta_{m})$ attains its maximum can be used as an estimate for
$\theta_{m}$ and the `volume' $\delta\theta_{m}$ is then interpreted as an
uncertainty. Then the integral of $p(X|m,\theta_{m})$ can be approximated by
the product $L_{\text{max}}\times\delta\theta_{m}$. Thus, in a very rough and
qualitative way the probability for the model given the data is
\begin{equation}
p(m|X)\propto\frac{L_{\text{max}}\times\delta\theta_{m}}{V_{m}}~.
\end{equation}
We can now interpret eq.(\ref{model selection}) as follows. Our preference for
a model will be dictated by how well the model fits the data; this is measured
by $\left[  p(X|m,\theta_{m})\right]  _{\text{max}}=L_{\text{max}}$. The
volume of the region of uncertainty $\delta\theta_{m}$ also contributes: if
more values of the parameters are consistent with the data, then there are
more ways the model agrees with the data, and the model is favored. Finally,
the larger the volume of possible parameter values $V_{m}$ the more the model
is penalized. Since a larger volume $V_{m}$ means a more complex model the
$1/V_{m}$ factor penalizes complexity. The preference for simpler models is
said to implement Occam's razor. This is a reference to the principle, stated
by William of Occam, a 13th century Franciscan monk, that one should not seek
a more complicated explanation when a simpler one will do. Such an
interpretation is satisfying but ultimately it is quite unnecessary. Occam's
principle does not need not be put in by hand: Bayes' theorem takes care of it
automatically in eq.(\ref{model selection})!

\subsection{Maximum Likelihood}

If one adopts the frequency interpretation of probabilities then most uses of
Bayes' theorem are not allowed. The reason is simple: it makes sense to assign
a probability distribution $p(x|\theta)$ to the data $X=\{X_{i}\}$ because the
$x$ are random variables but it is absolutely meaningless to talk about
probabilities for the parameters $\theta$ because they have no frequency
distributions, they are not \emph{random} variables, they are merely
\emph{unknown}. This means that many problems in science lie beyond the reach
of a frequentist probability theory.

To overcome this difficulty a new subject was invented: statistics. Within the
Bayesian approach the two subjects, statistics and probability theory, are
unified into the single field of inductive inference. In the frequentist
approach to statistics in order to infer an unknown quantity $\theta$ on the
basis of measurements of another quantity, the data $x$, one postulates the
existence of some function, called the `statistic', that relates the two,
$\theta=f(x)$. Since data are afflicted by experimental errors they are deemed
to be legitimate random variables to which frequentist probability concepts
can be applied. The problem is to estimate the unknown $\theta$ when the
sampling distribution $p(x|\theta)$ is known. The solution proposed by Fisher
was to select that value of $\theta$ that maximizes the probability of the
data that was actually obtained in the experiment. Since $p(x|\theta)$ is a
function of the variable $x$ and $\theta$ appears as a fixed parameter, Fisher
introduced a function of $\theta$, which he called the likelihood, where the
observed data $X$ appear as fixed parameters,
\begin{equation}
L\left(  \theta|X\right)  \overset{\operatorname*{def}}{=}p(X|\theta)~.
\end{equation}
Thus, this method of parameter estimation is called the method of `maximum
likelihood'. The likelihood function $L(\theta|X)$ is not a probability, it is
not normalized in any way, and it makes no sense to use it compute an average
or a variance, but the same intuition that leads one to propose maximization
of the likelihood to estimate $\theta$ also leads one to use the width of the
likelihood function as to estimate an error bar.

The Bayesian approach agrees with the method of maximum likelihood in the
special case where of prior is uniform,
\begin{equation}
p(\theta)=\operatorname*{const}\Rightarrow p(\theta|X)\propto p(\theta
)p(X|\theta)\propto p(X|\theta)~.
\end{equation}
This explains why the Bayesian discussion of this section has reproduced so
many of the standard results of the `orthodox' theory. But then there are
additional advantages. Unlike the likelihood, the posterior is a true
probability distribution that allows estimation not just of $\theta$ but of
any one of its moments. And, most important, there is no limitation to uniform
priors. If there is additional prior information that is relevant to a problem
the prior distribution provides a mechanism to take it into account.

\chapter[Entropy I: Carnot's Principle]{Entropy I: The Evolution of Carnot's
Principle}

An important problem that occupied the minds of many scientists in the 18th
century was either to devise a perpetual motion machine, or to prove its
impossibility from the established principles of mechanics. Both attempts
failed. Ever since the most rudimentary understanding of the laws of
thermodynamics was achieved in the 19th century no competent scientist would
waste time considering perpetual motion.\footnote{The science of
thermodynamics which led to statistical mechanics and eventually to
information theory was initially motivated by the desire to improve steam
engines. There seems to exist a curious historical parallel with the modern
day development of quantum information theory, which is being driven by the
desire to build quantum computers. The usefulness of thermodynamics far
outgrew its original aim. It is conceivable that the same will happen to
quantum information theory.} The other goal has also proved elusive; there
exist no derivations the Second Law from purely mechanical principles. It took
a long time, and for many the subject is still controversial, but the reason
has gradually become clear: entropy is not a physical quantity, it is a tool
for inference, a tool for reasoning in situations of incomplete information.
It is quite impossible that such a non-mechanical quantity could emerge from a
combination of mechanical notions. If anything it should be the other way around.

Much of the material including the title for this chapter is inspired by a
beautiful article by E. T. Jaynes [Jaynes 88]. I also borrowed from the
historical papers [Klein 70, 73, Uffink 04].

\section{Carnot: reversible engines}

Sadi Carnot was interested in improving the efficiency of steam engines, that
is, of maximizing the amount of useful work that can be extracted from an
engine per unit of burnt fuel. His work, published in 1824, was concerned with
whether appropriate choices of a working substance other than steam and of the
operating temperatures and pressures would improve the efficiency.

Carnot was quite convinced that perpetual motion was impossible even though he
had no proof. He could not have had a proof: thermodynamics had not been
invented yet. His conviction derived from the long list of previous attempts
that had ended in failure. His brilliant idea was to proceed anyway and to
\emph{postulate} what he knew was true but could not prove as the foundation
from which he would draw all sorts of other conclusions about
engines.\footnote{In his attempt to understand the undetectability of the
ether Einstein faced a similar problem: he knew that it was hopeless to seek
an understanding of the constancy of the speed of light on the basis of the
primitive physics of the atomic structure of solid rods that was available at
the time. Inspired by Carnot he deliberately followed the same strategy -- to
give up and declare victory -- and postulated the constancy of the speed of
light as the unproven but known truth which would serve as the foundation from
which other conclusions could be derived.
\par
{}}

At the time Carnot did his work the nature of heat as a form of energy had not
yet been understood. He adopted a model that was fashionable at the time --
the caloric model -- according to which heat is a substance that could be
transferred but neither created nor destroyed. For Carnot an engine used heat
to produce work in much the same way that falling water can turn a waterwheel
and produce work: the caloric would \textquotedblleft fall\textquotedblright%
\ from a higher temperature to a lower temperature thereby making the engine
turn. What was being transformed into work was not the caloric itself but the
energy acquired in the fall.

According to the caloric model the amount of heat extracted from the high
temperature source should be the same as the amount of heat discarded into the
low temperature sink. Later measurements showed that this was not true, but
Carnot was quite lucky. Although the model was seriously wrong, it did have a
great virtue: it suggested that the generation of work in a heat engine should
include not just the high temperature source from which heat is extracted (the
boiler) but also a low temperature sink (the condenser) into which heat is
discarded. Later, when heat was interpreted as a form of energy transfer it
was understood that for continued operation it was necessary that excess heat
be discarded into a low temperature sink so that the engine could complete
each cycle by returning to same initial state.

Carnot's caloric-waterwheel model was fortunate in yet another respect -- he
was not just lucky, he was very lucky -- a waterwheel engine can be operated
in reverse and used as a pump. This led him to consider a reversible heat
engine in which work would be used to draw heat from a cold source and `pump
it up' to deliver heat to the hot reservoir. The analysis of such reversible
heat engines led Carnot to the important conclusion

\noindent\textbf{Carnot's Principle:} \noindent\textquotedblleft\emph{No heat
engine }$E$ \emph{can be more efficient than a reversible one }$E_{R}%
$\emph{\ operating between the same temperatures}.\textquotedblright

The proof of Carnot's principle is quite straightforward but because he used
the caloric model Carnot's proof was not strictly correct -- the necessary
revisions were supplied by Clausius in 1850. As a side remark, it is
interesting that Carnot's notebooks, which were made public by his family
about 1870, long after his death, indicate that soon after 1824 Carnot came to
reject the caloric model and that he achieved the modern understanding of heat
as a form of energy transfer. This work -- which had preceded Joule's
experiments by about fifteen years -- was not published and therefore had no
influence on the history of thermodynamics [Wilson 81].

The following is Clausius' proof. In a standard cycle (Figure 3.1a) a heat engine $E$
extracts heat $q_{1}$ from a reservoir at high temperature $t_{1}$ and
partially converts it to useful work $w$. The difference $q_{1}-w=q_{2}$ is
wasted heat that is dumped into a reservoir at a lower temperature $t_{2}$.
The Carnot-Clausius argument is that if an engine $E_{S}$ exists that is more
efficient than a reversible engine $E_{R}$, then it is possible to build
perpetual motion machines. Since the latter do not exist Carnot's principle
follows: heat engines that are more efficient than reversible ones do not exist.

\begin{center}%
\vspace{0.15in}
\begin{figure}
[tbh]
\begin{center}
\includegraphics[
trim=1.203976in 0.456726in 0.000000in 0.764209in,
natheight=7.499600in,
natwidth=9.999800in,
height=3.1687in,
width=4.4261in
]%
{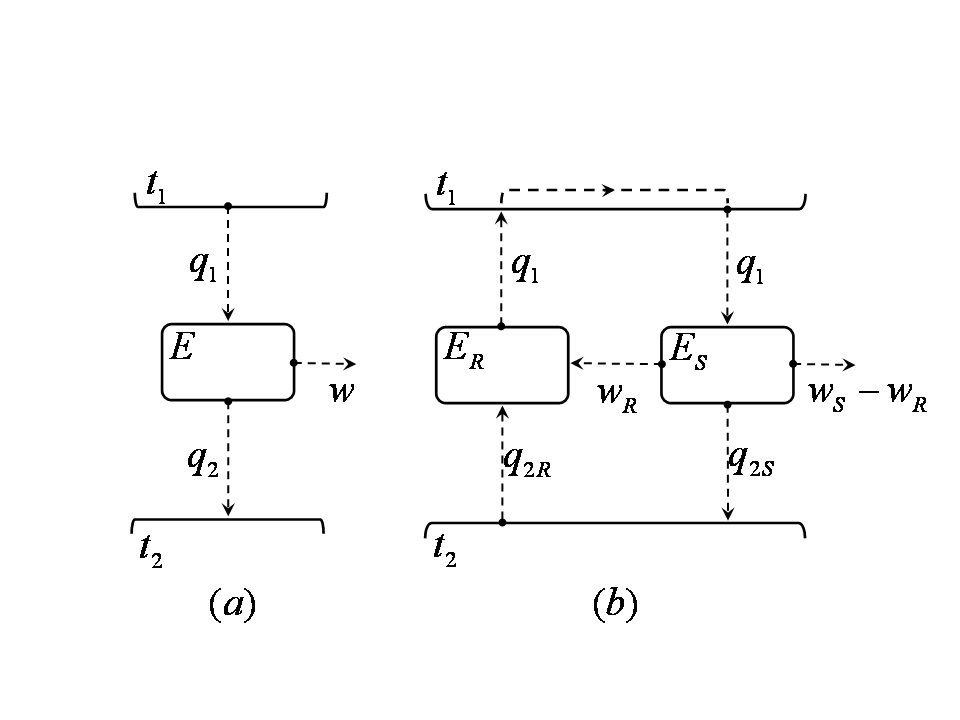}%
\caption{(a) A regular engine $E$ operating between heat reservoirs at
temperatures $t_{1}$ and $t_{2}$ generates work $w=q_{1}-q_{2}$. (b) A
(hypothetical) super-efficient engine $E_{S}$ linked to a reversed engine
$E_{R}$ would be a perpetual motion engine extracting heat from the cold
reservoir and converting it to work $w_{S}-w_{R}=q_{2R}-q_{2S}$. }%
\end{center}
\end{figure}

\end{center}

Consider two engines, one is super-efficient and the other is reversible,
$E_{S}$ and $E_{R}$, operating between the same hot and cold reservoirs. The
engine $E_{S}$ draws heat $q_{1}$ from the hot source, it generates work
$w_{S}$, and delivers the difference as heat $q_{2S}=q_{1}-w_{S}$ to the cold
sink (figure 3.1b). It is arranged that in its normal (forward) operation the reversible
engine $E_{R}$ draws the same heat $q_{1}$ from the hot source, it generates
work $w_{R}$, and discards the difference $q_{2R}=q_{1}-w_{R}$ to the cold
sink. Since $E_{S}$ is supposed to be more efficient than $E_{R}$ we have
$w_{S}>w_{R}$, it would be possible to use a part $w_{R}$ of the work produced
by $E_{S}$ to run $E_{R}$ in reverse. The result would be to extract heat
$q_{2R}$ from the cold source and pump the total heat $q_{2R}+w_{R}=q_{1}$
back up into the hot source. The remaining work $w_{S}-w_{R}$ produced by
$E_{S}$ would then be available for any other purposes. At the end of such
composite cycle the hot reservoir is left unchanged and the net result would
be to extract heat $q_{2R}-q_{2}>0$ from the cold reservoir and convert it to
work $w_{S}-w_{R}$ without any need for fuel. The conclusion is that the
existence of a super-efficient heat engine would allow the construction of a
perpetual motion engine.

The blank statement \emph{perpetual motion is not possible} is a true
principle but it does not tell the whole story. It blurs the important distinction between
perpetual motion engines that operate by violating energy conservation, which
are called machines of the \emph{first kind}, and perpetual motion engines
that do not violate energy conservation, which are thus called machines of the
\emph{second kind}. Carnot's conclusion deserves to be singled out as a new
principle because it is specific to the second kind of machine.

Other important conclusions obtained by Carnot are that all reversible engines
operating between the same temperatures are equally efficient; their
efficiency is a function of the temperatures only,
\begin{equation}
e\overset{\text{def}}{=}\frac{w}{q_{1}}=e(t_{1},t_{2})~,
\end{equation}
and is therefore independent of any and all other details of how the engine is
constructed and operated; that efficiency increases with the temperature
difference [see eq.(\ref{efficiency 2}) below]. Furthermore, the most
efficient heat engine cycle, now called the Carnot cycle, is one in which all
heat is absorbed at the high $t_{1}$ and all heat is discharged at the low
$t_{2}$. Thus, the Carnot cycle is defined by two isotherms and two adiabats.

The next important step, the determination of the universal function
$e(t_{1},t_{2})$, was accomplished by Kelvin.

\section{Kelvin: temperature}

After Joule's experiments in the 1840's on the conversion of work into heat
the caloric model had to be abandoned. Heat was finally recognized as a form
of energy and the additional relation $w=q_{1}-q_{2}$ was the ingredient that,
in the hands of Kelvin and Clausius, allowed Carnot's principle to be
developed into the next stage.

Suppose two reversible engines $E_{a}$ and $E_{b}$ are linked in series to
form a single more complex reversible engine $E_{c}$. The first operates
between temperatures $t_{1}$ and $t_{2}$, and the second between $t_{2}$ and
$t_{3}$. $E_{a}$ draws heat $q_{1}$ and discharges $q_{2}$, while $E_{b}$ uses
$q_{2}$ as input and discharges $q_{3}$. The efficiencies of the three engines
are
\begin{equation}
e_{a}=e\left(  t_{1},t_{2}\right)  =\frac{w_{a}}{q_{1}}\,,\quad e_{b}=e\left(
t_{2},t_{3}\right)  =\frac{w_{b}}{q_{2}}~,\label{efficiency 1}%
\end{equation}
and%
\begin{equation}
e_{c}=e\left(  t_{1},t_{3}\right)  =\frac{w_{a}+w_{b}}{q_{1}}%
~.\label{efficiency 2}%
\end{equation}
They are related by
\begin{equation}
e_{c}=e_{a}+\frac{w_{b}}{q_{2}}\frac{q_{2}}{q_{1}}=e_{a}+e_{b}\left(
1-\frac{w_{a}}{q_{1}}\right)  ~,
\end{equation}
or%
\begin{equation}
e_{c}=e_{a}+e_{b}-e_{a}e_{b}~,\label{efficiency 2a}%
\end{equation}
which is a functional equation for $e=e\left(  t_{1},t_{2}\right)  $. To find
the solution change variables to $x=\log\left(  1-e\right)  $, which
transforms eq.(\ref{efficiency 2a}) into
\begin{equation}
x_{c}\left(  t_{1},t_{3}\right)  =x_{a}\left(  t_{1},t_{2}\right)
+x_{b}\left(  t_{2},t_{3}\right)  ~,
\end{equation}
and then differentiate with respect to $t_{2}$ to get
\begin{equation}
\frac{\partial}{\partial t_{2}}x_{a}\left(  t_{1},t_{2}\right)  =-\frac
{\partial}{\partial t_{2}}x_{b}\left(  t_{2},t_{3}\right)  ~.
\end{equation}
The left hand side is independent of $t_{3}$ while the second is independent
of $t_{1}$, therefore $\partial x_{a}/\partial t_{2}$ must be some function
$g$ of $t_{2}$ only,
\begin{equation}
\frac{\partial}{\partial t_{2}}x_{a}\left(  t_{1},t_{2}\right)  =g(t_{2})~.
\end{equation}
Integrating gives $x(t_{1},t_{2})=F(t_{1})+G(t_{2})$ where the two functions
$F$ and $G$ are at this point unknown. The boundary condition $e\left(
t,t\right)  =0$ or equivalently $x(t,t)=0$ implies that we deal with merely
one unknown function: $G(t)=-F(t)$. Therefore
\begin{equation}
x(t_{1},t_{2})=F(t_{1})-F(t_{2})\quad\text{or}\quad e\left(  t_{1}%
,t_{2}\right)  =1-\frac{f(t_{2})}{f(t_{1})}~,\label{efficiency 3}%
\end{equation}
where $f=e^{-F}$. From eq.(\ref{efficiency 2}) we see that the efficiency
$e\left(  t_{1},t_{2}\right)  $ increases as the difference in temperature
increases, so that $f\left(  t\right)  $ must be a monotonically increasing function.

Kelvin recognized that there is nothing fundamental about the original
temperature scale $t$. It depends on the particular materials employed to
construct the thermometer. Kelvin realized that the freedom in
eq.(\ref{efficiency 3}) in the choice of the function $f$ corresponds to the
freedom of changing temperature scales by using different thermometric
materials. The only feature common to all thermometers that claim to rank
systems according to their `degree of hotness' is that they must agree that if
$A$ is hotter than $B$, and $B$ is hotter than $C$, then $A$ is hotter than
$C$. One can therefore \emph{regraduate} any old inconvenient $t$ scale by a
monotonic function to obtain a new scale $T$ chosen purely because it leads to
a more elegant formulation of the theory. From eq.(\ref{efficiency 3}) the
optimal choice is quite obvious, and thus Kelvin introduced the absolute scale
of temperature,
\begin{equation}
T=Cf\left(  t\right)  ~,
\end{equation}
where the arbitrary scale factor $C$ reflects the still remaining freedom to
choose the units. In the absolute scale the efficiency for the ideal
reversible heat engine is very simple,
\begin{equation}
e\left(  t_{1},t_{2}\right)  =1-\frac{T_{2}}{T_{1}}\ .\label{efficiency 4}%
\end{equation}

Carnot's principle that any heat engine $E^{\prime}$ must be less efficient
than the reversible one, $e^{\prime}\leq e$, is rewritten as
\begin{equation}
e^{\prime}=\frac{w}{q_{1}}=1-\frac{q_{2}}{q_{1}}\leq e=1-\frac{T_{2}}{T_{1}}~,
\end{equation}
or,
\begin{equation}
\frac{q_{1}}{T_{1}}-\frac{q_{2}}{T_{2}}\leq0~.
\end{equation}
It is convenient to redefine heat so that inputs are positive, $Q_{1}=q_{1}$,
and outputs are negative, $Q_{2}=-q_{2}$. Then,
\begin{equation}
\frac{Q_{1}}{T_{1}}+\frac{Q_{2}}{T_{2}}\leq0~,
\end{equation}
where the equality holds when and only when the engine is reversible.

The generalization to an engine or any system that undergoes a cyclic process
in which heat is exchanged with more than two reservoirs is straightforward.
If heat $Q_{i}$ is absorbed from the reservoir at temperature $T_{i}$ we
obtain the Kelvin form (1854) of Carnot's principle,
\begin{equation}%
{\textstyle\sum\limits_{i}}
\frac{Q_{i}}{T_{i}}\leq0~.\label{2nd law Kelvin}%
\end{equation}
which, in the hands of Clausius, led to the next non-trivial step, the
introduction of the concept of entropy.

\section{Clausius: entropy}

By about 1850 both Kelvin and Clausius had realized that two laws were
necessary as a foundation for thermodynamics. The somewhat awkward expressions
for the second law that they had adopted at the time were reminiscent of
Carnot's; they stated the impossibility of heat engines whose sole effect
would be to transform heat from a single source into work, or of refrigerators
that could pump heat from a cold to a hot reservoir without the input of
external work. It took Clausius until 1865 -- this is some fifteen years
later, which indicates that the breakthrough was not at all trivial -- before
he came up with a new compact statement of the second law that allowed
substantial further progress. [Cropper 86]

Clausius rewrote Kelvin's eq.(\ref{2nd law Kelvin}) for a cycle where the
system absorbs infinitesimal (positive or negative) amounts of heat $dQ$ from
a continuous sequence of reservoirs,
\begin{equation}%
{\textstyle\oint}
\frac{dQ}{T}\leq0~,\label{2nd law Clausius a}%
\end{equation}
where $T$ is the temperature of each reservoir. For a reversible process,
which is achieved when the system is slowly taken through a sequence of
equilibrium states and $T$ is the temperature of the system as well as the
reservoirs, the equality sign implies that the integral from any state $A$ to
any other state $B$ is independent of the path taken,
\begin{equation}%
{\textstyle\oint}
\frac{dQ}{T}=0\Rightarrow%
{\textstyle\int\limits_{R_{1}(A,B)}}
\frac{dQ}{T}=%
{\textstyle\int\limits_{R_{2}(A,B)}}
\frac{dQ}{T}~,
\end{equation}
where $R_{1}(A,B)$ and $R_{2}(A,B)$ denote any two reversible paths linking
the same initial state $A$ and final state $B$. Clausius saw that this implied
the existence of a function of the thermodynamic state, which he called the
entropy, and defined up to an additive constant by%
\begin{equation}
S_{B}=S_{A}+%
{\textstyle\int\limits_{R(A,B)}}
\frac{dQ}{T}~.\label{Clausius entropy}%
\end{equation}
At this stage in the development this entropy is `thermodynamic entropy', and
is defined only for equilibrium states.

Eq.(\ref{Clausius entropy}) seems like a mere reformulation of eqs.(
\ref{2nd law Kelvin}) and (\ref{2nd law Clausius a}) but it represents a major
advance because it allowed thermodynamics to reach beyond the study of cyclic
processes. Consider a possibly irreversible process in which a system is taken
from an initial state $A$ to a final state $B$, and suppose the system is
returned to the initial state along some other reversible path. Then, the more
general eq.(\ref{2nd law Clausius a}) gives
\begin{equation}%
{\textstyle\int\limits_{A,\operatorname{irrev}}^{B}}
\frac{dQ}{T}+%
{\textstyle\int\limits_{R(A,B)}}
\frac{dQ}{T}\leq0~.
\end{equation}
From eq.(\ref{Clausius entropy}) the second integral is $S_{A}-S_{B}$. In the
first integral $-dQ$ is the amount is the amount of heat absorbed by the
reservoirs at temperature $T$ and therefore it represents minus the change in
the entropy of the reservoirs, which in this case represent the rest of the
universe,
\begin{equation}
\left(  S_{A}^{\operatorname{res}}-S_{B}^{\operatorname{res}}\right)  +\left(
S_{A}-S_{B}\right)  \leq0\quad\text{or}\quad S_{B}^{\operatorname{res}}%
+S_{B}\geq S_{A}^{\operatorname{res}}+S_{A}~.
\end{equation}
Thus the second law can be stated in terms of the total entropy
$S^{\text{total}}=S^{\operatorname{res}}+S$ as
\begin{equation}
S_{\text{final}}^{\text{total}}\geq S_{\text{initial}}^{\text{total}%
}~,\label{2nd law Clausius b}%
\end{equation}
and Clausius could then summarize the laws of thermodynamics as
\textquotedblleft\emph{The energy of the universe is constant. The entropy of
the universe tends to a maximum}.\textquotedblright\ All restrictions to
cyclic processes have disappeared.

Clausius was also responsible for initiating another independent line of
research in this subject. His paper \textquotedblleft On the kind of motion we
call heat\textquotedblright\ (1857) was the first (failed!) attempt to deduce
the second law from purely mechanical principles applied to molecules. His
results referred to averages taken over all molecules, for example the kinetic
energy per molecule, and involved theorems in mechanics such as the virial
theorem. For him the increase of entropy was meant to be an absolute law and
not just a matter of overwhelming probability.

\section{Maxwell: probability}

We owe to Maxwell the introduction of probabilistic notions into fundamental
physics (1860). (Perhaps he was inspired by his earlier study of the rings of
Saturn which required reasoning about particles undergoing very complex
trajectories.) He realized the impossibility of keeping track of the exact
motion of all the molecules in a gas and pursued a less detailed description
in terms of the distribution of velocities. Maxwell interpreted his
distribution function as the fraction of molecules with velocities in a
certain range, and also as the \textquotedblleft probability\textquotedblright%
\ $P(\vec{v})d^{3}v$ that a molecule has a velocity $\vec{v}$ in a certain
range $d^{3}v$. It would take a long time to achieve a clearer understanding
of the meaning of the term `probability'. In any case, Maxwell concluded that
\textquotedblleft velocities are distributed among the particles according to
the same law as the errors are distributed in the theory of the `method of
least squares',\textquotedblright\ and on the basis of this distribution he
obtained a number of significant results on the transport properties of gases.

Over the years he proposed several derivations of his velocity distribution
function. The earlier one (1860) is very elegant. It involves two assumptions:
the first is a symmetry requirement, the distribution should only depend on
the actual magnitude $\left\vert \vec{v}\right\vert =v$ of the velocity and
not on its direction,
\begin{equation}
P(v)d^{3}v=P\left(  \sqrt{v_{x}^{2}+v_{y}^{2}+v_{z}^{2}}\right)  d^{3}v~.
\end{equation}
The second assumption is that velocities along orthogonal directions should be
independent
\begin{equation}
P(v)d^{3}v=p(v_{x})p(v_{y})p(v_{z})d^{3}v~.
\end{equation}
Therefore
\begin{equation}
P\left(  \sqrt{v_{x}^{2}+v_{y}^{2}+v_{z}^{2}}\right)  =p(v_{x})p(v_{y}%
)p(v_{z})~.
\end{equation}
Setting $v_{y}=v_{z}=0$ we get
\begin{equation}
P\left(  v_{x}\right)  =p(v_{x})p(0)p(0)~,
\end{equation}
so that we obtain a functional equation for $p$,
\begin{equation}
p\left(  \sqrt{v_{x}^{2}+v_{y}^{2}+v_{z}^{2}}\right)  p(0)p(0)=p(v_{x}%
)p(v_{y})p(v_{z})~,
\end{equation}
or
\begin{equation}
\log\left[  \frac{p\left(  \sqrt{v_{x}^{2}+v_{y}^{2}+v_{z}^{2}}\right)
}{p(0)}\right]  =\log\left[  \frac{p(v_{x})}{p(0)}\right]  +\log\left[
\frac{p(v_{y})}{p(0)}\right]  +\log\left[  \frac{p(v_{z})}{p(0)}\right]  ~,
\end{equation}
or, introducing the functions $G$,
\begin{equation}
G\left(  \sqrt{v_{x}^{2}+v_{y}^{2}+v_{z}^{2}}\right)  =G(v_{x})+G(v_{y}%
)+G(v_{z}).
\end{equation}
The solution is straightforward. Differentiate with respect to $v_{x}$ and to
$v_{y}$ to get
\begin{equation}
\frac{G^{\prime}\left(  \sqrt{v_{x}^{2}+v_{y}^{2}+v_{z}^{2}}\right)  }%
{\sqrt{v_{x}^{2}+v_{y}^{2}+v_{z}^{2}}}v_{x}=G^{\prime}(v_{x})\quad
\text{and}\quad\frac{G^{\prime}\left(  \sqrt{v_{x}^{2}+v_{y}^{2}+v_{z}^{2}%
}\right)  }{\sqrt{v_{x}^{2}+v_{y}^{2}+v_{z}^{2}}}v_{x}=G^{\prime}(v_{x})~,
\end{equation}
or
\begin{equation}
\frac{G^{\prime}(v_{x})}{v_{x}}=\frac{G^{\prime}(v_{y})}{v_{y}}=-2\alpha~,
\end{equation}
where $-2\alpha$ is a constant. Integrating gives
\begin{equation}
\log\left[  \frac{p(v_{x})}{p(0)}\right]  =G(v_{x})=-\alpha v_{x}%
^{2}+\operatorname*{const}~,
\end{equation}
so that
\begin{equation}
P(v)=\left(  \frac{\alpha}{\pi}\right)  ^{3/2}\exp\left[  -\alpha\left(
v_{x}^{2}+v_{y}^{2}+v_{z}^{2}\right)  \right]  ~,
\end{equation}
the same distribution as \textquotedblleft errors in the method of least
squares\textquotedblright.

Maxwell's distribution applies whether the molecule is part of a gas, a
liquid, or a solid and, with the benefit of hindsight, the reason is quite
easy to see. The probability that a molecule have velocity $\vec{v}$ and
position $\vec{x}$ is given by the Boltzmann distribution $\propto\exp-H/kT $.
For a large variety of situations the Hamiltonian for one molecule is of the
form $H=mv^{2}/2+V(\vec{x})$ where the potential $V(\vec{x})$ includes the
interactions, whether they be weak or strong, with all the other molecules. If
the potential $V(\vec{x})$ is independent of $\vec{v}$, then the distribution
for $\vec{v}$ and $\vec{x}$ factorizes. Velocity and position are
statistically independent, and the velocity distribution is Maxwell's.

Maxwell was the first to realize that the second law is not an absolute law
(this was expressed in his popular textbook `Theory of Heat' in 1871), that it
`has only statistical certainty' and indeed, that in fluctuation phenomena
`the second law is continually being violated'. Such phenomena are not rare:
just look out the window and you can see the sky is blue -- a consequence of
the scattering of light by density fluctuations in the atmosphere.

Maxwell introduced the notion of probability, but what did he actually mean by
the word `probability'? He used his distribution function as a velocity
distribution, the number of molecules with velocities in a certain range,
which betrays a frequentist interpretation. These probabilities are ultimately
mechanical properties of the gas. But he also used his distribution to
represent the lack of information we have about the precise microstate of the
gas. This latter interpretation is particularly evident in a letter he wrote
in 1867 where he argues that the second law could be violated by
\textquotedblleft a finite being who knows the paths and velocities of all
molecules by simple inspection but can do no work except open or close a
hole.\textquotedblright\ Such a \textquotedblleft demon\textquotedblright%
\ could allow fast molecules to pass through a hole from a vessel containing
hot gas into a vessel containing cold gas, and could allow slow molecules pass
in the opposite direction. The net effect being the transfer of heat from a
low to a high temperature, a violation of the second law. All that was
required was that the demon \textquotedblleft know\textquotedblright\ the
right information. [Klein 70]

\section{Gibbs: beyond heat}

Gibbs generalized the second law in two directions: to open systems and to
inhomogeneous systems. With the introduction of the concept of the chemical
potential, a quantity that regulates the transfer of particles in much the
same way that temperature regulates the transfer of heat, he could apply the
methods of thermodynamics to phase transitions, mixtures and solutions,
chemical reactions, and much else. His paper \textquotedblleft On the
Equilibrium of Heterogeneous Systems\textquotedblright\ [Gibbs 1875-78]\ is
formulated as the purest form of thermodynamics -- a phenomenological theory
of extremely wide applicability because its foundations do not rest on
particular models about the structure and dynamics of the microscopic constituents.

And yet, Gibbs was keenly aware of the significance of the underlying
molecular constitution -- he was familiar with Maxwell's writings and in
particular with his \textquotedblleft Theory of Heat\textquotedblright%
\ (indeed, he found mistakes in it). His discussion of the process of mixing
gases led him to analyze the \textquotedblleft paradox\textquotedblright\ that
bears his name. The entropy of two different gases increases when the gases
are mixed; but does the entropy also increase when two gases of the same
molecular species are mixed? Is this an irreversible process?

For Gibbs there never was a `paradox', much less one that would require some
esoteric new (quantum) physics for its resolution. For him it was quite clear
that thermodynamics was not concerned with microscopic details, but rather
with the changes from one macrostate to another. He explained that the mixing
of two gases of the same molecular species cannot be reversed because the
mixing does not lead to a different \textquotedblleft
thermodynamic\textquotedblright\ state:

\begin{description}
\item \qquad\textquotedblleft...we do not mean a state in which each particle
shall occupy more or less exactly the same position as at some previous epoch,
but only a state which shall be indistinguishable from the previous one in its
sensible properties. It is to states of systems thus incompletely defined that
the problems of thermodynamics relate.\textquotedblright\ [Gibbs\ 1875-78]
\end{description}

Gibbs' resolution of the paradox hinges on recognizing, as had Maxwell before
him, that the explanation of the second law cannot rest on purely mechanical
arguments, that probabilistic concepts are required. This led him to conclude:
\textquotedblleft In other words, the impossibility of an uncompensated
decrease of entropy seems to be reduced to improbability,\textquotedblright\ a
sentence that Boltzmann adopted as the motto for the second volume of his
\textquotedblleft Lectures on the Theory of Gases.\textquotedblright\ (For a
modern discussion of the Gibbs' paradox see section 4.12.)

Remarkably neither Maxwell nor Gibbs established a connection between
probability and entropy. Gibbs was very successful at showing what one can
accomplish by maximizing entropy but he did not address the issue of what
entropy is or what it means. The crucial steps in this direction were taken by Boltzmann.

But Gibbs' contributions did not end here. The ensemble theory introduced in
his \textquotedblleft Principles of Statistical Mechanics\textquotedblright%
\ in 1902 (it was Gibbs who coined the term `statistical mechanics') represent
a practical and conceptual step beyond Boltzmann's understanding of entropy.

\section{Boltzmann: entropy and probability}

It was Boltzmann who found the connection between entropy and probability, but
his path was long and tortuous [Klein 73, Uffink 04]. Over the years he
adopted several different interpretations of probability and, to add to the
confusion, he was not always explicit about which one he was using, sometimes
mixing them within the same paper, and even within the same equation. At
first, he defined the probability of a molecule having a velocity $\vec{v}$
within a small cell $d^{3}v$ as being proportional to the amount of time that
the particle spent within that particular cell, but he also defined that same
probability as the fraction of particles within the cell.

By 1868 he had managed to generalize the Maxwell distribution for point
particles and\ the theorem of equipartition of energy to complex molecules in
the presence of an external field. The basic argument, which led him to the
Boltzmann distribution, was that in equilibrium the distribution should be
stationary, that it should not change as a result of collisions among particles.

The collision argument only gave the distribution for individual molecules; it
was also in 1868 that he first applied probability to the system as a whole
rather than to the individual molecules. He identified the probability of the
system being in some region of the $N$-particle phase space (rather than the
space of molecular velocities) with the relative time the system would spend
in that region -- the so-called \textquotedblleft time\textquotedblright%
\ ensemble. Alternatively, probability was also defined at a given instant in
time as being proportional to the volume of the region. At first he did not
think it was necessary to comment on whether the two definitions are
equivalent or not, but eventually he realized that their `probable'
equivalence should be explicitly expressed as the hypothesis, which later came
to be known as the `ergodic hypothesis', that over a long time the trajectory
of the system would cover the whole region of phase space consistent with the
given value of the energy. At the time all these probabilities were still
conceived as mechanical properties of the gas.

In 1871 Boltzmann achieved a significant success in establishing a connection
between thermodynamic entropy and microscopic concepts such as the probability
distribution in phase space. In modern notation his argument was as follows.
The energy of $N$ interacting particles is given by
\begin{equation}
H=%
{\textstyle\sum\limits_{i}^{N}}
~\frac{p_{i}^{2}}{2m}+U\left(  x_{1},\ldots,x_{N}\right)  ~.
\end{equation}
The first non-trivial decision was to specify what quantity defined in purely
microscopic terms corresponds to the macroscopic internal energy. He opted for
the \textquotedblleft average\textquotedblright\
\begin{equation}
E=\left\langle H\right\rangle =%
{\textstyle\int}
dz_{N}\,P_{N}\,H~,
\end{equation}
where $dz_{N}=d^{3N}xd^{3N}p$ is the volume element in the $N$-particle phase
space, and $P_{N}$ is the $N$-particle distribution function,
\begin{equation}
P_{N}=\frac{\exp\left(  -\beta H\right)  }{Z}\quad\text{where}\quad Z=%
{\textstyle\int}
dz_{N}\ e^{-\beta H}~,
\end{equation}
and $\beta=1/kT$, so that,
\begin{equation}
E=\frac{3}{2}NkT+\left\langle U\right\rangle ~.\label{Boltz E}%
\end{equation}

The connection to the thermodynamic entropy requires a clear idea of the
nature of heat and how it differs from work. One needs to express heat in
purely microscopic terms, and this is quite subtle because at the molecular
level there is no distinction between thermal motions and just plain motions.
The distribution function is the crucial ingredient. In any infinitesimal
transformation the change in the internal energy separates into two
contributions,
\begin{equation}
\delta E=%
{\textstyle\int}
dz_{N}\ H\delta P_{N}~+%
{\textstyle\int}
dz_{N}\,P_{N}\delta H~.
\end{equation}
The second integral, which can be written as $\left\langle \delta
H\right\rangle =\left\langle \delta U\right\rangle $, arises purely from
changes in the potential function $U$, which depends among other things on the
volume of the vessel containing the gas. Now, a change in the potential is
precisely what one means by mechanical work $\delta W$, therefore, since
$\delta E=\delta Q+\delta W$, the first integral must represent the
transferred heat $\delta Q$,
\begin{equation}
\delta Q=\delta E-\left\langle \delta U\right\rangle ~.
\end{equation}
On the other hand, substituting $\delta E$ from eq.(\ref{Boltz E}), one gets%
\begin{equation}
\delta Q=\frac{3}{2}Nk\delta T+\delta\left\langle U\right\rangle -\left\langle
\delta U\right\rangle ~.
\end{equation}
This is not a complete differential, but dividing by the temperature yields
\begin{equation}
\frac{\delta Q}{T}=\delta\left[  \frac{3}{2}Nk\log T+\frac{\left\langle
U\right\rangle }{T}+k\log\left(
{\textstyle\int}
d^{3N}x\,e^{-\beta U}\right)  +\operatorname*{const}\right]  ~,
\end{equation}
which suggests that the expression in brackets should be identified with the
thermodynamic entropy $S$. Further rewriting leads to
\begin{equation}
S=\frac{E}{T}+k\log Z+\operatorname*{const}\,,
\end{equation}
which is recognized as the correct modern expression.

Boltzmann's path towards understanding the second law was guided by one notion
from which he never wavered: matter is an aggregate of molecules. Apart from
this the story of his progress is the story of the increasingly more important
role played by probabilistic notions, and ultimately, it is the story of the
evolution of his understanding of the notion of probability itself. By 1877
Boltzmann achieves his final goal and explains entropy purely in terms of
probability -- mechanical notions were by now reduced to the bare minimum
consistent with the subject matter: we are, after all, talking about
collections of molecules and their energy is conserved. His final achievement
hinges on the introduction of yet another way of thinking about probabilities.

He considered an idealized system consisting of $N$ particles whose
single-particle phase space is divided into $m$ cells each with energy
$\varepsilon_{n}$, $n=1,...,m$. The number of particles in the $n$th cell is
denoted $w_{n}$, and the distribution `function' is given by the set of
numbers $w_{1},\ldots,w_{m}$. In Boltzmann's previous work the determination
of the distribution function had been based on figuring out its time evolution
from the mechanics of collisions. Here he used a purely combinatorial
argument. A completely specified state, which he called a complexion, and we
call a microstate, is defined by specifying the cell of each individual
molecule. A macrostate is less completely specified by the distribution
function, $w_{1},\ldots,w_{m}.$ The number of microstates compatible with a
given macrostate, which he called the `permutability', and we call the
`multiplicity' is
\begin{equation}
W=\frac{N!}{w_{1}!\ldots w_{m}!}\ .
\end{equation}
Boltzmann's assumption was that the probability of the macrostate was
proportional to its multiplicity, to the number of ways in which it could be
achieved, which assumes each microstate is as likely as any other -- the
`equal a priori probability postulate'.

The most probable macrostate is that which maximizes $W$ subject to the
constraints of a fixed total number of particles $N$ and a fixed total energy
$E$,
\begin{equation}%
{\textstyle\sum\limits_{n=1}^{m}}
w_{n}=N\quad\text{and}\quad%
{\textstyle\sum\limits_{n=1}^{m}}
w_{n}\varepsilon_{n}=E.\label{Boltz constraints}%
\end{equation}
When the numbers $w_{n}$ are large enough that one can use Stirling's
approximation for the factorials, we have
\begin{align}
\log W  & =N\log N-N-%
{\textstyle\sum\limits_{n=1}^{m}}
\left(  w_{n}\log w_{n}-w_{n}\right) \\
& =-%
{\textstyle\sum\limits_{n=1}^{m}}
w_{n}\log w_{n}+\operatorname*{const}~,
\end{align}
or perhaps better%
\begin{equation}
\log W=-N%
{\textstyle\sum\limits_{n=1}^{m}}
\frac{w_{n}}{N}\log\frac{w_{n}}{N}%
\end{equation}
so that
\begin{equation}
\log W=-N%
{\textstyle\sum\limits_{n=1}^{m}}
f_{n}\log f_{n}\label{log W}%
\end{equation}
where $f_{n}=w_{n}/N$ is the fraction of molecules in the $n$th cell with
energy $\varepsilon_{n}$, or, alternatively the probability that a molecule is
in its $n$th state. The distribution that maximizes $\log W$ subject to the
constraints (\ref{Boltz constraints}) is such that
\begin{equation}
f_{n}=\frac{w_{n}}{N}\propto e^{-\beta\varepsilon_{n}}~,
\end{equation}
where $\beta$ is a Lagrange multiplier determined by the total energy. When
applied to a gas, the possible states of a molecule are cells in phase space.
Therefore
\begin{equation}
\log W=-N\int dz_{1}\,f(x,p)\log f(x,p)~,\label{log W b}%
\end{equation}
where $dz_{1}=d^{3}xd^{3}p$ and the most probable distribution is the
equilibrium distribution found earlier by Maxwell and generalized by Boltzmann.

In this approach probabilities are central. The role of dynamics is minimized
but it is not eliminated. The Hamiltonian enters the discussion in two places.
One is quite explicit: there is a conserved energy the value of which is
imposed as a constraint. The second is much more subtle; we saw above that the
probability of a macrostate could be taken proportional to the multiplicity
$W$ provided microstates are assigned equal probabilities, or equivalently,
equal volumes in phase space are assigned equal a priori weights. As always
equal probabilities must be justified in terms of some form of underlying
symmetry. In this case, the symmetry follows from Liouville's theorem -- under
a Hamiltonian time evolution a region in phase space will move around and its
shape will be distorted but its volume will be conserved; Hamiltonian time
evolution preserves volumes in phase space. The nearly universal applicability
of the `equal a priori postulate' can be traced to the fact that what is
needed is a Hamiltonian; any Hamiltonian would do.

It is very remarkable that although Boltzmann calculated the maximized value
$\log W$ for an ideal gas and knew that it agreed with the thermodynamical
entropy except for a scale factor, he never wrote the famous equation that
bears his name
\begin{equation}
S=k\log W~.
\end{equation}
This equation, as well as Boltzmann's constant $k$, were both first written by Planck.

There is, however, a problem with eq.(\ref{log W b}): it involves the
distribution function $f(x,p)$ in the one-particle phase space and therefore
it cannot take correlations into account. Indeed, eq.(\ref{log W b}) gives the
correct form of the entropy only for ideal gases of non-interacting particles.
The expression that applies to systems of interacting particles
is\footnote{For the moment we disregard the question of the distinguishability
of the molecules. The so-called Gibbs paradox and the extra factor of $1/N!$
will be discussed in detail in chapter 4.}%
\begin{equation}
\log W=-\int dz_{N}\,\,f_{N}\log f_{N}~,\label{log W c}%
\end{equation}
where $f_{N}=f_{N}(x_{1},p_{1},\ldots,x_{N},p_{N})$ is the probability
distribution in the $N$-particle phase space. This equation is usually
associated with the name of Gibbs who, in his \textquotedblleft Principles of
Statistical Mechanics\textquotedblright\ (1902), developed Boltzmann's
combinatorial arguments into a very powerful theory of ensembles. The
conceptual gap between eq.(\ref{log W b}) and (\ref{log W c}) is enormous; it
goes well beyond the issue of intermolecular interactions. The probability in
Eq.(\ref{log W b}) is the single-particle distribution, it can be interpreted
as a \textquotedblleft mechanical\textquotedblright\ property, namely, the
relative number of molecules in each cell. The entropy Eq.(\ref{log W b}) is a
mechanical property of the individual system. In contrast, eq.(\ref{log W c})
involves the $N$-particle distribution which is not a property of any single
individual system but a property of an ensemble of replicas of the system.
Gibbs was not very explicit about his interpretation of probability. He wrote

\begin{description}
\item \qquad\textquotedblleft The states of the bodies which we handle are
certainly not \emph{known} to us exactly. What we \emph{know} about a body can
generally be described most accurately and most simply by saying that it is
one taken at random from a great number (ensemble) of bodies which are
completely described.\textquotedblright\ [my italics, Gibbs 1902, p.163]
\end{description}

\noindent It is clear that for Gibbs probabilities represent a state of
knowledge, that the ensemble is a purely imaginary construction, just a tool
for handling incomplete information. On the other hand, it is also clear that
Gibbs still thinks of probabilities in terms of frequencies, and since the
actual replicas of the system do not exist, he is forced to imagine them.

This brings our story of entropy up to about 1900. In the next chapter we
start a more deliberate and systematic study of the connection between entropy
and information.

\section{Some remarks}

I end with a disclaimer: this chapter has historical overtones but it is not
history. Lines of research such as the Boltzmann equation and the ergodic
hypothesis that were historically very important have been omitted because
they represent paths that diverge from the central theme of this work, namely
how laws of physics can be derived from rules for handling information and
uncertainty. Our goal has been and will be to discuss thermodynamics and
statistical mechanics as the first historical example of such an
\emph{information} physics. At first I tried to write a `history as it should
have happened'. I wanted to trace the development of the concept of entropy
from its origins with Carnot in a manner that reflects the logical rather than
the actual evolution. But I found that this approach would not do; it
trivializes the enormous achievements of the 19th century thinkers and it
misrepresents the actual nature of research. Scientific research is not a tidy business.

I mentioned that this chapter was inspired by a beautiful article by E. T.
Jaynes with the same title [Jaynes 88]. I think Jaynes' article has great
pedagogical value but I disagree with him on how well Gibbs understood the
logical status of thermodynamics and statistical mechanics as examples of
inferential and probabilistic thinking. My own assessment runs in quite the
opposite direction: the reason why the conceptual foundations of
thermodynamics and statistical mechanics have been so controversial throughout
the 20th century is precisely because neither Gibbs nor Boltzmann were
particularly clear on the interpretation of probability. I think that we could
hardly expect them to have done much better; they did not benefit from the
writings of Keynes (1921), Ramsey (1931), de Finetti (1937), Jeffreys (1939),
Cox (1946), Shannon (1948), Polya (1954) and, of course, Jaynes himself
(1957). Indeed, whatever clarity Jaynes attributes to Gibbs, is not Gibbs'; it
is the hard-won clarity that Jaynes attained through his own efforts and after
absorbing much of the best the 20th century had to offer.\newpage

\thispagestyle{empty}

\bigskip

\chapter{Entropy II: Measuring Information}

What is information? Our central goal is to gain insight into the nature of
information, how one manipulates it, and the implications of such insights for
physics. In chapter 2 we provided a first partial answer. We might not yet
know precisely what information is, but we know it when we see it. For
example, it is clear that experimental data contains information, that it is
processed using Bayes' rule, and that this is very relevant to the empirical
aspect of science, namely, to data analysis. Bayes' rule is the machinery that
processes the information contained in data to update from a prior to a
posterior probability distribution. This suggests the following
generalization: \textquotedblleft information\textquotedblright\ is whatever
induces one to update from one state of belief to another. This is a notion
worth exploring and to which we will return later.

In this chapter we pursue another point of view that has turned out to be
extremely fruitful. We saw that the natural way to deal with uncertainty, that
is, with lack of information, is to introduce the notion of degrees of belief,
and that these measures of plausibility should be manipulated and calculated
using the ordinary rules of the calculus of probabilities. But with this
achievement we do not yet reach our final goal. The rules of probability
theory allow us to assign probabilities to some \textquotedblleft
complex\textquotedblright\ propositions on the basis of the probabilities that
have been previously assigned to other, perhaps more \textquotedblleft
elementary\textquotedblright\ propositions.

In this chapter we introduce a new inference tool designed specifically for
assigning those elementary probabilities. The new tool is Shannon's measure of
an \textquotedblleft amount of information\textquotedblright\ and the
associated method of reasoning is Jaynes' Method of Maximum Entropy, or
MaxEnt. [Shannon 48, Jaynes 57b, 83, 03]

\section{Shannon's information measure}

We appeal once more to the idea that if a general theory exists it must apply
to special cases. Consider a set of mutually exclusive and exhaustive
alternatives $i$, for example, the possible values of a variable, or the
possible states of a system. The state of the system is unknown. On the basis
of the incomplete information $I$ we have we can at best assign probabilities
$p(i|I)=p_{i}$. In order to select just one among the possible states more
information is required. The question we address here is how much more? Note
that we are not asking the more difficult question of which particular piece
of information is missing, but merely the quantity that is missing. It seems
reasonable that the amount of information that is missing in a sharply peaked
distribution is smaller than the amount missing in a broad distribution, but
how much smaller? Is it possible to quantify the notion of amount of
information? Can one find a unique quantity $S$ that is a function of the
$p_{i}$'s, that tends to be large for broad distributions and small for narrow ones?

Consider a discrete set of $n$ mutually exclusive and exhaustive discrete
alternatives $i$, each with probability $p_{i}$. According to Shannon, any
measure $S$ of the amount of information that is missing when all we know is a
probability distribution must satisfy three axioms. It is quite remarkable
that these conditions are sufficiently constraining to determine the quantity
$S$ uniquely. The first two axioms are deceptively simple.

\noindent\textbf{Axiom 1}. $S$ is a real continuous function of the
probabilities $p_{i}$, $S[p]=S\left(  p_{1},\ldots p_{n}\right)  $.

\noindent\emph{Remark:} It is explicitly assumed that $S[p]$ depends only on
the $p_{i}$ and on nothing else. What we seek here is an \emph{absolute}
measure of the amount of missing information in $p$. If the objective were to
update from a prior $q$ to a posterior distribution $p$ -- a problem that will
be later tackled in chapter 6 -- then we would require a functional $S[p,q]$
depending on both $q$ and $p$. Such $S[p,q]$ would at best be a
\emph{relative} measure: the information in $p$ relative to the reference
distribution $q$.

\noindent\textbf{Axiom 2}. If all the $p_{i}$'s are equal, $p_{i}=1/n$. Then
$S=S\left(  1/n,\ldots,1/n\right)  =F\left(  n\right)  $, where $F\left(
n\right)  $ is an increasing function of $n$.

\noindent\emph{Remark:} This means that it takes less information to pinpoint
one alternative among a few than among many and also that knowing the number
$n$ of available states is already a valuable piece of information. Notice
that the uniform distribution $p_{i}=1/n$ is singled out to play a very
special role. Indeed, although no reference distribution has been explicitly
mentioned, the uniform distribution will, in effect, provide the standard of
complete ignorance.

The third axiom is a consistency requirement and is somewhat less intuitive.
The entropy $S[p]$ measures the amount of additional information beyond the
incomplete information $I$ already codified in the $p_{i}$ that will be needed
to pinpoint the actual state of the system. Imagine that this missing
information were to be obtained not all at once, but in installments. The
consistency requirement is that the particular manner in which we obtain this
information should not matter. This idea can be expressed as follows.

Imagine the $n$ states are divided into $N$ groups labeled by $g=1,\ldots,N$.
The probability that the system is found in group $g$ is
\begin{equation}
P_{g}=\sum_{i\in g}\,p_{i}\,.\label{Pg}%
\end{equation}
Let $p_{i|g}$ denote the conditional probability that the system is in the
state $i\in g$ given it is in group $g$,
\begin{equation}
p_{i|g}=\frac{p_{i}}{P_{g}}\quad\text{for}\quad i\in g.\label{pi|g}%
\end{equation}
Suppose we were to obtain the desired information in two steps, the first of
which would allow us to single out one of the groups $g$ while the second
would allow us to decide on the actual $i$ within the selected group $g$. The
amount of information required in the first step is $S_{G}=S[P]$ where
$P=\{P_{g}\}$ with $g=1\ldots N$. Now suppose we did get this information, and
as a result we found, for example, that the system was in group $g_{1}$. Then
for the second step, to single out the state $i$ within the group $g_{1} $,
the amount of additional information needed would be $S_{g_{1}}=S[p_{\cdot
|g_{1}}]$. Similarly, information amounts $S_{g_{2}},S_{g_{3}},\ldots$ or
$S_{g_{N}}$ would be required had the selected groups turned out to be
$g_{2},g_{3},\ldots$ or $g_{N}$. But at the beginning of this process we do
not yet know which of the $g$'s is the correct one. The \emph{expected amount
of missing information} to take us from the $g$'s to the actual $i$'s is
$\sum_{g}P_{g}S_{g}$. The point is that it should not matter whether we get
the total missing information in one step, which completely determines $i $,
or in two steps, the first of which has low resolution and only determines one
of the groups, say $g$, while the second step provides the fine tuning that
determines $i$ within the given $g$. This gives us our third axiom: \ \ \

\noindent\textbf{Axiom 3}. For all possible groupings $g=1\ldots N$ of the
states $i=1\ldots n$ we must have
\begin{equation}
S=S_{G}+%
{\textstyle\sum\limits_{g}}
P_{g}S_{g}\,.\label{grouping property}%
\end{equation}
This is called the \textquotedblleft grouping\textquotedblright\ property.

\noindent\emph{Remark:} Given axiom 3 it might seem more appropriate to
interpret $S$ as a measure of the \emph{expected} rather than the \emph{actual
}amount of missing information, but if $S$ is the expected value of something,
it is not clear, at this point, what that something would be. We will return
to this below.

The solution to Shannon's constraints is obtained in two steps. First assume
that all states $i$ are equally likely, $p_{i}=1/n$. Also assume that the $N$
groups $g$ all have the same number of states, $m=n/N$, so that $P_{g}=1/N$
and $p_{i|g}=p_{i}/P_{g}=1/m$. Then by axiom 2,
\begin{equation}
S[p_{i}]=S\left(  1/n,\ldots,1/n\right)  =F\left(  n\right)  ,
\end{equation}%
\begin{equation}
S_{G}[P_{g}]=S\left(  1/N,\ldots,1/N\right)  =F\left(  N\right)  ,
\end{equation}
and
\begin{equation}
S_{g}[p_{i|g}]=S(1/m,\ldots,1/m)=F(m).
\end{equation}
Then, axiom 3 gives
\begin{equation}
F\left(  mN\right)  =F\left(  N\right)  +F\left(  m\right)  \text{
}.\label{eq for F}%
\end{equation}
This should be true for all integers $N$ and $m$. It is easy to see that one
solution of this equation is
\begin{equation}
F\left(  m\right)  =k\,\log\,m\text{ },\label{k log m}%
\end{equation}
where $k$ is any positive constant, but it is easy to see that
eq.(\ref{eq for F}) has infinitely many other solutions. Indeed, since any
integer $m$ can be uniquely decomposed as a product of prime numbers, $m=%
{\textstyle\prod_{r}}
q_{r}^{\alpha_{r}}$, where $\alpha_{i}$ are integers and $q_{r} $ are prime
numbers, using eq.(\ref{eq for F}) we have
\begin{equation}
F\left(  m\right)  =%
{\textstyle\sum_{r}}
\alpha_{r}F(q_{r})\label{eq for F b}%
\end{equation}
which means that eq.(\ref{eq for F}) can be satisfied by arbitrarily
specifying $F(q_{r})$ on the primes and then defining $F(m)$ for any other
integer through eq.(\ref{eq for F b}).

A unique solution is obtained when we impose the additional requirement that
$F(m)$ be monotonic increasing in $m$ (axiom 2). The following argument is
found in [Jaynes 03]. Consider any two integers $s$ and $t$ both larger than
$1$. The ratio of their logarithms can be approximated arbitrarily closely by
a rational number, i.e., we can find integers $\alpha$ and $\beta$ (with
$\beta$ arbitrarily large) such that
\begin{equation}
\frac{\alpha}{\beta}\leq\frac{\log s}{\log t}<\frac{\alpha+1}{\beta}%
\quad\text{or}\quad t^{\alpha}\leq r^{\beta}<t^{\alpha+1}~.\label{eq for F c}%
\end{equation}
But $F$ is monotonic increasing, therefore
\begin{equation}
F(t^{\alpha})\leq F(s^{\beta})<F(t^{\alpha+1})~,
\end{equation}
and using eq.(\ref{eq for F}),
\begin{equation}
\alpha F(t)\leq\beta F(s)<(\alpha+1)F(t)\quad\text{or}\quad\frac{\alpha}%
{\beta}\leq\frac{F(s)}{F(t)}<\frac{\alpha+1}{\beta}~.\label{eq for F d}%
\end{equation}
Which means that the ratio $F(r)/F(s)$ can be approximated by the same
rational number $\alpha/\beta$. Indeed, comparing eqs.(\ref{eq for F c}) and
(\ref{eq for F d}) we get
\begin{equation}
\left\vert \frac{F(s)}{F(t)}-\frac{\log s}{\log t}\right\vert \leq\frac
{1}{\beta}%
\end{equation}
or,
\begin{equation}
\left\vert \frac{F(s)}{\log s}-\frac{F(t)}{\log t}\right\vert \leq\frac
{F(t)}{\beta\log s}%
\end{equation}
We can make the right hand side arbitrarily small by choosing $\beta$
sufficiently large, therefore $F(s)/\log s$ must be a constant, which proves
(\ref{k log m}) is the unique solution.

In the second step of our derivation we will still assume that all $i$s are
equally likely, so that $p_{i}=1/n$ and $S[p]=F\left(  n\right)  $. But now we
assume the groups $g$ have different sizes, $m_{g}$, with $P_{g}=m_{g}/n$ and
$p_{i|g}=1/m_{g}$. Then axiom 3 becomes
\[
F\left(  n\right)  =S_{G}[P]+\sum_{g}P_{g}\,F(m_{g})\text{,}%
\]
Therefore,
\[
S_{G}[P]=F\left(  n\right)  -\sum_{g}P_{g}F\left(  m_{g}\right)  =\sum
_{g}P_{g}\, \left[  F\left(  n\right)  -F\left(  m_{g}\right)  \right]  \,.
\]
\newline Substituting our previous expression for $F$ we get
\[
S_{G}[P]=\sum_{g}P_{g}\,k\,\log\frac{n}{m_{g}}=-k\,\sum_{i=1}^{N}P_{g}\,\log
P_{g}\,.
\]
Therefore Shannon's quantitative measure of the amount of missing information,
the entropy of the probability distribution $p_{1},\ldots,p_{n} $ is
\begin{equation}
S[p]=-k\,%
{\textstyle\sum\limits_{i=1}^{n}}
p_{i}\,\log\,p_{i}\,.\label{Shannon S}%
\end{equation}

\subsection*{Comments}

Notice that for discrete probability distributions we have $\,p_{i}\leq1$ and
$\log\,p_{i}\leq0$. Therefore $S\geq0$ for $k>0$. As long as we interpret $S$
as the amount of uncertainty or of missing information it cannot be negative.
We can also check that in cases where there is no uncertainty we get $S=0$: if
any state has probability one, all the other states have probability zero and
every term in $S$ vanishes.

The fact that entropy depends on the available information implies that there
is no such thing as \emph{the} entropy of a system. The same system may have
many different entropies. Notice, for example, that already in the third axiom
we find an explicit reference to two entropies $S[p]$ and $S_{G}[P]$ referring
to two different descriptions of the same system. Colloquially, however, one
does refer to \emph{the} entropy of a system; in such cases the relevant
information available about the system should be obvious from the context. In
the case of thermodynamics what one means by \emph{the} entropy is the
particular entropy that one obtains when the only information available is
specified by the known values of those few variables that specify the
thermodynamic macrostate.

The choice of the constant $k$ is purely a matter of convention. A convenient
choice is $k=1$. In thermodynamics the choice is Boltzmann's constant $k_{B}=$
$1.38\times10^{-16}$erg/K which reflects the historical choice of units of
temperature. In communication theory and computer science, the conventional
choice is $k=1/\log_{e}2\approx1.4427$, so that
\begin{equation}
S[p]=-\,%
{\textstyle\sum\limits_{i=1}^{n}}
p_{i}\,\log_{2}\,p_{i}~.
\end{equation}
The base of the logarithm is $2$, and the entropy is said to measure
information in units called `bits'.

Now we turn to the question of interpretation. Earlier we mentioned that from
axiom 3 it seems more appropriate to interpret $S$ as a measure of the
\emph{expected} rather than the \emph{actual} amount of missing information.
If one adopts this interpretation, the actual amount of information that we
gain when we find that $i$ is the true alternative would have to be
$\log1/p_{i}$. But this is not quite satisfactory. Consider a variable that
takes just two values, $0$ with probability $p$ and $1$ with probability
$1-p$. For very small $p$, $\log1/p$ would be very large, while the
information that communicates the true alternative is conveyed by a very short
one bit message, namely \textquotedblleft$0$\textquotedblright. It appears
that it is not the \emph{actual amount} of information that $\log1/p$ seems to
measure but rather how unexpected or how surprising the piece of information
might be. Accordingly, $\log1/p_{i}$ is sometimes called the \textquotedblleft
surprise\textquotedblright\ of $i$.

It seems reasonable to expect that more information implies less uncertainty.
We have used the word `uncertainty' as roughly synonymous to `lack of
information'. The following example illustrates the potential pitfalls. I
normally keep my keys in my pocket. My state of knowledge about the location
of my keys is represented by a probability distribution that is sharply peaked
at my pocket and reflects a small uncertainty. But suppose I check and I find
that my pocket is empty. Then my keys could be virtually anywhere. My new
state of knowledge is represented by a very broad distribution that reflects a
high uncertainty. We have here a situation where more information has
increased the uncertainty rather than decreased it.

The point of these remarks is not to suggest that there is something wrong
with the mathematical derivation -- eq.(\ref{Shannon S}) does follow from the
axioms -- but to suggest caution when interpreting $S$. The notion of
information is at this point still vague. Any attempt to find its measure will
always be open to the objection that it is not clear what it is that is being
measured. Indeed, the first two of Shannon's axioms seem to be particularly
intuitive, but the third one, the grouping property, is not nearly as
compelling. Is entropy the only way to measure uncertainty? Doesn't the
variance also measure uncertainty? Shannon and Jaynes both argued that one
should not place too much significance on the axiomatic derivation of
eq.(\ref{Shannon S}), that its use can be \emph{fully} justified a posteriori
by its formal properties, for example, by the various inequalities it
satisfies. However, this position can be questioned on the grounds that it is
the axioms that confer meaning to the entropy; the disagreement is not about
the actual equations, but about what they mean and, ultimately, about how they
should be used. Other measures of uncertainty can be introduced and, indeed,
they have been introduced by Renyi and by Tsallis, creating a whole industry
of alternative theories. [Renyi 61, Tsallis 88] Whenever one can make an
inference using Shannon's entropy, one can make other inferences using any one
of the Renyi's entropies. Which, among all those alternatives, should one choose?

\subsection*{The two-state case}

To gain intuition about $S[p]$ consider the case of a variable that can take
two values. The proverbial example is a biased coin -- for example, a bent
coin -- for which the outcome `heads' is assigned probability $p$ and `tails'
probability $1-p$. The corresponding entropy is
\begin{equation}
S(p)=-p\,\log\,p-(1-p)\,\log\,(1-p)~\text{,}\label{twostateS}%
\end{equation}
where we chose $k=1$. It is easy to check that $S\geq0$ and that the maximum
uncertainty, attained for $p=1/2$, is $S_{\max}=\log2$.

An important set of properties of the entropy follows from the concavity of
the entropy which follows from the concavity of the logarithm. Suppose we
can't decide whether the actual probability of heads is $p_{1}$ or $p_{2}$. We
may decide to assign probability $q$ to the first alternative and probability
$1-q$ to the second. The actual probability of heads then is the mixture
$qp_{1}+(1-q)p_{2}$. The corresponding entropies satisfy the inequality
\begin{equation}
S\left(  qp_{1}+(1-q)p_{2}\right)  \geq qS\left(  p_{1}\right)  +\left(
1-q\right)  S\left(  p_{2}\right)  ~,\label{concavity}%
\end{equation}
with equality in the extreme cases where $p_{1}=p_{2}$, or $q=0$, or $q=1$.
Eq.(\ref{concavity}) says that however ignorant we might be when we invoke a
probability distribution, an uncertainty about the probabilities themselves
will introduce an even higher degree of ignorance.

\section{Relative entropy}

The following entropy-like quantity turns out to be useful
\begin{equation}
K[p,q]=+\,%
{\textstyle\sum\limits_{i}}
p_{i}\,\log\,\frac{p_{i}}{q_{i}}\,.\label{Ka}%
\end{equation}
Despite the positive sign $K$ is sometimes read as the `entropy of $p$
relative to $q$,' and thus called \textquotedblleft relative
entropy.\textquotedblright\ It is easy to see that in the special case when
$q_{i}$ is a uniform distribution then $K$ is essentially equivalent to the
Shannon entropy -- they differ by a constant. Indeed, for $q_{i}=1/n$,
eq.(\ref{Ka}) becomes
\begin{equation}
K[p,1/n]=\,%
{\textstyle\sum\limits_{i}^{n}}
p_{i}\,\left(  \log p_{i}+\log n\right)  =\log n-S[p]~.\label{Kb}%
\end{equation}

The relative entropy is also known by many other names including cross
entropy, information divergence, information for discrimination, and
Kullback-Leibler distance [Kullback 59] who recognized its importance for
applications in statistics, and studied many of its properties). However, the
expression (\ref{Ka}) has a much older history. It was already used by Gibbs
in his \emph{Elementary Principles of Statistical Mechanics} [Gibbs 1902].

It is common to interpret $K[p,q]$ as the amount of information that is gained
(thus the positive sign) when one thought the distribution that applies to a
random process is $q$ and one learns that the distribution is actually $p$.
The interpretation suffers from the same conceptual difficulties mentioned
earlier concerning the Shannon entropy. In the next chapter we will see that
the relative entropy turns out to be the fundamental quantity for inference --
indeed, more fundamental, more general, and therefore, more useful than
entropy itself -- and that the interpretational difficulties that afflict the
Shannon entropy can be avoided. (We will also redefine it with a negative
sign, $S[p,q]\overset{\operatorname*{def}}{=}-K[p,q]$, so that it really is a
true entropy.) In this chapter we just derive some properties and consider
some applications.

An important property of the relative entropy is the Gibbs inequality,
\begin{equation}
K[p,q]\geq0~,\label{Kc}%
\end{equation}
with equality if and only if $p_{i}=q_{i}$ for all $i$. The proof uses the
concavity of the logarithm,
\begin{equation}
\log x\leq x-1\quad\text{or}\quad\log\frac{q_{i}}{p_{i}}\leq\frac{q_{i}}%
{p_{i}}-1~,
\end{equation}
\noindent\noindent\noindent which implies
\begin{equation}
\,%
{\textstyle\sum\limits_{i}}
p_{i}\log\frac{q_{i}}{p_{i}}\leq%
{\textstyle\sum\limits_{i}}
\left(  q_{i}-p_{i}\right)  =0~.
\end{equation}

The Gibbs inequality provides some justification to the common interpretation
of $K[p,q]$ as a measure of the \textquotedblleft distance\textquotedblright%
\ between the distributions $p$ and $q$. Although useful, this language is not
quite correct because $K[p,q]\neq K[q,p]$ while a true distance $d$ is
required to be symmetric, $d[p,q]=d[q,p]$. However, as we shall later see, if
the two distributions are sufficiently close the relative entropy $K[p+\delta
p,p]$ satisfies all the requirements of a metric. Indeed, it turns out that up
to a constant factor, it is the only natural Riemannian metric on the manifold
of probability distributions. It is known as the Fisher-Rao metric or, perhaps
more appropriately, the information metric.

The two inequalities $S[p]\geq0$ and $K[p,q]\geq0$ together with eq.(\ref{Kb})
imply
\begin{equation}
0\leq S[p]\leq\log n~,\label{entropy range}%
\end{equation}
which establishes the range of the entropy between the two extremes of
complete certainty ($p_{i}=\delta_{ij}$ for some value $j$) and complete
uncertainty (the uniform distribution) for a variable that takes $n$ discrete values.

\section{Joint entropy, additivity, and subadditivity}

The entropy $S[p_{x}]$ reflects the uncertainty or lack of information about
the variable $x$ when our knowledge about it is codified in the probability
distribution $p_{x}$. It is convenient to refer to $S[p_{x}]$ directly as the
\textquotedblleft entropy of the variable $x$\textquotedblright\ and write
\begin{equation}
S_{x}\overset{\operatorname*{def}}{=}S[p_{x}]=-\,%
{\textstyle\sum\limits_{x}}
p_{x}\,\log\,p_{x}\ .
\end{equation}
The virtue of this notation is its compactness but one must keep in mind the
same symbol $x$ is used to denote both a variable $x$ and its values $x_{i}$.
To be more explicit,
\begin{equation}
-\,%
{\textstyle\sum\limits_{x}}
p_{x}\,\log\,p_{x}=-\,%
{\textstyle\sum\limits_{i}}
p_{x}(x_{i})\,\log\,p_{x}(x_{i})~.
\end{equation}

The uncertainty or lack of information about two (or more) variables $x$ and
$y$ is expressed by the joint distribution $p_{xy}$ and the corresponding
\emph{joint} entropy is
\begin{equation}
S_{xy}=-%
{\textstyle\sum\limits_{xy}}
p_{xy}\log p_{xy}~.\label{joint S}%
\end{equation}

When the variables $x$ and $y$ are independent, $p_{xy}=p_{x}p_{y}$, the joint
entropy is \emph{additive}
\begin{equation}
S_{xy}=-%
{\textstyle\sum\limits_{xy}}
p_{x}p_{y}\,\log(p_{x}p_{y})=S_{x}+S_{y}~,\label{additivity}%
\end{equation}
that is, the joint entropy of independent variables is the sum of the
entropies of each variable. This \emph{additivity} property also holds for the
other measure of uncertainty we had introduced earlier, namely, the variance,
\begin{equation}
\operatorname*{var}(x+y)=\operatorname*{var}(x)+\operatorname*{var}(y)~.
\end{equation}

In thermodynamics additivity is called \emph{extensivity}: the entropy of an
extended system is the sum of the entropies of its parts provided these parts
are independent. The thermodynamic entropy can be extensive only when the
interactions between various subsystems are sufficiently weak that
correlations between them can be neglected.

When the two variables $x$ and $y$ are not independent the equality
(\ref{additivity}) can be generalized into an inequality. Consider the joint
distribution $p_{xy}=p_{x}p_{y|x}=p_{y}p_{x|y}$. The relative entropy or
Kullback \textquotedblleft distance\textquotedblright\ of $p_{xy}$ to the
product distribution $p_{x}p_{y}$ that would represent uncorrelated variables
is given by
\begin{align}
K[p_{xy},p_{x}p_{y}]  & =\,%
{\textstyle\sum\limits_{xy}}
p_{xy}\,\log\,\frac{p_{xy}}{p_{x}p_{y}}\nonumber\\
& =-S_{xy}-%
{\textstyle\sum\limits_{xy}}
p_{xy}\,\log p_{x}-%
{\textstyle\sum\limits_{xy}}
p_{xy}\,\log p_{y}\nonumber\\
& =-S_{xy}+S_{x}+S_{y}~.\label{Kd}%
\end{align}
Therefore, using $K\geq0$ we get
\begin{equation}
S_{xy}\leq S_{x}+S_{y}~,\label{subadditivity}%
\end{equation}
with the equality holding when the two variables $x$ and $y$ are independent.
This inequality is called the \emph{subadditivity} property. Its
interpretation is clear: entropy increases when information about correlations
is discarded.

\section{Conditional entropy and mutual information}

Consider again two variables $x$ and $y$. We want to measure the amount of
uncertainty about one variable $x$ when we have some limited information about
another variable $y$. This quantity, called the conditional entropy, and
denoted $S_{x|y}$, is obtained by calculating the entropy of $x$ as if the
precise value of $y$ were known and then taking the expectation over the
possible values of $y$%
\begin{equation}
S_{x|y}=%
{\textstyle\sum\limits_{y}}
p_{y}S[p_{x|y}]=-\,%
{\textstyle\sum\limits_{y}}
p_{y}%
{\textstyle\sum\limits_{x}}
p_{x|y}\,\log\,p_{x|y}=-%
{\textstyle\sum\limits_{x,y}}
p_{xy}\,\log\,p_{x|y}~,
\end{equation}
where $p_{xy}$ is the joint distribution of $x$ and $y$.

The conditional entropy is related to the entropy of $x$ and the joint entropy
by the following \textquotedblleft chain rule.\textquotedblright\ Use the
product rule for the joint distribution
\begin{equation}
\log\,p_{xy}=\log\,p_{y}+\log\,p_{x|y}~,
\end{equation}
and take the expectation over $x$ and $y$ to get
\begin{equation}
S_{xy}=S_{y}+S_{x|y}~.\label{conditional chain rule}%
\end{equation}
In words: the entropy of two variables is the entropy of one plus the
conditional entropy of the other. Also, since $S_{y}$ is positive we see that
conditioning reduces entropy,
\begin{equation}
S_{xy}\geq S_{x|y}~.
\end{equation}

Another useful entropy-like quantity is the so-called \textquotedblleft mutual
information\textquotedblright\ of $x$ and $y$, denoted $M_{xy}$, which
\textquotedblleft measures\textquotedblright\ how much information $x$ and $y$
have in common. This is given by the relative entropy between the joint
distribution $p_{xy}$ and the product distribution $p_{x}p_{y}$ that discards
all information contained in the correlations. Using eq.(\ref{Kd}),
\begin{equation}
M_{xy}\overset{\operatorname*{def}}{=}K[p_{xy},p_{x}p_{y}]=S_{x}+S_{y}%
-S_{xy}\geq0~,
\end{equation}
which shows that it is symmetrical in $x$ and $y$. Using
eq.(\ref{conditional chain rule}) the mutual information is related to the
conditional entropies by
\begin{equation}
M_{xy}=S_{x}-S_{x|y}=S_{y}-S_{y|x}~.
\end{equation}

The relationships among these various entropies can be visualized by a figure
that resembles a Venn diagram. (The diagram is usually considered a purely
mnemonic aid, but recent work [Knuth 02-06] on the duality between assertions
and questions, and the corresponding duality between probabilities and
entropies suggests that the resemblance between the two types of Venn diagrams
is not accidental.)

\section{Continuous distributions}

Shannon's derivation of the expression for entropy, eq.(\ref{Shannon S}),
applies to probability distributions of discrete variables. The generalization
to continuous variables is not quite straightforward.

The discussion will be carried out for a one-dimensional continuous variable;
the generalization to more dimensions is trivial. The starting point is to
note that the expression
\begin{equation}
-%
{\textstyle\int}
dx\,p(x)\log\,p(x)\label{wrong S a}%
\end{equation}
is unsatisfactory. A change of variables $x\rightarrow y=y(x)$ changes the
probability density $p(x)$ to $p^{\prime}(y)$ but does not represent a loss or
gain of information. Therefore, the actual probabilities do not change,
$p(x)dx=p^{\prime}(y)dy$, and neither should the entropy. However, one can
check that (\ref{wrong S a}) is not invariant,%
\begin{align}%
{\textstyle\int}
dx\,p(x)\log\,p(x)  & =%
{\textstyle\int}
dy\,p^{\prime}(y)\log\left[  p^{\prime}(y)\left\vert \frac{dy}{dx}\right\vert
\right] \nonumber\\
& \neq%
{\textstyle\int}
dy\,p^{\prime}(y)\log\,p^{\prime}(y)\,.
\end{align}

We approach the continuous case as a limit from the discrete case. Consider a
continuous distribution $p(x)$ defined on an interval for $x_{a}\leq x\leq
x_{b}$. Divide the interval into equal intervals $\Delta x=\left(  x_{b}%
-x_{a}\right)  /N$. The distribution $p(x)$ can be approximated by a discrete
distribution
\begin{equation}
p_{n}=p(x_{n})\Delta x~,
\end{equation}
where $x_{n}=x_{a}+n\Delta x$ and $n$ is an integer. The discrete entropy is
\begin{equation}
S_{N}=-%
{\textstyle\sum\limits_{n=0}^{N}}
\Delta x\,p(x_{n})\,\log\left[  p(x_{n})\Delta x\right]  \,~,
\end{equation}
and as $N\rightarrow\infty$ we get
\begin{equation}
S_{N}\longrightarrow\log N-%
{\textstyle\int\limits_{x_{a}}^{x_{b}}}
dx\,p(x)\,\log\, \left[  \frac{p(x)}{1/\left(  x_{b}-x_{a}\right)  }\right]
\end{equation}
which diverges. This is quite to be expected: it takes a finite amount of
information to identify one discrete alternative within a finite set, but it
takes an infinite amount to single out one point in a continuum. The
difference $S_{N}-\log N$ has a well defined limit and we are tempted to
consider
\begin{equation}
-%
{\textstyle\int\limits_{x_{a}}^{x_{b}}}
dx\,p(x)\,\log\,\left[  \frac{p(x)}{1/\left(  x_{b}-x_{a}\right)  }\right]
\label{wrong S b}%
\end{equation}
as a candidate for the continuous entropy, until we realize that, except for
an additive constant, it coincides with the unacceptable expression
(\ref{wrong S a}) and should be discarded for precisely the same reason: it is
not invariant under changes of variables. Had we first changed variables to
$y=y(x)$ and then discretized into $N$ equal $\Delta y$ intervals we would
have obtained a different limit
\begin{equation}
-%
{\textstyle\int\limits_{y_{a}}^{y_{b}}}
dy\,p^{\prime}(y)\,\log\,\left[  \frac{p^{\prime}(y)}{1/\left(  y_{b}%
-y_{a}\right)  }\right]  ~.\label{wrong S c}%
\end{equation}
The problem is that the limiting procedure depends on the particular choice of
discretization; the limit depends on which particular set of intervals $\Delta
x$ or $\Delta y$ we have arbitrarily decided to call equal. Another way to
express the same idea is to note that the denominator $1/\left(  x_{b}%
-x_{a}\right)  $ in (\ref{wrong S b}) represents a probability density that is
uniform in the variable $x$, but not in $y$. Similarly, the density $1/\left(
y_{b}-y_{a}\right)  $ in (\ref{wrong S c}) is uniform in $y$, but not in $x$.

Having identified the origin of the problem we can now suggest a solution. On
the basis of our prior knowledge of the problem at hand we must decide on a
privileged set of equal intervals, or alternatively, on one preferred
probability distribution $\mu(x)$ we are willing to define as
\textquotedblleft uniform.\textquotedblright\ Then, and only then, it makes
sense to propose the following definition%
\begin{equation}
S[p,\mu]\overset{\operatorname*{def}}{=}-%
{\textstyle\int\limits_{x_{a}}^{x_{b}}}
dx\,p(x)\,\log\frac{p(x)}{\mu(x)}~.\label{continuum S}%
\end{equation}
It is easy to check that this is invariant,
\begin{equation}%
{\textstyle\int\limits_{x_{a}}^{x_{b}}}
dx\,p(x)\,\log\frac{p(x)}{\mu(x)}=%
{\textstyle\int\limits_{y_{a}}^{y_{b}}}
dy\,p^{\prime}(y)\,\log\frac{p^{\prime}(y)}{\mu^{\prime}(y)}~.
\end{equation}

Examples illustrating possible choices of the uniform $\mu(x)$ are the following.

\begin{enumerate}
\item When the variable $x$ refers to position in \textquotedblleft
physical\textquotedblright\ space, we can feel fairly comfortable with what we
mean by equal volumes: use Cartesian coordinates and choose $\mu
(x)=\operatorname*{constant}$.

\item In a curved space $D$-dimensional, with a known metric tensor $g_{ij}$,
i.e., the distance between neighboring points with coordinates $x^{i}$ and
$x^{i}+dx^{i}$ is given by $d\ell^{2}=g_{ij}dx^{i}dx^{j}$, the volume elements
are given by $\left(  \det g\right)  ^{1/2}d^{D}x$. In this case choose
$\mu(x)\propto\left(  \det g\right)  ^{1/2}$.

\item In classical statistical mechanics the Hamiltonian evolution in phase
space is, according to Liouville's theorem, such that phase space volumes are
conserved. This leads to a natural definition of equal intervals or equal
volumes. The corresponding choice of uniform $\mu$ is called the postulate of
\textquotedblleft equal a priori probabilities.\textquotedblright
\end{enumerate}

Notice that the expression in eq.(\ref{continuum S}) is a relative entropy
$-K[p,\mu]$. This is a hint for a theme that will be fully developed in
chapter 6: relative entropy is the more fundamental quantity. Strictly, there
is no Shannon entropy in the continuum -- not only do we have to subtract an
infinite constant and spoil its (already shaky) interpretation as an
information measure, but we have to appeal to prior knowledge and introduce
the measure $\mu$. On the other hand there are no difficulties in obtaining
the continuum relative entropy from its discrete version. We can check that
\begin{equation}
K_{N}=%
{\textstyle\sum\limits_{n=0}^{N}}
\,p_{n}\,\log\frac{\,p_{n}\,}{q_{n}}=%
{\textstyle\sum\limits_{n=0}^{N}}
\Delta x\,p(x_{n})\,\log\frac{\,p(x_{n})\Delta x\,}{q(x_{n})\Delta x}%
\end{equation}
has a well defined limit,
\begin{equation}
K[p,q]=%
{\textstyle\int\limits_{x_{a}}^{x_{b}}}
dx\,p(x)\,\log\frac{p(x)}{q(x)}~,
\end{equation}
which is explicitly invariant under coordinate transformations.

\section{Communication Theory}

Here we give the briefest introduction to some basic notions of communication
theory as originally developed by Shannon [Shannon 48, Shannon Weaver 49]. For
a more comprehensive treatment see [Cover Thomas 91].

Communication theory studies the problem of how a message that was selected at
some point of origin can be best reproduced at some later destination point.
The complete communication system includes an \emph{information source} that
generates a message composed of, say, words in English, or pixels on a
picture. A \emph{transmitter} translates the message into an appropriate
signal. For example, sound pressure is encoded into an electrical current, or
letters into a sequence of zeros and ones. The signal is such that it can be
transmitted over a \emph{communication channel}, which could be electrical
signals propagating in coaxial cables or radio waves through the atmosphere.
Finally, a \emph{receiver }reconstructs\emph{\ }the signal back into a message
that can be interpreted by an agent at the destination point.

From the engineering point of view the communication system must be designed
with only a limited information about the set of possible messages. In
particular, it is not known which specific messages will be selected for
transmission. The typical sort of questions one wishes to address concern the
minimal physical requirements needed to communicate the messages that could
potentially be generated by a particular information source. One wants to
characterize the sources, measure the capacity of the communication channels,
and learn how to control the degrading effects of noise. And after all this,
it is somewhat ironic but nevertheless true that \textquotedblleft information
theory\textquotedblright\ is completely unconcerned with whether any
\textquotedblleft information\textquotedblright\ is being communicated at all.
As far as the engineering goes, whether the messages convey some meaning or
not is completely irrelevant.

To illustrate the basic ideas consider the problem of \textquotedblleft data
compression.\textquotedblright\ A useful idealized model of an information
source is a sequence of random variables $x_{1},x_{2},\ldots$ which take
values from a finite alphabet of symbols. We will assume that the variables
are independent and identically distributed. (Eliminating these limitations is
both possible and important.) Suppose that we deal with a binary source in
which the variables $x_{i}$, which are usually called `bits', take the values
zero or one with probabilities $p$ or $1-p$ respectively. Shannon's idea was
to classify the possible sequences $x_{1},\ldots,x_{N}$ into \emph{typical}
and \emph{atypical} according to whether they have high or low probability.
For large $N$ the expected number of zeros and ones is $Np$ and $N(1-p)$
respectively. The probability of these \emph{typical} sequences is
\begin{equation}
P(x_{1},\ldots,x_{N})\approx p^{Np}(1-p)^{N(1-p)}~,
\end{equation}
so that
\begin{equation}
-\log P(x_{1},\ldots,x_{N})\approx-N[p\log p-(1-p)\log(1-p)]=NS(p)
\end{equation}
where $S(p)$ is the two-state entropy, eq.(\ref{twostateS}), the maximum value
of which is $S_{\max}=\log2$. Therefore, the probability of typical sequences
is roughly%
\begin{equation}
P(x_{1},\ldots,x_{N})\approx e^{-NS(p)}\,.\label{AEP a}%
\end{equation}
Since the total probability is less than one, we see that the number of
typical sequences has to be less than about $e^{NS(p)}$ which for large $N$ is
considerably less than the total number of possible sequences, $2^{N}%
=e^{N\log2}$. This fact is very significant. Transmitting an arbitrary
sequence irrespective of whether it is typical or not requires a long message
of $N$ bits, but we do not have to waste resources in order to transmit all
sequences. We only need to worry about the far fewer typical sequences because
the atypical sequences are too rare. The number of typical sequences is about
\begin{equation}
e^{NS(p)}=2^{NS(p)/\log2}=2^{NS(p)/S_{\max}}%
\end{equation}
and therefore we only need about $NS(p)/S_{\max}$ bits to identify each one of
them. Thus, it must be possible to compress the original long message into a
much shorter one. The compression might imply some small probability of error
because the actual message might conceivably turn out to be atypical but one
can, if desired, avoid any such errors by using one additional bit to flag the
sequence that follows as typical and short or as atypical and long. Actual
schemes for implementing the data compression are discussed in [Cover Thomas 91].

Next we state these intuitive notions in a mathematically precise way.

\noindent\textbf{Theorem:} The Asymptotic Equipartition Property (AEP). If
$x_{1},\ldots,x_{N}$ are independent variables with the same probability
distribution $p(x)$, then
\begin{equation}
-\frac{1}{N}\log P(x_{1},\ldots,x_{N})\longrightarrow S[p]\quad\text{in
probability.}\label{AEP b}%
\end{equation}

\noindent\textbf{Proof:} If the variables $x_{i}$ are independent, so are
their logarithms, $\log p(x_{i})$,
\begin{equation}
-\frac{1}{N}\log P(x_{1},\ldots,x_{N})=-\frac{1}{N}%
{\textstyle\sum\limits_{i}^{N}}
\log p(x_{i})\,,
\end{equation}
and the law of large numbers (see section 2.8) gives
\begin{equation}
\lim_{N\rightarrow\infty}\operatorname*{Prob}\left[  \left\vert -\frac{1}%
{N}\log P(x_{1},\ldots,x_{N})+\left\langle \log p(x)\right\rangle \right\vert
\leq\varepsilon\right]  =1\,,\label{AEP c}%
\end{equation}
where
\begin{equation}
-\left\langle \log p(x)\right\rangle =S[p]\,\,.
\end{equation}
This concludes the proof.

We can elaborate on the AEP idea further. The typical sequences are those for
which eq.(\ref{AEP a}) or (\ref{AEP b}) is satisfied. To be precise let us
define the typical set $A_{N,\varepsilon}$ as the set of sequences with
probability $P(x_{1},\ldots,x_{N})$ such that
\begin{equation}
e^{-N\left[  S(p)+\varepsilon\right]  }\leq P(x_{1},\ldots,x_{N})\leq
e^{-N\left[  S(p)-\varepsilon\right]  }\,.\label{typical seq}%
\end{equation}
\newpage

\noindent\textbf{Theorem of typical sequences:}

\begin{enumerate}
\item[\textbf{(1)}] For $N$ sufficiently large $\operatorname*{Prob}%
[A_{N,\varepsilon}]>1-\varepsilon$.

\item[\textbf{(2)}] $\left\vert A_{N,\varepsilon}\right\vert \leq e^{N\left[
S(p)+\varepsilon\right]  }$ where $\left\vert A_{N,\varepsilon}\right\vert $
is the number of sequences in $A_{N,\varepsilon}$.

\item[\textbf{(3)}] For $N$ sufficiently large $\left\vert A_{N,\varepsilon
}\right\vert \geq(1-\varepsilon)e^{N\left[  S(p)-\varepsilon\right]  }$.
\end{enumerate}

\noindent In words: the typical set has probability near one, typical
sequences are nearly equally probable (thus the `equipartition'), and there
are about $e^{NS(p)}$ of them. To summarize:

\begin{description}
\item \emph{Almost all events are almost equally likely}.
\end{description}

\noindent\textbf{Proof:} \textbf{\ }Eq.(\ref{AEP c}) states that for fixed
$\varepsilon$, for any given $\delta$ there is an $N_{\delta}$ such that for
all $N>N_{\delta}$, we have
\begin{equation}
\operatorname*{Prob}\left[  \left\vert -\frac{1}{N}\log P(x_{1},\ldots
,x_{N})+S[p]\right\vert \leq\varepsilon\right]  \geq1-\delta\,.
\end{equation}
Thus, the probability that the sequence $(x_{1},\ldots,x_{N})$ is
$\varepsilon$-typical tends to one, and therefore so must
$\operatorname*{Prob}[A_{N,\varepsilon}]$. Setting $\delta=\varepsilon$ yields
part (\textbf{1)}. To prove \textbf{(2)} write
\begin{align}
1  & \geq\operatorname*{Prob}[A_{N,\varepsilon}]=%
{\textstyle\sum\limits_{(x_{1},\ldots,x_{N})\in A_{N,\varepsilon}}}
P(x_{1},\ldots,x_{N})\nonumber\\
& \geq%
{\textstyle\sum\limits_{(x_{1},\ldots,x_{N})\in A_{N,\varepsilon}}}
e^{-N\left[  S(p)+\varepsilon\right]  }=e^{-N\left[  S(p)+\varepsilon\right]
}\left\vert A_{N,\varepsilon}\right\vert \,.
\end{align}
Finally, from part \textbf{(1)},\textbf{\ }%
\begin{align}
1-\varepsilon & <\operatorname*{Prob}[A_{N,\varepsilon}]=%
{\textstyle\sum\limits_{(x_{1},\ldots,x_{N})\in A_{N,\varepsilon}}}
P(x_{1},\ldots,x_{N})\nonumber\\
& \leq%
{\textstyle\sum\limits_{(x_{1},\ldots,x_{N})\in A_{N,\varepsilon}}}
e^{-N\left[  S(p)-\varepsilon\right]  }=e^{-N\left[  S(p)-\varepsilon\right]
}\left\vert A_{N,\varepsilon}\right\vert \,,
\end{align}
which proves \textbf{(3)}.

We can now quantify the extent to which messages generated by an information
source of entropy $S[p]$ can be compressed. A scheme that produces compressed
sequences that are more than $NS(p)/S_{\max}$ bits is capable of
distinguishing among all the typical sequences. The compressed sequences can
be reliably decompressed into the original message. Conversely, schemes that
yield compressed sequences of fewer than $NS(p)/S_{\max}$ bits cannot describe
all typical sequences and are not reliable. This result is known as
\emph{Shannon's} \emph{noiseless channel coding theorem}.

\section{Assigning probabilities: MaxEnt}

Probabilities are introduced to deal with lack of information. The notion that
entropy $S[p]$ can be interpreted as a quantitative measure of the amount of
missing information has one remarkable consequence: it provides us with a
method to assign probabilities. \noindent The idea is simple:

\emph{Among all possible probability distributions that agree with whatever we
know select that particular distribution that reflects maximum ignorance about
everything else. Since ignorance is measured by entropy, the method is
mathematically implemented by selecting the distribution that maximizes
entropy subject to the constraints imposed by the available information. This
method of reasoning is called the method of Maximum Entropy, and is often
abbreviated as MaxEnt.}

Ultimately, the method of maximum entropy is based on an ethical principle of
intellectual honesty that demands that one should not assume information one
does not have. The idea is quite compelling but its justification relies
heavily on interpreting entropy as a measure of missing information and
therein lies its weakness: to what extent are we sure that entropy is the
unique measure of information or of uncertainty?

As a simple illustration of the MaxEnt method in action consider a variable
$x$ about which absolutely nothing is known except that it can take $n$
discrete values $x_{i}$ with $i=1\ldots n$. The distribution that represents
the state of maximum ignorance is that which maximizes the entropy subject to
the single constraint that the probabilities be normalized, $%
{\textstyle\sum\nolimits_{i}}
p_{i}=1$. Introducing a Lagrange multiplier $\alpha$ to handle the constraint,
the variation $p_{i}\rightarrow p_{i}+\delta p_{i}$ gives
\begin{equation}
0=\delta\left(  S[p]-\alpha%
{\textstyle\sum\limits_{i}}
p_{i}\right)  =-\,%
{\textstyle\sum\limits_{i=1}^{n}}
\left(  \log\,p_{i}+1+\alpha\right)  \delta p_{i}~,
\end{equation}
so that the selected distribution is
\begin{equation}
p_{i}=e^{-1-\alpha}\quad\text{or}\quad p_{i}=\frac{1}{n}~,\label{uniform a}%
\end{equation}
where the multiplier $\alpha$ has been determined from the normalization
constraint. We can check that the maximum value attained by the entropy,
\begin{equation}
S_{\text{max}}=-%
{\textstyle\sum\nolimits_{i}}
\,\frac{1}{n}\log\frac{1}{n}=\log n\,,\label{uniform b}%
\end{equation}
agrees with eq.(\ref{entropy range}).

\noindent\textbf{Remark: }The distribution of maximum ignorance turns out to
be uniform. It coincides with what we would have obtained using Laplace's
Principle of Insufficient Reason. It is sometimes asserted that the MaxEnt
method provides a proof of Laplace's principle but such a claim is
questionable. As we saw earlier, the privileged status of the uniform
distribution was imposed through the Shannon's axioms from the very beginning.

\section{Canonical distributions}

The available information constrains the possible probability distributions.
Although the constraints can take any form whatsoever, in this section we
develop the MaxEnt formalism for the special case of constraints that are
linear in the probabilities. The most important applications are to situations
of thermodynamic equilibrium where the relevant information is given in terms
of the expected values of those few macroscopic variables such as energy,
volume, and number of particles over which one has some experimental control.
(In the next chapter we revisit this problem more explicitly.)

The goal is to select the distribution of maximum entropy from within the
family of all distributions for which the expectations of some functions
$f^{k}(x)$, $k=1,2,\ldots$ have known numerical values $F^{k}$,%
\begin{equation}
\left\langle f^{k}\right\rangle =%
{\textstyle\sum\limits_{i}}
p_{i}f_{i}^{k}=F^{k}~,\label{constraint k}%
\end{equation}
where we set $f^{k}(x_{i})=f_{i}^{k}$ to simplify the notation. In addition
there is a normalization constraint, $%
{\textstyle\sum}
p_{i}=1$. Introducing the necessary multipliers, the entropy maximization is
achieved setting
\begin{align}
0  & =\delta\left(  S[p]-\alpha%
{\textstyle\sum\limits_{i}}
p_{i}-\lambda_{k}\left\langle f^{k}\right\rangle \right) \nonumber\\
& =-%
{\textstyle\sum\limits_{i}}
\left(  \log\,p_{i}+1+\alpha+\lambda_{k}f_{i}^{k}\right)  \delta p_{i}~,
\end{align}
where we adopt the summation convention that repeated upper and lower indices
are summed over. The solution is the so-called `canonical' distribution,
\begin{equation}
p_{i}=\exp-(\lambda_{0}+\lambda_{k}f_{i}^{k})~,\label{canon dist}%
\end{equation}
where we have set $1+\alpha=\lambda_{0}$. The normalization constraint
determines $\lambda_{0}$,
\begin{equation}
e^{\lambda_{0}}=%
{\textstyle\sum\limits_{i}}
\exp(-\lambda_{k}f_{i}^{k})\overset{\operatorname*{def}}{=}Z\left(
\lambda_{1},\lambda_{2},\ldots\right) \label{partition Z}%
\end{equation}
where we have introduced the partition function $Z$. Substituting
eqs.(\ref{canon dist}) and (\ref{partition Z}) into the other constraints,
eqs.(\ref{constraint k}), gives a set of equations that implicitly determine
the remaining multipliers,
\begin{equation}
-\frac{\partial\log Z}{\partial\lambda_{k}}=F^{k}~,\label{expected F}%
\end{equation}
and substituting into $S[p]=-%
{\textstyle\sum}
{}p_{i}\log p_{i}$ we obtain the maximized value of the entropy,
\begin{equation}
S_{\text{max}}=%
{\textstyle\sum\limits_{i}}
p_{i}(\lambda_{0}+\lambda_{k}f_{i}^{k})=\lambda_{0}+\lambda_{k}F^{k}%
~.\label{maximum S}%
\end{equation}

Equations (\ref{canon dist}-\ref{expected F}) are a generalized form of the
\textquotedblleft canonical\textquotedblright\ distributions first discovered
by Maxwell, Boltzmann and Gibbs. Strictly, the calculation above only shows
that the entropy is stationary, $\delta S=0$. To complete the argument we must
show that (\ref{maximum S}) is the absolute maximum rather than just a local
extremum or a stationary point.

Consider any other distribution $q_{i}$ that satisfies precisely the same
constraints in eqs.(\ref{constraint k}). According to the basic Gibbs
inequality for the relative entropy of $q$ and the canonical $p$,
\begin{equation}
K(q,p)=\,%
{\textstyle\sum\limits_{i}}
q_{i}\,\log\,\frac{q_{i}}{p_{i}}\geq0~,
\end{equation}
or%
\begin{equation}
S[q]\leq-%
{\textstyle\sum\limits_{i}}
q_{i}\,\log\,p_{i}~.
\end{equation}
Substituting eq.(\ref{canon dist}) gives
\begin{equation}
S[q]\leq%
{\textstyle\sum\limits_{i}}
q_{i}(\lambda_{0}+\lambda_{k}f_{i}^{k})=\lambda_{0}+\lambda_{k}F^{k}~.
\end{equation}
Therefore
\begin{equation}
S[q]\leq S[p]=S_{\text{max}}~.
\end{equation}
In words: within the family of all distributions $q$ that satisfy the
constraints (\ref{constraint k}) the distribution that achieves the maximum
entropy is the canonical distribution $p$ given in eq.(\ref{canon dist}).

Having found the maximum entropy distribution we can now develop the MaxEnt
formalism along lines that closely parallel the formalism of statistical
mechanics. Each distribution within the family (\ref{canon dist}) can be
thought of as a point in a continuous space -- the manifold of canonical
distributions. Each specific choice of expected values $(F^{1},F^{2},\ldots) $
determines a unique point within the space, and therefore the $F^{k}$ play the
role of coordinates. To each point $(F^{1},F^{2},\ldots)$ we can associate a
number, the value of the maximized entropy. Therefore, $S_{\text{max}}$ is a
scalar field which we denote $S(F^{1},F^{2},\ldots)=S(F)$. In thermodynamics
it is conventional to drop the suffix `max' and to refer to $S(F)$ as
\emph{the entropy} of the system. This language is inappropriate because it
can be misleading. We should constantly remind ourselves that $S(F)$ is just
one out of many possible entropies. $S(F)$ is that particular entropy that
measures the amount of missing information of a subject whose knowledge
consists of the numerical values of the $F$s and nothing else. The multiplier
\begin{equation}
\lambda_{0}=\log Z(\lambda_{1},\lambda_{2},\ldots)=\log Z(\lambda)
\end{equation}
is sometimes called the \textquotedblleft free energy\textquotedblright%
\ because it is closely related to the thermodynamic free energy,
\begin{equation}
S(F)=\log Z(\lambda)+\lambda_{k}F^{k}.\label{Legendre transf}%
\end{equation}
The quantities $S(F)$ and $\log Z(\lambda)$ contain the same information; the
equation above shows that they are Legendre transforms of each other. Just as
the $F$s are obtained from $\log Z(\lambda)$ from eq.(\ref{expected F}), the
$\lambda$s can be obtained from $S(F)$
\begin{equation}
\frac{\partial S(F)}{\partial F^{k}}=\frac{\partial\log Z(\lambda)}%
{\partial\lambda_{j}}\frac{\partial\lambda_{j}}{\partial F^{k}}+\frac
{\partial\lambda_{j}}{\partial F^{k}}F^{j}+\lambda_{k}~,
\end{equation}
or, using eq.(\ref{expected F}),
\begin{equation}
\frac{\partial S(F)}{\partial F^{k}}=\lambda_{k}~,\label{lambda = dS/dF}%
\end{equation}
which shows that the multipliers $\lambda_{k}$ are the components of the
gradient of the entropy $S(F)$ on the manifold of canonical distributions.

A useful extension of the formalism is the following. It is common that the
functions $f^{k}$ are not fixed but depend on one or more parameters $v$ that
can be externally manipulated, $f_{i}^{k}=f^{k}(x_{i},v)$. For example
$f_{i}^{k}$ could refer to the energy of the $i$th state of the system, and
the parameter $v$ could be the volume of the system or an externally applied
magnetic field.

Then a general change in the expected value $F^{k}$ induced by changes in both
$f^{k}$ and $\lambda_{k}$, is expressed as
\begin{equation}
\delta F^{k}=\delta\left\langle f^{k}\right\rangle =%
{\textstyle\sum\limits_{i}}
\left(  p_{i}\delta f_{i}^{k}+f_{i}^{k}\delta p_{i}\right)
\,,\label{delta Fk}%
\end{equation}
The first term on the right is
\begin{equation}
\left\langle \delta f^{k}\right\rangle =%
{\textstyle\sum\limits_{i}}
p_{i}\frac{\partial f_{i}^{k}}{\partial v}\delta v=\left\langle \frac{\partial
f^{k}}{\partial v}\right\rangle \delta v~.
\end{equation}
When $F^{k}$ represents the internal energy then $\left\langle \delta
f^{k}\right\rangle $ is a small energy transfer that can be controlled through
an external parameter $v$. This suggests that $\left\langle \delta
f^{k}\right\rangle $ represents a kind of \textquotedblleft generalized
work,\textquotedblright\ $\delta W^{k}$, and the expectations $\left\langle
\partial f^{k}/\partial v\right\rangle $ are analogues of pressure or
susceptibility,
\begin{equation}
\delta W^{k}\overset{\operatorname*{def}}{=}\left\langle \delta f^{k}%
\right\rangle =\left\langle \frac{\partial f^{k}}{\partial v}\right\rangle
\delta v\,.
\end{equation}
The second term in eq.(\ref{delta Fk}),
\begin{equation}
\delta Q^{k}\overset{\operatorname*{def}}{=}%
{\textstyle\sum\limits_{i}}
f_{i}^{k}\delta p_{i}=\delta\left\langle f^{k}\right\rangle -\left\langle
\delta f^{k}\right\rangle ~\label{gen heat}%
\end{equation}
is a kind of \textquotedblleft generalized heat\textquotedblright, and
\begin{equation}
\delta F^{k}=\delta W^{k}+\delta Q^{k}\label{gen 1st law}%
\end{equation}
is a \textquotedblleft generalized first law.\textquotedblright\

The corresponding change in the entropy is obtained from
eq.(\ref{Legendre transf}),
\begin{align}
\delta S  & =\delta\log Z(\lambda)+\delta(\lambda_{k}F^{k})\nonumber\\
& =-\frac{1}{Z}%
{\textstyle\sum\limits_{i}}
\left[  \delta\lambda_{k}f_{i}^{k}+\lambda_{k}\delta f_{i}^{k}\right]
\,e^{-\lambda_{k}f_{i}^{k}}+\delta\lambda_{k}F^{k}+\lambda_{k}\delta
F^{k}\nonumber\\
& =\lambda_{k}\left(  \delta\left\langle f^{k}\right\rangle -\left\langle
\delta f^{k}\right\rangle \right)  ,
\end{align}
which, using eq.(\ref{gen heat}), gives%

\begin{equation}
\delta S=\lambda_{k}\delta Q^{k}~.\label{entropy change}%
\end{equation}
It is easy to see that this is equivalent to eq.(\ref{lambda = dS/dF}) where
the partial derivatives are derivatives at constant $v$.

Thus the entropy remains constant in infinitesimal \textquotedblleft
adiabatic\textquotedblright\ processes -- those with $\delta Q^{k}=0$. From
the information theory point of view [see eq.(\ref{gen heat})] this result is
a triviality: the amount of information in a distribution cannot change when
the probabilities do not change,
\begin{equation}
\delta p_{i}=0\Rightarrow\delta Q^{k}=0\Rightarrow\delta S=0~.
\end{equation}
.

\section{On constraints and relevant information}

The method of maximum entropy has been successful in many applications, but
there are cases where it has failed. Are these symptoms of irreparable flaws
or mere examples of misuses of the method? MaxEnt is a method for processing
information: what information are we talking about? The importance of this
issue cannot be overestimated. Here we collect a few remarks; this is a topic
to which we will return repeatedly.

One point that must be made is that questions about how information is
processed -- and this is the problem that MaxEnt is supposed to address --
should not be confused with questions about how the information was obtained
in the first place. These are two separate issues.

Here is an example of a common error. Once we accept that certain constraints
might refer to the expected values of certain variables, how do we decide
their numerical magnitudes? The numerical values of expectations are seldom
known and one might be tempted to replace expected values by sample averages
because it is the latter that are directly available from experiment. But the
two are not the same: \emph{Sample averages are experimental data. Expected
values are not experimental data.}

For very large samples such a replacement can be justified by the law of large
numbers -- there is a high probability that sample averages will approximate
the expected values. However, for small samples using one as an approximation
for the other can lead to incorrect inferences. It is important to realize
that these incorrect inferences do not represent an intrinsic flaw of the
MaxEnt method; they are merely a warning of how the MaxEnt method should not
be used.

There are many other objections that have been raised against the logic behind
the MaxEnt method. We make no attempt to survey them all; many have already
received adequate answers (see, e.g., [Jaynes 83] and [Jaynes 03],
particularly section 11.8). But some objections remain that are quite
legitimate and demand our attention. They revolve around the following
question: Once we accept that constraints will be in the form of the expected
values of certain variables, how do we decide which variables to choose?

When using the MaxEnt method to obtain, say, the canonical Boltzmann
distribution ($p_{i}\propto e^{-\beta E_{i}}$) it has been common to adopt the
following language:

\begin{description}
\item[\textbf{(A)}] We seek the probability distribution that codifies the
information we actually have (say, the expected energy) and is maximally
unbiased (\emph{i.e.} maximally ignorant or maximum entropy) about all the
other information we do not possess.\
\end{description}

\noindent Many authors find this justification unsatisfactory. Indeed, they
might argue, for example, that

\begin{description}
\item[\textbf{(B1)}] The observed spectrum of black body radiation is whatever
it is, independently of the information that happens to be available to us.
\end{description}

\noindent We prefer to phrase the objection differently:

\begin{description}
\item[\textbf{(B2)}] In most realistic situations the expected value of the
energy is not a quantity we happen to know; how, then, can we justify using it
as a constraint?
\end{description}

\noindent Alternatively, even when the expected values of some quantities
happen to be known, according to (A) what MaxEnt provides is the best possible
inferences given the limited information that is available. This is no mean
feat, but there is no guarantee that the resulting inferences will be any good
at all. The predictions of statistical mechanics are spectacularly accurate:
how can we hope to achieve equally spectacular predictions in other fields?

\begin{description}
\item[\textbf{(B3)}] We need some understanding of which are the
\textquotedblleft correct\textquotedblright\ quantities the expectation values
of which codify the relevant information for the problem at hand.
\end{description}

\noindent Merely that some particular expected value happens to be known is
neither an adequate nor a sufficient explanation.

A partial answer to these objections starts with the observation that whether
the value of the expected energy is known or not, it is nevertheless still
true that maximizing entropy subject to the energy constraint leads to the
indisputably correct \textit{family} of thermal equilibrium distributions
(e.g., the black-body spectral distribution). The justification behind
imposing a constraint on the expected energy cannot be that this is a quantity
that happens to be known -- because of the brute fact that it is not known --
but rather that the\emph{\ }expected energy is the quantity that \emph{should
}be\emph{\ }known. Even when its actual numerical value is unknown, we
recognize it as the \emph{relevant} information without which no successful
predictions are possible. (In the next chapter we revisit this important question.)

Therefore we allow MaxEnt to proceed as if this crucial information were
available which leads us to a \emph{family }of distributions containing the
temperature as a free parameter. The actual value of this parameter will have
to be inferred from the experiment itself either directly, using a
thermometer, or indirectly by Bayesian analysis from other empirical data.

To summarize: \emph{It is not just what you happen to know; you have to know
the right thing.} The constraints that should be imposed are those that codify
the information that is relevant to the problem under consideration. Between
one extreme of ignorance (we know neither which variables are relevant nor
their expected values), and the other extreme of useful knowledge (we know
which variables are relevant and we also know their expected values), there is
\emph{an intermediate state of knowledge} -- and this is the rule rather than
the exception -- in which the relevant variables have been correctly
identified but their actual expected values remain unknown. In this
intermediate state, the information about which are the relevant variables is
taken into account using MaxEnt to select a parametrized family of probability
distributions, while the actual expected values must then be inferred
independently either by direct measurement or inferred indirectly using Bayes'
rule from other experimental data.

Achieving this `intermediate state of knowledge' is the difficult problem
presented by (B3). Historically progress has been achieved in individual cases
mostly by \textquotedblleft intuition,\textquotedblright\ that is, trial and
error. Perhaps the seeds for a more systematic \textquotedblleft theory of
relevance\textquotedblright\ can already be seen in the statistical theories
of model selection and of non-parametric density estimation.

\bigskip

\chapter{Statistical Mechanics}

Among the various theories that make up what we call physics, thermodynamics
holds a very special place because it provided the first example of a
fundamental theory that could be interpreted as a procedure for processing
relevant information. Our goal in this chapter is to provide a more explicit
discussion of statistical mechanics as a theory of inference. We show that
several notoriously controversial topics such as the Second Law of
thermodynamics, irreversibility, reproducibility, and the Gibbs paradox can be
considerably clarified when viewed from the information/inference perspective.

Since the success of any problem of inference hinges on identifying the
relevant information we start by providing some background on the dynamical
evolution of probability distributions -- Liouville's theorem -- and then we
justify why in situations of thermal equilibrium the relevant constraint is
encapsulated into the expected value of the energy (and/or other such
conserved quantities).

\section{Liouville's theorem}

Perhaps the most \emph{relevant}, and therefore, most \emph{important} piece
of information that has to be incorporated into any inference about physical
systems is that their time evolution is constrained by equations of motion.
Whether these equations -- those of Newton, Maxwell, Yang and Mills, or
Einstein -- can themselves be derived as examples of inference are questions
which will not concern us at this point.

To be specific, in this section we will limit ourselves to discussing
classical systems such as fluids. In this case there is an additional crucial
piece of relevant information: these systems are composed of molecules. For
simplicity we will assume that the molecules have no internal structure, that
they are described by their positions and momenta, and that they behave
according to classical mechanics.

The import of these remarks is that the proper description of the
\emph{microstate} of a fluid of $N$ particles in a volume $V$ is in terms of a
\textquotedblleft vector\textquotedblright\ in the $N$-particle phase space,
$z=(\vec{x}_{1},\vec{p}_{1},\ldots\vec{x}_{N},\vec{p}_{N})$. The time
evolution is given by Hamilton's equations,
\begin{equation}
\frac{d\vec{x}_{i}}{dt}=\frac{\partial H}{\partial\vec{p}_{i}}\quad
\text{and}\quad\frac{d\vec{p}_{i}}{dt}=-\frac{\partial H}{\partial\vec{x}_{i}%
}\,,\label{Hamilton's eqs}%
\end{equation}
where $H$ is the Hamiltonian,
\begin{equation}
H=%
{\textstyle\sum\limits_{i=1}^{N}}
\frac{p_{i}^{2}}{2m}+U(\vec{x}_{1},\ldots\vec{x}_{N},V)~.
\end{equation}
But the actual positions and momenta of the molecules are unknown and thus the
\emph{macrostate} of the fluid is described by a probability density in phase
space, $f(z,t)$. When the system evolves continuously according to Hamilton's
equations there is no information loss and the probability flow satisfies a
local conservation equation,
\begin{equation}
\frac{\partial}{\partial t}f(z,t)=-\nabla_{z}\cdot
J(z,t)~,\label{conserved prob}%
\end{equation}
where the probability current $J$ is a vector given by
\begin{equation}
J(z,t)=f(z,t)\dot{z}=\left\{  f(z,t)\frac{d\vec{x}_{i}}{dt},\,f(z,t)\frac
{d\vec{p}_{i}}{dt}\right\}  ~.\label{conserved prob b}%
\end{equation}
Evaluating the divergence explicitly using (\ref{Hamilton's eqs}) gives
\begin{align}
\frac{\partial f}{\partial t}  & =-%
{\textstyle\sum\limits_{i=1}^{N}}
\left[  \frac{\partial}{\partial\vec{x}_{i}}\cdot\left(  f(z,t)\frac{d\vec
{x}_{i}}{dt}\right)  +\frac{\partial}{\partial\vec{p}_{i}}\cdot\left(
f(z,t)\frac{d\vec{p}_{i}}{dt}\right)  \right] \nonumber\\
& =-%
{\textstyle\sum\limits_{i=1}^{N}}
\left(  \frac{\partial f}{\partial\vec{x}_{i}}\cdot\frac{\partial H}%
{\partial\vec{p}_{i}}-\frac{\partial f}{\partial\vec{p}_{i}}\cdot
\frac{\partial H}{\partial\vec{x}_{i}}\right)  ~.
\end{align}
Thus the time derivative of $f(z,t)$ at a fixed point $z$ is given by the
Poisson bracket with the Hamiltonian $H$,
\begin{equation}
\frac{\partial f}{\partial t}=\{H,f\}\overset{\text{def}}{=}%
{\textstyle\sum\limits_{i=1}^{N}}
\left(  \frac{\partial H}{\partial\vec{x}_{i}}\cdot\frac{\partial f}%
{\partial\vec{p}_{i}}-\frac{\partial H}{\partial\vec{p}_{i}}\cdot
\frac{\partial f}{\partial\vec{x}_{i}}\right)  ~.\label{Liouville a}%
\end{equation}
This is called the Liouville equation.

Two important corollaries are the following. Instead of focusing on the change
in $f(z,t)$ at a fixed point $z$ we can study the change in $f\left(
z(t),t\right)  $ at a point $z(t)$ that is being carried along by the flow.
This defines the so-called \textquotedblleft convective\textquotedblright%
\ time derivative,
\begin{equation}
\frac{d}{dt}f\left(  z(t),t\right)  =\frac{\partial}{\partial t}f(z,t)+%
{\textstyle\sum\limits_{i=1}^{N}}
\left(  \frac{\partial f}{\partial\vec{x}_{i}}\cdot\frac{d\vec{x}_{i}}%
{dt}+\frac{\partial f}{\partial\vec{p}_{i}}\cdot\frac{d\vec{p}_{i}}%
{dt}\right)  ~.
\end{equation}
Using Hamilton's equations shows that the second term is $-\{H,f\}$ and
cancels the first, therefore
\begin{equation}
\frac{d}{dt}f\left(  z(t),t\right)  =0~,\label{Liouville b}%
\end{equation}
which means that $f$ is constant along a flow line. Explicitly,
\begin{equation}
f\left(  z(t),t\right)  =f\left(  z(t^{\prime}),t^{\prime}\right)  ~.
\end{equation}

Next consider a small volume element $\Delta z(t)$ that is being carried along
by the fluid flow. Since trajectories cannot cross each other (because
Hamilton's equations are first order in time) they cannot cross the boundary
of the evolving volume $\Delta z(t)$ and therefore the total probability
within $\Delta z(t)$ is conserved,
\begin{equation}
\frac{d}{dt}\operatorname*{Prob}[\Delta z(t)]=\frac{d}{dt}[\Delta z(t)f\left(
z(t),t\right)  ]=0~.
\end{equation}
But $f$ itself is constant, eq.(\ref{Liouville b}), therefore
\begin{equation}
\frac{d}{dt}\Delta z(t)=0~,\label{Liouville c}%
\end{equation}
which means that the shape of a region of phase space may get deformed by time
evolution but its volume remains invariant.

\section{Derivation of Equal a Priori Probabilities}

Earlier, in section 4.5, we pointed out that a proper definition of entropy in
a continuum, eq.(\ref{continuum S}), requires that one specify a privileged
background measure $\mu(z)$,
\begin{equation}
S[f,\mu]=-%
{\textstyle\int}
dz\,f(z)\log\frac{f(z)}{\mu(z)}~,
\end{equation}
where $dz=d^{3N}xd^{3N}p$. The choice of $\mu(z)$ is important: it determines
what we mean by a uniform or maximally ignorant distribution.

It is customary to set $\mu(z)$ equal to a constant which we might as well
choose to be one. This amounts to \emph{postulating} that equal volumes of
phase space are assigned the same a priori probabilities. Ever since the
introduction of Boltzmann's ergodic hypothesis there have been many failed
attempts to derive it from purely dynamical considerations. In this section we
want to \emph{determine} $\mu(z)$ by proving the following theorem

\noindent\textbf{The Equal a Priori Probability Theorem: }\emph{Since}
Hamiltonian dynamics involves no loss of information, \emph{if the entropy
}$S[f,\mu]$\emph{\ is to be interpreted as the measure of amount of
information}, then $\mu(z)$ must be a constant in phase space.

\noindent\textbf{Remark: }In chapter 6 the requirement that the entropy $S$
must be interpreted as a measure of information will be removed and thus the
logic of statistical mechanics as a theory of inference will be considerably strengthened.

\noindent\textbf{Proof:} The main non-dynamical hypothesis is that entropy
measures information. The \emph{information }entropy of the time-evolved
distribution $f(z,t)$ is
\begin{equation}
S(t)=-%
{\textstyle\int}
dz\,f(z,t)\log\frac{f(z,t)}{\mu(z)}~.\label{S[f,mu]}%
\end{equation}
The first input from Hamiltonian dynamics is that information is not lost and
therefore we must require that $S(t)$ be constant,
\begin{equation}
\frac{d}{dt}S(t)=0~.\label{dS/dt a}%
\end{equation}
Therefore,
\begin{equation}
\frac{d}{dt}S(t)=-%
{\textstyle\int}
dz\,\left[  \frac{\partial f(z,t)}{\partial t}\log\frac{f(z,t)}{\mu(z)}%
+\frac{\partial f(z,t)}{\partial t}\right]  ~.\label{dS/dt a2}%
\end{equation}
The second term vanishes,
\begin{equation}%
{\textstyle\int}
dz\,\frac{\partial f(z,t)}{\partial t}=\frac{d}{dt}%
{\textstyle\int}
dz\,f(z,t)=0~.
\end{equation}
A second input from Hamiltonian dynamics is that probabilities are not merely
conserved, they are locally conserved, which is expressed by
eqs.(\ref{conserved prob}) and (\ref{conserved prob b}). The first term of
eq.(\ref{dS/dt a2}) can be rewritten,
\begin{equation}
\frac{d}{dt}S(t)=%
{\textstyle\int}
dz\,\nabla_{z}\cdot J(z,t)\log\frac{f(z,t)}{\mu(z)}~,
\end{equation}
so that integrating by parts (the surface term vanishes) gives
\begin{align}
\frac{d}{dt}S(t)  & =-%
{\textstyle\int}
dz\,f(z,t)\dot{z}\cdot\nabla_{z}\log\frac{f(z,t)}{\mu(z)}\nonumber\\
& =%
{\textstyle\int}
dz\,\left[  -\dot{z}\cdot\nabla_{z}f(z,t)+f(z,t)\dot{z}\cdot\nabla_{z}\log
\mu(z)\right]  ~.
\end{align}
Hamiltonian dynamics enters here once again: the first term vanishes by
Liouville's equation (\ref{Liouville a}),
\begin{equation}
-%
{\textstyle\int}
dz\,\dot{z}\cdot\nabla_{z}f(z,t)=%
{\textstyle\int}
dz\left\{  H,f(z,t)\right\}  =%
{\textstyle\int}
dz\,\frac{\partial f(z,t)}{\partial t}=0~,
\end{equation}
and therefore, imposing (\ref{dS/dt a}),
\begin{equation}
\frac{d}{dt}S(t)=%
{\textstyle\int}
dz\,f(z,t)\dot{z}\cdot\nabla_{z}\log\mu(z)=0~.
\end{equation}
This integral must vanish for any arbitrary choice of the distribution
$f(z,t)$, therefore
\begin{equation}
\dot{z}\cdot\nabla_{z}\log\mu(z)=0~.
\end{equation}
Furthermore, we have considerable freedom about the particular Hamiltonian
operating on the system. We could choose to change the volume in any
arbitrarily prescribed way by pushing on a piston, or we could choose to vary
an external magnetic field. Either way we can change $H(t)$ and therefore
$\dot{z}$ at will. The time derivative $dS/dt$ must still vanish irrespective
of the particular choice of the vector $\dot{z}$. We conclude that
\begin{equation}
\nabla_{z}\log\mu(z)=0\quad\text{or}\quad\mu(z)=\operatorname*{const}.
\end{equation}

To summarize: the requirement that information is not lost in Hamiltonian
dynamics implies that the measure of information must be a constant of the
motion,
\begin{equation}
\frac{d}{dt}S(t)=0~,\label{dS/dt b}%
\end{equation}
and this singles out the Gibbs entropy,
\begin{equation}
S(t)=-%
{\textstyle\int}
dz\,f(z,t)\log f(z,t)~,\label{Gibbs entropy}%
\end{equation}
as the correct \emph{information} entropy.

It is sometimes asserted that (\ref{dS/dt b}) implies that the Gibbs entropy
cannot be identified with the \emph{thermodynamic} entropy because this would
be in contradiction to the second law. As we shall see below, this is not
true; in fact, it is quite the opposite.

\section{The relevant constraints}

Thermodynamics is concerned with situations of thermal equilibrium. What is
the relevant information needed to make inferences that apply to these special
cases? The first condition we must impose on $f\left(  z,t\right)  $ to
describe equilibrium is that it be independent of time. Thus we require that
$\{H,f\}=0$ and $f$ must be a function of conserved quantities such as energy,
momentum, angular momentum, or number of particles. But we do not want $f$ to
be merely stationary, as say, for a rotating fluid, we want it to be truly
static. We want $f$ to be invariant under time reversal. For these problems it
turns out that it is not necessary to impose that the total momentum and total
angular momentum vanish; these constraints will turn out to be satisfied
automatically. To simplify the situation even more we will only consider
problems where the number of particles is held fixed. Processes where
particles are exchanged as in the equilibrium between a liquid and its vapor,
or where particles are created and destroyed as in chemical reactions,
constitute an important but straightforward extension of the theory.

It thus appears that it is sufficient to impose that $f$ be some function of
the energy. According to the formalism developed in section 4.8 and the
remarks in 4.9 this is easily accomplished: the constraints codifying the
information that could be relevant to problems of thermal equilibrium should
be the expected values of functions $\phi(E)$ of the energy. For example,
$\langle\phi(E)\rangle$ could include various moments, $\langle E\rangle$,
$\langle E^{2}\rangle$,\ldots\ or perhaps more perhaps complicated functions.
The remaining question is which functions $\phi(E)$ and how many of them.

To answer this question we look at thermal equilibrium from the point of view
leading to what is known as the \emph{microcanonical formalism}. Let us
enlarge our description to include the system of interest $A$ and its
environment, that is, the thermal bath $B$ with which it is in equilibrium.
The advantage of this broader view is that the composite system $C=A+B$ can be
assumed to be isolated and \emph{we know that its energy }$E_{c}$\emph{\ is
some fixed constant}. This is highly relevant information: when the value of
$E_{c}$ is known, not only do we know $\left\langle E_{c}\right\rangle $ but
we know the expected values $\langle\phi(E_{c})\rangle$ for absolutely all
functions $\phi(E_{c}).$ In other words, in this case we have succeeded in
identifying the relevant information and we are finally ready to assign
probabilities using the MaxEnt method. (When the value of $E_{c}$ is not known
we are in that state of \textquotedblleft intermediate\textquotedblright%
\ knowledge described in section 4.9.)

To simplify the notation it is convenient to divide phase space into discrete
cells of equal volume. For system $A$ let the (discretized) microstate $z_{a}$
have energy $E_{a}$. For the thermal bath $B$ a much less detailed description
is sufficient. Let the number of bath microstates with energy $E_{b}$ be
$\Omega_{B}(E_{b})$. Our relevant information includes the fact that $A$ and
$B$ interact very weakly, just barely enough to attain equilibrium, and thus
the known total energy $E_{c}$ constrains the allowed microstates of $A+B$ to
the subset that satisfies
\begin{equation}
E_{a}+E_{b}=E_{c}~.
\end{equation}
The total number of such microstates is
\begin{equation}
\Omega(E_{c})=%
{\textstyle\sum\nolimits_{a}}
\Omega_{B}(E_{c}-E_{a})~.
\end{equation}

We are in a situation where we know absolutely nothing beyond the fact that
the composite system $C$ can be in any one of its $\Omega(E_{c})$ allowed
microstates. This is precisely the problem tackled in section 4.7: the maximum
entropy distribution is uniform, eq.(\ref{uniform a}), and the probability of
any microstate of $C$ is $1/\Omega(E_{c})$. More importantly, the probability
that system $A$ is in the particular microstate $a$ when it is in thermal
equilibrium with the bath $B$ is
\begin{equation}
p_{a}=\frac{\Omega_{B}(E_{c}-E_{a})}{\Omega(E_{c})}~.\label{thermal eq a}%
\end{equation}
This is the result we sought; now we need to interpret it. It is convenient to
rewrite $p_{a}$ in terms of the the entropy of the bath $S_{B}=k\log
\,\Omega_{B}$,
\begin{equation}
p_{a}\propto\exp\frac{1}{k}S_{B}(E_{c}-E_{a})~.
\end{equation}
There is one final piece of relevant information we can use: the thermal bath
$B$ is much larger than system $A$, $E_{c}\gg E_{a}$, and we can Taylor
expand
\begin{equation}
S_{B}(E_{c}-E_{a})=S_{B}(E_{c})-\frac{E_{a}}{T}+\ldots\,,
\end{equation}
where the temperature $T$ of the bath has been introduced according to the
standard thermodynamic definition,
\begin{equation}
\left.  \frac{\partial S_{B}}{\partial E_{b}}\right\vert _{E_{c}}%
\overset{\operatorname*{def}}{=}\frac{1}{T}~.
\end{equation}
The term $S_{B}(E_{c})$ is a constant independent of the label $a$ which can
be absorbed into the normalization. We conclude that the distribution that
codifies the relevant information about equilibrium is
\begin{equation}
p_{a}\propto\exp\left(  -\frac{E_{a}}{kT}\right)  ~,\label{thermal eq b}%
\end{equation}
which we recognize as having the canonical form of eq.(\ref{canon dist}).

Our goal in this section was to identify the relevant variables. Here is the
answer: the relevant information about thermal equilibrium can be summarized
by the expected value of the energy $\left\langle E\right\rangle $ because
someone who just knows $\left\langle E\right\rangle $ and is maximally
ignorant about everything else is led to assign probabilities according to
eq.(\ref{canon dist}) which coincides with (\ref{thermal eq b}).

But our analysis has also disclosed an important limitation.
Eq.(\ref{thermal eq a}) shows that in general the distribution for a system in
equilibrium with a bath depends in a complicated way on the properties of the
bath. The information in $\left\langle E\right\rangle $ is adequate only when
the system and the bath interact weakly and the bath is so much larger than
the system that its effects can be represented by a single parameter, the
temperature $T$. Conversely, if these conditions are not met, then more
information is needed. For example, the system might be sufficiently isolated
that within the time scales of interest it can only reach thermal equilibrium
with the few degrees of freedom in its very immediate vicinity. Then the
surrounding bath need not be large and the information contained in the
expected value $\left\langle E\right\rangle $ while still useful and relevant
might just not be sufficient; more will be needed.

\noindent\textbf{Remark:} The notion of relevance is relative. A particular
piece of information might be relevant to one specific question and irrelevant
to another. In the discussion above the system is in equilibrium, but we have
not been sufficiently explicit about what specific questions one wants to
address. It is implicit in this whole approach that one refers to the typical
questions addressed in thermodynamics.

\section{The canonical formalism}

We consider a system (say, a fluid) in thermal equilibrium. The energy of the
(conveniently discretized) microstate $z_{a}$ is $E_{a}=E_{a}(V)$ where $V$ is
the volume of the system. We assume further that the expected value of the
energy is known, $\left\langle E\right\rangle =\bar{E}$.

Maximizing the (discretized) Gibbs entropy,
\begin{equation}
S[p]=-k\,%
{\textstyle\sum\limits_{a}}
p_{a}\,\log\,p_{a}\quad\text{where}\quad p_{a}=f(z_{a})\Delta
z~,\label{Gibbs entropy b}%
\end{equation}
subject to constraints on normalization $\left\langle 1\right\rangle =1$ and
energy $\left\langle E\right\rangle =\bar{E}$ yields, eq.(\ref{canon dist}),
\begin{equation}
p_{a}=\frac{1}{Z}e^{-\beta E_{a}}\label{canon dist b}%
\end{equation}
where the Lagrange multiplier $\beta$ is determined from
\begin{equation}
-\frac{\partial\log Z}{\partial\beta}=\bar{E}\quad\text{and}\quad Z(\beta,V)=%
{\textstyle\sum\nolimits_{a}}
e^{-\beta E_{a}}~.\label{E and Z}%
\end{equation}

The maximized value of the Gibbs entropy is, eq.(\ref{maximum S}),
\begin{equation}
S(\bar{E},V)=k\log Z+k\beta\bar{E}~.\label{S(E,V)}%
\end{equation}
Differentiating with respect to $\bar{E}$ we obtain the analogue of
eq.(\ref{lambda = dS/dF}),
\begin{equation}
\left(  \frac{\partial S}{\partial\bar{E}}\right)  _{V}=k\frac{\partial\log
Z}{\partial\beta}\frac{\partial\beta}{\partial\bar{E}}+k\frac{\partial\beta
}{\partial\bar{E}}\bar{E}+k\beta\ =k\beta~,
\end{equation}
where eq.(\ref{E and Z}) has been used to cancel the first two terms. In
thermodynamics temperature is defined by
\begin{equation}
\left(  \frac{\partial S}{\partial\bar{E}}\right)  _{V}\overset
{\operatorname*{def}}{=}\frac{1}{T}~,
\end{equation}
therefore,
\begin{equation}
\beta=\frac{1}{kT}\ .
\end{equation}

\emph{The connection between the formalism above and thermodynamics hinges on
a suitable identification of work and heat.} A small change in the internal
energy $\delta E$ can be induced by small changes in $T$ and $V$,
\begin{equation}
\delta\bar{E}=%
{\textstyle\sum\nolimits_{a}}
p_{a}\delta E_{a}+%
{\textstyle\sum\nolimits_{a}}
E_{a}\delta p_{a}\,.\label{dW and dQ}%
\end{equation}
Since $E_{a}=E_{a}(V)$ the first term $\left\langle \delta E\right\rangle $ is
an energy change that can be induced by small changes in volume,
\begin{equation}
\left\langle \delta E\right\rangle =%
{\textstyle\sum\nolimits_{a}}
p_{a}\frac{\partial E_{a}}{\partial V}\delta V=\left\langle \frac{\partial
E}{\partial V}\right\rangle \delta V~,
\end{equation}
this suggests that we can identify it with the mechanical work,
\begin{equation}
\left\langle \delta E\right\rangle =\delta W=-P\delta V~,
\end{equation}
and therefore, the pressure is given by
\begin{equation}
P=-\left\langle \frac{\partial E}{\partial V}\right\rangle ~.\label{pressure}%
\end{equation}
This is the microscopic definition of pressure.

The second term in eq.(\ref{dW and dQ}) must therefore represent heat,
\begin{equation}
\delta Q=\delta\bar{E}-\delta W=\delta\left\langle E\right\rangle
-\left\langle \delta E\right\rangle .\label{first law a}%
\end{equation}
The corresponding change in entropy is obtained from eq.(\ref{S(E,V)}),
\begin{align}
\frac{\delta S}{k}  & =\delta\log Z+\delta(\beta\bar{E})\nonumber\\
& =-\frac{1}{Z}%
{\textstyle\sum\nolimits_{a}}
e^{-\beta E_{a}}\left(  E_{a}\delta\beta+\beta\delta E_{a}\right)  +\bar
{E}\delta\beta+\beta\delta\bar{E}\nonumber\\
& =\beta(\delta\bar{E}-\left\langle \delta E\right\rangle )~,
\end{align}
therefore,
\begin{equation}
\delta S=\frac{\delta Q}{T}~.\label{dS=dQ/T}%
\end{equation}
This result is important. It proves that

\begin{description}
\item \qquad\emph{The maximized Gibbs entropy, }$S(\bar{E},V)$\emph{, is
identical to the thermodynamic entropy originally defined by Clausius}.
\end{description}

\noindent Substituting into eq.(\ref{first law a}), yields the
\emph{fundamental thermodynamic identity},
\begin{equation}
\delta\bar{E}=T\delta S-P\delta V~.\label{first law b}%
\end{equation}
Incidentally, it shows that the \textquotedblleft natural\textquotedblright%
\ variables for energy are $S$ and $V$, that is, $\bar{E}=\bar{E}(S,V)$.
Similarly, writing%
\begin{equation}
\delta S=\frac{1}{T}\delta\bar{E}+\frac{P}{T}\delta V\label{first law c}%
\end{equation}
confirms that $S=S(\bar{E},V)$.

The free energy $F$ is defined by
\begin{equation}
Z=e^{-\beta F}\quad\text{or}\quad F=-kT\log Z(T,V)~.\label{free energy a}%
\end{equation}
Eq.(\ref{S(E,V)}) then leads to
\begin{equation}
F=\bar{E}-TS~,\label{free energy b}%
\end{equation}
so that
\begin{equation}
\delta F=-S\delta T-P\delta V~,\label{free energy c}%
\end{equation}
which shows that $F=F(T,V)$.

Several useful thermodynamic relations can be easily obtained from
eqs.(\ref{first law b}), (\ref{first law c}), and (\ref{free energy c}). For
example, the identities
\begin{equation}
\left(  \frac{\partial F}{\partial T}\right)  _{V}=-S\quad\text{and}%
\quad\left(  \frac{\partial F}{\partial V}\right)  _{V}=-P\,,
\end{equation}
can be read directly from eq.(\ref{free energy c}).

\section{The Second Law of Thermodynamics}

We saw that in 1865 Clausius summarized the two laws of thermodynamics into
\textquotedblleft The energy of the universe is constant. The entropy of the
universe tends to a maximum.\textquotedblright\ We can be a bit more explicit
about the Second Law: \textquotedblleft In an adiabatic non-quasi-static
process that starts and ends in equilibrium the total entropy increases; if
the process is adiabatic and quasi-static process the total entropy remains
constant.\textquotedblright\ The Second Law was formulated in a somewhat
stronger form by Gibbs (1878) \textquotedblleft For irreversible processes not
only does the entropy tend to increase, but it does increase to the maximum
value allowed by the constraints imposed on the system.\textquotedblright\

We are now ready to prove the Second Law. The proof below proposed by E. T.
Jaynes in 1965 is mathematically very simple, but it is also conceptually
subtle [Jaynes 65]. It may be useful to recall some of our previous results.
The entropy mentioned in the Second Law is the \textquotedblleft
thermodynamic\textquotedblright\ entropy $S_{T}$. It is defined only for
equilibrium states by the Clausius relation,
\begin{equation}
S_{T}(B)-S_{T}(A)=%
{\textstyle\int\limits_{A}^{B}}
\frac{dQ}{T}~,
\end{equation}
where the integral is along a reversible path of intermediate equilibrium
states. But as we saw in the previous section, in thermal equilibrium the
\emph{maximized} Gibbs entropy $S_{G}^{\text{can}}$ -- that is, the entropy
computed from the canonical distribution -- satisfies the same relation,
eq.(\ref{dS=dQ/T}),
\begin{equation}
\delta S_{G}^{\text{can}}=\frac{\delta Q}{T}\Rightarrow S_{G}^{\text{can}%
}(B)-S_{G}^{\text{can}}(A)=%
{\textstyle\int\limits_{A}^{B}}
\frac{dQ}{T}~.
\end{equation}
If the arbitrary additive constant is adjusted so $S_{G}^{\text{can}}$ matches
$S_{T}$ for one equilibrium state they will be equal for all equilibrium
states. Therefore, if at any time $t$ the system is in thermal equilibrium and
its relevant macrovariables agree with expected values, say $X_{t}$,
calculated using the canonical distribution then,
\begin{equation}
S_{T}(t)=S_{G}^{\text{can}}(t)~.\label{2nd law a}%
\end{equation}

The system, which is assumed to be thermally insulated from its environment,
is allowed (or forced) to evolve according to a certain Hamiltonian. The
evolution could, for example, be the free expansion of a gas into vacuum, or
it could be given by the time-dependent Hamiltonian that describes some
externally prescribed influence, say, a moving piston or an imposed field.
Eventually a new equilibrium is reached at some later time $t^{\prime}$. Such
a process is adiabatic; no heat was exchanged with the environment. Under
these circumstances the initial canonical distribution $f_{\text{can}}(t)$,
e.g. eq.(\ref{canon dist}) or (\ref{canon dist b}), evolves according to
Liouville's equation, eq.(\ref{Liouville a}),
\begin{equation}
f_{\text{can}}(t)\overset{H(t)}{\longrightarrow}f(t^{\prime})~,
\end{equation}
and, according to eq.(\ref{dS/dt b}), the corresponding Gibbs entropy remains
constant,
\begin{equation}
S_{G}^{\text{can}}(t)=S_{G}(t^{\prime})~.\label{2nd law b}%
\end{equation}
Since the Gibbs entropy remains constant it is sometimes argued that this
contradicts the Second Law but note that the time-evolved $S_{G}(t^{\prime}) $
is not the thermodynamic entropy because $f(t^{\prime})$ is not necessarily of
the canonical form, eq.(\ref{canon dist}).

From the new distribution $f(t^{\prime})$ we can, however, compute the
expected values $X_{t^{\prime}}$ that apply to the state of equilibrium at
$t^{\prime}$. Of all distributions agreeing with the new values $X_{t^{\prime
}}$ the canonical distribution $f_{\text{can}}(t^{\prime})$ is that which has
maximum Gibbs entropy, $S_{G}^{\text{can}}(t^{\prime})$. Therefore
\begin{equation}
S_{G}(t^{\prime})\leq S_{G}^{\text{can}}(t^{\prime})~.\label{2nd law c}%
\end{equation}
But $S_{G}^{\text{can}}(t^{\prime})$ coincides with the thermodynamic entropy
of the new equilibrium state,
\begin{equation}
S_{G}^{\text{can}}(t^{\prime})=S_{T}(t^{\prime})~.\label{2nd law d}%
\end{equation}
Collecting all these results, eqs.(\ref{2nd law a})-(\ref{2nd law d}), we
conclude that the thermodynamic entropy has increased,
\begin{equation}
S_{T}(t)\leq S_{T}(t^{\prime})~,\label{2nd law e}%
\end{equation}
which is the Second Law. The equality applies when the time evolution is
quasistatic so that throughout the process the distribution is always
canonical; in particular, $f(t^{\prime})=f_{\text{can}}(t^{\prime})$. The
argument above can be generalized considerably by allowing heat exchanges or
by introducing uncertainties into the actual Hamiltonian dynamics.

To summarize, the chain of steps is
\begin{equation}
S_{T}(t)\underset{\text{(1)}}{=}S_{G}^{\text{can}}(t)\underset{\text{(2)}}%
{=}S_{G}(t^{\prime})\underset{\text{(3)}}{\leq}S_{G}^{\text{can}}(t^{\prime
})\underset{\text{(4)}}{=}S_{T}(t^{\prime})~.\label{2nd law f}%
\end{equation}
Steps (1) and (4) hinge on identifying the maximized Gibbs entropy with the
thermodynamic entropy -- which works provided we have correctly identified the
relevant macrovariables for the particular problem at hand. Step (2) follows
from the constancy of the Gibbs entropy under Hamiltonian evolution -- this is
the least controversial step. Of course, if we did not have complete knowledge
about the exact Hamiltonian $H(t)$ acting on the system an inequality would
have been introduced already at this point. The crucial inequality, however,
is introduced in step (3) where \emph{information is discarded}. The
distribution $f(t^{\prime})$ contains information about the macrovariables
$X_{t^{\prime}}$ at time $t^{\prime}$, and since the Hamiltonian is known, it
also contains information about the values $X_{t}$ the macrovariables had at
the initial time $t$. In contrast, a description in terms of the distribution
$f_{\text{can}}(t^{\prime})$ contains information about the macrovariables
$X_{t^{\prime}}$ at time $t^{\prime}$ \emph{and nothing else}. In a
thermodynamic description all memory of the history of the system is lost.

The Second Law refers to thermodynamic entropies only. These entropies measure
the amount of information available to someone with only macroscopic means to
observe and manipulate the system. \emph{The irreversibility implicit in the
Second Law arises from this restriction to thermodynamic descriptions.}

It is important to emphasize what has just been proved: in an adiabatic
process from one state of equilibrium to another the \emph{thermodynamic}
entropy increases. This is the Second Law. Many questions remain unanswered:
We have assumed that the system tends towards and finally reaches an
equilibrium; how do we know that this happens? What are the relaxation times,
transport coefficients, etc.? There are all sorts of aspects of
non-equilibrium irreversible processes that remain to be explained but this
does not detract from what Jaynes' explanation did in fact accomplish, namely,
it explained the Second Law, no more and, most emphatically, no less.

\section{The thermodynamic limit}

If the Second Law \textquotedblleft has only statistical
certainty\textquotedblright\ (Maxwell, 1871) and any violation
\textquotedblleft seems to be reduced to improbability\textquotedblright%
\ (Gibbs, 1878)\ how can thermodynamic predictions attain so much certainty?
Part of the answer hinges on restricting the kind of questions we are willing
to ask to those concerning the few macroscopic variables over which we have
some control. Most other questions are not \textquotedblleft
interesting\textquotedblright\ and thus they are never asked. For example,
suppose we are given a gas in equilibrium within a cubic box, and the question
is where will particle \#23 be found. The answer is that we expect the
particle to be at the center of the box but with a very large standard
deviation -- the particle can be anywhere in the box. The answer is not
particularly impressive. On the other hand, if we ask for the energy of the
gas at temperature $T$, or how it changes as the volume is changed by $\delta
V$, then the answers are truly impressive.

Consider a system in thermal equilibrium in a macrostate described by a
canonical distribution $f(z)$ assigned on the basis of constraints on the
values of certain macrovariables $X$. For simplicity we will assume $X$ is a
single variable, the energy, $X=\left\langle E\right\rangle =\bar{E}$. The
microstates $z$ can be divided into typical and atypical microstates. The
typical microstates are all contained within a \textquotedblleft high
probability\textquotedblright\ region $\mathcal{R}_{\varepsilon}$ to be
defined below that has total probability $1-\varepsilon$, where $\varepsilon$
is a small positive number, and within which $f(z)$ is greater than some lower
bound. The \textquotedblleft phase\textquotedblright\ volume of the typical
region is
\begin{equation}
\operatorname*{Vol}(\mathcal{R}_{\varepsilon})=%
{\textstyle\int\nolimits_{\mathcal{R}_{\varepsilon}}}
dz=W_{\varepsilon}~.
\end{equation}
Our goal is to establish that the thermodynamic entropy and the volume of the
region $\mathcal{R}_{\varepsilon}$ are related through Boltzmann's equation,
\begin{equation}
S_{T}\approx k\log W_{\varepsilon}~.
\end{equation}
The surprising feature is that the result is essentially independent of
$\varepsilon$. The following theorems which are adaptations of the Asymptotic
Equipartition Property (section 4.6) state this result in a mathematically
precise way.

\noindent\textbf{Theorem:} Let $f(z)$ be the canonical distribution and
$kS=S_{G}=S_{T}$ the corresponding entropy,
\begin{equation}
f(z)=\frac{e^{-\beta E(z)}}{Z}\quad\text{and}\quad S=\beta\bar{E}+\log
Z~.~\label{canon dist x}%
\end{equation}
Then as $N\rightarrow\infty$,
\begin{equation}
-\frac{1}{N}\log f(z)\longrightarrow\frac{S}{N}\quad\text{in probability, }%
\end{equation}
provided that the system is such that the energy fluctuations increase slower
than $N$, that is, $\lim_{N\rightarrow\infty}\Delta E/N=0$. ($\Delta$ denotes
the standard deviation.)

The theorem roughly means that

\begin{description}
\item \emph{The accessible microstates are essentially equally likely}.
\end{description}

\noindent Microstates $z$ for which $(-\log f(z))/N$ differs substantially
from $S/N$ have either too low probability and are deemed \textquotedblleft
inaccessible,\textquotedblright\ or they might individually have a high
probability but are too few to contribute significantly.

\noindent\textbf{Remark: }The word `essentially' is tricky because $f(z)$ may
differ from $e^{-S}$ by a huge factor, but $\log f(z)$ differs from $-S$ by an
unimportant amount that grows less rapidly than $N$.

\noindent\textbf{Remark:} Note that the theorem applies only to those systems
with interparticle interactions such that the energy fluctuations are
sufficiently well behaved. Typically this requires that as $N$ and $V$ tend to
infinity with $N/V$ constant, the spatial correlations fall sufficiently fast
that distant particles are uncorrelated. Under these circumstances energy and
entropy are extensive quantities.

\noindent\textbf{Proof:} Apply the Tchebyshev inequality (see section 2.8),
\begin{equation}
P\left(  \left\vert x-\langle x\rangle\right\vert \geq\varepsilon\right)
\leq\left(  \frac{\Delta x}{\varepsilon}\right)  ^{2}\text{ ,}%
\end{equation}
to the variable
\begin{equation}
x=\frac{-1}{N}\log f(z)~.
\end{equation}
The mean is the entropy per particle,
\begin{align}
\left\langle x\right\rangle  & =\frac{-1}{N}\left\langle \log f\right\rangle
\nonumber\\
& =\frac{S}{N}=\frac{1}{N}\left(  \beta\bar{E}+\log Z\right)  ~.
\end{align}
To calculate the variance,
\begin{equation}
(\Delta x)^{2}=\frac{1}{N^{2}}\left[  \left\langle (\log f)^{2}\right\rangle
-\left\langle \log f\right\rangle ^{2}\right]  ~,
\end{equation}
use
\begin{align}
\left\langle \left(  \log f\right)  ^{2}\right\rangle  & =\left\langle \left(
\beta E+\log Z\right)  ^{2}\right\rangle \nonumber\\
& =\beta^{2}\left\langle E^{2}\right\rangle +2\beta\left\langle E\right\rangle
\log Z+\left(  \log Z\right)  ^{2}~,
\end{align}
so that
\begin{equation}
(\Delta x)^{2}=\frac{\beta^{2}}{N^{2}}\left(  \left\langle E^{2}\right\rangle
-\left\langle E\right\rangle ^{2}\right)  =\left(  \frac{\beta\Delta E}%
{N}\right)  ^{2}~.
\end{equation}
Collecting these results gives
\begin{equation}
\operatorname*{Prob}\left[  \left\vert -\frac{1}{N}\log f(z)-\frac{S}%
{N}\right\vert \geq\varepsilon\right]  \leq\left(  \frac{\beta\Delta
E}{N\varepsilon}\right)  ^{2}~.
\end{equation}
For systems such that the relative energy fluctuations $\Delta E/\bar{E}$ tend
to $0$ as $N^{-1/2}$ when $N\rightarrow\infty$, and the energy is an extensive
quantity, $\bar{E}\propto N$, the limit on the right is zero, $\Delta
E/N\rightarrow0$, therefore,%

\begin{equation}
\lim_{N\rightarrow\infty}\operatorname*{Prob}\left[  \left\vert -\frac{1}%
{N}\log f(z)-\frac{S}{N}\right\vert \geq\varepsilon\right]
=0~,\label{typ microstates a}%
\end{equation}
which concludes the proof.

The following theorem elaborates on these ideas further. To be precise let us
define the typical region $\mathcal{R}_{\varepsilon}$ as the set of
microstates with probability $f(z)$ such that
\begin{equation}
e^{-S-N\varepsilon}\leq f(z)\leq e^{-S+N\varepsilon}%
\,,\label{typ microstates b}%
\end{equation}
or, using eq.(\ref{canon dist x}),
\begin{equation}
\frac{1}{Z}e^{-\beta\bar{E}-N\varepsilon}\leq f(z)\leq\frac{1}{Z}e^{-\beta
\bar{E}+N\varepsilon}~.
\end{equation}
This last expression shows that typical microstates are those for which the
energy per particle $E(z)/N$ lies within a narrow interval $2\varepsilon kT$
about the expected value $\bar{E}/N$.

\noindent\textbf{Remark:} Even though some states $z$ (namely those with
energy $E(z)<\bar{E}$) can individually be more probable than the typical
states it turns out (see below) that they are too few and their volume is
negligible compared to $W_{\epsilon}$.

\noindent\textbf{Theorem of typical microstates:} For $N$ sufficiently large

\begin{enumerate}
\item[\textbf{(1)}] $\operatorname*{Prob}[\mathcal{R}_{\varepsilon
}]>1-\varepsilon$

\item[\textbf{(2)}] $\operatorname*{Vol}(\mathcal{R}_{\varepsilon
})=W_{\varepsilon}\leq e^{S+N\varepsilon}$.

\item[\textbf{(3)}] $W_{\varepsilon}\geq(1-\varepsilon)e^{S-N\varepsilon} $.

\item[\textbf{(4)}] $\lim_{N\rightarrow\infty}(\log W_{\varepsilon}-S)/N=0 $.
\end{enumerate}

\noindent In words:

\begin{description}
\item \qquad\emph{The typical region has probability close to one; typical
microstates are almost equally probable; the phase volume they occupy is about
}$e^{S_{T}/k}$\emph{, that is, }$S_{T}=k\log W$\emph{.}
\end{description}

\noindent The Gibbs entropy is a measure of the logarithm of the phase volume
of typical states and for large $N$ it does not much matter what we mean by
typical (i.e., what we choose for $\varepsilon$). Incidentally, note that it
is the Gibbs entropy that satisfies the Boltzmann formula $S_{G}=k\log W$.

\noindent\textbf{Proof:} Eq.(\ref{typ microstates a}) states that for fixed
$\varepsilon$, for any given $\delta$ there is an $N_{\delta}$ such that for
all $N>N_{\delta}$, we have
\begin{equation}
\operatorname*{Prob}\left[  \left\vert -\frac{1}{N}\log f(z)-\frac{S}%
{N}\right\vert \leq\varepsilon\right]  \geq1-\delta\,.
\end{equation}
Thus, the probability that the microstate $z$ is $\varepsilon$-typical tends
to one, and therefore so must $\operatorname*{Prob}[\mathcal{R}_{\varepsilon
}]$. Setting $\delta=\varepsilon$ yields part (\textbf{1)}. This also shows
that the total probability of the set of states with $E(z)<\bar{E}$ is
negligible -- they must occupy a negligible volume. To prove \textbf{(2)}
write
\begin{align}
1  & \geq\operatorname*{Prob}[\mathcal{R}_{\varepsilon}]=%
{\textstyle\int\nolimits_{\mathcal{R}_{\varepsilon}}}
dz\,f(z)\nonumber\\
& \geq e^{-S-N\varepsilon}%
{\textstyle\int\nolimits_{\mathcal{R}_{\varepsilon}}}
dz\,=e^{-S-N\varepsilon}W_{\varepsilon}\,.
\end{align}
Similarly, to prove \textbf{(3)} use \textbf{(1)},\textbf{\ }%
\begin{align}
1-\varepsilon & <\operatorname*{Prob}[\mathcal{R}_{\varepsilon}]=%
{\textstyle\int\nolimits_{\mathcal{R}_{\varepsilon}}}
dz\,f(z)\nonumber\\
& \leq e^{-S+N\varepsilon}%
{\textstyle\int\nolimits_{\mathcal{R}_{\varepsilon}}}
dz=e^{-S+N\varepsilon}W_{\varepsilon}\,,
\end{align}
Finally, from (2) and (3),
\begin{equation}
(1-\varepsilon)e^{S-N\varepsilon}\leq W_{\varepsilon}\leq e^{S+N\varepsilon}~,
\end{equation}
which is the same as
\begin{equation}
\frac{S}{N}-\varepsilon+\frac{\log(1-\varepsilon)}{N}\leq\frac{\log
W_{\varepsilon}}{N}\leq\frac{S}{N}+\varepsilon~,
\end{equation}
and proves \textbf{(4)}.

\noindent\textbf{Remark:} The theorems above can be generalized to situations
involving several macrovariables $X^{k}$ in addition to the energy. In this
case, the expected value of $\log f(z)$ is
\begin{equation}
\left\langle -\log f\right\rangle =S=\lambda_{k}\left\langle X^{k}%
\right\rangle +\log Z~,
\end{equation}
and its variance is
\begin{equation}
\left(  \Delta\log f\right)  ^{2}=\lambda_{k}\lambda_{m}\left(  \left\langle
X^{k}X^{m}\right\rangle -\left\langle X^{k}\right\rangle \left\langle
X^{m}\right\rangle \right)  ~.
\end{equation}

\section[Interpretation of the Second Law]{Interpretation of the Second Law:
Reproducibility}

We saw that the Gibbs entropy is a measure of the logarithm of the phase
volume of typical states. In the proof of the Second Law (section 4.11.1) we
started with a system at time $t$ in a state of thermal equilibrium defined by
the macrovariables $X_{t}$. We saw (section 4.11.2) that within the typical
region $\mathcal{R}(t)$ fluctuations of the $X_{t}$ are negligible: all
microstates are characterized by the same values of $X$. Furthermore, the
typical region $\mathcal{R}(t)$ includes essentially all possible initial
states compatible with the initial $X_{t}$.

The volume $W(t)=e^{S_{T}(t)/k}$ of the typical region can be interpreted in
two ways. On one hand it is a measure of our ignorance as to the true
microstate when all we know are the macrovariables $X_{t}$. On the other hand,
the volume $W(t)$ is also a measure of the extent that we can control the
actual microstate of the system when the $X_{t}$ are the only parameters we
can manipulate.

Having been prepared in equilibrium at time $t$ the system is then subjected
to an adiabatic process and it eventually attains a new equilibrium at time
$t^{\prime}$. The Hamiltonian evolution deforms the initial region
$\mathcal{R}(t)$ into a new region $\mathcal{R}(t^{\prime})$ with exactly the
same volume $W(t)=W(t^{\prime})$; the macrovariables evolve from their initial
values $X_{t}$ to new values $X_{t^{\prime}}$.

Now suppose we adopt a thermodynamic description for the new equilibrium; the
preparation history is forgotten, and all we know are the new values
$X_{t^{\prime}}$. The new typical region $\mathcal{R}^{\prime}(t^{\prime})$
has a volume $W^{\prime}(t^{\prime})$ and it includes all microstates
compatible with the information $X_{t^{\prime}}$.

After these preliminaries we come to the crux of the argument: With the
limited experimental means at our disposal we can guarantee that the initial
microstate will be somewhere within $W(t)$ and therefore that in due course of
time it will be within $W(t^{\prime})$. In order for the process
$X_{t}\rightarrow X_{t^{\prime}}$ to be experimentally reproducible it must be
that all microstates in $W(t^{\prime})$ will also be within $W^{\prime
}(t^{\prime})$ which means that $W(t)=W(t^{\prime})\leq W^{\prime}(t^{\prime
})$. Conversely, if it were true that $W(t)>W^{\prime}(t^{\prime})$ we would
sometimes observe that an initial microstate within $W(t)$ would evolve into a
final microstate lying outside $W^{\prime}(t^{\prime})$ that is, sometimes we
would observe $X_{t}\nrightarrow X_{t^{\prime}}$. Thus, when $W(t)>W^{\prime
}(t^{\prime})$ the experiment is not reproducible.

A new element has been introduced into the discussion of the Second Law:
\emph{reproducibility}. [Jaynes 65] Thus, we can express the Second Law in the
somewhat tautological form:

\begin{description}
\item \qquad\emph{In a reproducible adiabatic process the thermodynamic
entropy cannot decrease.}
\end{description}

We can address this question from a different angle: How do we know that the
chosen constraints $X$ are the relevant macrovariables that provide an
\emph{adequate} thermodynamic description? In fact, what do we mean by an
\emph{adequate} description? Let us rephrase these questions differently:
Could there exist additional unknown physical constraints $Y$ that
significantly restrict the microstates compatible with the initial macrostate
and which therefore provide an even better description? The answer is that
such variables can, of course, exist but that including them in the
description does not necessarily lead to an improvement. If the process
$X_{t}\rightarrow X_{t^{\prime}}$ is reproducible when no particular care has
been taken to control the values of $Y$ we can expect that to the extent that
we are only interested in the $X$'s the $Y$'s are irrelevant; keeping track of
them will not yield a better description. \emph{Reproducibility is the
criterion whereby we can decide whether a particular thermodynamic description
is adequate or not. }

\section{Remarks on irreversibility}

A considerable source of confusion on the question of reversibility is that
the same word `reversible' is used with several different meanings [Uffink 01]:

\noindent(a) \emph{Mechanical or microscopic reversibility} refers to the
possibility of reversing the velocities of every particle. Such reversals
would allow the system not just to retrace its steps from the final macrostate
to the initial macrostate but it would also allow it to retrace its detailed
microstate trajectory as well.

\noindent(b) \emph{Carnot or macroscopic reversibility} refers to the
possibility of retracing the history of \newline macrostates of a system in
the opposite direction. The required amount of control over the system can be
achieved by forcing the system along a prescribed path of intermediate
macroscopic equilibrium states that are infinitesimally close to each other.
Such a reversible process is normally and appropriately called
\emph{quasi-static}. There is no implication that the trajectories of the
individual particles will be retraced.

\noindent(c) \emph{Thermodynamic reversibility} refers to the possibility of
starting from a final macrostate and completely recovering the initial
macrostate without any other external changes. There is no need to retrace the
intermediate macrostates in reverse order. In fact, rather than
`reversibility' it may be more descriptive to refer to `\emph{recoverability}%
'. Typically a state is irrecoverable when there is friction, decay, or
corruption of some kind.

Notice that when one talks about the \textquotedblleft
irreversibility\textquotedblright\ of the Second Law and about the
\textquotedblleft reversibility\textquotedblright\ of mechanics there is no
inconsistency or contradiction: the word `reversibility' is being used with
two entirely different meanings.

Classical thermodynamics assumes that isolated systems approach and eventually
attain a state of equilibrium. The state of equilibrium is, by definition, a
state that, once attained, will not spontaneously change in the future. On the
other hand, it is understood that changes might have happened in the past.
Classical thermodynamics introduces a time asymmetry: it treats the past and
the future differently.

The situation with statistical mechanics is, however, somewhat different. Once
equilibrium has been attained fluctuations are possible. In fact, if we wait
long enough we can expect that large fluctuations can be expected to happen in
the future, just as they might have happened in the past. The situation is
quite symmetric. The interesting asymmetry arises when we realize that for a
large fluctuation to happen spontaneously in the future might require an
extremely long time while we just happen to know that a similarly large
\textquotedblleft fluctuation\textquotedblright\ was observed in the very
recent past. This might seem strange because the formalism of statistical
mechanics does not introduce any time asymmetry. The solution to the puzzle is
that the large \textquotedblleft fluctuation\textquotedblright\ in the recent
past most likely did not happen spontaneously but was quite deliberately
brought about by human (or otherwise) intervention. The system was prepared in
some unusual state by applying appropriate constraints which were subsequently
removed -- we do this all the time.

\section{Entropies, descriptions and the Gibbs paradox}

Under the generic title of \textquotedblleft Gibbs Paradox\textquotedblright%
\ one usually considers a number of related questions in both phenomenological
thermodynamics and in statistical mechanics: (1) The entropy change when two
distinct gases are mixed happens to be independent of the nature of the gases.
Is this in conflict with the idea that in the limit as the two gases become
identical the entropy change should vanish? (2) Should the thermodynamic
entropy of Clausius be an extensive quantity or not? (3) Should two
microstates that differ only in the exchange of identical particles be counted
as two or just one microstate?

The conventional wisdom asserts that the resolution of the paradox rests on
quantum mechanics but this analysis is unsatisfactory; at best it is
incomplete. While it is true that the exchange of identical quantum particles
does not lead to a new microstate this approach ignores the case of classical,
and even non-identical particles. For example, nanoparticles in a colloidal
suspension or macromolecules in solution are both classical and non-identical.
Several authors (e.g., [Grad 61, Jaynes 92]) have recognized that quantum
theory has no bearing on the matter; indeed, as remarked in section 3.5, this
was already clear to Gibbs.

Our purpose here is to discuss the Gibbs paradox from the point of view of
information theory. The discussion follows [Tseng Caticha 01]. Our conclusion
will be that the paradox is resolved once it is realized that there is no such
thing as \emph{the} entropy of a system, that there are \emph{many} entropies.
The choice of entropy is a choice between a description that treats particles
as being distinguishable and a description that treats them as
indistinguishable; which of these alternatives is more convenient depends on
the resolution of the particular experiment being performed.

The \textquotedblleft grouping\textquotedblright\ property of entropy,
eq.(\ref{grouping property}),
\[
S[p]=S_{G}[P]+%
{\textstyle\sum\nolimits_{g}}
P_{g}S_{g}[p_{\cdot|g}]
\]
plays an important role in our discussion. It establishes a relation between
two different descriptions and refers to three different entropies. One can
describe the system with high resolution as being in a microstate $i$ (with
probability $p_{i}$), or alternatively, with lower resolution as being in one
of the groups $g$ (with probability $P_{g}$). Since the description in terms
of the groups $g$ is less detailed we might refer to them as `mesostates'. A
thermodynamic description, on the other hand, corresponds to an even lower
resolution that merely specifies the equilibrium macrostate. For simplicity,
we will define the macrostate with a single variable, the energy. Including
additional variables is easy and does not modify the gist of the argument.

The standard connection between the thermodynamic description in terms of
macrostates and the description in terms of microstates is established in
section 4.10.4. If the energy of microstate $a$ is $E_{a}$, to the macrostate
of energy $\bar{E}=\left\langle E\right\rangle $ we associate that canonical
distribution (\ref{canon dist b})
\begin{equation}
p_{a}=\frac{e^{-\beta E_{a}}}{Z_{H}}\,,\label{can dist H}%
\end{equation}
where the partition function $Z_{H}$ and the Lagrange multiplier $\beta$ are
determined from eqs.(\ref{E and Z}),
\begin{equation}
Z_{H}=%
{\textstyle\sum\limits_{i}}
e^{-\beta E_{i}}\quad\text{and\quad}\frac{\partial\log Z_{H}}{\partial\beta
}=-\bar{E}~.
\end{equation}
The corresponding entropy, eq.(\ref{S(E,V)}) is (setting $k=1$)
\begin{equation}
S_{H}=\beta\bar{E}+\log Z_{H}\,,
\end{equation}
measures the amount of information required to specify the microstate when all
we know is the value $\bar{E}$ .

\subsection*{Identical particles}

Before we compute and interpret the probability distribution over mesostates
and its corresponding entropy we must be more specific about which mesostates
we are talking about. Consider a system of $N$ classical particles that are
exactly identical. The interesting question is whether these identical
particles are also \textquotedblleft distinguishable.\textquotedblright\ By
this we mean the following: we look at two particles now and we label them. We
look at the particles later. Somebody might have switched them. Can we tell
which particle is which? The answer is: it depends. Whether we can distinguish
identical particles or not depends on whether we were able and willing to
follow their trajectories.

A slightly different version of the same question concerns an $N$-particle
system in a certain state. Some particles are permuted. Does this give us a
different state? As discussed earlier the answer to this question requires a
careful specification of what we mean by a state.

Since by a \emph{microstate} we mean a point in the $N$-particle phase space,
then a permutation does indeed lead to a new microstate. On the other hand,
our concern with permutations suggests that it is useful to introduce the
notion of a \emph{mesostate} defined as the group of those $N!$ microstates
that are obtained as permutations of each other. With this definition it is
clear that a permutation of the identical particles does not lead to a new mesostate.

Now we can return to discussing the connection between the thermodynamic
macrostate description and the description in terms of mesostates using, as
before, the Method of Maximum Entropy. Since the particles are (sufficiently)
identical, all those $N!$ microstates $i$ within the same mesostate $g$ have
the same energy, which we will denote by $E_{g}$ (\emph{i.e.}, $E_{i}=E_{g}$
for all $i\in g$). To the macrostate of energy $\bar{E}=\left\langle
E\right\rangle $ we associate the canonical distribution,
\begin{equation}
P_{g}=\frac{e^{-\beta E_{g}}}{Z_{L}}\,,\label{can dist L}%
\end{equation}
where
\begin{equation}
Z_{L}=%
{\textstyle\sum\limits_{g}}
e^{-\beta E_{g}}\quad\text{and\quad}\frac{\partial\log Z_{L}}{\partial\beta
}=-\bar{E}\,.
\end{equation}
The corresponding entropy, eq.(\ref{S(E,V)}) is (setting $k=1$)
\begin{equation}
S_{L}=\beta\bar{E}+\log Z_{L}\,,
\end{equation}
measures the amount of information required to specify the mesostate when all
we know is $\bar{E}$.

Two different entropies $S_{H}$ and $S_{L}$ have been assigned to the same
macrostate $\bar{E}$; they measure the different amounts of additional
information required to specify the state of the system to a high resolution
(the microstate) or to a low resolution (the mesostate).

The relation between $Z_{H}$ and $Z_{L}$ is obtained from
\begin{equation}
Z_{H}=%
{\textstyle\sum\limits_{i}}
e^{-\beta E_{i}}=N!%
{\textstyle\sum\limits_{g}}
e^{-\beta E_{g}}=N!Z_{L}\quad\text{or}\quad Z_{L}=\frac{Z_{H}}{N!}%
\,.\label{ZLZH}%
\end{equation}
The relation between $S_{H}$ and $S_{L}$ is obtained from the
\textquotedblleft grouping\textquotedblright\ property,
eq.(\ref{grouping property}), with $S=S_{H}$ and $S_{G}=S_{L}$, and
$p_{i|g}=1/N!$. The result is
\begin{equation}
S_{L}=S_{H}-\log N!\,.\label{SLSH}%
\end{equation}
Incidentally, note that
\begin{equation}
S_{H}=-%
{\textstyle\sum\nolimits_{a}}
p_{a}\log p_{a}=-%
{\textstyle\sum\nolimits_{g}}
P_{g}\log P_{g}/N!~.
\end{equation}
Equations (\ref{ZLZH}) and (\ref{SLSH}) both exhibit the Gibbs $N!$
\textquotedblleft corrections.\textquotedblright\ Our analysis shows (1) that
the justification of the $N!$ factor is not to be found in quantum mechanics,
and (2) that the $N!$ does not correct anything. The $N!$ is not a fudge
factor that fixes a wrong (possibly nonextensive) entropy $S_{H}$ into a
correct (possibly extensive) entropy $S_{L}$. Both entropies $S_{H}$ and
$S_{L}$ are correct. They differ because they measure different things: one
measures the information to specify the microstate, the other measures the
information to specify the mesostate.

An important goal of statistical mechanics is to provide a justification, an
explanation of thermodynamics. Thus, we still need to ask which of the two
statistical entropies, $S_{H}$ or $S_{L}$, should be identified with the
thermodynamic entropy of Clausius $S_{T}$. Inspection of eqs.(\ref{ZLZH}) and
(\ref{SLSH}) shows that, as long as one is not concerned with experiments that
involve changes in the number of particles, the same thermodynamics will
follow whether we set $S_{H}=S_{T}$ or $S_{L}=S_{T}$.

But, of course, experiments involving changes in $N$ are very important (for
example, in the equilibrium between different phases, or in chemical
reactions). Since in the usual thermodynamic experiments we only care that
some number of particles has been exchanged, and we do not care which were the
actual particles exchanged, we expect that the correct identification is
$S_{L}=S_{T}$. Indeed, the quantity that regulates the equilibrium under
exchanges of particles is the chemical potential defined by
\begin{equation}
\mu=-kT\left(  \frac{\partial S_{T}}{\partial N}\right)  _{E,V,\ldots}%
\end{equation}
The two identifications $S_{H}=S_{T}$ or $S_{L}=S_{T}$, lead to two different
chemical potentials, related by
\begin{equation}
\mu_{L}=\mu_{H}-NkT\,.
\end{equation}
It is easy to verify that, under the usual circumstances where surface effects
can be neglected relative to the bulk, $\mu_{L}$ has the correct functional
dependence on $N$: it is intensive and can be identified with the
thermodynamic $\mu$. On the other hand, $\mu_{H}$ is not an intensive quantity
and cannot therefore be identified with $\mu$.

\subsection*{Non-identical particles}

We saw that classical identical particles can be treated, depending on the
resolution of the experiment, as being distinguishable or indistinguishable.
Here we go further and point out that even non-identical particles can be
treated as indistinguishable. Our goal is to state explicitly in precisely
what sense it is up to the observer to decide whether particles are
distinguishable or not.

We defined a mesostate as a subset of $N!$ microstates that are obtained as
permutations of each other. With this definition it is clear that a
permutation of particles does not lead to a new mesostate even if the
exchanged particles are not identical. This is an important extension because,
unlike quantum particles, classical particles cannot be expected to be exactly
identical down to every minute detail. In fact in many cases the particles can
be grossly different -- examples might be colloidal suspensions or solutions
of organic macromolecules. A high resolution device, for example an electron
microscope, would reveal that no two colloidal particles or two macromolecules
are exactly alike. And yet, for the purpose of modelling most of our
macroscopic observations it is not necessary to take account of the myriad
ways in which two particles can differ.

Consider a system of $N$ particles. We can perform rather crude macroscopic
experiments the results of which can be summarized with a simple
phenomenological thermodynamics where $N$ is one of the relevant variables
that define the macrostate. Our goal is to construct a statistical foundation
that will explain this macroscopic model, reduce it, so to speak, to
\textquotedblleft first principles.\textquotedblright\ The particles might
ultimately be non-identical, but the crude phenomenology is not sensitive to
their differences and can be explained by postulating mesostates $g$ and
microstates $i$ with energies $E_{i}\approx E_{g}$, for all $i\in g$, as if
the particles were identical. As in the previous section this statistical
model gives%

\begin{equation}
Z_{L}=\frac{Z_{H}}{N!}\quad\text{with\quad}Z_{H}=\sum_{i}e^{-\beta E_{i}}\,,
\end{equation}
and the connection to the thermodynamics is established by postulating
\begin{equation}
S_{T}=S_{L}=S_{H}-\log N!\,.
\end{equation}

Next we consider what happens when more sophisticated experiments are
performed. The examples traditionally offered in discussions of this sort
refer to the new experiments that could be made possible by the discovery of
membranes that are permeable to some of the $N$ particles but not to the
others. Other, perhaps historically more realistic examples, are afforded by
the availability of new experimental data, for example, more precise
measurements of a heat capacity as a function of temperature, or perhaps
measurements in a range of temperatures that had previously been inaccessible.

Suppose the new phenomenology can be modelled by postulating the existence of
two kinds of particles. (Experiments that are even more sophisticated might
allow us to detect three or more kinds, perhaps even a continuum of different
particles.) What we previously thought were $N$ identical particles we will
now think as being $N_{a}$ particles of type $a$ and $N_{b}$ particles of type
$b$. The new description is in terms of macrostates defined by $N_{a}$ and
$N_{b}$ as the relevant variables.

To construct a statistical explanation of the new phenomenology from `first
principles' we need to revise our notion of mesostate. Each new mesostate will
be a group of microstates which will include all those microstates obtained by
permuting the $a$ particles among themselves, and by permuting the $b$
particles among themselves, but will not include those microstates obtained by
permuting $a$ particles with $b$ particles. The new mesostates, which we will
label $\hat{g}$ and to which we will assign energy $\varepsilon_{\hat{g}}$,
will be composed of $N_{a}!N_{b}!$ microstates $\hat{\imath}$, each with a
well defined energy $E_{\hat{\imath}}=E_{\hat{g}}$, for all $\hat{\imath}%
\in\hat{g}$. The new statistical model gives%

\begin{equation}
\hat{Z}_{L}=\frac{\hat{Z}_{H}}{N_{a}!N_{b}!}\quad\text{with\quad}\hat{Z}%
_{H}=\sum_{\hat{\imath}}e^{-\beta E_{\hat{\imath}}}\,,
\end{equation}
and the connection to the new phenomenology is established by postulating
\begin{equation}
\hat{S}_{T}=\hat{S}_{L}=\hat{S}_{H}-\log N_{a}!N_{b}!\,.
\end{equation}

In discussions of this topic it is not unusual to find comments to the effect
that in the limit as particles $a$ and $b$ become identical one expects that
the entropy of the system with two kinds of particles tends to the entropy of
a system with just one kind of particle. The fact that this expectation is not
met is one manifestation of the Gibbs paradox.

From the information theory point of view the paradox does not arise because
there is no such thing as \emph{the entropy of the system}, there are several
entropies. It is true that as $a\rightarrow b$ we will have $\hat{Z}%
_{H}\rightarrow Z_{H}$, and accordingly $\hat{S}_{H}\rightarrow S_{H}$, but
there is no reason to expect a similar relation between $\hat{S}_{L}$ and
$S_{L}$ because these two entropies refer to mesostates $\hat{g}$ and $g$ that
remain different even as $a$ and $b$ became identical. In this limit the
mesostates $\hat{g}$, which are useful for descriptions that treat particles
$a$ and $b$ as indistinguishable among themselves but distinguishable from
each other, lose their usefulness.

\subsection*{Conclusion}

The Gibbs paradox in its various forms arises from the widespread
misconception that entropy is a real physical quantity and that one is
justified in talking about \emph{the entropy} of the system. The thermodynamic
entropy is not a property of the system. Entropy is a property of our
description of the system, it is a property of the macrostate. More
explicitly, it is a function of the macroscopic variables used to define the
macrostate. To different macrostates reflecting different choices of variables
there correspond different entropies for the very same system.

But this is not the complete story: entropy is not just a function of the
macrostate. Entropies reflect a relation between two descriptions of the same
system: in addition to the macrostate, we must also specify the set of
microstates, or the set of mesostates, as the case might be. Then, having
specified the macrostate, an entropy can be interpreted as the amount of
additional information required to specify the microstate or mesostate. We
have found the `grouping' property very valuable precisely because it
emphasizes the dependence of entropy on the choice of micro or
mesostates.\newpage

\thispagestyle{empty}

\bigskip

\chapter{Entropy III: Updating Probabilities}

The general problem of inductive inference is to update from a prior
probability distribution to a posterior distribution when new information
becomes available. The challenge is to develop updating methods that are both
systematic and objective. In Chapter 2 we saw that Bayes' rule is the natural
way to update when the information is in the form of data. We also saw that
Bayes' rule could not be derived just from the requirements of consistency
implicit in the sum and product rules of probability theory. An additional
Principle of Minimal Updating (PMU) was necessary: \emph{Prior information is
valuable and should not be discarded; beliefs should be revised only to the
extent required by the data. }A few interesting questions were just barely
hinted at: How do we update when the information is not in the form of data?
If the information is not data, what else could it possibly be? Indeed what,
after all, is information?

Then in Chapter 4 we saw that the method of maximum entropy, MaxEnt, allowed
one to deal with information in the form of constraints on the allowed
probability distributions. So here we have a partial answer to one of our
questions: in addition to data information can take the form of constraints.
However, MaxEnt is not a method for updating; it is a method for assigning
probabilities on the basis of the constraint information, but it does not
allow us to take into account the information contained in prior distributions.

Thus, Bayes' rule allows for the information contained in arbitrary priors and
in data, but not in arbitrary constraints,\footnote{Bayes' rule can handle
constraints when they are expressed in the form of data that can be plugged
into a likelihood function. Not all constraints are of this kind.} while on
the other hand, MaxEnt can handle arbitrary constraints but not arbitrary
priors. In this chapter we bring those two methods together: by generalizing
the PMU we show how the MaxEnt method can be extended beyond its original
scope, as a rule to assign probabilities, to a full-fledged method for
inductive inference, that is, a method for updating from arbitrary priors
given information in the form of arbitrary constraints. It should not be too
surprising that the extended Maximum Entropy method, which we will henceforth
abbreviate as ME, includes both MaxEnt and Bayes' rule as special cases.

Historically the ME method is a direct descendant of MaxEnt. As we saw in
chapter 4 within the MaxEnt method entropy is interpreted through the Shannon
axioms as a measure of the amount of uncertainty\ or of the amount of
information\ that is missing\ in a probability distribution. We discussed some
limitations of this approach. The Shannon axioms refer to probabilities of
discrete variables; for continuous variables the entropy is not defined. But a
more serious objection was raised: even if we grant that the Shannon axioms do
lead to a reasonable expression for the entropy, to what extent do we believe
the axioms themselves? Shannon's third axiom, the grouping property, is indeed
sort of reasonable, but is it necessary? Is entropy the only consistent
measure of uncertainty or of information? What is wrong with, say, the
standard deviation? Indeed, there exist examples in which the Shannon entropy
does not seem to reflect one's intuitive notion of information [Uffink 95].
Other entropies, justified by a\ different choice of axioms, can be introduced
(prominent examples are [Renyi 61, Tsallis 88]).

From our point of view the real limitation is that neither Shannon nor Jaynes
were concerned with updating probabilities. Shannon was analyzing the capacity
of communication channels and characterizing the potential diversity of
messages generated by information sources (section 4.6). His entropy makes no
reference to prior distributions. On the other hand, as we already mentioned,
Jaynes conceived MaxEnt as a method to assign probabilities on the basis of
constraint information and a fixed underlying measure, not an arbitrary prior.
He never meant to update from one probability distribution to another.

Considerations such as these motivated several attempts to develop ME directly
as a method for updating probabilities without invoking questionable measures
of uncertainty. Prominent among them are [Shore and Johnson 80, Skilling
88-90, Csiszar 91]. The important contribution by Shore and Johnson was the
realization that one could axiomatize the updating method itself rather than
the information measure. Their axioms are justified on the basis of a
fundamental principle of consistency -- if a problem can be solved in more
than one way the results should agree -- but the axioms themselves and other
assumptions they make have raised some objections [Karbelkar 86, Uffink 95]).
Despite such criticism Shore and Johnson's pioneering papers have had an
enormous influence; they identified the correct goal to be achieved.

Another approach to entropy was proposed by Skilling. His axioms are clearly
inspired by those of Shore and Johnson but his approach is different in
several important aspects. in particular Skilling did not explore the
possibility of using his induction method for the purpose for inductive
\emph{inference}, that is, for updating from prior to posterior probabilities.

The primary goal of this chapter is to apply Skilling's method of eliminative
induction to Shore and Johnson's problem of updating probabilities and, in the
process, to overcome the objections that can be raised against either. The
presentation below follows [Caticha 03, Caticha Giffin 06, Caticha 07].

As we argued earlier when developing the theory of degrees of belief, our
general approach differs from the way in which many physical theories have
been developed in the past. The more traditional approach consists of first
setting up the mathematical formalism and then seeking an acceptable
interpretation. The drawback of this procedure is that questions can always be
raised about the uniqueness of the proposed interpretation, and about the
criteria that makes it acceptable or not.

In contrast, here we proceed in the opposite order: we first decide what we
are talking about, what goal we want to achieve, and only then we design a
suitable mathematical formalism. The advantage is that the issue of meaning
and interpretation is resolved from the start. The preeminent example of this
approach is Cox's algebra of probable inference (discussed in chapter 2) which
clarified the meaning and use of the notion of probability: after Cox it was
no longer possible to raise doubts about the legitimacy of the degree of
belief interpretation. A second example is special relativity: the actual
physical significance of the $x$ and $t$ appearing in the mathematical
formalism of Lorentz and Poincare was a matter of controversy until Einstein
settled the issue by deriving the formalism,that is, the Lorentz
transformations, from more basic principles. Yet a third example is the
derivation of the mathematical formalism of quantum theory. [Caticha 98] In
this chapter we explore a fourth example: the concept of relative entropy is
introduced as a tool for reasoning which reduces to the usual entropy in the
special case of uniform priors. There is no need for an interpretation in
terms of heat, multiplicity of states, disorder, uncertainty, or even in terms
of an amount of information. In this approach we find an explanation for why
the search for the meaning of entropy has turned out to be so elusive:
\emph{Entropy needs no interpretation}. \noindent We do not need to know what
`entropy' means; we only need to know how to use it.

Since the PMU is the driving force behind both Bayesian and ME updating it is
worthwhile to investigate the precise relation between the two. We show that
Bayes' rule can be derived as a special case of the ME method.\footnote{This
result was first obtained by Williams (see [Williams 80, Diaconis 82]) long
before the logical status of the ME method -- and therefore the full extent of
its implications -- had been sufficiently clarified.} The virtue of our
derivation, which hinges on translating information in the form of data into a
constraint that can be processed using ME, is that it is particularly clear.
It throws light on Bayes' rule and demonstrates its complete compatibility
with ME updating. A slight generalization of the same ideas shows that
Jeffrey's updating rule (section 2.10.2) is also a special case of the ME
method. Thus, within the ME framework maximum entropy and Bayesian methods are
unified into a single consistent theory of inference.

There is a second function that the ME method must perform in order to fully
qualify as a method of inductive inference: once we have decided that the
distribution of maximum entropy is to be preferred over all others the
following question arises immediately: the maximum of the entropy function is
never infinitely sharp, are we really confident that distributions with
entropy very close to the maximum are totally ruled out? We must find a
quantitative way to assess the extent to which distributions with lower
entropy are ruled out. This matter is addressed following the treatment in
[Caticha 00].

\section{What is information?}

It is not unusual to hear that systems \textquotedblleft
carry\textquotedblright\ or \textquotedblleft contain\textquotedblright%
\ information and that \textquotedblleft information is
physical\textquotedblright. This mode of expression can perhaps be traced to
the origins of information theory in Shannon's theory of communication. We say
that we have received information when among the vast variety of messages that
could conceivably have been generated by a distant source, we discover which
particular message was actually sent. It is thus that the message
\textquotedblleft carries\textquotedblright\ information. The analogy with
physics is straightforward: the set of all possible states of a physical
system can be likened to the set of all possible messages, and the actual
state of the system corresponds to the message that was actually sent. Thus,
the system \textquotedblleft conveys\textquotedblright\ a message: the system
\textquotedblleft carries\textquotedblright\ information about its own state.
Sometimes the message might be difficult to read, but it is there nonetheless.

This language -- information is physical\ -- useful as it has turned out to
be, does not exhaust the meaning of the word `information'. The goal of
information theory, or better, communication theory, is to characterize the
sources of information, to measure the capacity of the communication channels,
and to learn how to control the degrading effects of noise. It is somewhat
ironic but nevertheless true that this \textquotedblleft
information\textquotedblright\ theory\ is unconcerned with the central
Bayesian issue of how the message affects the beliefs of a rational agent.

\emph{A fully Bayesian information theory demands an explicit account of the
relation between information and beliefs.}

The notion that the theory for reasoning with incomplete information is the
theory of degrees of rational belief led us to tackle two different
problems.\footnote{We mentioned earlier, and emphasize again here, that the
qualifier `rational' is crucial: we are interested in the reasoning of an
idealized rational agent and not of real imperfect humans.} The first was to
understand the conditions required to achieve consistency within a web of
interconnected beliefs. This problem was completely solved: degrees of belief
are consistent when they obey the rules of probability theory, which led us to
conclude that \emph{rational degrees of belief are probabilities}.

The second problem is that of updating probabilities when new information
becomes available. The desire and need to update our beliefs is driven by the
conviction that not all probability assignments are equally good. This bears
on the issue of whether probabilities are subjective, objective, or somewhere
in between. We argued earlier that what makes one probability assignment
better than another is that it better reflects some \textquotedblleft
objective\textquotedblright\ feature of the world, that is, it provides a
better guide to the \textquotedblleft truth\textquotedblright\ -- whatever
this might mean. Therefore objectivity is a desirable goal. It is their
(partial) objectivity that makes probabilities useful. Indeed, what we seek
are updating mechanisms that allow us to process information and incorporate
its objective features into our beliefs.

Bayes' rule behaves precisely in this way. We saw in section 2.10 that as more
and more data are taken into account the original (possibly subjective) prior
becomes less and less relevant, and all rational agents become more and more
convinced of the \emph{same} truth. This is crucial: were it not this way
Bayesian reasoning would not be deemed acceptable.

We are now ready to answer the question `What, after all, is information?' The
result of being confronted with new information\ should be a restriction on
our options as to what we are honestly and rationally allowed to believe.
This, I propose, is the defining characteristic of information. By
information, in its most general form, I mean a set of constraints on the
family of acceptable posterior distributions. Thus,

\begin{description}
\item \emph{Information is whatever constrains rational beliefs}.
\end{description}

\noindent We can phrase this idea somewhat differently. Since our objective is
to update from a prior distribution to a posterior when new information
becomes available we can state that

\begin{description}
\item \emph{Information is what forces a change of beliefs. }\noindent
\end{description}

\noindent An important aspect of this notion is that for a rational agent the
updating is not optional: it is a moral imperative.

Our definition captures an idea of information that is directly related to
changing our minds: information is the driving force behind the process of
learning. Note also that although there is no need to talk about amounts of
information, whether measured in units of bits or otherwise, our notion of
information allows precise quantitative calculations. Indeed, constraints on
the acceptable posteriors are precisely the kind of information the method of
maximum entropy (see below) is designed to handle.

The constraints that convey, or rather, that are information can take a wide
variety of forms. For example, they can represent data (see section 6.5
below), or they can be in the form of expected values (as in statistical
mechanics, see chapter 5). Although one cannot directly measure expected
values or probabilities one can still use them to convey information. This is
what we do, for example, when we specify a prior or the likelihood function --
this is not something that one can measure but by constraining our beliefs
they certainly are valuable information. Constraints can also be specified
through geometrical relations (see section 6.7 and also [Caticha 01, Caticha
Cafaro 07]).

It may be worthwhile to point out an analogy with dynamics -- the study of
change. In Newtonian dynamics the state of motion of a system is described in
terms of momentum -- the \textquotedblleft quantity\textquotedblright\ of
motion -- while the change from one state to another is explained in terms of
an applied force. Similarly, in Bayesian inference a state of belief is
described in terms of probabilities -- a \textquotedblleft
quantity\textquotedblright\ of belief -- and the change from one state to
another is due to information. Just as a force is defined as that which
induces a change from one state of motion to another, so \emph{information is
that which induces a change from one state of belief to another}.

What about prejudices and superstitions? What about divine revelations? Do
they constitute information? Perhaps they lie outside our chosen subject of
ideally rational beliefs, but to the extent that their effects are
indistinguishable from those of other sorts of information, namely, they
affect beliefs, they qualify as information too. Whether the sources of such
information are reliable or not is quite another matter. False information is
information too and even ideally rational agents are affected by false information.

What about limitations in our computational power? They influence our
inferences. Should they be considered information? No. Limited computational
resources may affect the numerical approximation to the value of, say, an
integral, but they do not affect the actual value of the integral. Similarly,
limited computational resources may affect the approximate imperfect reasoning
of real humans and real computers but they do not affect the reasoning of
those ideal rational agents that are the subject of our present concerns.

\section{Entropy as a tool for updating probabilities}

Consider a variable $x$ the value of which is uncertain. The variable can be
discrete or continuous, in one or in several dimensions. For example, $x$
could represent the possible microstates of a physical system, a point in
phase space, or an appropriate set of quantum numbers. The uncertainty about
$x$ is described by a probability distribution $q(x)$. Our goal is to update
from the prior distribution $q(x)$ to a posterior distribution $p(x)$ when new
information -- by which we mean a set of constraints -- becomes available. The
information can be given in terms of expected values but this is not
necessary. The question is: of all those distributions within the family
defined by the constraints, what distribution $p(x)$ should we select?

To select the posterior one could proceed by attempting to place all candidate
distributions in increasing \emph{order of preference}. [Skilling 88]
Irrespective of what it is that makes one distribution preferable over another
it is clear that any ranking according to preference must be transitive: if
distribution $p_{1}$ is preferred over distribution $p_{2}$, and $p_{2}$ is
preferred over $p_{3}$, then $p_{1}$ is preferred over $p_{3}$. Such
transitive rankings are implemented by assigning to each $p(x)$ a real number
$S[p]$ in such a way that if $p_{1}$ is preferred over $p_{2}$, then
$S[p_{1}]>S[p_{2}]$. The selected distribution (one or possibly many, for
there may be several equally preferred distributions) will be that which
maximizes the functional $S[p]$ which we will call the entropy of $p$. We are
thus led to a method of Maximum Entropy (ME) that is a variational method
involving entropies which are real numbers. These are features imposed by
design; they are dictated by the function that the ME method is supposed to perform.

Next, to define the ranking scheme, we must decide on the functional form of
$S[p]$. First, \emph{the purpose of the method is to update from priors to
posteriors.} The ranking scheme must depend on the particular prior $q$ and
therefore the entropy $S$ must be a functional of both $p$ and $q$. The
entropy $S[p,q]$ describes a ranking of the distributions $p$ \emph{relative}
to the given prior $q$. $S[p,q]$ is the entropy of $p$ \emph{relative} to $q$,
and accordingly $S[p,q]$ is commonly called \emph{relative entropy}. This is
appropriate and sometimes we will follow this practice. However, as discussed
in section 4.5, even the `regular' Shannon entropy is relative, it is the
entropy of $p$ relative to an underlying uniform distribution. Since all
entropies are relative to some prior, the qualifier `relative' and is
redundant can be dropped. This is somewhat analogous to the situation with
energy: all energies are relative to some origin or to some reference frame
but we do not feel compelled to constantly refer to the `relative energy'. It
is just taken for granted.

Second, since we deal with incomplete information the method, by its very
nature, cannot be deductive: \emph{the} \emph{method must be inductive}. The
best we can do is generalize from those few special cases where we know what
the preferred distribution should be to the much larger number of cases where
we do not. In order to achieve its purpose, we must assume that $S[p,q]
$\emph{\ }is of \emph{universal }applicability. There is no justification for
this universality beyond the usual pragmatic justification of induction: in
order to avoid the paralysis of not generalizing at all we must risk making
wrong generalizations. An induction method must be allowed to induce.

We will apply the \textbf{Principle of Eliminative Induction }introduced in
chapter 1:

\begin{description}
\item \qquad If a general theory exists it must apply to special cases.

\item \qquad If special examples are known then all candidate theories that
fail to reproduce the known examples are discarded.

\item \qquad If a sufficient number of special examples are known then the
general theory might be completely determined.
\end{description}

\noindent The best we can do is use those special cases where we know what the
preferred distribution should be to eliminate those entropy functionals
$S[p,q]$ that fail to provide the right update. The known\ special cases will
be called (perhaps inappropriately) the \emph{axioms} of the theory. They play
a crucial role: they define what makes one distribution preferable over another.

The three axioms below are chosen to reflect the moral conviction that
information collected in the past and codified into the prior distribution is
very valuable and should not be frivolously discarded. This attitude is
radically conservative: the only aspects of one's beliefs that should be
updated are those for which new evidence has been supplied. This is important
and it is worthwhile to consider it from a different angle. Degrees of belief,
probabilities, are said to be subjective: two different individuals might not
share the same beliefs and could conceivably assign probabilities differently.
But subjectivity does not mean arbitrariness. It is not a blank check allowing
the rational agent to change its mind for no good reason. Valuable prior
information should not be discarded until new information renders it obsolete.

Furthermore, since the axioms do not tell us what and how to update, they
merely tell us what not to update, they have the added bonus of maximizing
objectivity -- there are many ways to change something but only one way to
keep it the same. Thus, we adopt the

\textbf{Principle of Minimal Updating} (PMU): \emph{Beliefs should be updated
only to the extent required by the new information.}

\noindent The three axioms, the motivation behind them, and their consequences
for the functional form of the entropy functional are given below. As will
become immediately apparent the axioms do not refer to merely three cases; any
induction from such a weak foundation would hardly be reliable. The reason the
axioms are convincing and so constraining is that they refer to three
infinitely large classes of known special cases. Detailed proofs are deferred
to the next section.

\textbf{Axiom 1: Locality}. \emph{Local information has local effects.}

\noindent Suppose the information to be processed does not refer to a
particular subdomain $\mathcal{D}$ of the space $\mathcal{X}$ of $x$s. In the
absence of any new information about $\mathcal{D}$ the PMU demands we do not
change our minds about $\mathcal{D}$. Thus, we design the inference method so
that $q(x|\mathcal{D})$, the prior probability of $x$ conditional on
$x\in\mathcal{D}$, is not updated. The selected conditional posterior is
$P(x|\mathcal{D})=q(x|\mathcal{D})$. We emphasize: the point is not that we
make the unwarranted assumption that keeping $q(x|\mathcal{D})$ is guaranteed
to lead to correct inferences. It need not. Induction is risky. The point is,
rather, that in the absence of any evidence to the contrary there is no reason
to change our minds and the prior information takes priority.

The consequence of axiom 1 is that non-overlapping domains of $x$ contribute
additively to the entropy,
\begin{equation}
S[p,q]=\int dx\,F\left(  p(x),q(x),x\right)  \ ,\label{axiom1}%
\end{equation}
where $F$ is some unknown function -- not a functional, just a regular
function of three arguments.

\textbf{Axiom 2: Coordinate invariance.} \emph{The system of coordinates
carries no information. }

\noindent The points $x$ can be labeled using any of a variety of coordinate
systems. In certain situations we might have explicit reasons to believe that
a particular choice of coordinates should be preferred over others. This
information might have been given to us in a variety of ways, but unless the
evidence was in fact given we should not assume it: the ranking of probability
distributions should not depend on the coordinates used.

To grasp the meaning of this axiom it may be useful to recall some facts about
coordinate transformations. Consider a change from old coordinates $x$ to new
coordinates $x^{\prime}$ such that $x=\Gamma(x^{\prime})$. The new volume
element $dx^{\prime}$ includes the corresponding Jacobian,
\begin{equation}
dx=\gamma(x^{\prime})dx^{\prime}\quad\text{where}\quad\gamma(x^{\prime
})=\left\vert \frac{\partial x}{\partial x^{\prime}}\right\vert
.\label{coord jacobian}%
\end{equation}
Let $m(x)$ be any density; the transformed density $m^{\prime}(x^{\prime})$ is
such that $m(x)dx=m^{\prime}(x^{\prime})dx^{\prime}$. This is true, in
particular, for probability densities such as $p(x)$ and $q(x)$, therefore
\begin{equation}
m(x)=\frac{m^{\prime}(x^{\prime})}{\gamma(x^{\prime})}~,\quad p(x)=\frac
{p^{\prime}(x^{\prime})}{\gamma(x^{\prime})}\quad\text{and}\quad
q(x)=\frac{q^{\prime}(x^{\prime})}{\gamma(x^{\prime})}%
\,.\label{coord trans dens}%
\end{equation}
The coordinate transformation gives%

\begin{align}
S[p,q]  & =\int dx\,F\left(  p(x),q(x),x\right) \nonumber\\
& =\int\gamma(x^{\prime})dx^{\prime}\,F\left(  \frac{p^{\prime}(x^{\prime}%
)}{\gamma(x^{\prime})},\frac{q^{\prime}(x^{\prime})}{\gamma(x^{\prime}%
)},\Gamma(x^{\prime})\right)  ,\label{sa}%
\end{align}
which is a mere change of variables. The identity above is valid always, for
all $\Gamma$ and for all $F$; it imposes absolutely no constraints on
$S[p,q]$. The real constraint arises from realizing that we could have
\emph{started} in the $x^{\prime}$ coordinate frame, in which case we would
have have ranked the distributions using the entropy
\begin{equation}
S[p^{\prime},q^{\prime}]=\int dx^{\prime}\,F\left(  p^{\prime}(x^{\prime
}),q^{\prime}(x^{\prime}),x^{\prime}\right)  \,,\label{sb}%
\end{equation}
but this should have no effect on our conclusions. This is the nontrivial
content of axiom 2. It is not that we can change variables, we can always do
that; but rather that the two rankings, the one according to $S[p,q]$ and the
other according to $S[p^{\prime},q^{\prime}]$ must coincide. This requirement
is satisfied if, for example, $S[p,q]$ and $S[p^{\prime},q^{\prime}]$ turn out
to be numerically equal, but this is not necessary.

The consequence of axiom 2 is that $S[p,q]$ can be written in terms of
coordinate invariants such as $dx\,m(x)$ and $p(x)/m(x)$, and $q(x)/m(x)$:
\begin{equation}
S[p,q]=\int dx\,m(x)\Phi\left(  \frac{p(x)}{m(x)},\frac{q(x)}{m(x)}\right)
~.\label{axiom2}%
\end{equation}
Thus the unknown function $F$ which had three arguments has been replaced by a
still unknown function $\Phi$ with two arguments plus an unknown density
$m(x)$.

Next we determine the density $m(x)$ by invoking the locality axiom 1 once
again. A situation in which no new information is available is dealt by
allowing the domain $\mathcal{D}$ to cover the whole space $\mathcal{X}$. The
requirement that in the absence of any new information the prior conditional
probabilities $q(x|\mathcal{D})=q(x|\mathcal{X})=pqx)$ should not be updated,
can be expressed as

\textbf{Axiom 1 (special case): }\emph{When there is no new information there
is no reason to change one's mind. }

\noindent When there are no constraints the selected posterior distribution
should coincide with the prior distribution, that is, $P(x)=q(x)$. The
consequence of this second use of locality is that the arbitrariness in the
density $m(x)$ is removed: up to normalization $m(x)$ must be the prior
distribution $q(x)$, and therefore at this point we have succeeded in
restricting the entropy to functionals of the form
\begin{equation}
S[p,q]=\int dx\,q(x)\Phi\left(  \frac{p(x)}{q(x)}\right) \label{axiom1b}%
\end{equation}

\textbf{Axiom 3:\ Consistency for independent subsystems}. \emph{When a system
is composed of subsystems that are \textbf{known} to be independent it should
not matter whether the inference procedure treats them separately or jointly.
}

This axiom is perhaps subtler than it appears at first sight. Two points must
be made clear. The first point concerns how the information about independence
is to be handled as a constraint. Consider a system composed of two (or more)
subsystems which we know are independent. This means that both the prior and
the posterior are products. If the subsystem priors are $q_{1}(x_{1})$ and
$q_{2}(x_{2})$, then the prior for the whole system is the product
\begin{equation}
q(x_{1},x_{2})=q_{1}(x_{1})q_{2}(x_{2})~,\label{indep a}%
\end{equation}
while the joint posterior is constrained within the family
\begin{equation}
p(x_{1},x_{2})=p_{1}(x_{1})p_{2}(x_{2})~.\label{indep b}%
\end{equation}
Further suppose that new information is acquired, say constraints
$\mathcal{C}_{1}$ such that $q_{1}(x_{1})$ is updated to $P_{1}(x_{1})$, and
constraints $\mathcal{C}_{2}$ such that $q_{2}(x_{2})$ is updated to
$P_{2}(x_{2})$. Axiom 3 is implemented as follows: First we treat the two
subsystems separately. For subsystem $1$ we maximize
\begin{equation}
S[p_{1},q_{1}]=\int dx_{1}\,q_{1}(x_{1})\Phi\left(  \frac{p_{1}(x_{1})}%
{q_{1}(x_{1})}\right)  ,
\end{equation}
subject to constraints $\mathcal{C}_{1}$ on the marginal distribution
$p_{1}(x_{1})=\int dx_{2}\,p(x_{1},x_{2})$ to select the posterior
$P_{1}(x_{1})$. The constraints $\mathcal{C}_{1}$ could, for example, include
normalization, or they could involve the known expected value of a function
$f_{1}(x_{1})$,
\begin{equation}
\int dx_{1}f_{1}(x_{1})p_{1}(x_{1})=\int dx_{1}dx_{2}\,f_{1}(x_{1}%
)p(x_{1},x_{2})=F_{1}~.\label{C1}%
\end{equation}
Similarly, for subsystem $2$ we maximize the corresponding $S[p_{2},q_{2}]$
subject to constraints $\mathcal{C}_{2}$ on $p_{2}(x_{2})=\int dx_{1}%
\,p(x_{1},x_{2})$ to select the posterior $P_{2}(x_{2})$.

Next the subsystems are treated jointly. Since we are concerned with those
special examples where we have the information that the subsystems are
independent, we are \emph{required} to search for the posterior within the
restricted family of joint distributions that take the form of the product
(\ref{indep b}); this is an \emph{additional} constraint over and above the
original $\mathcal{C}_{1}$ and $\mathcal{C}_{2}$. The new constraint
$p=p_{1}p_{2}$ is easily implemented by direct substitution. Instead of
maximizing the joint entropy, $S[p,q_{1}q_{2}]$, we now maximize
\begin{equation}
S[p_{1}p_{2},q_{1}q_{2}]=\int dx_{1}dx_{2}\,q_{1}(x_{1})q_{2}(x_{2}%
)\Phi\left(  \frac{p_{1}(x_{1})p_{2}(x_{2})}{q_{1}(x_{1})q_{2}(x_{2})}\right)
,\label{joint S[P1P2]}%
\end{equation}
under independent variations $\delta p_{1}$ and $\delta p_{2}$ subject to the
same constraints $\mathcal{C}_{1}$ and $\mathcal{C}_{2}$. The function $\Phi$
is then determined -- or at least constrained -- by demanding that the
selected posterior be $P_{1}(x_{1})P_{2}(x_{2})$.

The second point is that the axiom applies to \emph{all instances }of systems
that happen to be independent -- this is why it is so powerful. The axiom
applies to situations where we deal with just two systems -- as in the
previous paragraph -- and it also applies when we deal with many, whether just
a few or a very large number. The axiom applies when the independent
subsystems are identical, and also when they are not.

The final conclusion is that probability distributions $p(x)$ should be ranked
relative to the prior $q(x)$ according to the relative entropy,
\begin{equation}
S[p,q]=-\int dx\,p(x)\log\frac{p(x)}{q(x)}.\label{S[p,q]}%
\end{equation}

The\ lengthy proof leading to (\ref{S[p,q]}) is given in the next section. It
involves three steps. First we show (subsection 6.3.4) that applying Axiom 3
to subsystems that happen to be identical restricts the entropy functional to
a member of the one-parameter family of entropies $S_{\eta}[p,q]$ parametrized
by an \textquotedblleft inference parameter\textquotedblright\ $\eta$,
\begin{equation}
S_{\eta}[p,q]=\frac{1}{\eta(\eta+1)}\left(  1-\int dx\,p^{\eta+1}q^{-\eta
}\right)  ~.\label{S-eta}%
\end{equation}
It is easy to see that there are no singularities for $\eta=0$ or $-1$. The
limits $\eta\rightarrow0$ and $\eta\rightarrow-1$ are well behaved. In
particular, to take $\eta\rightarrow0$ use
\begin{equation}
y^{\eta}=\exp(\eta\log y)\approx1+\eta\log y~,
\end{equation}
which leads to the usual logarithmic entropy, $S_{0}[p,q]=S[p,q]$ given in
eq.(\ref{S[p,q]}). Similarly, for $\eta\rightarrow-1$ we get $S_{-1}%
[p,q]=S[q,p]$.

In the second step (subsection 6.3.5) axiom 3 is applied to two independent
systems that are not identical and could in principle be described by
different parameters $\eta_{1}$ and $\eta_{2}$. The consistency demanded by
axiom 3 implies that the two parameters must be equal, $\eta_{1}=\eta_{2}$,
and since this must hold for all pairs of independent systems we conclude that
$\eta$ must be a universal constant. In the final step the value of this
constant -- which turns out to be $\eta=0$ -- is determined (subsection 6.3.5)
by demanding that axiom 3 apply to $N$ identical subsystems where $N$ is very large.

We can now summarize our overall conclusion:

\begin{description}
\item[\textbf{The ME method}: ] \emph{We want to update from a prior
distribution }$q(x)$\emph{\ to a posterior distribution }$p(x)$\emph{\ when
information in the form of a constraint that specifies the allowed posteriors
becomes available. The posterior selected by induction from special cases that
implement locality, coordinate invariance and consistency for independent
subsystems, is that which maximizes the relative entropy }$S[p,q]$%
\emph{\ subject to the available constraints. No interpretation for }%
$S[p,q]$\emph{\ is given and none is needed.}
\end{description}

This extends the method of maximum entropy beyond its original purpose as a
rule to assign probabilities from a given underlying measure (MaxEnt) to a
method for updating probabilities from any arbitrary prior (ME). Furthermore,
the logic behind the updating procedure does not rely on any particular
meaning assigned to the entropy, either in terms of information, or heat, or
disorder. Entropy is merely a tool for inductive inference; we do not need to
know what it means; we only need to know how to use it.

The derivation above has singled out \emph{a unique }$S[p,q]$\emph{\ to be
used in inductive inference}. Other `entropies' could turn out to be useful
for other purposes -- perhaps as measures of information, or of ecological
diversity, or something else -- but they are not an induction from the special
cases set down in the axioms.

\section{The proofs}

In this section we establish the consequences of the three axioms leading to
the final result eq.(\ref{S[p,q]}). The details of the proofs are important
not just because they lead to our final conclusions, but also because the
translation of the verbal statement of the axioms into precise mathematical
form is a crucial part of unambiguously specifying what the axioms actually say.

\subsection{Axiom 1: Locality}

Here we prove that axiom 1 leads to the expression eq.(\ref{axiom1}) for
$S[p,q]$. The requirement that probabilities be normalized is handled by
imposing normalization as one among so many other constraints that one might
wish to impose. To simplify the proof we consider the case of a discrete
variable, $p_{i}$ with $i=1\ldots n$, so that $S[p,q]=S(p_{1}\ldots
p_{n},q_{1}\ldots q_{n})$. The generalization to a continuum is straightforward.

Suppose the space of states $\mathcal{X}$ is partitioned into two
non-overlapping domains $\mathcal{D}$ and $\mathcal{D}^{\prime}$ with
$\mathcal{D\cup D}^{\prime}=\mathcal{X}$, and that the information to be
processed is in the form of a constraint that refers to the domain
$\mathcal{D}^{\prime}$,
\begin{equation}%
{\textstyle\sum\limits_{j\in\mathcal{D}^{\prime}}}
a_{j}p_{j}=A\text{ .}\label{locality a}%
\end{equation}
Axiom 1 states that the constraint on $\mathcal{D}^{\prime}$ does not have an
influence on the \emph{conditional} probabilities $p_{i|\mathcal{D}}$. It may
however influence the probabilities $p_{i}$ within $\mathcal{D}$ through an
overall multiplicative factor. To deal with this complication consider then a
special case where the overall probabilities of $\mathcal{D}$ and
$\mathcal{D}^{\prime}$ are constrained too,
\begin{equation}%
{\textstyle\sum\limits_{i\in\mathcal{D}}}
p_{i}=P_{\mathcal{D}}\quad\text{and \quad}%
{\textstyle\sum\limits_{j\in\mathcal{D}^{\prime}}}
p_{j}=P_{\mathcal{D}^{\prime}}\text{ ,}\label{locality b}%
\end{equation}
with $P_{\mathcal{D}}+P_{\mathcal{D}^{\prime}}=1$. Under these special
circumstances constraints on $\mathcal{D}^{\prime}$ will not influence $p_{i}%
$s within $\mathcal{D}$, and vice versa.

To obtain the posterior maximize $S[p,q]$ subject to these three constraints,
\begin{align*}
0  & =\left[  \delta S-\lambda\left(
{\textstyle\sum\limits_{i\in\mathcal{D}}}
p_{i}-P_{\mathcal{D}}\right)  +\right. \\
& -\left.  \lambda^{\prime}\left(
{\textstyle\sum\limits_{j\in\mathcal{D}^{\prime}}}
p_{i}-P_{\mathcal{D}^{\prime}}\right)  +\mu\left(
{\textstyle\sum\limits_{j\in\mathcal{D}^{\prime}}}
a_{j}p_{j}-A\right)  \right]  ~,
\end{align*}
leading to
\begin{align}
\frac{\partial S}{\partial p_{i}}  & =\lambda\text{\quad for\quad}%
i\in\mathcal{D}~,\label{locality c}\\
\frac{\partial S}{\partial p_{j}}  & =\lambda^{\prime}+\mu a_{j}\text{\quad
for\quad}j\in\mathcal{D}^{\prime}~.\label{locality d}%
\end{align}
Eqs.(\ref{locality a}-\ref{locality d}) are $n+3$ equations we must solve for
the $p_{i}$s and the three Lagrange multipliers. Since $S=S(p_{1}\ldots
p_{n},q_{1}\ldots q_{n})$ its derivative
\begin{equation}
\frac{\partial S}{\partial p_{i}}=f_{i}(p_{1}\ldots p_{n},q_{1}\ldots q_{n})
\end{equation}
could in principle also depend on all $2n$ variables. But this violates the
locality axiom because any arbitrary change in $a_{j}$ within $\mathcal{D}%
^{\prime}$ would influence the $p_{i}$s within $\mathcal{D}$. The only way
that probabilities within $\mathcal{D}$ can be shielded from arbitrary changes
in the constraints pertaining to $\mathcal{D}^{\prime}$ is that the functions
$f_{i}$ with $i\in\mathcal{D}$ depend only on $p_{i}$s while the functions
$f_{j}$ depend only on $p_{j}$s. Furthermore, this must hold not just for one
particular partition of $\mathcal{X}$ into domains $\mathcal{D}$ and
$\mathcal{D}^{\prime}$, it must hold for all conceivable partitions. Therefore
$f_{i}$ can depend only on $p_{i}$ and, at this point, on any of the $q$s,
\begin{equation}
\frac{\partial S}{\partial p_{i}}=f_{i}(p_{i},q_{1}\ldots q_{n}%
)~.\label{locality e}%
\end{equation}

But the power of the locality axiom is not exhausted yet. The information to
be incorporated into the posterior can enter not just through constraints but
also through the prior. Suppose that the local information about domain
$\mathcal{D}^{\prime}$ is altered by changing the prior within $\mathcal{D}%
^{\prime}$. Let $q_{j}\rightarrow q_{j}+\delta q_{j}$ for $j\in\mathcal{D}%
^{\prime}$. Then (\ref{locality e}) becomes
\begin{equation}
\frac{\partial S}{\partial p_{i}}=f_{i}(p_{i},q_{1}\ldots q_{j}+\delta
q_{j}\ldots q_{n})
\end{equation}
which shows that $p_{i}$ with $i\in\mathcal{D}$ will be influenced by
information about $\mathcal{D}^{\prime}$ unless $f_{i}$ with $i\in\mathcal{D}$
is independent of all the $q_{j}$s for $j\in\mathcal{D}^{\prime}$. Again, this
must hold for all partitions into $\mathcal{D}$ and $\mathcal{D}^{\prime}$,
and therefore,
\begin{equation}
\frac{\partial S}{\partial p_{i}}=f_{i}(p_{i},q_{i})\text{\quad for all\quad
}i\in\mathcal{X}~.
\end{equation}
Integrating, one obtains%
\begin{equation}
S[p,q]=%
{\textstyle\sum\limits_{i}}
F_{i}(p_{i},q_{i})+\operatorname{constant}\text{.}%
\end{equation}
for some undetermined functions $F_{i}$. The corresponding expression for a
continuous variable $x$ is obtained replacing $i$ by $x$, and the sum over $i
$ by an integral over $x$ leading to eq.(\ref{axiom1}).

\subsection{Axiom 2: Coordinate invariance}

Next we prove eq.(\ref{axiom2}) It is convenient to introduce a function
$m(x)$ which transforms as a density and rewrite the expression (\ref{axiom1})
for the entropy in the form
\begin{align}
S[p,q]  & =\int dx\,m(x)\frac{1}{m(x)}F\left(  \frac{p(x)}{m(x)}%
m(x),\frac{q(x)}{m(x)}m(x),x\right) \\
& =\int dx\,m(x)\Phi\left(  \frac{p(x)}{m(x)},\frac{q(x)}{m(x)},m(x),x\right)
,
\end{align}
where the function $\Phi$ is defined by
\begin{equation}
\Phi(\alpha,\beta,m,x)\overset{\operatorname*{def}}{=}\frac{1}{m}F(\alpha
m,\beta m,m,x).
\end{equation}

Next, we consider a special situation where the new information are
constraints which do not favor one coordinate system over another. For example
consider the constraint
\begin{equation}
\int dx\,p(x)a(x)=A
\end{equation}
where $a(x)$ is a scalar, \emph{i.e.}, invariant under coordinate changes,
\begin{equation}
a(x)\rightarrow a^{\prime}(x^{\prime})=a(x).\label{ci-a}%
\end{equation}
The usual normalization condition $\int dx\,p(x)=1$ is a simple example of a
scalar constraint.

Maximizing $S[p,q]$ subject to the constraint,
\begin{equation}
\delta\left[  S[p,q]+\lambda\left(  \int dx\,p(x)a(x)-A\right)  \right]  =0,
\end{equation}
gives
\begin{equation}
\dot{\Phi}\left(  \frac{p(x)}{m(x)},\frac{q(x)}{m(x)},m(x),x\right)  =\lambda
a(x)~,\label{ci-b}%
\end{equation}
where the dot represents the derivative with respect to the first argument,
\begin{equation}
\dot{\Phi}\left(  \alpha,\beta,m,x\right)  \overset{\operatorname*{def}}%
{=}\frac{\partial\Phi\left(  \alpha,\beta,m,x\right)  }{\partial\alpha}%
\end{equation}
But we could have started using the primed coordinates,
\begin{equation}
\dot{\Phi}\left(  \frac{p^{\prime}(x^{\prime})}{m^{\prime}(x^{\prime})}%
,\frac{q^{\prime}(x^{\prime})}{m^{\prime}(x^{\prime})},m^{\prime}(x^{\prime
}),x^{\prime}\right)  =\lambda^{\prime}a^{\prime}(x^{\prime}),
\end{equation}
or, using (\ref{coord trans dens}) and (\ref{ci-a}),
\begin{equation}
\dot{\Phi}\left(  \frac{p(x)}{m(x)},\frac{q^{\prime}(x^{\prime})}{m^{\prime
}(x^{\prime})},m(x)\gamma(x^{\prime}),x^{\prime}\right)  =\lambda^{\prime
}a(x).\label{ci-c}%
\end{equation}
Dividing (\ref{ci-c}) by (\ref{ci-b}) we get
\begin{equation}
\frac{\dot{\Phi}\left(  \alpha,\beta,m\gamma,x^{\prime}\right)  }{\dot{\Phi
}\left(  \alpha,\beta,m,x\right)  }=\frac{\lambda^{\prime}}{\lambda
}.\label{ci-d}%
\end{equation}
This identity should hold for any transformation $x=\Gamma(x^{\prime})$. On
the right hand side the multipliers $\lambda$ and $\lambda^{\prime}$ are just
constants; the ratio $\lambda^{\prime}/\lambda$ might depend on the
transformation $\Gamma$ but it does not depend on $x$. Consider the special
case of a transformation $\Gamma$ that has unit determinant everywhere,
$\gamma=1$, and differs from the identity transformation only within some
arbitrary region $\mathcal{D}$. Since for $x$ outside this region
$\mathcal{D}$ we have $x=x^{\prime}$, the left hand side of eq.(\ref{ci-d})
equals 1. Thus, for this particular $\Gamma$ the ratio is $\lambda^{\prime
}/\lambda=1$; but $\lambda^{\prime}/\lambda=\operatorname{constant}$, so
$\lambda^{\prime}/\lambda=1$ holds within $\mathcal{D}$ as well. Therefore,
for $x$ within $\mathcal{D}$,
\begin{equation}
\dot{\Phi}\left(  \alpha,\beta,m,x^{\prime}\right)  =\dot{\Phi}\left(
\alpha,\beta,m,x\right)  .
\end{equation}
Since the choice of $\mathcal{D}$ is arbitrary we conclude is that the
function $\dot{\Phi}$ cannot depend on its third argument, $\dot{\Phi}%
=\dot{\Phi}\left(  \alpha,\beta,m\right)  $.

Having eliminated the third argument, let us go back to eq.(\ref{ci-d}),
\begin{equation}
\frac{\dot{\Phi}\left(  \alpha,\beta,m\gamma\right)  }{\dot{\Phi}\left(
\alpha,\beta,m\right)  }=\frac{\lambda^{\prime}}{\lambda}\ ,
\end{equation}
and consider a different transformation $\Gamma$, one with unit determinant
$\gamma=1$ only outside the region $\mathcal{D}$. Therefore the constant ratio
$\lambda^{\prime}/\lambda$ is again equal to $1$, so that
\begin{equation}
\dot{\Phi}\left(  \alpha,\beta,m\gamma\right)  =\dot{\Phi}\left(  \alpha
,\beta,m\right)  .
\end{equation}
But within $\mathcal{D}$ the transformation $\Gamma$ is quite arbitrary, it
could have any arbitrary Jacobian $\gamma\neq1$. Therefore the function
$\dot{\Phi}$ cannot depend on its second argument either, and therefore
$\dot{\Phi}=\dot{\Phi}(\alpha,\beta)$. Integrating with respect to $\alpha$
gives $\Phi=\Phi(\alpha,\beta)+\operatorname{constant}$. The additive
constant, which could depend on $\beta$, has no effect on the maximization and
can be dropped. This completes the proof of eq.(\ref{axiom2}).

\subsection{Axiom 1 again}

The locality axiom implies that when there are no constraints the selected
posterior distribution should coincide with the prior distribution. This
provides us with an interpretation of the density $m(x)$ that had been
artificially introduced. The argument is simple: maximize $S[p,q]$ in
(\ref{axiom2}) subject to the single requirement of normalization,
\begin{equation}
\delta\left[  S[p,q]+\lambda\left(  \int dx\,p(x)-1\right)  \right]  =0,
\end{equation}
to get
\begin{equation}
\dot{\Phi}\left(  \frac{p(x)}{m(x)},\frac{q(x)}{m(x)}\right)  =\lambda
.\label{sc-a}%
\end{equation}
Since $\lambda$ is a constant, the left hand side must be independent of $x$
for arbitrary choices of the prior $q(x)$. This could, for example, be
accomplished if the function $\dot{\Phi}(\alpha,\beta)$ were itself a
constant, independent of its arguments $\alpha$ and $\beta$. But this gives
\begin{equation}
\Phi(\alpha,\beta)=c_{1}\alpha+c_{2}%
\end{equation}
where $c_{1}$ and $c_{2}$ are constants and leads to the unacceptable form
$S[p,q]\propto\int dx\,p(x)+\operatorname*{constant}$.

If the dependence on $x$ cannot be eliminated by an appropriate choice of
$\dot{\Phi}$, we must secure it by a choice of $m(x)$. Eq.(\ref{sc-a}) is an
equation for $p(x)$. In the absence of new information the selected posterior
distribution must coincide with the prior, $P(x)=q(x)$. The obvious way to
secure that (\ref{sc-a}) be independent of $x$ is to choose $m(x)\propto
q(x)$. Therefore $m(x)$ must, except for an overall normalization, be chosen
to coincide with the prior distribution.

\subsection{Axiom 3: Consistency for identical independent subsystems}

In this subsection we show that applying axiom 3 to subsystems that happen to
be identical restricts the entropy functional to a member of the one-parameter
family of $\eta$-entropies $S_{\eta}[p,q]$ parametrized by $\eta$. For
$\eta=0$ one obtains the standard logarithmic entropy, eq.(\ref{S[p,q]}),
\begin{equation}
S_{0}[p,q]=-\int dx\,p(x)\log\frac{p(x)}{q(x)}~.\label{S eta a}%
\end{equation}
For $\eta=-1$ one obtains
\begin{equation}
S_{-1}[p,q]=\int dx\,q(x)\log\frac{p(x)}{q(x)}~,\label{S eta b}%
\end{equation}
which coincides with $S_{0}[q,p]$ with the arguments switched. Finally, for a
generic value of $\eta\neq-1,0$ the result is
\begin{equation}
S_{\eta}[p,q]=-\int dx\,p(x)\left(  \frac{p(x)}{q(x)}\right)  ^{\eta
}~.\label{S eta c}%
\end{equation}

It is worthwhile to recall that the objective of this whole exercise is to
rank probability distributions according to preference and therefore different
entropies that induce the same ranking scheme are effectively equivalent. This
is very convenient as it allows considerable simplifications by an appropriate
choice of additive and multiplicative constants. Taking advantage of this
freedom we can, for example, combine the three expressions (\ref{S eta a}),
(\ref{S eta b}), and (\ref{S eta c}) into the single expression
\begin{equation}
S_{\eta}[p,q]=\frac{1}{\eta(\eta+1)}\left(  1-\int dx\,p^{\eta+1}q^{-\eta
}\right)  ~,
\end{equation}
that we met earlier in eq.(\ref{S-eta}).

The proof below is fairly lengthy and may be skipped on a first reading. It
follows the treatment in [Caticha Giffin 06] and is based upon and extends a
previous proof by Karbelkar who showed that belonging to the family of $\eta
$-entropies is a sufficient condition to satisfy the consistency axiom for
\emph{identical} systems. He conjectured but did not prove that this was
perhaps also a necessary condition. [Karbelkar 86] Although necessity was not
essential to his argument it is crucial for ours. We show below that for
identical subsystems there are no acceptable entropies outside the $S_{\eta} $ family.

First we treat the subsystems separately. For subsystem $1$ we maximize the
entropy $S[p_{1},q_{1}]$ subject to normalization and the constraint
$\mathcal{C}_{1}$ in eq.(\ref{C1}). Introduce Lagrange multipliers $\alpha
_{1}$ and $\lambda_{1}$,
\begin{equation}
\delta\left[  S[p_{1},q_{1}]-\lambda_{1}\left(  \int dx_{1}f_{1}P_{1}%
-F_{1}\right)  -\alpha_{1}\left(  \int dx_{1}\,P_{1}-1\right)  \right]  =0,
\end{equation}
which gives
\begin{equation}
\Phi^{\prime}\left(  \frac{p_{1}(x_{1})}{q_{1}(x_{1})}\right)  =\lambda
_{1}f_{1}(x_{1})+\alpha_{1}~,\label{phi 1}%
\end{equation}
where the prime indicates a derivative with respect to the argument,
$\Phi^{\prime}(y)=d\Phi(y)/dy$. For subsystem $2$ we need only consider the
extreme situation where the constraints $\mathcal{C}_{2}$ determine the
posterior completely: $p_{2}(x_{2})=$ $P_{2}(x_{2})$.

Next we treat the subsystems jointly. The constraints $\mathcal{C}_{2}$ are
easily implemented by direct substitution and thus, we maximize the entropy
$S[p_{1}P_{2},q_{1}q_{2}]$ by varying over $p_{1}$ subject to normalization
and the constraint $\mathcal{C}_{1}$ in eq.(\ref{C1}). Introduce Lagrange
multipliers $\alpha$ and $\lambda$,
\begin{equation}
\delta\left[  S[p_{1}P_{2},q_{1}q_{2}]-\lambda\left(  \int dx_{1}f_{1}%
p_{1}-F_{1}\right)  -\alpha\left(  \int dx_{1}\,p_{1}-1\right)  \right]  =0,
\end{equation}
which gives
\begin{equation}
\int dx_{2}\,p_{2}\Phi^{\prime}\left(  \frac{p_{1}P_{2}}{q_{1}q_{2}}\right)
=\lambda\lbrack P_{2},q_{2}]f_{1}(x_{1})+\alpha\lbrack P_{2},q_{2}%
]~,\label{phi 2}%
\end{equation}
where the multipliers $\lambda$ and $\alpha$ are independent of $x_{1}$ but
could in principle be functionals of $P_{2}$ and $q_{2}$.

The consistency condition that constrains the form of $\Phi$ is that if the
solution to eq.(\ref{phi 1}) is $P_{1}(x_{1})$ then the solution to
eq.(\ref{phi 2}) must also be $P_{1}(x_{1})$, and this must be true
irrespective of the choice of $P_{2}(x_{2})$. Let us then consider a small
change $P_{2}\rightarrow P_{2}+\delta P_{2}$ that preserves the normalization
of $P_{2}$. First introduce a Lagrange multiplier $\alpha_{2}$ and rewrite
eq.(\ref{phi 2}) as
\begin{equation}
\int dx_{2}\,p_{2}\Phi^{\prime}\left(  \frac{P_{1}P_{2}}{q_{1}q_{2}}\right)
-\alpha_{2}\left[  \int dx_{2}Pp_{2}-1\right]  =\lambda\lbrack P_{2}%
,q_{2}]f_{1}(x_{1})+\alpha\lbrack P_{2},q_{2}]~,
\end{equation}
where we have replaced $p_{1}$ by the known solution $P_{1}$ and thereby
effectively transformed eqs.(\ref{phi 1}) and (\ref{phi 2}) into an equation
for $\Phi$. The $\delta P_{2}$ variation gives,
\begin{equation}
\Phi^{\prime}\left(  \frac{P_{1}P_{2}}{q_{1}q_{2}}\right)  +\frac{P_{1}P_{2}%
}{q_{1}q_{2}}\Phi^{\prime\prime}\left(  \frac{P_{1}P_{2}}{q_{1}q_{2}}\right)
=\frac{\delta\lambda}{\delta P_{2}}f_{1}(x_{1})+\frac{\delta\alpha}{\delta
P_{2}}+\alpha_{2}~.
\end{equation}
Next use eq.(\ref{phi 1}) to eliminate $f_{1}(x_{1})$,
\begin{equation}
\Phi^{\prime}\left(  \frac{P_{1}P_{2}}{q_{1}q_{2}}\right)  +\frac{P_{1}P_{2}%
}{q_{1}q_{2}}\Phi^{\prime\prime}\left(  \frac{P_{1}P_{2}}{q_{1}q_{2}}\right)
=A[P_{2},q_{2}]\Phi^{\prime}\left(  \frac{P_{1}}{q_{1}}\right)  +B[P_{2}%
,q_{2}]~,\label{phi 3}%
\end{equation}
where
\begin{equation}
A[P_{2},q_{2}]=\frac{1}{\lambda_{1}}\frac{\delta\lambda}{\delta P_{2}}%
\quad\text{and}\quad B[P_{2},q_{2}]=-\frac{\delta\lambda}{\delta P_{2}}%
\frac{\alpha_{1}}{\lambda_{1}}+\frac{\delta\alpha}{\delta P_{2}}+\alpha_{2}~,
\end{equation}
are at this point unknown functionals of $P_{2}$ and $q_{2}$. Differentiating
eq.(\ref{phi 3}) with respect to $x_{1}$ the $B$ term drops out and we get
\begin{equation}
A[P_{2},q_{2}]=\left[  \frac{d}{dx_{1}}\Phi^{\prime}\left(  \frac{P_{1}}%
{q_{1}}\right)  \right]  ^{-1}\frac{d}{dx_{1}}\left[  \Phi^{\prime}\left(
\frac{P_{1}P_{2}}{q_{1}q_{2}}\right)  +\frac{P_{1}P_{2}}{q_{1}q_{2}}%
\Phi^{\prime\prime}\left(  \frac{P_{1}P_{2}}{q_{1}q_{2}}\right)  \right]  ~,
\end{equation}
which shows that $A$ is not a functional of $P_{2}$ and $q_{2}$ but a mere
function of $P_{2}/q_{2}$. Substituting back into eq.(\ref{phi 3}) we see that
the same is true for $B$. Therefore eq.(\ref{phi 3}) can be written as
\begin{equation}
\Phi^{\prime}\left(  y_{1}y_{2}\right)  +y_{1}y_{2}\Phi^{\prime\prime}\left(
y_{1}y_{2}\right)  =A(y_{2})\Phi^{\prime}\left(  y_{1}\right)  +B(y_{2}%
)~,\label{phi 4}%
\end{equation}
where $y_{1}=P_{1}/q_{1}$, $y_{2}=P_{2}/q_{2}$, and $A(y_{2})$, $B(y_{2})$ are
unknown functions of $y_{2}$.

Now we specialize to identical subsystems. Then we can exchange the labels
$1\leftrightarrow2$, and we get
\begin{equation}
A(y_{2})\Phi^{\prime}\left(  y_{1}\right)  +B(y_{2})=A(y_{1})\Phi^{\prime
}\left(  y_{2}\right)  +B(y_{1})~.\label{phi 5}%
\end{equation}
To find the unknown functions $A$ and $B$ differentiate with respect to
$y_{2}$,
\begin{equation}
A^{\prime}(y_{2})\Phi^{\prime}\left(  y_{1}\right)  +B^{\prime}(y_{2}%
)=A(y_{1})\Phi^{\prime\prime}\left(  y_{2}\right) \label{b}%
\end{equation}
and then with respect to $y_{1}$ to get
\begin{equation}
\frac{A^{\prime}(y_{1})}{\Phi^{\prime\prime}\left(  y_{1}\right)  }%
=\frac{A^{\prime}(y_{2})}{\Phi^{\prime\prime}\left(  y_{2}\right)
}=a=\operatorname{const}~.
\end{equation}
Integrate to get
\begin{equation}
A(y_{1})=a\Phi^{\prime}\left(  y_{1}\right)  +b~,
\end{equation}
then substitute back into eq.(\ref{b}) and integrate again to get
\begin{equation}
B^{\prime}(y_{2})=b\Phi^{\prime\prime}\left(  y_{2}\right)  \quad
\text{and}\quad B(y_{2})=b\Phi^{\prime}\left(  y_{2}\right)  +c~,
\end{equation}
where $b$ and $c$ are constants. We can check that $A(y)$ and $B(y)$ are
indeed solutions of eq.(\ref{phi 5}). Substituting into eq.(\ref{phi 4})
gives
\begin{equation}
\Phi^{\prime}\left(  y_{1}y_{2}\right)  +y_{1}y_{2}\Phi^{\prime\prime}\left(
y_{1}y_{2}\right)  =a\Phi^{\prime}\left(  y_{1}\right)  \Phi^{\prime}\left(
y_{2}\right)  +b\left[  \Phi^{\prime}\left(  y_{1}\right)  +\Phi^{\prime
}\left(  y_{2}\right)  \right]  +c~.\label{phi 6}%
\end{equation}
This is a peculiar differential equation. We can think of it as one
differential equation for $\Phi^{\prime}\left(  y_{1}\right)  $ for each given
constant value of $y_{2}$ but there is a complication in that the various
(constant) coefficients $\Phi^{\prime}\left(  y_{2}\right)  $ are themselves
unknown. To solve for $\Phi$ choose a fixed value of $y_{2}$, say $y_{2}=1$,
\begin{equation}
y\Phi^{\prime\prime}\left(  y\right)  -\eta\Phi^{\prime}\left(  y\right)
-\kappa=0~,\label{phi 7}%
\end{equation}
where $\eta=a\Phi^{\prime}\left(  1\right)  +b-1$ and $\kappa=b\Phi^{\prime
}\left(  1\right)  +c$. To eliminate the constant $\kappa$ differentiate with
respect to $y$,
\begin{equation}
y\Phi^{\prime\prime\prime}+\left(  1-\eta\right)  \Phi^{\prime\prime
}=0~,\label{phi 8}%
\end{equation}
which is a linear homogeneous equation and is easy to integrate.

For generic values of $\eta\neq-1,0$ the solution is
\begin{equation}
\Phi^{\prime\prime}(y)\propto y^{\eta-1}\Rightarrow\Phi^{\prime}(y)=\alpha
y^{\eta}+\beta~.
\end{equation}
The constants $\alpha$ and $\beta$ are chosen so that this is a solution of
eq.(\ref{phi 6}) for all values of $y_{2}$ (and not just for $y_{2}=1$).
Substituting into eq.(\ref{phi 6}) and equating the coefficients of various
powers of $y_{1}y_{2}$, $y_{1}$, and $y_{2}$ gives three conditions on the two
constants $\alpha$ and $\beta$,
\begin{equation}
\alpha(1+\eta)=a\alpha^{2},\quad0=a\alpha\beta+b\alpha,\quad\beta=a\beta
^{2}+2b\beta+c~.
\end{equation}
The nontrivial ($\alpha\neq0$) solutions are $\alpha=(1+\eta)/a$ and
$\beta=-b/a$, while the third equation gives $c=b(1-b)/4a$. We conclude that
for generic values of $\eta$ the solution of eq.(\ref{phi 6}) is
\begin{equation}
\Phi(y)=\frac{1}{a}y^{\eta+1}-\frac{b}{a}y+C~,\label{sol a}%
\end{equation}
where $C$ is a new constant. Substituting into eq.(\ref{axiom1b}) yields
\begin{equation}
S_{\eta}[p,q]=\frac{1}{a}\int dx\,p(x)\left(  \frac{p(x)}{q(x)}\right)
^{\eta}-\frac{b}{a}\int dx\,p(x)+C\int dx\,q(x)~.
\end{equation}
This complicated expression can be simplified considerably by exploiting the
freedom to choose additive and multiplicative constants. We can drop the last
two terms and choose $a=-1$ so that the preferred distribution is that which
maximizes entropy. This reproduces eq.(\ref{S eta c}).

For $\eta=0$ we return to eq.(\ref{phi 8}) and integrate twice to get
\begin{equation}
\Phi(y)=a^{\prime}y\log y+b^{\prime}y+c^{\prime}~,
\end{equation}
for some new constants $a^{\prime}$, $b^{\prime}$, and $c^{\prime}$.
Substituting into eq.(\ref{axiom1b}) yields
\begin{equation}
S_{0}[p,q]=a^{\prime}\int dx\,p(x)\log\frac{p(x)}{q(x)}+b^{\prime}\int
dx\,p(x)+c^{\prime}\int dx\,q(x)~.
\end{equation}
Again, choosing $a^{\prime}=-1$ and dropping the last two terms does not
affect the ranking scheme. This yields the standard expression for relative
entropy, eq.(\ref{S eta a}).

Finally, for $\eta=-1$ integrating eq.(\ref{phi 8}) twice gives
\begin{equation}
\Phi(y)=a^{\prime\prime}\log y+b^{\prime\prime}y+c^{\prime\prime}~,
\end{equation}
for some new constants $a^{\prime\prime}$, $b^{\prime\prime}$, and
$c^{\prime\prime}$. Substituting into eq.(\ref{axiom1b}) yields
\begin{equation}
S_{0}[p,q]=a^{\prime\prime}\int dx\,q(x)\log\frac{p(x)}{q(x)}+b^{\prime\prime
}\int dx\,p(x)+c^{\prime\prime}\int dx\,q(x)~.
\end{equation}
Again, choosing $a^{\prime\prime}=1$ and dropping the last two terms yields
eq.(\ref{S eta b}). This completes our derivation.

\subsection{Axiom 3: Consistency for non-identical subsystems}

Let us summarize our results so far. The goal is to update probabilities by
ranking the distributions according to an entropy $S$ that is of general
applicability. The allowed functional forms of the entropy $S$ have been
constrained down to a member of the one-dimensional family $S_{\eta}$. One
might be tempted to conclude that there is no $S$ of universal applicability;
that inferences about different systems could to be carried out with different
$\eta$-entropies. But we have not yet exhausted the full power of the
consistency axiom 3. Consistency is universally desirable; there is no reason
why it should be limited to identical systems.

To proceed further we ask: What is $\eta$? Is it a property of the individual
carrying out the inference or of the system under investigation? The former is
unacceptable; we insist that the updating must be objective in that different
individuals with the same prior and with the same constraints must make the
same inference. Therefore the \textquotedblleft inference
parameter\textquotedblright\ $\eta$ can only be a property of the system.

Consider two different systems characterized by $\eta_{1}$ and $\eta_{2}$. Let
us further suppose that these systems are known to be independent (perhaps
system $\#1$ lives here on Earth while system $\#2$ lives in a distant galaxy)
so that they fall under the jurisdiction of axiom 3. Separate inferences about
systems $\#1$ and $\#2$ are carried out with $S_{\eta_{1}}[p_{1},q_{1}]$ and
$S_{\eta_{2}}[p_{2},q_{2}]$ respectively. For the combined system we are also
required to use an $\eta$-entropy, say $S_{\eta}[p_{1}p_{2},q_{1}q_{2}]$. The
question is what $\eta$s do we choose that will lead to consistent inferences
whether we treat the systems separately or jointly. The results of the
previous subsection indicate that a joint inference with $S_{\eta}[p_{1}%
p_{2},q_{1}q_{2}]$ is equivalent to separate inferences with $S_{\eta}%
[p_{1},q_{1}]$ and $S_{\eta}[p_{2},q_{2}] $. Therefore we must choose
$\eta=\eta_{1}$ and also $\eta=\eta_{2}$ which is possible only if we had
$\eta_{1}=\eta_{2}$ from the start.

But this is not all: consider a third system $\#3$ that also lives here on
Earth. We do not know whether system $\#3$ is independent from system $\#1$ or
not but we can confidently assert that it will certainly be independent of the
system $\#2$ living in the distant galaxy. The argument of the previous
paragraph leads us to conclude that $\eta_{3}=\eta_{2}$, and therefore that
$\eta_{3}=\eta_{1}$ even when systems $\#1$ and $\#3$ are not known to be
independent! We conclude that \emph{all systems must be characterized by the
same parameter }$\eta$ whether they are independent or not because we can
always find a common reference system that is sufficiently distant to be
independent of any two of them. The inference parameter $\eta$ is a universal
constant, the value of which is at this point still unknown.

The power of a consistency argument resides in its universal applicability: if
an entropy $S[p,q]$ exists then it must be one chosen from among the $S_{\eta
}[p,q]$. The remaining problem is to determine this universal constant $\eta$.
Here we give one argument; in the next subsection we give another one.

One possibility is to regard $\eta$ as a quantity to be determined
experimentally. Are there systems for which inferences based on a known value
of $\eta$ have repeatedly led to success? The answer is yes; they are quite common.

As we discussed in Chapter 5 statistical mechanics and thus thermodynamics are
theories of inference based on the value $\eta=0$. The relevant entropy, which
is the Boltzmann-Gibbs-Shannon entropy, can be interpreted as the special case
of the ME when one updates from a uniform prior. It is an experimental fact
\emph{without any known exceptions} that inferences about \emph{all} physical,
chemical and biological systems that are in thermal equilibrium or close to it
can be carried out by assuming that $\eta=0$. Let us emphasize that this is
not an obscure and rare example of purely academic interest; these systems
comprise essentially all of natural science. (Included is every instance where
it is useful to introduce a notion of temperature.)

In conclusion: consistency for non-identical systems requires that $\eta$ be a
universal constant and there is abundant experimental evidence for its value
being $\eta=0$. Other $\eta$-entropies may turn out to be useful for other
purposes but \emph{the logarithmic entropy }$S[p,q]$\emph{\ in
eq.(\ref{S[p,q]}) provides the only consistent ranking criterion for updating
probabilities that can claim general applicability.}

\subsection{Axiom 3: Consistency with the law of large numbers}

Here we offer a second argument, also based on a broader application of axiom
3, that the value of the universal constant $\eta$ must be $\eta=0$. We
require consistency for large numbers of independent identical subsystems. In
such cases the weak law of large numbers is sufficient to make the desired inferences.

Let the state for each individual system be described by a discrete variable
$i=1\ldots m$.

First we treat the individual systems separately. The identical priors for the
individual systems are $q_{i}$ and the available information is that the
potential posteriors $p_{i}$ are subject, for example, to an expectation value
constraint such as $\langle a\rangle=A$, where $A$ is some specified value and
$\langle a\rangle=%
{\textstyle\sum}
a_{i}p_{i}$. The preferred posterior $P_{i}$ is found maximizing the $\eta
$-entropy $S_{\eta}[p,q]$ subject to $\langle a\rangle=A$.

To treat the systems jointly we let the number of systems found in state $i$
be $n_{i}$, and let $f_{i}=n_{i}/N$ be the corresponding frequency. The two
descriptions are related by the law of large numbers: for large $N$ the
frequencies $f_{i}$ converge (in probability) to the desired posterior $P_{i}
$ while the sample average $\bar{a}=%
{\textstyle\sum}
a_{i}f_{i}$ converges (also in probability) to the expected value $\langle
a\rangle=A$.

Now we consider the set of $N$ systems treated jointly. The probability of a
particular frequency distribution $f=(f_{1}\ldots f_{n})$ generated by the
prior $q$ is given by the multinomial distribution,
\begin{equation}
Q_{N}\left(  f|q\right)  =\frac{N!}{n_{1}!\ldots n_{m}!}q_{1}^{n_{1}}\ldots
q_{m}^{n_{m}}\quad\text{with}\quad%
{\textstyle\sum\limits_{i=1}^{m}}
n_{i}=N~.
\end{equation}
When the $n_{i}$ are sufficiently large we can use Stirling's approximation,
\begin{equation}
\log n!=n\log n-n+\log\sqrt{2\pi n}+O(1/n)~.
\end{equation}
Then
\begin{align}
\log Q_{N}\left(  f|q\right)   & \approx N\log N-N+\log\sqrt{2\pi
N}\nonumber\\
& -%
{\textstyle\sum\limits_{i}}
\left(  n_{i}\log n_{i}-n_{i}+\log\sqrt{2\pi n_{i}}-n_{i}\log q_{i}\right)
\nonumber\\
& =-N%
{\textstyle\sum\limits_{i}}
\frac{n_{i}}{N}\log\frac{n_{i}}{Nq_{i}}-%
{\textstyle\sum\limits_{i}}
\log\sqrt{\frac{n_{i}}{N}}-(N-1)\log\sqrt{2\pi N}\nonumber\\
& =NS[f,q]-%
{\textstyle\sum\limits_{i}}
\log\sqrt{f_{i}}-(N-1)\log\sqrt{2\pi N}~,
\end{align}
where $S[f,q]$ is the $\eta=0$ entropy given by eq.(\ref{S[p,q]}). Therefore
for large $N$ can be written as
\begin{equation}
Q_{N}\left(  f|q\right)  \approx C_{N}(%
{\textstyle\prod\limits_{i}}
f_{i})^{-1/2}\exp(NS[f,q])
\end{equation}
where $C_{N}$ is a normalization constant. The Gibbs inequality $S[f,q]\leq0
$, eq.(\ref{Kc}), shows that for large $N$ the probability $Q_{N}\left(
f|q\right)  $ shows an exceedingly sharp peak. The most likely frequency
distribution is numerically equal to the probability distribution $q_{i}$.
This is the weak law of large numbers. Equivalently, we can rewrite it as
\begin{equation}
\frac{1}{N}\log Q_{N}\left(  f|q\right)  \approx S[f,q]+r_{N}~,
\end{equation}
where $r_{N}$ is a correction that vanishes as $N\rightarrow\infty$. This
means that finding the most probable frequency distribution is equivalent to
maximizing the entropy $S[f,q]$.

The most probable frequency distribution that satisfies the constraint
$\bar{a}=A$ is the distribution that maximizes $Q_{N}\left(  f|q\right)  $
subject to the constraint $\bar{a}=A$, which is equivalent to maximizing the
entropy $S[f,q]$ subject to $\bar{a}=A$. In the limit of large $N$ the
frequencies $f_{i}$ converge (in probability) to the desired posterior $P_{i}$
while the sample average $\bar{a}=%
{\textstyle\sum}
a_{i}f_{i}$ converges (also in probability) to the expected value $\langle
a\rangle=A$.

The two procedures agree only when we choose $\eta=0$. Inferences carried out
with with $\eta\neq0$ are not consistent with inferences from the law of large
numbers. This is the \emph{Principle of Eliminative Induction} in action:\ it
is the successful falsification of all rival $\eta$-entropies that
corroborates the surviving entropy with $\eta=0$. The reason the competing
$\eta$-entropies are discarded is clear: $\eta\neq0$ is inconsistent with the
law of large numbers.

[Csiszar 84] and [Grendar 01] have argued that the asymptotic argument above
provides by itself a valid justification for the ME method of updating. An
agent whose prior is $q$ receives the information $\left\langle a\right\rangle
=A$ which can be reasonably interpreted as a sample average $\bar{a}=A$ over a
large ensemble of $N$ trials. The agent's beliefs are updated so that the
posterior $P$ coincides with the most probable $f$ distribution. This is quite
compelling but, of course, as a justification of the ME method it is
restricted to situations where it is natural to think in terms of ensembles
with large $N$. This justification is not nearly as compelling for singular
events for which large ensembles either do not exist or are too unnatural and
contrived. From our point of view the asymptotic argument above does not by
itself provide a fully convincing justification for the universal validity of
the ME method but it does provide considerable inductive support. It serves as
a valuable consistency check that must be passed by any inductive inference
procedure that claims to be of \emph{general} applicability.

\section{Random remarks}

\subsection{On deductive vs. inductive systems}

In a deductive axiomatic system certain statements are chosen as axioms and
other statements called theorems are derived from them. The theorems can be
asserted to be true only when conditions are such that the axioms hold true.
Within a deductive axiomatic system it makes no sense to make assertions that
go beyond the reach of applicability of the axioms. In contrast the purpose of
eliminative induction is precisely to venture into regions beyond those known
special cases -- the axioms -- and accordingly, the truth of the resulting
inferences -- the theorems -- is not guaranteed.

A second interesting difference is that in a deductive system there is a
certain preference for minimizing the number of axioms as this clarifies the
relations among various elements of the system and the structure of the whole.
In contrast when doing induction one strives to maximize the number of axioms
as it is much safer to induce from many known instances than from just a few.

\subsection{On priors}

\emph{All entropies are relative entropies}. In the case of a discrete
variable, if one assigns equal a priori probabilities, $q_{i}=1$, one obtains
the Boltzmann-Gibbs-Shannon entropy, $S[p]=-\sum_{i}\,p_{i}\log p_{i}\,$. The
notation $S[p]$ has a serious drawback: it misleads one into thinking that $S$
depends on $p$ only. In particular, we emphasize that whenever $S[p]$ is used,
the prior measure $q_{i}=1$ has been implicitly assumed. In Shannon's axioms,
for example, this choice is implicitly made in his first axiom, when he states
that the entropy is a function of the probabilities $S=S(p_{1}...p_{n})$ and
nothing else, and also in his second axiom when the uniform distribution
$p_{i}=1/n$ is singled out for special treatment.

The absence of an explicit reference to a prior $q_{i}$ may erroneously
suggest that prior distributions have been rendered unnecessary and can be
eliminated. It suggests that it is possible to transform information
(\emph{i.e.}, constraints) directly into posterior distributions in a totally
objective and unique way. This was Jaynes' hope for the MaxEnt program. If
this were true the old controversy, of whether probabilities are subjective or
objective, would have been resolved -- probabilities would ultimately be
totally objective. But the prior $q_{i}=1$ is implicit in $S[p]$; the
postulate of equal a priori probabilities or Laplace's \textquotedblleft
Principle of Insufficient Reason\textquotedblright\ still plays a major,
though perhaps hidden, role. Any claims that probabilities assigned using
maximum entropy will yield absolutely objective results are unfounded; not all
subjectivity has been eliminated. \emph{Just as with Bayes' theorem, what is
objective here is the manner in which information is processed to update from
a prior to a posterior, and not the prior probabilities themselves.}

Choosing the prior density $q(x)$ can be tricky. Sometimes symmetry
considerations can be useful in fixing the prior (three examples were given in
section 4.5) but otherwise there is no fixed set of rules to translate
information into a probability distribution except, of course, for Bayes'
theorem and the ME method themselves.

What if the prior $q(x)$ vanishes for some values of $x$? $S[p,q]$ can be
infinitely negative when $q(x)$ vanishes within some region $\mathcal{D}$. In
other words, the ME method confers an overwhelming preference on those
distributions $p(x)$ that vanish whenever $q(x)$ does. One must emphasize that
this is as it should be; it is not a problem. A similar situation\ also arises
in the context of Bayes' theorem where a vanishing prior represents a
tremendously serious commitment because no amount of data to the contrary
would allow us to revise it. In both ME and Bayes updating we should recognize
the implications of assigning a vanishing prior. Assigning a very low but
non-zero prior represents a safer and less prejudiced representation of one's beliefs.

For more on the choice of priors see the review [Kass Wasserman 96]; in
particular for entropic priors see [Rodriguez 90-03, Caticha Preuss 04]

\subsection{Comments on other axiomatizations}

One feature that distinguishes the axiomatizations proposed by various authors
is how they justify maximizing a functional. In other words, why \emph{maximum
entropy}? In the approach of Shore and Johnson this question receives no
answer; it is just one of the axioms. Csiszar provides a better answer. He
derives the `maximize a functional' rule from reasonable axioms of regularity
and locality [Csiszar 91]. In Skilling's and in the approach developed here
the rule is not derived, but it does not go unexplained either: it is imposed
by design, it is justified by the function that $S$ is supposed to perform, to
achieve a transitive ranking.

Both Shore and Johnson and Csiszar require, and it is not clear why, that
updating from a prior must lead to a unique posterior, and accordingly, there
is a restriction that the constraints define a convex set. In Skilling's
approach and in the one advocated here there is no requirement of uniqueness,
we are perfectly willing to entertain situations where the available
information points to several equally preferable distributions.

There is another important difference between the axiomatic approach presented
by Csiszar and the present one. Since our ME method is a method for induction
we are justified in applying the method as if it were of universal
applicability. As with all inductive procedures, in any particular instance of
induction can turn out to be wrong -- because, for example, not all relevant
information has been taken into account -- but this does not change the fact
that ME is still the unique inductive inference method that generalizes from
the special cases chosen as axioms. Csiszar's version of the MaxEnt method is
not designed to generalize beyond the axioms. His method was developed for
linear constraints and therefore he does not feel justified in carrying out
his \emph{deductions} beyond the cases of linear constraints. In our case, the
application to non-linear constraints is precisely the kind of
\emph{induction} the ME method was designed to perform.

It is interesting that if instead of axiomatizing the inference process, one
axiomatizes the entropy itself by specifying those properties expected of a
measure of separation between (possibly unnormalized) distributions one is led
to a continuum of $\eta$-entropies,\ [Amari 85]\
\begin{equation}
S_{\eta}[p,q]=\frac{1}{\eta(\eta+1)}\int dx\left[  (\eta+1)p-\eta q-p^{\eta
+1}q^{-\eta}\right]  ~,\label{S-eta b}%
\end{equation}
labelled by a parameter $\eta$. These entropies are equivalent, for the
purpose of updating, to the relative Renyi entropies [Renyi 61, Aczel 75]. The
shortcoming\ of this approach is that it is not clear when and how such
entropies are to be used, which features of a probability distribution are
being updated and which preserved, or even in what sense do these entropies
measure an amount of information. Remarkably, if one further requires that
$S_{\eta}$ be additive over independent sources of uncertainty, as any
self-respecting measure ought to be, then the continuum in $\eta$ is
restricted to just the two values $\eta=0$ and $\eta=-1$ which correspond to
the entropies $S[p,q]$ and $S[q,p]$. \

For the special case when $p$ is normalized and a uniform prior $q=1$ we get
(dropping the integral over $q$)%
\begin{equation}
S_{\eta}=\frac{1}{\eta}\left(  1-\frac{1}{\eta+1}\int dx\,p^{\eta}\right)  ~.
\end{equation}
A related entropy
\begin{equation}
S_{\eta}^{\prime}=\frac{1}{\eta}\left(  1-\int dx\,p^{\eta+1}\right)
\end{equation}
has been proposed in [Tsallis 88] and forms the foundation of his
non-extensive statistical mechanics. Clearly these two entropies are
equivalent in that they generate equivalent variational problems -- maximizing
$S_{\eta}$ is equivalent to maximizing $S_{\eta}^{\prime}$. To conclude our
brief remarks on the entropies $S_{\eta}$ we point out that quite apart from
the difficulty of achieving consistency with the law of large numbers, some
the probability distributions obtained maximizing $S_{\eta}$ may also be
derived through a more standard use of MaxEnt or ME as advocated in these
lectures. [Plastino 94]

\section{Bayes' rule as a special case of ME}

Since the ME method and Bayes' rule are both designed for updating
probabilities, and both invoke a Principle of Minimal Updating, it is
important to explore the relations between them. In particular we would like
to know if the two are mutually consistent or not. [Caticha Giffin 06]

As described in section 2.10 the goal is to update our beliefs about
$\theta\in\Theta$ ($\theta$ represents one or many parameters) on the basis of
three pieces of information: (1) the prior information codified into a prior
distribution $q(\theta)$; (2) the data $x\in\mathcal{X}$ (obtained in one or
many experiments); and (3) the known relation between $\theta$ and $x$ given
by the model as defined by the sampling distribution or likelihood,
$q(x|\theta)$. The updating consists of replacing the \emph{prior} probability
distribution $q(\theta)$ by a \emph{posterior} distribution $P(\theta)$ that
applies after the data has been processed.

The crucial element that will allow Bayes' rule to be smoothly incorporated
into the ME scheme is the realization that before the data information is
available not only we do not know $\theta$, we do not know $x$ either. Thus,
the relevant space for inference is not $\Theta$ but the product space
$\Theta\times\mathcal{X}$ and the relevant joint prior is $q(x,\theta
)=q(\theta)q(x|\theta)$. We should emphasize that the information about how
$x$ is related to $\theta$ is contained in the \emph{functional form} of the
distribution $q(x|\theta)$ -- for example, whether it is a Gaussian or a
Cauchy distribution or something else -- and not in the actual values of the
arguments $x$ and $\theta$ which are, at this point, still unknown.

Next we collect data and the observed values turn out to be $X$. We must
update to a posterior that lies within the family of distributions
$p(x,\theta)$ that reflect the fact that $x$ is now known,
\begin{equation}
p(x)=%
{\textstyle\int}
d\theta\,p(\theta,x)=\delta(x-X)~.\label{data constraint a}%
\end{equation}
This data information constrains but is not sufficient to determine the joint
distribution
\begin{equation}
p(x,\theta)=p(x)p(\theta|x)=\delta(x-X)p(\theta|X)~.
\end{equation}
Any choice of $p(\theta|X)$ is in principle possible. So far the formulation
of the problem parallels section 2.10 exactly. We are, after all, solving the
same problem. Next we apply the ME method and show that we get the same answer.

According to the ME method the selected joint posterior $P(x,\theta)$ is that
which maximizes the entropy,%
\begin{equation}
S[p,q]=-%
{\textstyle\int}
dxd\theta~p(x,\theta)\log\frac{p(x,\theta)}{q(x,\theta)}~,~\label{entropy}%
\end{equation}
subject to the appropriate constraints. Note that the information in the data,
eq.(\ref{data constraint a}), represents an \emph{infinite} number of
constraints on the family $p(x,\theta)$: for each value of $x$ there is one
constraint and one Lagrange multiplier $\lambda(x)$. Maximizing $S$,
(\ref{entropy}), subject to (\ref{data constraint a}) and normalization,
\begin{equation}
\delta\left\{  S+\alpha\left[
{\textstyle\int}
dxd\theta~p(x,\theta)-1\right]  +%
{\textstyle\int}
dx\,\lambda(x)\left[
{\textstyle\int}
d\theta~p(x,\theta)-\delta(x-X)\right]  \right\}  =0~,
\end{equation}
yields the joint posterior,
\begin{equation}
P(x,\theta)=q(x,\theta)\,\frac{e^{\lambda(x)}}{Z}~,
\end{equation}
where $Z$ is a normalization constant, and the multiplier $\lambda(x)$ is
determined from (\ref{data constraint a}),
\begin{equation}%
{\textstyle\int}
d\theta~q(x,\theta)\frac{\,e^{\lambda(x)}}{Z}=q(x)\frac{\,e^{\lambda(x)}}%
{Z}=\delta(x-X)~,
\end{equation}
so that the joint posterior is%
\begin{equation}
P(x,\theta)=q(x,\theta)\frac{\,\delta(x-X)}{q(x)}=\delta(x-X)q(\theta|x)~,
\end{equation}
The corresponding marginal posterior probability $P(\theta)$ is
\begin{equation}
P(\theta)=%
{\textstyle\int}
dx\,P(\theta,x)=q(\theta|X)=q(\theta)\frac{q(X|\theta)}{q(X)}%
~,\label{Bayes rule c}%
\end{equation}
which is recognized as Bayes' rule, eq.(\ref{Bayes rule}). Thus Bayes' rule is
consistent with, and indeed, is a special case of the ME method.

To summarize: the prior $q(x,\theta)=q(x)q(\theta|x)$ is updated to the
posterior $P(x,\theta)=P(x)P(\theta|x)$ where $P(x)=\delta(x-X)$ is fixed by
the observed data while $P(\theta|X)=q(\theta|X)$ remains unchanged. Note that
in accordance with the philosophy that drives the ME method \emph{one only
updates those aspects of one's beliefs for which corrective new evidence has
been supplied}.

I conclude with a few simple examples that show how the ME allows
generalizations of Bayes' rule. The background for these generalized Bayes
problems is the familiar one: We want to make inferences about some variables
$\theta$ on the basis of information about other variables $x$. As before, the
prior information consists of our prior knowledge about $\theta$ given by the
distribution $q(\theta)$ and the relation between $x$ and $\theta$ is given by
the likelihood $q(x|\theta)$; thus, the prior joint distribution $q(x,\theta)$
is known. But now the information about $x$ is much more limited.

\subsubsection*{Bayes updating with uncertain data\noindent}

The data is uncertain: $x$ is not known. The marginal posterior $p(x)$ is no
longer a sharp delta function but some other known distribution,
$p(x)=P_{D}(x)$. This is still an infinite number of constraints
\begin{equation}
p(x)=%
{\textstyle\int}
d\theta\,p(\theta,x)=P_{D}(x)~,\label{data constraint b}%
\end{equation}
that are easily handled by ME. Maximizing $S$, (\ref{entropy}), subject to
(\ref{data constraint b}) and normalization, leads to
\begin{equation}
P(x,\theta)=P_{D}(x)q(\theta|x)~.
\end{equation}
The corresponding marginal posterior,
\begin{equation}
P(\theta)=%
{\textstyle\int}
dx\,P_{D}(x)q(\theta|x)=q(\theta)%
{\textstyle\int}
dx\,P_{D}(x)\frac{q(x|\theta)}{q(x)}~,\label{Jeffrey}%
\end{equation}
is known as Jeffrey's rule which we met earlier in section 2.10.

\subsubsection*{Bayes updating with information about $x$ moments}

Now we have even less information: $p(x)$ is not known. All we know about
$p(x)$ is an expected value
\begin{equation}
\left\langle f\right\rangle =%
{\textstyle\int}
dx\,p(x)f(x)=F~.\label{data constraint c}%
\end{equation}
Maximizing $S$, (\ref{entropy}), subject to (\ref{data constraint c}) and
normalization,
\begin{equation}
\delta\left\{  S+\alpha\left[
{\textstyle\int}
dxd\theta~p(x,\theta)-1\right]  +\lambda%
{\textstyle\int}
dxd\theta~p(x,\theta)f(x)-F\right\}  =0~,
\end{equation}
yields the joint posterior,
\begin{equation}
P(x,\theta)=q(x,\theta)\,\frac{e^{\lambda f(x)}}{Z}~,
\end{equation}
where the normalization constant $Z$ and the multiplier $\lambda$ are obtained
from
\begin{equation}
Z=%
{\textstyle\int}
dx~q(x)e^{\lambda f(x)}\quad\text{and}\quad\frac{d\log Z}{d\lambda}=F~.
\end{equation}
The corresponding marginal posterior is
\begin{equation}
P(\theta)=q(\theta)%
{\textstyle\int}
dx\,\,\frac{e^{\lambda f(x)}}{Z}q(x|\theta)~.\label{Bayes ME}%
\end{equation}
These two examples (\ref{Jeffrey}) and (\ref{Bayes ME}) are sufficiently
intuitive that one could have written them down directly without deploying the
full machinery of the ME\ method, but they do serve to illustrate the
essential compatibility of Bayesian and Maximum Entropy methods. Next we
consider a slightly less trivial example.

\subsubsection*{Updating with data and information about $\theta$ moments}

Here we follow [Giffin Caticha 07]. In addition to data about $x$ we have
additional information about $\theta$ in the form of a constraint on the
expected value of some function $f(\theta)$,
\begin{equation}%
{\textstyle\int}
dxd\theta\,P(x,\theta)f(\theta)=\left\langle f(\theta)\right\rangle
=F~.\label{<f>}%
\end{equation}

In the standard Bayesian practice it is possible to impose constraint
information at the level of the prior, but this information need not be
preserved in the posterior. What we do here that differs from the standard
Bayes' rule is that we can require that the constraint (\ref{<f>}) be
satisfied by the posterior distribution.

Maximizing the entropy (\ref{entropy}) subject to normalization, the data
constraint (\ref{data constraint a}), and the moment constraint (\ref{<f>})
yields the joint posterior,%
\begin{equation}
P(x,\theta)=q(x,\theta)\frac{e^{\lambda(x)+\beta f(\theta)}}{z}~,
\end{equation}
where $z$ is a normalization constant,%
\begin{equation}
z=%
{\textstyle\int}
dxd\theta\,e^{\lambda(x)+\beta f(\theta)}q(x,\theta)\,.
\end{equation}
The Lagrange multipliers $\lambda(x)$ are determined from the data constraint,
(\ref{data constraint}),%
\begin{equation}
\frac{e^{\lambda(x)}}{z}=\frac{\delta(x-X)}{Zq(X)}\quad\text{where}\quad
Z(\beta,X)=%
{\textstyle\int}
d\theta\,e^{\beta f(\theta)}q(\theta|X)~,
\end{equation}
so that the joint posterior becomes%
\begin{equation}
P(x,\theta)=\delta(x-X)q(\theta|X)\frac{e^{\beta f(\theta)}}{Z}%
~.\label{joint posterior}%
\end{equation}
The remaining Lagrange multiplier $\beta$ is determined by imposing that the
posterior $P(x,\theta)$ satisfy the constraint (\ref{<f>}). This yields an
implicit equation for $\beta$,
\begin{equation}
\frac{\partial\log Z}{\partial\beta}=F~.\label{F}%
\end{equation}
Note that since $Z=Z(\beta,X)$ the resultant $\beta$ will depend on the
observed data $X$. Finally, the new marginal distribution for $\theta$ is%
\begin{equation}
P(\theta)=q(\theta|X)\frac{e^{\beta f(\theta)}}{Z}=q(\theta)\frac{q(X|\theta
)}{q(X)}\frac{e^{\beta f(\theta)}}{Z}~.\label{main result}%
\end{equation}
For $\beta=0$ (no moment constraint) we recover Bayes' rule. For $\beta\neq0$
Bayes' rule is modified by a \textquotedblleft canonical\textquotedblright%
\ exponential factor.

\section{Commuting and non-commuting constraints}

The ME method allows one to process information in the form of constraints.
When we are confronted with several constraints we must be particularly
cautious. In what order should they be processed? Or should they be processed
together? The answer depends on the problem at hand. (Here we follow [Giffin
Caticha 07].)

We refer to constraints as \emph{commuting} when it makes no difference
whether they are handled simultaneously or sequentially. The most common
example is that of Bayesian updating on the basis of data collected in
multiple experiments: for the purpose of inferring $\theta$ it is well-known
that the order in which the observed data $x^{\prime}=\{x_{1}^{\prime}%
,x_{2}^{\prime},\ldots\}$ is processed does not matter. (See section 2.10.3.)
The proof that ME is completely compatible with Bayes' rule implies that data
constraints implemented through $\delta$ functions, as in
(\ref{data constraint a}), commute. It is useful to see how this comes about.

When an experiment is repeated it is common to refer to the value of $x$ in
the first experiment and the value of $x$ in the second experiment. This is a
dangerous practice because it obscures the fact that we are actually talking
about \emph{two} separate variables. We do not deal with a single $x$ but with
a composite $x=(x_{1},x_{2})$ and the relevant space is $\mathcal{X}_{1}%
\times\mathcal{X}_{2}\times\Theta$. After the first experiment yields the
value $X_{1}$, represented by the constraint $c_{1}:P(x_{1})=\delta
(x_{1}-X_{1})$, we can perform a second experiment that yields $X_{2}$ and is
represented by a second constraint $c_{2}:P(x_{2})=\delta(x_{2}-X_{2})$. These
constraints $c_{1}$ and $c_{2}$ commute because they refer to \emph{different}
variables $x_{1}$ and $x_{2}$. An experiment, once performed and its outcome
observed, cannot be \emph{un-performed} and its result cannot be
\emph{un-observed} by a second experiment. Thus, imposing the second
constraint does not imply a revision of the first.

In general constraints need not commute and when this is the case the order in
which they are processed is critical. For example, suppose the prior is $q $
and we receive information in the form of a constraint, $C_{1}$. To update we
maximize the entropy $S[p,q]$ subject to $C_{1}$ leading to the posterior
$P_{1}$ as shown in Figure 6.1. Next we receive a second piece of information
described by the constraint $C_{2}$. At this point we can proceed in
essentially two different ways:

\noindent\textbf{(a) Sequential updating.} Having processed $C_{1}$, we use
$P_{1}$ as the current prior and maximize $S[p,P_{1}]$ subject to the new
constraint $C_{2}$. This leads us to the posterior $P_{a}$.

\noindent\textbf{(b)\ Simultaneous updating.} Use the original prior $q$ and
maximize $S[p,q]$ subject to both constraints $C_{1}$ and $C_{2}$
simultaneously. This leads to the posterior $P_{b}$.\footnote{At first sight
it might appear that there exists a third possibility of\ simultaneous
updating: (c) use $P_{1}$ as the current prior and maximize $S[p,P_{1}]$
subject to both constraints $C_{1}$ and $C_{2}$ simultaneously. Fortunately,
and this is a valuable check for the consistency of the ME method, it is easy
to show that case (c) is equivalent to case (b). Whether we update from $q$ or
from $P_{1}$ the selected posterior is $P_{b}$.}%

\begin{figure}
[tbh]
\begin{center}
\includegraphics[
trim=0.000000in 1.002697in 0.000000in 1.003447in,
natheight=7.499600in,
natwidth=9.999800in,
height=2.7752in,
width=5.028in
]%
{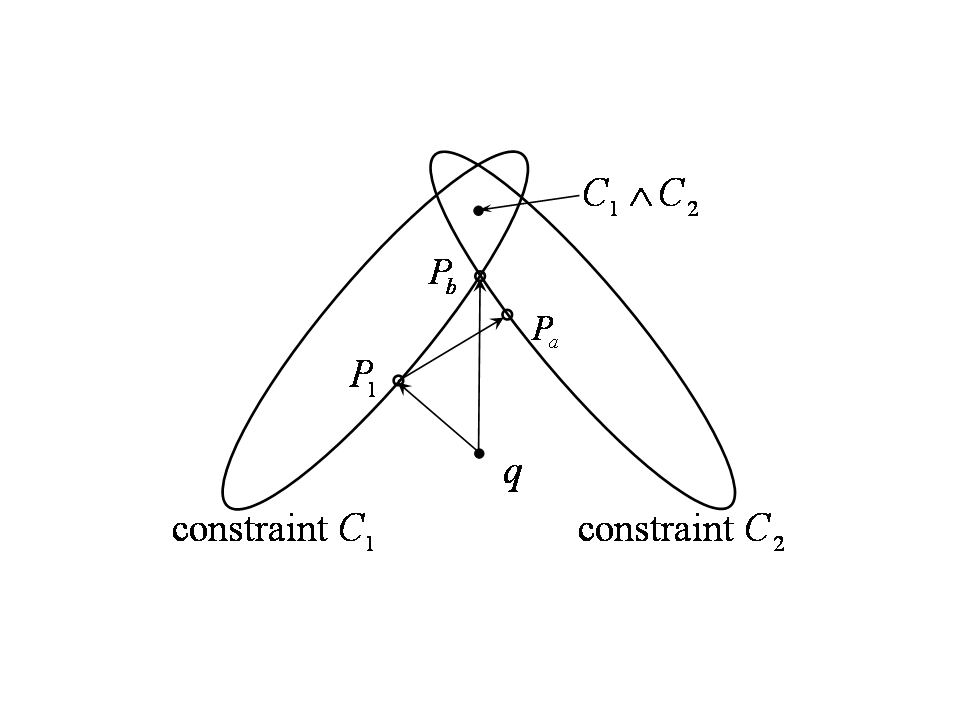}%
\caption{Illustrating the difference between processing two constraints
$C_{1}$ and $C_{2}$ sequentially ($q\rightarrow P_{1}\rightarrow P_{a}$) and
simultaneously ($q\rightarrow P_{b} $ or $q\rightarrow P_{1}\rightarrow P_{b}%
$). }%
\end{center}
\end{figure}

To decide which path (a) or (b) is appropriate we must be clear about how the
ME method handles constraints. The ME machinery interprets a constraint such
as $C_{1}$ in a very mechanical way: all distributions satisfying $C_{1}$ are
in principle allowed and all distributions violating $C_{1}$ are ruled out.
Updating to a posterior $P_{1}$ consists precisely in revising those aspects
of the prior $q$ that disagree with the new constraint $C_{1}$. However, there
is nothing final about the distribution $P_{1}$. It is just the best we can do
in our current state of knowledge and it may happen that future information
will require us to revise it further. Indeed, when new information $C_{2}$ is
received we must reconsider whether the original constraint $C_{1}$ remains
valid or not. Are \emph{all} distributions satisfying the new $C_{2}$ really
allowed, even those that violate $C_{1}$? If we decide that this is the case
then the new $C_{2}$ takes over and we update from the current $P_{1}$ to the
final posterior $P_{a}$. The old constraint $C_{1}$ may still exert some
limited influence on the final posterior $P_{a}$ through its effect on the
intermediate posterior $P_{1}$, but from now on $C_{1}$ is considered obsolete
and will be ignored.

Alternatively, we may decide that the old constraint $C_{1}$ retains its
validity. The new $C_{2}$ is not meant to replace the old $C_{1}$ but to
provide an additional refinement of the family of allowed posteriors. If this
is the case, then the constraint that correctly reflects the new information
is not $C_{2}$ but the more restrictive $C_{1}\wedge C_{2}$. The two
constraints $C_{1}$ and $C_{2}$ should be processed simultaneously to arrive
at the correct posterior $P_{b}$.

To summarize: sequential updating is appropriate when old constraints become
obsolete and are superseded by new information; simultaneous updating is
appropriate when old constraints remain valid. The two cases refer to
different states of information and therefore \emph{we expect} that they will
result in different inferences. These comments are meant to underscore the
importance of understanding what information is being processed; failure to do
so will lead to errors that do not reflect a shortcoming of the ME method but
rather a misapplication of it.

\section{Information geometry}

This section provides a very brief introduction to an important subject that
deserves a much more extensive treatment. [Amari 85, Amari Nagaoka 00]

Consider a family of distributions $p(x|\theta)$ labelled by a finite number
of parameters $\theta^{i}$, $i=1\ldots n$. It is usually possible to think of
the family of distributions $p(x|\theta)$ as a manifold -- an $n$-dimensional
space that is locally isomorphic to $\mathbb{R}^{n}$.\footnote{Of course it is
possible to conceive of sufficiently singular families of distributions that
are not smooth manifolds. This does not detract from the value of the methods
of information geometry any more than the existence of spaces with complicated
geometries detracts from the general value of geometry itself.} The
distributions $p(x|\theta)$ are points in this \textquotedblleft statistical
manifold\textquotedblright\ with coordinates given by the parameters
$\theta^{i}$. We can introduce the idea of a distance between two such points
-- that is, a `distance' between probability distributions. The distance
$d\ell$ between two neighboring points $\theta$ and $\theta+d\theta$ is given
by a generalization of Pythagoras' theorem in terms of a metric tensor
$g_{ij}$,\footnote{The use of superscripts rather than subscripts for the
indices labelling coordinates is a standard and very convenient notational
convention in differential geometry. We adopt the standard Einstein convention
of summing over repeated indices when one appears as a superscript and the
other as a subscript.}
\begin{equation}
d\ell^{2}=g_{ij}d\theta^{i}d\theta^{j}~.
\end{equation}
The singular importance of the metric tensor $g_{ij}$ derives from a most
remarkable theorem due to \v{C}encov that we mention without proof. [Cencov
81, Campbell 86] The theorem states that the metric $g_{ij}$ on the manifold
of probability distributions is unique: there is only one metric that takes
into account the fact that these are not distances between simple
structureless dots but between probability distributions. Up to a scale
factor, which merely reflects a choice of units, the unique distance is given
by the information metric which we introduce below in three independent but
intuitively appealing ways.

\subsection{Derivation from distinguishability}

We seek a quantitative measure of the extent that two distributions
$p(x|\theta)$ and $p(x|\theta+d\theta)$ can be distinguished. The following
argument is intuitively appealing. Consider the relative difference,
\begin{equation}
\frac{p(x|\theta+d\theta)-p(x|\theta)}{p(x|\theta)}=\frac{\partial\log
p(x|\theta)}{\partial\theta^{{}i}}\,d\theta^{{}i}.
\end{equation}
The expected value of the relative difference might seem a good candidate, but
it does not work because it vanishes identically,
\begin{equation}
\int dx\,p(x|\theta)\,\frac{\partial\log p(x|\theta)}{\partial\theta^{{}i}%
}\,d\theta^{{}i}=d\theta^{{}i}\,\frac{\partial}{\partial\theta^{{}i}}\int
dx\,p(x|\theta)=0.
\end{equation}
However, the variance does not vanish,
\begin{equation}
d\ell^{2}=\int dx\,p(x|\theta)\,\frac{\partial\log p(x|\theta)}{\partial
\theta^{{}i}}\,\frac{\partial\log p(x|\theta)}{\partial\theta^{{}j}}%
\,d\theta^{{}i}d\theta^{{}j}\,\,.
\end{equation}
This is the measure of distinguishability we seek; a small value of $d\ell
^{2}$ means the points $\theta$ and $\theta+d\theta$ are difficult to
distinguish. It suggests introducing the matrix $g_{ij}$
\begin{equation}
g_{ij}\overset{\text{def}}{=}\int dx\,p(x|\theta)\,\frac{\partial\log
p(x|\theta)}{\partial\theta^{{}i}}\,\frac{\partial\log p(x|\theta)}%
{\partial\theta^{{}j}}\label{info metric a}%
\end{equation}
called the Fisher information \emph{matrix} [Fisher 25], so that
\begin{equation}
d\ell^{2}=g_{ij}\,d\theta^{{}i}d\theta^{{}j}\,\,.\label{info metric b}%
\end{equation}

Up to now no notion of distance has been introduced. Normally one says that
the reason it is difficult to distinguish two points in say, the three
dimensional space we seem to inhabit, is that they happen to be too close
together. It is very tempting to invert the logic and assert that the two
points $\theta$ and $\theta+d\theta$ must be very close together because they
are difficult to distinguish. Furthermore, note that being a variance
$d\ell^{2}$ is positive and vanishes only when $d\theta$ vanishes. Thus it is
natural to interpret $g_{ij}$ as the metric tensor of a Riemannian space [Rao
45]. It is known as the \emph{information metric}. The recognition by Rao that
$g_{ij}$ is a metric in the space of probability distributions gave rise to
the subject of information geometry [Amari 85], namely, the application of
geometrical methods to problems in inference and in information theory. A
disadvantage of this heuristic argument is that it does not make explicit a
crucial property mentioned above that except for an overall multiplicative
constant this metric is unique. [Cencov 81, Campbell 86]

The coordinates $\theta$ are quite arbitrary; one can freely relabel the
points in the manifold. It is then easy to check that $g_{ij}$ are the
components of a tensor and that the distance $d\ell^{2}$ is an invariant, a
scalar under coordinate transformations.

\subsection{Derivation from a Euclidean metric}

Consider a discrete variable $a=1\ldots n$. The restriction to discrete
variables is not a serious limitation, we can choose $n$ sufficiently large to
approximate a continuous distribution to any desired degree. The possible
probability distributions of $a$ can be labelled by the probability values
themselves: a probability distribution can be specified by a point $p$ with
coordinates $(p^{1}\ldots p^{n})$. The corresponding statistical manifold is
the simplex $\mathcal{S}_{n-1}=\{p=(p^{1}\ldots p^{n}):%
{\textstyle\sum_{a}}
p^{a}=1\}$.

Next we change to new coordinates $\psi^{a}=\left(  p^{a}\right)  ^{1/2}$. In
these new coordinates the equation for the simplex $\mathcal{S}_{n-1}$ -- the
normalization condition -- reads $%
{\textstyle\sum}
\left(  \psi^{a}\right)  ^{2}=1$, which we recognize as the equation of an
$(n-1)$-sphere embedded in an $n$-dimensional Euclidean space $\mathbb{R}^{n}%
$, \emph{provided} the $\psi^{a} $ are interpreted as Cartesian coordinates.
This suggests that we assign the simplest possible metric: the distance
between the distribution $p(\psi)$ and its neighbor $p(\psi+d\psi)$ is the
Euclidean distance in $\mathbb{R}^{n}$,
\begin{equation}
d\ell^{2}=%
{\textstyle\sum\limits_{a}}
\left(  d\psi^{a}\right)  ^{2}=\delta_{ab}d\psi^{a}d\psi^{b}~.
\end{equation}
Distances between more distant distributions are merely angles defined on the
surface of the sphere $\mathcal{S}_{n-1}$.

Except for an overall constant this is the same information metric
(\ref{info metric b}) we defined earlier! Indeed, consider an $m$-dimensional
subspace ($m\leq n-1$) of the sphere $\mathcal{S}_{n-1}$ defined by $\psi
=\psi(\theta^{1},\ldots,\theta^{m})$. The parameters $\theta^{i}$, $i=1\ldots
m$, can be used as coordinates on the subspace. The Euclidean metric on
$\mathbb{R}^{n}$ induces a metric on the subspace. The distance between
$p(\theta)$ and $p(\theta+d\theta)$ is
\begin{align}
d\ell^{2}  & =\delta_{ab}d\psi^{a}d\psi^{b}=\delta_{ab}\frac{\partial\psi^{a}%
}{\partial\theta^{i}}d\theta^{i}\frac{\partial\psi^{b}}{\partial\theta^{j}%
}d\theta^{j}\nonumber\\
& =\frac{1}{4}%
{\textstyle\sum\limits_{a}}
p^{a}\frac{\partial\log p^{a}}{\partial\theta^{i}}\frac{\partial\log p^{a}%
}{\partial\theta^{j}}d\theta^{i}d\theta^{j}~,
\end{align}
which (except for the factor 1/4) we recognize as the discrete version of
(\ref{info metric a}) and (\ref{info metric b}). This interesting result does
not constitute a \textquotedblleft derivation.\textquotedblright\ There is a
priori no reason why the coordinates $\psi$ should be singled out as special
and attributed a Euclidean metric. But perhaps it helps to lift the veil of
mystery that might otherwise surround the strange expression
(\ref{info metric a}).

\subsection{Derivation from relative entropy}

The \textquotedblleft derivation\textquotedblright\ that follows has the merit
of drawing upon our intuition about relative entropy. Consider the entropy of
one distribution $p(x|\theta^{\prime})$ relative to another $p(x|\theta)$,
\begin{equation}
S(\theta^{\prime},\theta)=-\int dx\,p(x|\theta^{\prime})\log\frac
{p(x|\theta^{\prime})}{p(x|\theta)}~.
\end{equation}
We study how this entropy varies when $\theta^{\prime}=\theta+d\theta$ is in
the close vicinity of a given $\theta$. As we had seen in section 4.2 --
recall the Gibbs inequality $S(\theta^{\prime},\theta)\leq0$ with equality if
and only if $\theta^{\prime}=\theta$ -- the entropy $S(\theta^{\prime}%
,\theta)$ attains an absolute maximum at $\theta^{\prime}=\theta$ . Therefore,
the first nonvanishing term in the Taylor expansion about $\theta$ is second
order in $d\theta$%
\begin{equation}
S(\theta+d\theta,\theta)=\left.  \frac{1}{2}\frac{\partial S(\theta^{\prime
},\theta)}{\partial\theta^{\prime i}\partial\theta^{\prime j}}\right\vert
_{\theta^{\prime}=\theta}d\theta^{i}d\theta^{j}+\ldots\leq0~,
\end{equation}
and we use this quadratic form to define the information metric,
\begin{equation}
g_{ij}\overset{\operatorname*{def}}{=}\left.  -\frac{\partial S(\theta
^{\prime},\theta)}{\partial\theta^{\prime i}\partial\theta^{\prime j}%
}\right\vert _{\theta^{\prime}=\theta}~,\label{info metric c}%
\end{equation}
so that
\begin{equation}
S(\theta+d\theta,\theta)=-\frac{1}{2}d\ell^{2}~.\label{info metric d}%
\end{equation}
It is straightforward to show that (\ref{info metric c}) coincides with
(\ref{info metric a}).

\subsection{Volume elements in curved spaces}

Having decided on a measure of distance we can now also measure angles, areas,
volumes and all sorts of other geometrical quantities. Here we only consider
calculating the $m$-dimensional volume of the manifold of distributions
$p(x|\theta)$ labelled by parameters $\theta^{i}$ with $i=1\ldots m$.

The parameters $\theta^{i}$ are coordinates for the point $p$ and in these
coordinates it may not be obvious how to write down an expression for a volume
element $dV$. But within a sufficiently small region -- which is what a volume
element is -- any curved space looks flat. Curved spaces are `locally flat'.
The idea then is rather simple: within that very small region we should use
Cartesian coordinates and the metric takes a very simple form, it is the
identity matrix, $\delta_{ab}$. In locally Cartesian coordinates $\phi^{a}$
the volume element is simply given by the product
\begin{equation}
dV=d\phi^{1}d\phi^{2}\ldots d\phi^{m}~,
\end{equation}
which, in terms of the old coordinates $\theta^{i}$, is
\begin{equation}
dV=\left\vert \frac{\partial\phi}{\partial\theta}\right\vert d\theta
^{1}d\theta^{2}\ldots d\theta^{m}=\left\vert \frac{\partial\phi}%
{\partial\theta}\right\vert d^{m}\theta~.
\end{equation}
This is the volume we seek written in terms of the coordinates $\theta$. Our
remaining problem consists in calculating the Jacobian $\left\vert
\partial\phi/\partial\theta\right\vert $ of the transformation that takes the
metric $g_{ij}$ into its Euclidean form $\delta_{ab}$.

Let the locally Cartesian coordinates be defined by $\phi^{a}=\Phi^{a}%
(\theta^{1},\ldots\theta^{m})$. A small change in $d\theta$ corresponds to a
small change in $d\phi$,
\begin{equation}
d\phi^{a}=X_{i}^{a}d\theta^{i}\quad\text{where}\quad X_{i}^{a}\overset
{\operatorname*{def}}{=}\frac{\partial\phi^{a}}{\partial\theta^{i}}~,
\end{equation}
and the Jacobian is given by the determinant of the matrix $X_{i}^{a}$,
\begin{equation}
\left\vert \frac{\partial\phi}{\partial\theta}\right\vert =\left\vert
\det\left(  X_{i}^{a}\right)  \right\vert ~.
\end{equation}
The distance between two neighboring points is the same whether we compute it
in terms of the old or the new coordinates,
\begin{equation}
d\ell^{2}=g_{ij}d\theta^{i}d\theta^{j}=\delta_{ab}d\phi^{a}d\phi^{b}%
\end{equation}
Thus the relation between the old and the new metric is%
\begin{equation}
g_{ij}=\delta_{ab}X_{i}^{a}X_{j}^{b}~.
\end{equation}
The right hand side represents the product of three matrices. Taking the
determinant we get
\begin{equation}
g\overset{\operatorname*{def}}{=}\det(g_{ab})=\left[  \det\left(  X_{i}%
^{a}\right)  \right]  ^{2}~,
\end{equation}
so that
\begin{equation}
\left\vert \det\left(  X_{a}^{\alpha}\right)  \right\vert =g^{1/2}~.
\end{equation}
We have succeeded in expressing the volume element totally in terms of the
coordinates $\theta$ and the known metric $g_{ij}(\theta)$. The answer is
\begin{equation}
dV=g^{1/2}(\theta)d^{m}\theta~.
\end{equation}
The volume of any extended region on the manifold is
\begin{equation}
V=\int dV=\int g^{1/2}(\theta)d^{n}\theta~.~
\end{equation}

After this technical detour we are now ready to return to the main subject of
this chapter -- updating probabilities -- and derive one last and very
important feature of the ME method.

\section{Deviations from maximum entropy}

There is one last issue that must be addressed before one can claim that the
design of the ME\ method is more or less complete. Higher entropy represents
higher preference but there is nothing in the previous arguments to tell us by
how much. Does twice the entropy represent twice the preference or four times
as much? We can rank probability distributions $p$ relative to a prior $q$
according to the relative entropy $S[p,q]$ but any monotonic function of the
relative entropy will accomplish the same goal. Once we have decided that the
distribution of maximum entropy is to be preferred over all others the
following question arises: Suppose the maximum of the entropy function is not
particularly sharp, are we really confident that distributions with entropy
close to the maximum are totally ruled out? Can we quantify `preference'? We
want a quantitative measure of the extent to which distributions with lower
entropy are ruled out. The discussion below follows [Caticha 00].

Suppose we have maximized the entropy $S[p,q]$ subject to certain constraints
and obtain a probability distribution $p_{0}(x)$. The question we now address
concerns the extent to which $p_{0}(x)$ should be preferred over other
distributions with lower entropy. Consider a family of distributions
$p(x|\theta)$ labelled by a finite number of parameters $\theta^{i}$,
$i=1\ldots n$. We assume that the $p(x|\theta)$ satisfy the same constraints
that led us to select $p_{0}(x)$ and that $p_{0}(x)$ itself is included in the
family. Further we choose the parameters $\theta$ so that $p_{0}%
(x)=p(x|\theta=0)$. The question about the extent that $p(x|\theta=0)$ is to
be preferred over $p(x|\theta\neq0)$ is a question about the probability
$p(\theta)$ of various values of $\theta$: what is the rational degree of
belief that the selected value should be $\theta$? The original problem which
led us to design the maximum entropy method was to assign a probability to
$x$; we now see that the full problem is to assign probabilities to both $x$
and $\theta$.\ We are concerned not just with $p(x)$ but rather with the joint
distribution $P(x,\theta)$; the universe of discourse has been expanded from
$\mathcal{X}$ (the space of $x$s) to $\mathcal{X}\times\Theta$ ($\Theta$ is
the space of parameters $\theta$).

To determine the joint distribution $P(x,\theta)$ we make use of essentially
the only method at our disposal -- the ME method itself -- but this requires
that we address the standard two preliminary questions: First, what is the
prior distribution, what do we know about $x$ and $\theta$ before we receive
information about the constraints? And second, what is this new information
that constrains the allowed $P(x,\theta)$?

This first question is the more subtle one: when we know absolutely nothing
about the $\theta$s we know neither their physical meaning nor whether there
is any relation to the $x$s. A prior that reflects this lack of correlations
is a product, $q(x,\theta)=q(x)\mu(\theta)$. We will assume that the prior
over $x$ is known -- it is the prior we had used when we updated from $q(x)$
to $p_{0}(x)$. Since we are totally ignorant about $\theta$ we would like to
choose $\mu(\theta)$ so that it reflects a uniform distribution but here we
stumble upon a problem: uniform means that equal volumes in $\Theta$ are
assigned equal probabilities and knowing nothing about the $\theta$s we do not
yet know what \textquotedblleft equal\textquotedblright\ volumes in $\Theta$
could possibly mean. We need some additional information.

Suppose next that we are told that the $\theta$s represent probability
distributions, they are parameters labeling some unspecified distributions
$p(x|\theta)$. We do not yet know the functional form of $p(x|\theta)$, but if
the $\theta$s derive their meaning solely from the $p(x|\theta)$ then there
exists a natural measure of distance in the space $\Theta$. It is the
information metric $g_{ij}(\theta)$ introduced in the previous section and the
corresponding volume elements are given by $g^{1/2}(\theta)d^{n}\theta$, where
$g(\theta)$ is the determinant of the metric. The uniform prior for $\theta$,
which assigns equal probabilities to equal volumes, is proportional to
$g^{1/2}(\theta)$ and therefore we choose $\mu(\theta)=g^{1/2}(\theta)$.

Next we tackle the second question: what are the constraints on the allowed
joint distributions $p(x|\theta)$? Consider the space of all joint
distributions. To each choice of the functional form of $p(x|\theta)$ (whether
we talk about Gaussians, Boltzmann-Gibbs distributions, or something else)
there corresponds a different subspace defined by distributions of the form
$P(x,\theta)=p(\theta)p(x|\theta)$. The crucial constraint is that which
specifies the subspace, that is, the particular functional form for
$p(x|\theta)$. This defines the meaning to the $\theta$s -- for example, the
$\theta$s could be the mean and variance in Gaussian distributions, or
Lagrange multipliers in Boltzmann-Gibbs distributions. It also fixes the prior
$\mu(\theta)$ on the relevant subspace. Notice that the kind of constraint
that we impose here is very different from those that appear in usual
applications of maximum entropy method, which are in the form of expectation values.

To select the preferred distribution $P(x,\theta)$ we maximize the entropy
$S[P|g^{1/2}q]$ over all distributions of the form $P(x,\theta)=p(\theta
)p(x|\theta)$ by varying with respect to $p(\theta)$ with $p(x|\theta)$ fixed.
It is convenient to write the entropy as
\begin{align}
S[P,g^{1/2}q]  & =-\int dx\,d\theta\,p(\theta)p(x|\theta)\,\log\frac
{p(\theta)p(x|\theta)}{g^{1/2}(\theta)q(x)}\nonumber\\
& =S[p,g^{1/2}]+\int d\theta\,p(\theta)S(\theta),\label{S[joint]}%
\end{align}
where
\begin{equation}
S[p,g^{1/2}]=-\int\,d\theta\,p(\theta)\log\frac{p(\theta)}{g^{1/2}(\theta)}%
\end{equation}
and
\begin{equation}
S(\theta)=-\int\,dx\,p(x|\theta)\log\frac{p(x|\theta)}{q(x)}.\label{Stheta}%
\end{equation}
The notation shows that $S[p,g^{1/2}]$ is a functional of $p(\theta)$ while
$S(\theta)$ is a function of $\theta$ (it is also a functional of
$p(x|\theta)$). Maximizing (\ref{S[joint]}) with respect to variations $\delta
p(\theta)$ such that $\int d\theta\,p(\theta)=1$, yields
\begin{equation}
0=\int\,d\theta\left(  -\log\frac{p(\theta)}{g^{1/2}(\theta)}+S(\theta
)+\log\zeta\right)  \,\delta p(\theta)\,,
\end{equation}
where the required Lagrange multiplier has been written as $1-\log\zeta$.
Therefore the probability that the value of $\theta$ should lie within the
small volume $g^{1/2}(\theta)d^{n}\theta$ is
\begin{equation}
p(\theta)d^{n}\theta=\frac{1}{\zeta}\,\,e^{S(\theta)}g^{1/2}(\theta
)d^{n}\theta\quad\text{with\quad}\zeta=\int d^{n}\theta\,g^{1/2}%
(\theta)\,e^{S(\theta)}.\label{main}%
\end{equation}
Equation (\ref{main}) is the result we seek. It tells us that, as expected,
the preferred value of $\theta$ is that which maximizes the entropy
$S(\theta)$, eq.(\ref{Stheta}), because this maximizes the scalar probability
density $\exp S(\theta)$. But it also tells us the degree to which values of
$\theta$ away from the maximum are ruled out. For macroscopic systems the
preference for the ME distribution can be overwhelming. Eq.(\ref{main}) agrees
with the Einstein thermodynamic fluctuation theory and extends it beyond the
regime of small fluctuations -- in the next section we deal with fluctuations
as an illustration. Note also that the density $\exp S(\theta)$ is a scalar
function and the presence of the Jacobian factor $g^{1/2}(\theta)$ makes
Eq.(\ref{main}) manifestly invariant under changes of the coordinates
$\theta^{i}$ in the space $\Theta$.

We conclude this section by pointing out that there are a couple of
interesting points of analogy between the pair \{maximum likelihood
method/Bayes' rule\} on one hand and the corresponding pair \{MaxEnt/ME\}
methods on the other hand. Note that maximizing the likelihood function
$L(\theta|x)\overset{\operatorname*{def}}{=}p(x|\theta)$ selects a single
preferred value of $\theta$ but no measure is given of the extent to which
other values of $\theta$ are ruled out. The method of maximum likelihood does
not provide us with a distribution for $\theta$ -- the likelihood function
$L(\theta|x)$ is not a probability distribution for $\theta$. Similarly,
maximizing entropy as prescribed by the MaxEnt method yields a single
preferred value of the label $\theta$ but MaxEnt fails to address the question
of the extent to which other values of $\theta$ are ruled out. Neither Bayes'
rule nor the ME method suffer from this limitation.

The second point of analogy is that neither the maximum likelihood nor the
MaxEnt methods are capable of handling information contained in prior
distributions, while both Bayesian and ME methods can. This limitation of
maximum likelihood and MaxEnt is not surprising since neither method was
designed for updating probabilities.

\section{An application to fluctuations}

The starting point for the standard formulation of the theory of fluctuations
in thermodynamic systems (see [Landau 77, Callen 85]) is Einstein's inversion
of Boltzmann's formula $S=k\log W$ to obtain the probability of a fluctuation
in the form $W\sim\exp S/k$. A careful justification, however, reveals a
number of approximations which, for most purposes, are legitimate and work
very well. A re-examination of fluctuation theory from the point of view of ME
is, however, valuable. Our general conclusion is that the ME point of view
allows exact formulations; in fact, it is clear that deviations from the
canonical predictions can be expected, although in general they will be
negligible. Other advantages of the ME\ approach include the explicit
covariance under changes of coordinates, the absence of restrictions to the
vicinity of equilibrium or to large systems, and the conceptual ease with
which one deals with fluctuations of both the extensive as well as their
conjugate intensive variables. [Caticha 00]

This last point is an important one: within the canonical formalism (section
4.8) the extensive variables such as energy are uncertain while the intensive
ones such as the temperature or the Lagrange multiplier $\beta$ are fixed
parameters, they do not fluctuate. There are, however, several contexts in
which it makes sense to talk about fluctuations of the conjugate variables. We
discuss the standard scenario of an open system that can exchange say, energy,
with its environment.

Consider the usual setting of a thermodynamical system with microstates
labelled by $x$. Let $m(x)dx$ be the number of microstates within the range
$dx$. According to the postulate of \textquotedblleft equal a priori
probabilities\textquotedblright\ we choose a uniform prior distribution
proportional to the density of states $m(x)$. The canonical ME distribution
obtained by maximizing $S[p,m]$ subject to constraints on the expected values
$\left\langle f^{k}\right\rangle =F^{k}$ of relevant variables $f^{k}(x)$, is%
\begin{equation}
p(x|F)=\frac{1}{Z(\lambda)}\,m(x)\,e^{-\lambda_{{}k}f^{{}k}(x)}\quad
\text{with}\quad Z(\lambda)=%
{\textstyle\int}
dx\,m(x)\,e^{-\lambda_{{}k}f^{{}k}(x)}~,\label{fluct canon dist}%
\end{equation}
and the corresponding entropy is
\begin{equation}
S(F)=\log Z(\lambda)+\lambda_{k}F^{k}~.\label{entropy of can dist}%
\end{equation}

Fluctuations of the variables $f^{{}k}(x)$ or of any other function of the
microstate $x$ are usually computed in terms of the various moments of
$p(x|F)$. Within this context all expected values such as the constraints
$\left\langle f^{{}k}\right\rangle =$ $F^{{}k}$ and the entropy $S(F)$ itself
are fixed; they do not fluctuate. The corresponding conjugate variables, the
Lagrange multipliers $\lambda_{{}k}=\partial S/\partial F^{{}k}$,
eq.(\ref{lambda = dS/dF}), do not fluctuate either.

The standard way to make sense of $\lambda$ fluctuations is to couple the
system of interest to a second system, a bath, and allow exchanges of the
quantities $f^{{}k}$. All quantities referring to the bath will be denoted by
primes: the microstates are $x^{\prime}$, the density of states is $m^{\prime
}(x^{\prime})$, and the variables are $f^{\prime k}(x^{\prime})$, etc. Even
though the overall expected value $\left\langle f^{{}k}+f^{\prime
k}\right\rangle =F_{T}^{{}k}$ of the combined system plus bath is fixed, the
individual expected values $\left\langle f^{{}k}\right\rangle =F^{{}k}$ and
$\left\langle f^{\prime k}\right\rangle =F^{\prime k}=F_{T}^{{}k}-F^{{}k}$ are
allowed to fluctuate. The ME distribution $p_{0}(x,x^{\prime})$ that best
reflects the prior information contained in $m(x)$ and $m^{\prime}(x^{\prime
})$ updated by information on the total $F_{T}^{{}k}$ is
\begin{equation}
p_{0}(x,x^{\prime})=\frac{1}{Z_{0}}\,m(x)m^{\prime}(x^{\prime})\,e^{-\lambda
_{0\alpha}\left(  f^{{}k}(x)+f^{\prime k}(x^{\prime})\right)  }.
\end{equation}
But less than ME distributions are not totally ruled out; to explore the
possibility that the quantities $F_{T}^{k}$ are distributed between the two
systems in a less than optimal way we consider distributions $p(x,x^{\prime
},F)$ constrained to the form
\begin{equation}
P(x,x^{\prime},F)=p(F)p(x|F)p(x^{\prime}|F_{T}-F),\label{pi.pp}%
\end{equation}
where $p(x|F)$ is the canonical distribution in eq.(\ref{fluct canon dist}),
its entropy is eq.(\ref{entropy of can dist}) and analogous expressions hold
for the primed quantities.

We are now ready to write down the probability that the value of $F$
fluctuates into a small volume $g^{1/2}(F)dF$. From eq.(\ref{main}) we have
\begin{equation}
p(F)dF=\frac{1}{\zeta}\,\,e^{S_{T}(F)}g^{1/2}(F)dF,
\end{equation}
where $\zeta$ is a normalization constant and the entropy $S_{T}(F)$ of the
system plus the bath is
\begin{equation}
S_{T}(F)=S(F)+S^{\prime}(F_{T}-F).
\end{equation}
The formalism simplifies considerably when the bath is large enough that
exchanges of $F$ do not affect it, and $\lambda^{\prime}$ remains fixed at
$\lambda_{0}$. Then
\begin{equation}
S^{\prime}(F_{T}-F)=\log Z^{\prime}(\lambda_{0})+\lambda_{0k}\left(  F_{T}%
^{{}k}-F^{{}k}\right)  =\operatorname{const}-\lambda_{0k}F^{{}k}%
\text{.}\label{bath entropy}%
\end{equation}

It remains to calculate the determinant $g(F)$ of the information metric given
by eq.(\ref{info metric c}),
\begin{equation}
g_{ij}=-\frac{\partial^{2}S_{T}(\dot{F},F)}{\partial\dot{F}^{i}\partial\dot
{F}^{j}}=-\frac{\partial^{2}}{\partial\dot{F}^{i}\partial\dot{F}^{j}}\left[
S(\dot{F},F)+S^{\prime}(F_{T}-\dot{F},F_{T}-F)\right]
\end{equation}
where the dot indicates that the derivatives act on the first argument. The
first term on the right is
\begin{align}
\frac{\partial^{2}S(\dot{F},F)}{\partial\dot{F}^{i}\partial\dot{F}^{j}}  &
=-\frac{\partial^{2}}{\partial\dot{F}^{i}\partial\dot{F}^{j}}\int
\,dx\,p(x|\dot{F})\log\frac{p(x|\dot{F})}{m(x)}\frac{m(x)}{p(x|F)}\nonumber\\
& =\frac{\partial^{2}S(F)}{\partial F^{i}\partial F^{j}}+\int\,dx\,\frac
{\partial^{2}p(x|F)}{\partial F^{i}\partial F^{j}}\log\frac{p(x|F)}{m(x)}~.
\end{align}
To calculate the integral on the right use
\begin{equation}
\log\frac{p(x|F)}{m(x)}=-\log Z(\lambda)-\lambda_{{}k}f^{{}k}(x)
\end{equation}
(from eq.(\ref{fluct canon dist}) so that the integral vanishes,
\begin{equation}
-\log Z(\lambda)\frac{\partial^{2}}{\partial F^{i}\partial F^{j}}%
\int\,dx\,p(x|F)\,-\lambda_{{}k}\frac{\partial^{2}}{\partial F^{i}\partial
F^{j}}\int\,dx\,p(x|F)f^{{}k}(x)=0~.
\end{equation}
Similarly
\begin{align}
\frac{\partial^{2}}{\partial\dot{F}^{i}\partial\dot{F}^{j}}S^{\prime}%
(F_{T}-\dot{F},F_{T}-F)  & =\frac{\partial^{2}S^{\prime}(F_{T}-F)}{\partial
F^{i}\partial F^{j}}\\
& +\int\,dx^{\prime}\,\frac{\partial^{2}p(x^{\prime}|F_{T}-F)}{\partial
F^{i}\partial F^{j}}\log\frac{p(x^{\prime}|F_{T}-F)}{m^{\prime}(x^{\prime}%
)}\nonumber
\end{align}
and here, using eq.(\ref{bath entropy}), both terms vanish. Therefore
\begin{equation}
g_{ij}=-\frac{\partial^{2}S(F)}{\partial F^{i}\partial F^{j}}~.
\end{equation}

We conclude that the probability that the value of $F$ fluctuates into a small
volume $g^{1/2}(F)dF$ becomes
\begin{equation}
p(F)dF=\frac{1}{\zeta}\,\,e^{S(F)-\lambda_{0k}F^{{}k}}g^{1/2}%
(F)dF~.\label{fluctuations}%
\end{equation}
This equation is exact.

An important difference with the usual theory stems from the presence of the
Jacobian factor $g^{1/2}(F)$. This is required by coordinate invariance and
can lead to small deviations from the canonical predictions. The quantities
$\left\langle \lambda_{{}k}\right\rangle $ and $\left\langle F^{{}%
k}\right\rangle $ may be close but will not in general coincide with the
quantities $\lambda_{0k}$ and $F_{0}^{{}k}$ at the point where the scalar
probability density attains its maximum. For most thermodynamic systems
however the maximum is very sharp. In its vicinity the Jacobian can be
considered constant, and one obtains the usual results [Landau 77], namely,
that the probability distribution for the fluctuations is given by the
exponential of a Legendre transform of the entropy.

The remaining difficulties are purely computational and of the kind that can
in general be tackled systematically using the method of steepest descent to
evaluate the appropriate generating function. Since we are not interested in
variables referring to the bath we can integrate Eq.(\ref{pi.pp}) over
$x^{\prime}$, and use the distribution $P(x,F)=p(F)p(x|F)$ to compute various
moments. As an example, the correlation between $\delta\lambda_{{}i}%
=\lambda_{{}i}-\left\langle \lambda_{{}i}\right\rangle $ and $\delta f^{{}%
j}=f^{{}j}-\left\langle f^{{}j}\right\rangle $ or $\delta F^{{}j}=F^{{}%
j}-\left\langle F^{{}j}\right\rangle $ is
\begin{equation}
\left\langle \delta\lambda_{{}i}\delta f^{{}j}\right\rangle =\left\langle
\delta\lambda_{{}i}\delta F^{{}j}\right\rangle =-\frac{\partial\left\langle
\lambda_{{}i}\right\rangle }{\partial\lambda_{0j}}+\left(  \lambda
_{0i}-\left\langle \lambda_{{}i}\right\rangle \right)  \left(  F_{0}^{{}%
j}-\left\langle F^{{}j}\right\rangle \right)  .
\end{equation}
When the differences $\lambda_{0i}-\left\langle \lambda_{{}i}\right\rangle $
or $F_{0}^{{}j}-\left\langle F^{{}j}\right\rangle $ are negligible one obtains
the usual expression,
\begin{equation}
\left\langle \delta\lambda_{{}i}\delta f^{{}j}\right\rangle \approx-\delta
_{{}i}^{{}j}~.
\end{equation}

\section{Conclusion}

Any Bayesian account of the notion of information cannot ignore the fact that
Bayesians are concerned with the beliefs of rational agents. The relation
between information and beliefs must be clearly spelled out. The definition we
have proposed -- that information is that which constrains rational beliefs
and therefore forces the agent to change its mind -- is convenient for two
reasons. First, the information/belief relation very explicit, and second, the
definition is ideally suited for quantitative manipulation using the ME\ method.

Dealing with uncertainty requires that one solve two problems. First, one must
represent a state of knowledge as a consistent web of interconnected beliefs.
The instrument to do it is probability. Second, when new information becomes
available the beliefs must be updated. The instrument for this is relative
entropy. It is the only candidate for an updating method that is of universal
applicability and obeys the moral injunction that one should not change one's
mind frivolously. Prior information is valuable and should not be revised
except when demanded by new evidence, in which case the revision is no longer
optional but obligatory. The resulting general method -- the ME\ method -- can
handle arbitrary priors and arbitrary constraints; it includes MaxEnt and
Bayes' rule as special cases; and it provides its own criterion to assess the
extent that non maximum-entropy distributions are ruled out.\footnote{For
possible developments and applications of these ideas, which we hope will be
the subject of future additions to these lectures, see the \textquotedblleft
Suggestions for further reading.\textquotedblright}

To conclude I cannot help but to express my continued sense of wonder and
astonishment that the method for reasoning under uncertainty -- which should
presumably apply to the whole of science -- turns out to rest upon an ethical
foundation of intellectual honesty. The moral imperative is to uphold those
beliefs and only those beliefs that obey very strict constraints of
consistency; the allowed beliefs must be consistent among themselves, and they
must be consistent with the available information. Just imagine the implications!

\bigskip

\chapter*{References%
\addcontentsline{toc}{chapter}{References}%
}

\begin{description}
\item[{[Aczel 75]}] J. Acz\'{e}l and Z. Dar\'{o}czy, \emph{On Measures of
Information and their Characterizations} (Academic Press, New York 1975).

\item[{[Amari 85]}] S. Amari, \emph{Differential-Geometrical Methods in
Statistics} (Springer-Verlag, 1985).

\item[{[Amari Nagaoka 00]}] S. Amari and H. Nagaoka, \emph{Methods of
Information Geometry} (Am. Math. Soc./Oxford U. Press, Providence, 2000).

\item[{[Brillouin 52]}] L. Brillouin, \emph{Science and Information Theory}
(Academic Press, New York, 1952).

\item[{[Callen 85]}] H. B. Callen, \emph{Thermodynamics and an Introduction to
Thermostatistics} (Wiley, New York, 1985).

\item[{[Campbell 86]}] L. L. Campbell: Proc. Am. Math. Soc. \textbf{98}, 135 (1986).

\item[{[Caticha 03]}] A. Caticha, \textquotedblleft Relative Entropy and
Inductive Inference,\textquotedblright\ \emph{Bayesian Inference and Maximum
Entropy Methods in Science and Engineering}, ed. by G. Erickson and Y. Zhai,
AIP Conf. Proc. \textbf{707}, 75 (2004) (arXiv.org/abs/physics/0311093).

\item[{[Caticha 06]}] A. Caticha and A. Giffin, \textquotedblleft Updating
Probabilities,\textquotedblright\ in \emph{Bayesian Inference and Maximum
Entropy Methods in Science and Engineering}, ed. by A. Mohammad-Djafari, AIP
Conf. Proc. \textbf{872}, 31 (2006) (arXiv.org/ abs/physics/0608185).

\item[{[Caticha 07]}] A. Caticha, \textquotedblleft Information and
Entropy,\textquotedblright\ in \emph{Bayesian Inference and Maximum Entropy
Methods in Science and Engineering}, ed. by K. Knuth \emph{et al.}, AIP Conf.
Proc. \textbf{954}, 11 (2007) (arXiv.org/abs/0710.1068).

\item[{[Caticha Preuss 04]}] A. Caticha and R. Preuss, `Maximum entropy and
\linebreak Bayesian data analysis: entropic prior distributions,' Phys. Rev.
\textbf{E70}, 046127 (2004) (arXiv.org/abs/physics/0307055).

\item[{[Cencov 81]}] N. N. \v{C}encov: \emph{Statistical Decision Rules and
Optimal Inference}, Transl. Math. Monographs, vol. 53, Am. Math. Soc.
(Providence, 1981).

\item[{[Cover Thomas 91]}] T. Cover and J. Thomas, \emph{Elements of
Information Theory }(Wiley,1991).

\item[{[Cox 46]}] R. T. Cox, `Probability, Frequency and Reasonable
Expectation', Am. J. Phys. \textbf{14}, 1 (1946); \emph{The Algebra of
Probable Inference} (Johns Hopkins, Baltimore, 1961).

\item[{[Cropper 86]}] W. H. Cropper, \textquotedblleft Rudolf Clausius and the
road to entropy,\textquotedblright\ Am. J. Phys. \textbf{54}, 1068 (1986).

\item[{[Csiszar 84]}] I. Csiszar, \textquotedblleft Sanov property,
generalized $I$-projection and a conditional limit theorem,\textquotedblright%
\ Ann. Prob. \textbf{12}, 768 (1984).

\item[{[Csiszar 85]}] I. Csisz\'{a}r \textquotedblleft An extended Maximum
Entropy Principle and a \linebreak Bayesian justification,\textquotedblright%
\ in \emph{Bayesian Statistics 2}, p.83, ed. by J. M. \linebreak Bernardo. M.
H. de Groot, D. V. Lindley, and A. F. M. Smith (North Holland, 1985);
\textquotedblleft MaxEnt, mathematics and information
theory,\textquotedblright\ \emph{Maximum Entropy and Bayesian Methods}, p. 35,
ed. by K. M. Hanson and R. N.Silver (Kluwer, 1996).

\item[{[Csiszar 91]}] I. Csisz\'{a}r, \textquotedblleft Why least squares and
maximum entropy: an axiomatic approach to inference for linear inverse
problems,\textquotedblright\ Ann. Stat. \textbf{19}, 2032 (1991).

\item[{[Diaconis 82]}] P. Diaconis and S. L. Zabell, \textquotedblleft
Updating Subjective Probabilities,\textquotedblright\ J. Am. Stat. Assoc.
\textbf{77}, 822 (1982).

\item[{[Earman 92]}] J. Earman, \emph{Bayes or Bust?: A Critical Examination
of Bayesian Confirmation Theory} (MIT Press, Cambridge, 1992).

\item[{[Fisher 25]}] R. A. Fisher: Proc. Cambridge Philos. Soc. \textbf{122},
700 (1925).

\item[{[Gibbs 1875-78]}] J. W. Gibbs, \textquotedblleft On the Equilibrium of
Heterogeneous Substances,\textquotedblright\ Trans. Conn. Acad. III (1875-78),
reprinted in \emph{The Scientific Papers of J. W. Gibbs} (Dover, NY, 1961).

\item[{[Gibbs 1902]}] J. W. Gibbs, \emph{Elementary Principles in Statistical
Mechanics} (Yale U. Press, New Haven, 1902; reprinted by Ox Bow Press,
Connecticut, 1981).

\item[{[Giffin Caticha 07]}] A. Giffin and A. Caticha, \textquotedblleft
Updating Probabilities with Data and Moments,\textquotedblright\ in
\emph{Bayesian Inference and Maximum Entropy Methods in Science and
Engineering}, ed. by K. Knuth \emph{et al.}, AIP Conf. Proc. \textbf{954},
74\textbf{\ }(2007) (arXiv.org/abs/0708.1593).

\item[{[Grad 61]}] H. Grad, \textquotedblleft The Many Faces of
Entropy,\textquotedblright\ Comm. Pure and Appl. Math. \textbf{14}, 323
(1961), and \textquotedblleft Levels of Description in Statistical Mechanics
and Thermodynamics\textquotedblright\ in \emph{Delaware Seminar in the
Foundations of Physics}, ed. by M. Bunge (Springer-Verlag, New York, 1967).
\end{description}

\thispagestyle{fancy} \fancyhf{} \fancyhead[LE,RO]{\thepage}
\fancyhead[LO,RE]{\emph{REFERENCES}} \renewcommand{\headrulewidth}{0pt}

\begin{description}
\item[{[Gregory 05]}] P. C. Gregory, \emph{Bayesian Logical Data Analysis for
the Physical Sciences} (Cambridge U. Press, 2005).

\item[{[Grendar 03]}] M. Grendar, Jr. and M. Grendar \textquotedblleft Maximum
Probability and Maximum Entropy Methods: Bayesian
interpretation,\textquotedblright\ in \emph{Bayesian Inference and Maximum
Entropy Methods in Science and Engineering}, ed. by G. Erickson and Y. Zhai,
AIP Conf. Proc. \textbf{707}, p. 490 (2004) (arXiv.org/abs/physics/0308005).

\item[{[Hacking 01]}] I. Hacking, \emph{An Introduction to Probability and
Inductive Logic} (Cambridge U. P., 2001)

\item[{[Howson Urbach 93]}] C. Howson and P. Urbach, \emph{Scientific
Reasoning, the Bayesian Approach} (Open Court, Chicago, 1993).

\item[{[James 1907]}] W. James, \emph{Pragmatism} (Dover, 1995) and \emph{The
Meaning of Truth} (Prometheus, 1997).

\item[{[Jeffrey 04]}] R. Jeffrey, \emph{Subjective Probability, the Real
Thing} (Cambridge U.P., 2004).

\item[{[Jeffreys 39]}] H. Jeffreys, \emph{Theory of Probability} (Oxford U.
P., 1939).

\item[{[Jaynes 57a]}] E. T. Jaynes, \textquotedblleft How does the Brain do
Plausible Reasoning\textquotedblright\ Stanford Univ. Microwave Lab. report
421 (1957), also published in \emph{Maximum Entropy and Bayesian Methods in
Science and Engineering}, G. J. Erickson and C. R. Smith (eds.) (Kluwer, 1988)
and online at http://bayes.wustl.edu.

\item[{[Jaynes 57b]}] E. T. Jaynes, \textquotedblleft Information Theory and
Statistical Mechanics\textquotedblright\ Phys. Rev. \textbf{106}, 620 and
\textbf{108}, 171 (1957).

\item[{[Jaynes 65]}] E. T. Jaynes, \textquotedblleft Gibbs vs. Boltzmann
Entropies,\textquotedblright\ Am. J. Phys. \textbf{33}, 391 (1965) (online at http://bayes.wustl.edu).

\item[{[Jaynes 83]}] \emph{E. T. Jaynes: Papers on Probability, Statistics and
Statistical Physics} edited by R. D. Rosenkrantz (Reidel, Dordrecht, 1983),
and papers online at http://bayes.wustl.edu.

\item[{[Jaynes 85]}] E. T. Jaynes, \textquotedblleft Bayesian Methods: General
Background\textquotedblright, in \emph{Maximum Entropy and Bayesian Methods in
Applied Statistics}, J. H. Justice (ed.) (Cambridge U. P., 1985) and at http://bayes.wustl.edu.

\item[{[Jaynes 88]}] E. T. Jaynes, "The Evolution of Carnot's Principle," pp.
267-281 in \emph{Maximum Entropy and Bayesian Methods in Science and
Engineering} ed. by G. J. Erickson and C. R. Smith (Kluwer, 1988) (online at http://bayes.wustl.edu).

\item[{[Jaynes 92] }] E. T. Jaynes, \textquotedblleft The Gibbs
Paradox\textquotedblright\ in \emph{Maximum Entropy and Bayesian Methods}, ed.
by C. R. Smith, G. J. Erickson and P. O. Neudorfer (Kluwer, Dordrecht, 1992).
\end{description}

\thispagestyle{fancy} \fancyhf{} \fancyhead[LE,RO]{\thepage}
\fancyhead[LO,RE]{\emph{REFERENCES}} \renewcommand{\headrulewidth}{0pt}

\begin{description}
\item[{[Jaynes 03]}] E. T. Jaynes, \emph{Probability Theory: The Logic of
Science} edited by G. L. Bretthorst (Cambridge U. Press, 2003).

\item[{[Karbelkar 86]}] S. N. Karbelkar, \textquotedblleft On the axiomatic
approach to the maximum entropy principle of inference,\textquotedblright%
\ Pramana -- J. Phys. \textbf{26}, 301 (1986).

\item[{[Kass Wasserman 96]}] R. E. Kass and L. Wasserman, J. Am. Stat. Assoc.
\textbf{91}, 1343 (1996).

\item[{[Klein 70]}] M. J. Klein, \textquotedblleft Maxwell, His Demon, and the
Second Law of Thermodynamics,\textquotedblright\ American Scientist
\textbf{58}, 84 (1970).

\item[{[Klein 73]}] M. J. Klein, \textquotedblleft The Development of
Boltzmann's Statistical Ideas\textquotedblright\ in \emph{The Boltzmann
Equation} ed. by E. G. D. Cohen and W. Thirring, (Springer Verlag, 1973).

\item[{[Kullback 59]}] S. Kullback, \emph{Information Theory and Statistics}
(Wiley, New York 1959).

\item[{[Landau 77]}] L. D. Landau and E. M. Lifshitz, \emph{Statistical
Physics} (Pergamon, New York, 1977).

\item[{[Lucas 70]}] J. R. Lucas, \emph{The Concept of Probability} (Clarendon
Press, Oxford, 1970).

\item[{[Mehra 98]}] J. Mehra, \textquotedblleft Josiah Willard Gibbs and the
Foundations of Statistical Mechanics,\textquotedblright\ Found.
Phys.\textbf{\ 28}, 1785 (1998).

\item[{[von Mises 57]}] R. von Mises, \emph{Probability, Statistics and Truth}
(Dover, 1957).

\item[{[Plastino 94]}] A. R. Plastino and A. Plastino, Phys. Lett.
\textbf{A193}, 140 (1994).

\item[{[Rao 45]}] C. R. Rao: Bull. Calcutta Math. Soc. \textbf{37}, 81 (1945).

\item[{[Renyi 61]}] A. Renyi, \textquotedblleft On measures of entropy and
information,\textquotedblright\ \emph{Proc. 4th Berkeley Symposium on
Mathematical Statistics and Probability}, Vol 1, p. 547 (U. of California
Press, 1961).

\item[{[Rodriguez 90]}] C. C. Rodr\'{\i}guez, \textquotedblleft Objective
Bayesianism and geometry\textquotedblright\ in \emph{Maximum Entropy and
Bayesian Methods}, P. F. Foug\`{e}re (ed.) (Kluwer, Dordrecht, 1990).

\item[{[Rodriguez 91]}] C. C. Rodr\'{\i}guez, \textquotedblleft Entropic
priors\textquotedblright\ in \emph{Maximum Entropy and Bayesian Methods},
edited by W. T. Grandy Jr. and L. H. Schick (Kluwer, Dordrecht, 1991).

\item[{[Rodriguez 02]}] C. C. Rodr\'{\i}guez: `Entropic Priors for Discrete
Probabilistic Networks and for Mixtures of Gaussian Models'. In:
\emph{Bayesian Inference and Maximum Entropy Methods in Science and
Engineering}, ed. by R. L. Fry, AIP Conf. Proc. \textbf{617}, 410 (2002) (arXiv.org/abs/physics/0201016).
\end{description}

\thispagestyle{fancy} \fancyhf{} \fancyhead[LE,RO]{\thepage}
\fancyhead[LO,RE]{\emph{REFERENCES}} \renewcommand{\headrulewidth}{0pt}

\begin{description}
\item[{[Rodriguez 03]}] C. C. Rodr\'{\i}guez, \textquotedblleft A Geometric
Theory of Ignorance\textquotedblright\ (omega.\linebreak albany.edu:8008/ignorance/ignorance03.pdf).

\item[{[Savage 72]}] L. J. Savage, \emph{The Foundations of Statistics}
(Dover, 1972).

\item[{[Shannon 48]}] C. E. Shannon, \textquotedblleft The Mathematical Theory
of Communication,\textquotedblright\ Bell Syst. Tech. J. \textbf{27}, 379 (1948).

\item[{[Shannon Weaver 49]}] C. E. Shannon and W. Weaver, \emph{The
Mathematical Theory of Communication}, (U. Illinois Press, Urbana 1949).

\item[{[Shore Johnson 80]}] J. E. Shore and R. W. Johnson, \textquotedblleft
Axiomatic derivation of the Principle of Maximum Entropy and the Principle of
Minimum Cross-Entropy,\textquotedblright\ IEEE Trans. Inf. Theory
\textbf{IT-26}, 26 (1980); \textquotedblleft Properties of Cross-Entropy
Minimization,\textquotedblright\ IEEE Trans. Inf. Theory \textbf{IT-27}, 26 (1981).

\item[{[Sivia Skilling 06]}] D. S. Sivia and J. Skilling, \emph{Data Analysis:
a Bayesian tutorial} (Oxford U. Press, 2006).

\item[{[Skilling 88]}] J. Skilling, \textquotedblleft The Axioms of Maximum
Entropy\textquotedblright\ in \emph{Maximum-Entropy and Bayesian Methods in
Science and Engineering}, G. J. Erickson and C. R. Smith (eds.) (Kluwer,
Dordrecht, 1988).

\item[{[Skilling 89]}] J. Skilling, \textquotedblleft Classic Maximum
Entropy\textquotedblright\ in \emph{Maximum Entropy and Bayesian Methods}, ed.
by J. Skilling (Kluwer, Dordrecht, 1989).

\item[{[Skilling 90]}] J. Skilling, \textquotedblleft Quantified Maximum
Entropy\textquotedblright\ in \emph{Maximum Entropy and Bayesian Methods}, ed.
by P. F. Foug\`{e}re (Kluwer, Dordrecht, 1990).

\item[{[Stapp 72]}] H. P. Stapp, \textquotedblleft The Copenhagen
Interpretation\textquotedblright\ Am. J. Phys. \textbf{40}, 1098 (1972).

\item[{[Tribus 69]}] M. Tribus, \emph{Rational Descriptions, Decisions and
Designs} (Pergamon, New York, 1969).

\item[{[Tsallis 88]}] C. Tsallis, J. Stat. Phys. \textbf{52}, 479 (1988).

\item[{[Tseng Caticha 01]}] C.-Y. Tseng and A. Caticha, \textquotedblleft Yet
another resolution of the Gibbs paradox: an information theory
approach,\textquotedblright\ in \emph{Bayesian Inference and Maximum Entropy
Methods in Science and Engineering}, ed. by R. L. Fry, A.I.P. Conf. Proc. Vol.
617, p. 331 (2002) (arXiv.org/abs/cond-mat/0109324).

\item[{[Uffink 95]}] J. Uffink, \textquotedblleft Can the Maximum Entropy
Principle be explained as a consistency Requirement?\textquotedblright%
\ Studies in History and Philosophy of Modern Physics \textbf{26B}, 223 (1995).

\item[{[Uffink 03]}] J. Uffink, \textquotedblleft Irreversibility and the
Second Law of Thermodynamics,\textquotedblright\ in \emph{Entropy}, ed. by A.
Greven et al. (Princeton U. Press, 2003).
\end{description}

\thispagestyle{fancy} \fancyhf{} \fancyhead[LE,RO]{\thepage}
\fancyhead[LO,RE]{\emph{REFERENCES}} \renewcommand{\headrulewidth}{0pt}

\begin{description}
\item[{[Uffink 04]}] J. Uffink, \textquotedblleft Boltzmann's Work in
Statistical Physics\textquotedblright\ in \emph{The Stanford Encyclopedia of
Philosophy} (http://plato.stanford.edu).

\item[{[Williams 80]}] P. M. Williams, Brit. J. Phil. Sci. \textbf{31}, 131 (1980).

\item[{[Wilson 81]}] S. S. Wilson,\textquotedblleft Sadi
Carnot,\textquotedblright\ Scientific American, August 1981, p. 134.
\end{description}

\section*{Suggestions for further reading}

Here is a very incomplete and very biased list of references on topics that we
plan to include in future editions of these lectures. The topics range form
inference proper -- the assignment of priors, information geometry, model
selection, inductive inquiry, evolutionary Bayes -- to the applications of all
these ideas to the foundations of quantum, classical, statistical, and
gravitational physics.

\begin{description}
\item[{[Caticha 98a]}] A. Caticha, \textquotedblleft Consistency and Linearity
in Quantum Theory,\textquotedblright\ Phys. Lett. \textbf{A244}, 13 (1998) (arXiv.org/abs/quant-ph/9803086).

\item[{[Caticha 98b]}] A. Caticha, \textquotedblleft Consistency, Amplitudes
and Probabilities in Quantum Theory,\textquotedblright\ Phys. Rev.
\textbf{A57}, 1572 (1998) (arXiv.org/abs/quant-ph\linebreak/9804012).

\item[{[Caticha 98c]}] A. Caticha, \textquotedblleft Insufficient reason and
entropy in quantum theory,\textquotedblright\ Found. Phys. \textbf{30}, 227
(2000) (arXiv.org/abs/quant-ph/9810074).

\item[{[Caticha 00]}] A. Caticha, `Maximum entropy, fluctuations and priors',
in \linebreak\emph{Bayesian Methods and Maximum Entropy in Science and
Engineering}, ed. by A. Mohammad-Djafari, AIP Conf. Proc. \textbf{568}, 94
(2001) (arXiv.org/\linebreak abs/math-ph/0008017).

\item[{[Caticha 01]}] A. Caticha, `Entropic Dynamics', in \emph{Bayesian
Methods and Maximum Entropy in Science and Engineering}, ed. by R. L. Fry,
A.I.P. Conf. Proc. \textbf{617 }(2002) (arXiv.org/abs/gr-qc/0109068).

\item[{[Caticha 04]}] A. Caticha \textquotedblleft Questions, Relevance and
Relative Entropy,\textquotedblright\ in \linebreak\emph{Bayesian Inference and
Maximum Entropy Methods in Science and Engineering}, R. Fischer \emph{et al.}
A.I.P. Conf. Proc. Vol. \textbf{735}, (2004) (arXiv.org/\linebreak abs/gr-qc/0409175).

\item[{[Caticha 05]}] A. Caticha, \textquotedblleft The Information Geometry
of Space and Time\textquotedblright\ in \linebreak\emph{Bayesian Inference and
Maximum Entropy Methods in Science and Engineering}, ed. by K. Knuth \emph{et
al. }AIP Conf. Proc. \textbf{803}, 355 (2006) (arXiv.org/abs/gr-qc/0508108).

\item[{[Caticha Cafaro 07]}] A. Caticha and C. Cafaro, `From Information
Geometry to Newtonian Dynamics', in \emph{Bayesian Inference and Maximum
Entropy Methods in Science and Engineering}, ed. by K. Knuth \emph{et al.},
AIP Conf. Proc. \textbf{954}, 165\textbf{\ }(2007) (arXiv.org/abs/0710.1071).
\end{description}

\thispagestyle{fancy} \fancyhf{} \fancyhead[LE,RO]{\thepage}
\fancyhead[LO,RE]{\emph{REFERENCES}} \renewcommand{\headrulewidth}{0pt}

\begin{description}
\item[{[CatichaN Kinouchi 98]}] N. Caticha and O. Kinouchi, \textquotedblleft
Time ordering in the evolution of information processing and modulation
systems,\textquotedblright\ Phil. Mag. \textbf{B 77}, 1565 (1998).

\item[{[CatichaN Neirotti 06]}] N. Caticha and J. P. Neirotti,
\textquotedblleft The evolution of learning systems: to Bayes or not to
be,\textquotedblright\ in \emph{Bayesian Inference and Maximum Entropy Methods
in Science and Engineering}, ed. by A. Mohammad-Djafari, AIP Conf. Proc.
\textbf{872}, 203 (2006).

\item[{[Dewar 03]}] R. Dewar, \textquotedblleft Information theory explanation
of the fluctuation theorem, maximum entropy production and self-organized
criticality in non-equilibrium stationary states,\textquotedblright\ J. Phys.
A: Math. Gen. \textbf{36} 631 (2003).

\item[{[Dewar 05]}] R. Dewar, \textquotedblleft Maximum entropy production and
the fluctuation theorem,\textquotedblright\ J. Phys. A: Math. Gen. \textbf{38}
L371 (2003).

\item[{[Jaynes 79]}] E. T. Jaynes, \textquotedblleft Where do stand on maximum
entropy?\textquotedblright\ in \emph{The Maximum Entropy Principle} ed. by R.
D. Levine and M. Tribus (MIT Press 1979).

\item[{[Knuth 02]}] K. H. Knuth, \textquotedblleft What is a
question?\textquotedblright\ in \emph{Bayesian Inference and Maximum Entropy
Methods in Science and Engineering}, ed. by C. Williams, AIP Conf. Proc.
\textbf{659}, 227 (2002).

\item[{[Knuth 03]}] K. H. Knuth, \textquotedblleft Deriving laws from ordering
relations\textquotedblright\ in \emph{Bayesian Inference and Maximum Entropy
Methods in Science and Engineering}, ed. by G.J. Erickson and Y. Zhai, AIP
Conf. Proc. \textbf{707}, 204 (2003).

\item[{[Knuth 05]}] K. H. Knuth, \textquotedblleft Lattice duality: The origin
of probability and entropy,\textquotedblright\ Neurocomputing \textbf{67C},
245 (2005).

\item[{[Knuth 06]}] K. H. Knuth, \textquotedblleft Valuations on lattices and
their application to information theory,\textquotedblright\ Proc. IEEE World
Congress on Computational Intelligence (2006).

\item[{[Neirotti CatichaN 03]}] J. P. Neirotti and N. Caticha,
\textquotedblleft Dynamics of the evolution of learning algorithms by
selection\textquotedblright\ Phys. Rev. \textbf{E 67}, 041912 (2003).

\item[{[Rodriguez 89]}] C. C. Rodr\'{\i}guez, \textquotedblleft The metrics
generated by the Kullback number\textquotedblright\ in \emph{Maximum Entropy
and Bayesian Methods}, J. Skilling (ed.) (Kluwer, Dordrecht, 1989) (omega.albany.edu:8008).

\item[{[Rodriguez 98]}] C. C. Rodr\'{\i}guez, \textquotedblleft Are we
cruising a hypothesis space?\textquotedblright\ (arxiv.\linebreak org/abs/physics/9808009).

\item[{[Rodriguez 04]}] C. C. Rodr\'{\i}guez, \textquotedblleft The Volume of
Bitnets\textquotedblright\ (omega.albany.edu:\linebreak8008/bitnets/bitnets.pdf).

\item[{[Rodriguez 05]}] C. C. Rodr\'{\i}guez, \textquotedblleft The ABC of
model selection: AIC, BIC and the new CIC\textquotedblright\ (omega.albany.edu:8008/CIC/me05.pdf).
\end{description}

\thispagestyle{fancy} \fancyhf{} \fancyhead[LE,RO]{\thepage}
\fancyhead[LO,RE]{\emph{REFERENCES}} \renewcommand{\headrulewidth}{0pt}

\begin{description}
\item[{[Tseng Caticha 04]}] C. Y. Tseng and A. Caticha, \textquotedblleft
Maximum Entropy and the Variational Method in Statistical Mechanics: an
Application to Simple Fluids\textquotedblright\ (arXiv.org/abs/cond-mat/0411625).
\end{description}

\thispagestyle{fancy} \fancyhf{} \fancyhead[LE,RO]{\thepage}
\fancyhead[LO,RE]{\emph{REFERENCES}} \renewcommand{\headrulewidth}{0pt}

\end{document}